%% file: main.tex
\documentclass[a4paper,11pt]{article}
\pdfoutput=1 % if your are submitting a pdflatex (i.e. if you have
\usepackage{jheppub} % for details on the use of the package, please see the JINST-author-manual
\usepackage{lineno}
\usepackage{slashed}
\usepackage{subcaption}
\usepackage{mathtools}
\usepackage{commath}
\usepackage{multirow}
\usepackage{hyperref}
\usepackage{cleveref}
\usepackage{booktabs}
\usepackage{nicefrac} 
\usepackage{comment}
\usepackage{graphicx}
\usepackage{float}
\usepackage{enumitem}
\usepackage{physics}

\title{Domain walls in the Two-Higgs-Doublet Model and their charge and CP-violating interactions with Standard Model fermions
}

\author[a]{Mohamed Younes Sassi}
\author[a,b]{Gudrid Moortgat-Pick}

\affiliation[a]{II. Institut für Theoretische Physik,
University of Hamburg,\\Luruper Chaussee 149, 22761 Hamburg, Germany}
\affiliation[b]{Deutsches Elektronen-Synchrotron DESY, Notkestr. 85, 22607 Hamburg, Germany}

\emailAdd{mohamed.younes.sassi@desy.de}
\emailAdd{gudrid.moortgat-pick@desy.de}

\preprint{DESY-23-134}

\abstract{Discrete symmetries play an important role in several extensions of the Standard Model (SM) of particle physics. For instance, in order to avoid flavor changing neutral currents, a discrete $Z_2$ symmetry is imposed on the Two-Higgs-Doublet Model (2HDM). This can lead to the formation of domain walls (DW) as the $Z_2$ symmetry gets spontaneously broken during electroweak symmetry breaking in the early universe and domain walls form between regions whose vacua are related by the discrete symmetry.
Due to this simultaneous spontaneous breaking of both the discrete symmetry and the electroweak symmetry, the vacuum manifold consists of two disconnected 3-spheres. Such a non-trivial disconnected vacuum manifold leads to several choices for the vacua at two adjacent regions, in contrast to models where only the discrete symmetry gets spontaneously broken and the vacuum manifold consists of several disconnected points. Due to this, we end up with several classes of DW solutions having different properties localized inside the wall, such as charge and/or CP violating vacua. We discuss the properties of these different classes of DW solutions as well as the interaction of SM fermions with such topological defects leading to different exotic phenomena such as, for example, the top quark being transmitted or reflected off the wall as a bottom quark.}

\begin{document} 

\maketitle
\flushbottom

\input{intro}

\input{2hdm}

\input{DWin2HDM}

\input{FermionScattering}

\input{Summary_and_conclusions}

\appendix

\acknowledgments
We would like to thank Luis Hellmich, Thomas Konstandin, Wilfried Buchmüller and the members of the DESY 2HDM working group for useful discussions. This work is funded by the Deutsche Forschungsgemeinschaft (DFG) through Germany’s Excellence Strategy
– EXC 2121 “Quantum Universe” — 390833306. Figures presented in this work were generated using \texttt{MatPlotLib} \cite{Hunter:2007} and \texttt{NumPy} \cite{2020NumPy-Array}.

\bibliography{references.bib}

\end{document}

%% file: intro.tex
\section{Introduction}
Domain walls are a type of topological defects that arise after the spontaneous symmetry breaking (SSB) of a discrete symmetry. The formation of topological defects on a cosmological scale after SSB was already hypothesized in the last century by Zel'dovich\cite{Zeldovich:1974uw}, Kibble \cite{Kibble:1976sj} and Zurek \cite{Zurek:1985qw}. In the case that such a spontaneous symmetry breaking occurred in the early universe, causally disconnected regions may end up with different vacua located in the different sectors of the vacuum manifold of the theory. The boundaries between these regions are then called domain walls, which are sheet-like two-dimensionful surfaces where the discrete symmetry is restored. The vacuum manifold of the Standard Model (SM), for instance, is a 3-sphere which is topologically trivial. Therefore, no topological defects arise and beyond standard model physics with extra symmetries are needed in order to produce topological defects at the early universe.  

Extensions of the SM scalar sector may lead to the existence of topological defects such as cosmic strings and domain walls. Such defects appear for example in axion models \cite{Servant:2023mwt, Dine:2023qsq, Dine:1982ah,Sikivie:1982qv}, supersymmetric models in the context of the spontaneous breaking of the R symmetry \cite{Takahashi:2008mu} and the $Z_3$ discrete symmetry in the NMSSM \cite{Maniatis:2009re}, discrete flavor symmetries \cite{PhysRevLett.72.9, Gelmini:2020bqg} as well as in the Two-Higgs-Doublet model (2HDM) \cite{Battye:2011jj,Battye:2020sxy,Law:2021ing}. In this work we investigate the domain walls solutions in the 2HDM, where the SM scalar sector is extended with a second Higgs doublet of $\text{SU}(2)_L$ and charged under $\text{U}_Y(1)$. Such an extension allows several discrete symmetries in the model \cite{Battye:2011jj}, whose spontaneous breaking after the electroweak symmetry breaking (EWSB) at the early universe can lead to the formation of domain walls. Recently, it was  found in \cite{Battye:2020sxy,Law:2021ing} that the domain walls in the 2HDM can have non trivial structures inside them. In particular, it was demonstrated that one dimensional domain wall solutions (usually denoted as kink solutions in the literature) exhibit CP and charge-violating vacua inside the defect. The spontaneous breaking of $\text{SU}_L(2)\times \text{U}(1)_Y$ alongside the discrete symmetry $Z_2$ leads to a degeneracy in the choice of the boundary conditions that one can impose on the vacua of different domains. This will then lead to several classes of kink solutions with different internal structures \cite{vachaspati_2023}. Such effects were already investigated for domain walls solutions arising in Grand Unified Theories such as $\text{SU}(5)\times Z_2$ \cite{Vachaspati:2001pw,Pogosian:2000xv, vachaspati_2023}.   

In this work we expand the analysis done in \cite{Law:2021ing} to also include the variation of all Goldstone and hypercharge angles of the $\text{SU}(2)_L\times \text{U}(1)_Y$ symmetry, as well as to study the evolution of the 1D kink solution when using von Neumann boundary conditions. We then discuss the dependence of the kink solutions on the physical parameters of the 2HDM such as the masses of the extra Higgs bosons and the parameter $\text{tan}(\beta)$.
We also study the scattering of SM fermions off the different types of domain walls. This is done by solving the Dirac equation on a domain wall background that includes charge and CP-violating vacua, demonstrating the CP violating scattering of the fermions off the wall. We finally discuss some interesting phenomena such as top quarks being transmitted through the wall or reflected off the wall as a bottom quark. Such effects are charge violating and the charge difference will be absorbed by gauge fields living on the wall. 

Our paper is organized as follows: First we briefly introduce the 2HDM, its potential and its different types of vacua. In section 3 we study the kink solutions in the model. We discuss in detail the different types of kink solutions and the effects of the variation of hypercharge and goldstone modes on the properties of the defect's core. We also discuss the profile of the kink solutions for different parameter points of the model. In section 4 we discuss the scattering of fermions off these kink solutions and show that the scattering exhibits CP and charge violating phenomena. Our summary and conclusions are given in section 5.

%% file: 2hdm.tex
\section{The Two-Higgs-Doublet Model}
In this section we briefly introduce the general 2HDM and the used notation in this work. In the 2HDM the Standard Model Higgs sector is extended by an extra doublet charged under $\text{SU(2)}_L \times \text{U(1)}_Y$. The general renormalizable scalar potential invariant under the SM symmetries is given by:
\begin{align}
    \notag V_{\text{2HDM}} &= m^2_{11}\Phi^{\dagger}_1\Phi_1 + m^2_{22}\Phi^{\dagger}_2\Phi_2 +  m^2_{12}(\Phi^{\dagger}_1\Phi_2 + h.c) \\ \notag
    & + \frac{\lambda_1}{2}\bigl(\Phi^{\dagger}_1\Phi_1\bigr)^2 + \frac{\lambda_2}{2}\bigl(\Phi^{\dagger}_2\Phi_2\bigr)^2   + \lambda_3\bigl(\Phi^{\dagger}_1\Phi_1\bigr)\bigl(\Phi^{\dagger}_2\Phi_2\bigr)  + \lambda_4\bigl(\Phi_1^{\dagger} \Phi_2\bigr)\bigl(\Phi_2^{\dagger} \Phi_1\bigr) \\ 
    & 
    +\biggl[\frac{\lambda_5}{2}\bigl(\Phi_1^{\dagger} \Phi_2\bigr)^2 + \lambda_6 \bigl(\Phi^{\dagger}_1\Phi_1\bigr)\bigl(\Phi^{\dagger}_1\Phi_2\bigr) + \lambda_7 \bigl(\Phi^{\dagger}_2\Phi_2\bigr)\bigl(\Phi^{\dagger}_1\Phi_2\bigl)  + h.c\biggr].
\end{align}
Depending on the choice of the parameters, the potential can also be invariant under various discrete or continuous symmetries relating the Higgs doublets $\Phi_1$ and $\Phi_2$ \cite{Battye:2011jj}. 
The general Yukawa sector of the theory is then given by \cite{Branco:2011iw}:
\begin{equation}
    \mathcal{L}_{Yukawa} = y^1_{ij}\bar{\psi_i} \psi_j\Phi_1 + y^2_{ij}\bar{\psi_i} \psi_j\Phi_2,
\end{equation}
where $\psi_i$ and $\psi_j$ denote the different fermion generations.
It is generally not possible to diagonalize the Yukawa couplings of fermions when having them couple to both Higgs doublets, this will then lead to the Yukawa couplings $y^1_{ij}$ and $y^2_{ij}$ being not simultaneously diagonalizable. Therefore, couplings between quarks of different flavor are then possible and this will lead to flavor changing neutral currents (FCNCs) at tree level \cite{Branco:2011iw}. Such phenomena are, however, strongly constrained experimentally, which makes it necessary to forbid them in the 2HDM. To avoid this problem one can impose that fermions with the same quantum numbers couple to the same Higgs doublet while other fermions couple to the second one. This can be achieved by imposing a $Z_2$ discrete symmetry on the Yukawa sector according to which the scalar doublets tranform in this way:
\begin{align}
   & \Phi_1 \xrightarrow{Z_2} \Phi_1, & \Phi_2 \xrightarrow{Z_2} -\Phi_2.
\end{align}
There are therefore 4 types of 2HDMs depending on the choice of scalar doublets that fermions couple to \cite{Muhlleitner:2016mzt}:
\begin{table}[h]
\centering
\begin{tabular}{ |c||c|c|c||c|c|c|c|c|  }
 \hline
 & u-type & d-type & leptons & Q & $u_R$ & $d_R$ & L & $l_R$\\
 \hline \hline
Type 1  & $\Phi_2$  & $\Phi_2$ & $\Phi_2$ & +  & - & - & + & - \\
Type 2 & $\Phi_2$ & $\Phi_1$  & $\Phi_1$ & + & - & + & + & - \\
Type 3 (lepton specific)  & $\Phi_2$  & $\Phi_2$ & $\Phi_1$ & +  & - & + & + & - \\
Type 4 (Flipped)  & $\Phi_2$   & $\Phi_1$ & $\Phi_2$ & +  & - & - & + & + \\
\hline
\end{tabular}
\caption{Types of Yukawa couplings between the fermions and the scalars in the 2HDM and the charges of the fermions under the $Z_2$ symmetry \cite{Branco:2011iw}. $Q$ and $L$ denote left-handed quark and lepton $SU(2)_L$ doublets while $u_R$, $d_R$ and $l_R$ denote $SU(2)_L$ right-handed singlets.}
\label{Tab:2hdmtypes}
\end{table}
\\
The scalar potential which respects $\text{SU(2)}_L \times \text{U(1)}_Y \times Z_2$ is then given by:
\begin{align}
  \notag V_{\text{2HDM}} &=  m^2_{11}\Phi^{\dagger}_1\Phi_1 + m^2_{22}\Phi^{\dagger}_2\Phi_2 + \frac{\lambda_1}{2}\bigl(\Phi^{\dagger}_1\Phi_1\bigr)^2 + \frac{\lambda_2}{2}\bigl(\Phi^{\dagger}_2\Phi_2\bigr)^2   + \lambda_3\bigl(\Phi^{\dagger}_1\Phi_1\bigr)\bigl(\Phi^{\dagger}_2\Phi_2\bigr) \\ 
    & + \lambda_4\bigl(\Phi_1^{\dagger} \Phi_2\bigr)\bigl(\Phi_2^{\dagger} \Phi_1\bigr)  
    +\biggl[\frac{\lambda_5}{2}\bigl(\Phi_1^{\dagger} \Phi_2\bigr)^2 + h.c\biggr].
\end{align}
After electroweak symmetry breaking, the Higgs doublets acquire a vacuum expectation value. The 2HDM includes 8 scalar degrees of freedom. In our work we adopt the non-linear representation \cite{Law:2021ing,Battye:2020sxy} to parameterize the vacua: 
\begin{align}
  &   \Phi_1 = \text{U} \tilde{\Phi}_1 = \text{U} \dfrac{1}{\sqrt{2}}
    \begin{pmatrix}
          0 \\      v_1
     \end{pmatrix},
\label{eq:nonlinrep1}      
\\ & \Phi_2 = \text{U} \tilde{\Phi}_2 = \text{U} \dfrac{1}{\sqrt{2}}
      \begin{pmatrix}
     v_+ \\
     v_2e^{i\xi}
      \end{pmatrix} ,  
\label{eq:nonlinrep2}      
\end{align}
where U is an element of the $\text{SU(2)}_L\times\text{U(1)}_Y$ group that is given by:
\begin{equation}
        \text{U}(x) = e^{i\theta(x)} \text{exp}\biggl(i\dfrac{\tilde{g}_i(x)\sigma_i}{2v_{sm}}\biggl),
\label{eq:EWmatrix}        
\end{equation}
where $\theta$, $\tilde{g}_{1,2,3}$ are respectively the hypercharge angle and Goldstone modes, $\sigma_i$ denote the Pauli matrices, the generators of SU(2) and $v_{sm}$ is the vacuum expectation value of the SM Higgs doublet. Using this representation, we can separate the Goldstone and hypercharge modes from the physical vacuum parameters $v_1$, $v_2$, $v_+$ and $\xi$.\\
There are 3 possible types of vacua in the 2HDM: charge-breaking, CP-breaking and neutral:
\begin{itemize}[noitemsep,topsep=-8pt]
\item The most general one occurs when $v_+$ is non zero and the vacuum is therefore charge breaking:
\begin{align}
    &   \tilde{\Phi}_1 = \dfrac{1}{\sqrt{2}}
    \begin{pmatrix}
          0 \\      v_1
     \end{pmatrix}, & \tilde{\Phi}_2 =  \dfrac{1}{\sqrt{2}}
      \begin{pmatrix}
     v_+ \\
     v_2e^{i\xi}
      \end{pmatrix}. 
\label{eq:chargedvacuum}      
\end{align}
Such a vacuum is obviously non-physical as it will give the photon a mass and allow charge breaking interactions such as the decay electrons into neutrinos or the decay of top quarks into bottom quarks via new decay channels \cite{Lahanas:1998wf}. \\
\item The CP-breaking vacuum occurs when the phase $\xi$ between the two doublets is non zero:
\begin{align}
    &   \tilde{\Phi}_1 = \dfrac{1}{\sqrt{2}}
    \begin{pmatrix}
          0 \\      v_1
     \end{pmatrix}, & \tilde{\Phi}_2 =  \dfrac{1}{\sqrt{2}}
      \begin{pmatrix}
     0 \\
     v_2e^{i\xi}
      \end{pmatrix}.  
\label{eq:CPviolatingvacuum}      
\end{align}
this type of vacuum leads to CP-breaking Yukawa couplings and can be useful in the context of baryogenesis to generate the needed CP-violation. \\
\item Neutral vacuum where $\xi = 0$ and $v_+ = 0$ :
\begin{align}
    &   \tilde{\Phi}_1 = \dfrac{1}{\sqrt{2}}
    \begin{pmatrix}
          0 \\      v_1
     \end{pmatrix}, & \tilde{\Phi}_2 =  \dfrac{1}{\sqrt{2}}
      \begin{pmatrix}
     0 \\
     v_2
      \end{pmatrix}.   
\label{eq:neutralvac}      
\end{align}
such a vaccum can accommodate the SM vacuum expectation value (VEV) when $\sqrt{v^2_1 + v^2_2}= v_{sm} = 246 \text{ GeV}$. \\
\end{itemize}
In \cite{Barroso:2005sm,Ivanov:2006yq}, it was shown that if a parameter point leads to a neutral minimum, then such a minimum of the potential will always lie above any possible charge or CP breaking minima.
Throughout this work we will only consider that all regions of the universe ended up with a neutral vacuum after electroweak spontaneous symmetry breaking.  
\begin{figure}[h]
     \centering
     \begin{subfigure}[b]{0.49\textwidth}
         \centering
         \includegraphics[width=\textwidth]{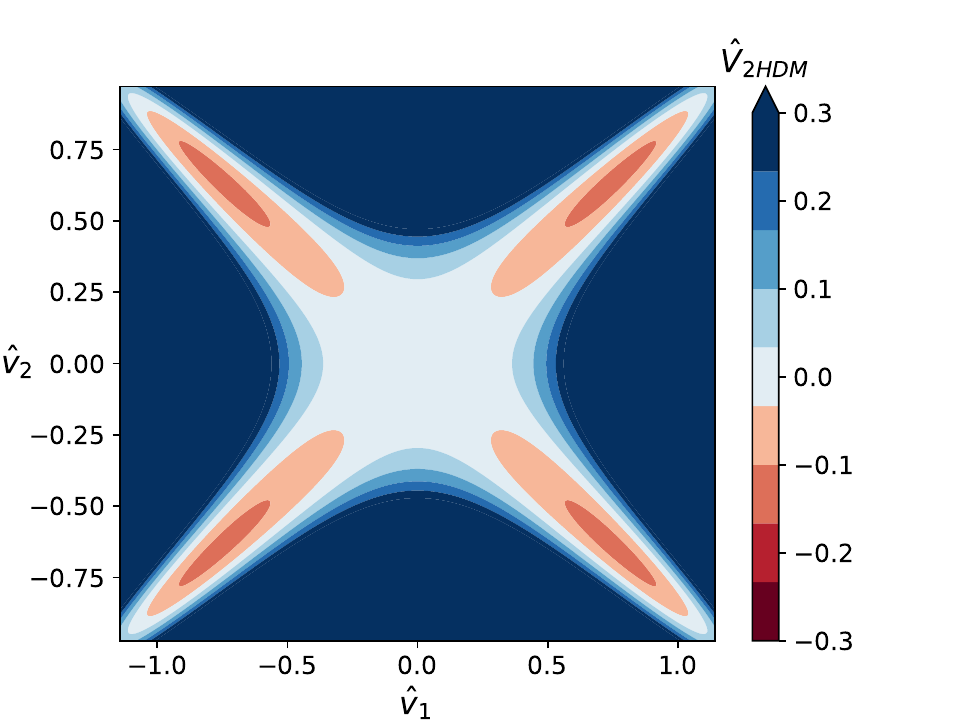}
         \subcaption{}\label{subfig:potential}
     \end{subfigure}
     \hfill
     \begin{subfigure}[b]{0.49\textwidth}
         \centering
         \includegraphics[width=\textwidth]{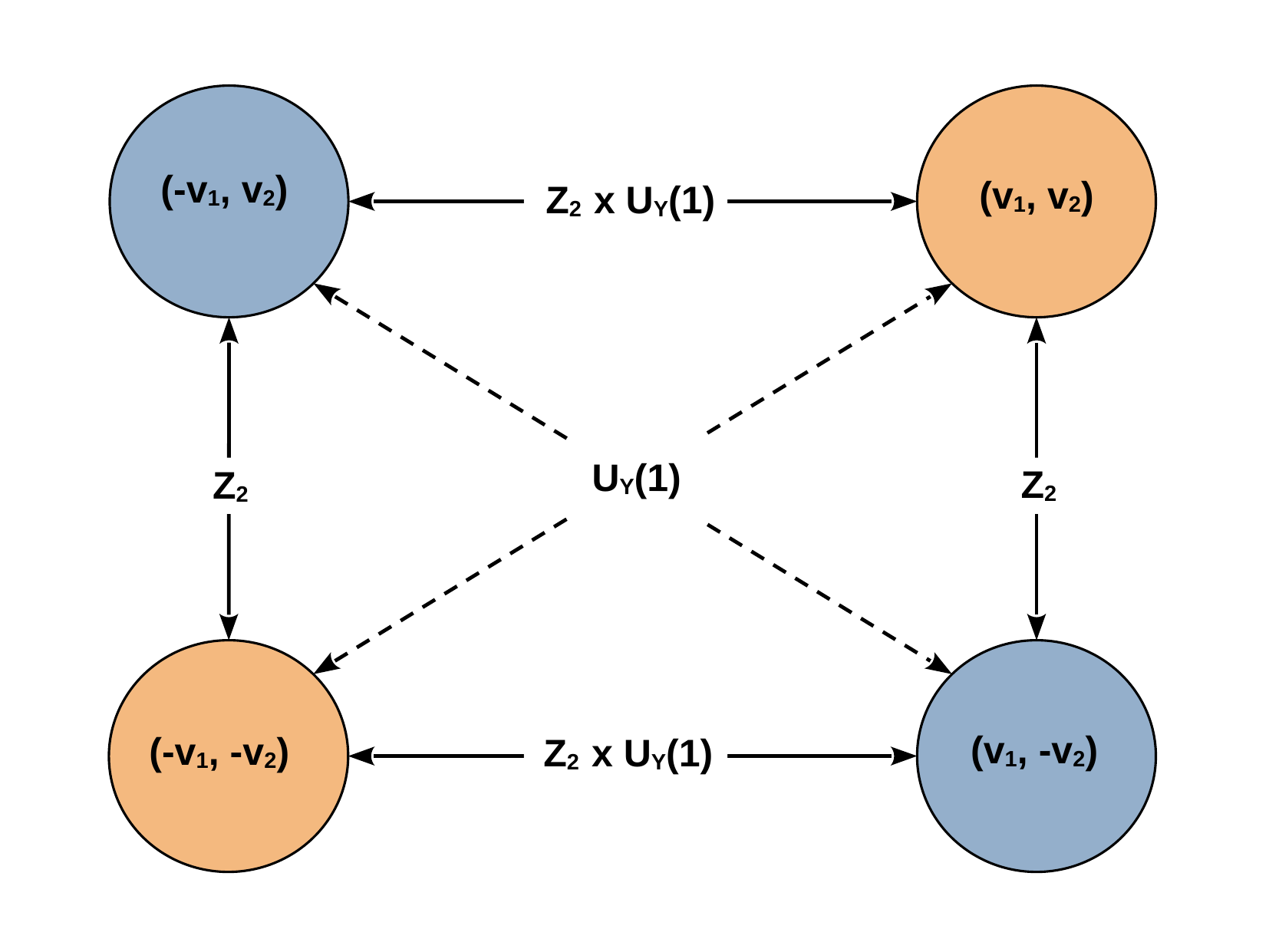}
         \subcaption{}\label{subfig:relationsvacua}
     \end{subfigure}
\caption{(a) Potential of the 2HDM in dimensionless units using PP I (\ref{eq:PP1}). (b) Symmetry relations between the minima in the potential.}
\label{fig:potential}
\end{figure}
Figure \ref{fig:potential} shows the dimensionless potential $\hat{V}_{\text{2HDM}} = V_{\text{2HDM}}/(m^2_hv^2_{sm})$ with rescaled parameters $\hat{v}_i = v_i/v_{sm}$. One distinguishes degenerate vacua that can be related by a $Z_2$ transformation, as for example, $(\hat{v}_1, -\hat{v}_2)$ and $(\hat{v}_1, \hat{v}_2)$ as well as degenerate minima that are related by a hypercharge transformation $\text{U}_Y(1)$ such as $(-\hat{v}_1, -\hat{v}_2)$ and $(\hat{v}_1, \hat{v}_2)$. We also have a multitude of other degenerate vacua that can be obtained from the latter ones, by using a gauge transformation of $\text{SU}(2)\times \text{U}_Y(1)$.\\
The particle content of the CP-conserving 2HDM includes 5 physical Higgs scalars: two CP-even with masses $m_h$ and $m_H$, one CP-odd with a mass $m_A$ and two charged Higgs bosons with a degenerate mass $m_C$. Using the mass parametrization, one can trade the parameters in the scalar potential with physical parameters such as the masses of the physical scalars, the ratio between the 2 vevs of the doublets $\text{tan}(\beta) = \frac{v_2}{v_1}$, the standard model vev $v_{sm} = 246 \text{ GeV}$ and the mixing angle $\alpha$. The potential parameters are therefore given by:
\allowdisplaybreaks
\begin{align}
 \lambda_{1} &= \frac{1}{v^2_1}\biggl(-m^2_{12}\text{tan}(\beta) +  m^2_h \text{cos}^2(\alpha) + m^2_H \text{sin}^2(\alpha) \biggl), \\
 \lambda_2 &=   \frac{1}{v^2_2}\biggl(-m^2_{12}\text{tan}(\beta) +  m^2_h \text{sin}^2(\alpha) + m^2_H \text{cos}^2(\alpha) \biggr), \\
 \lambda_3 &= \frac{1}{v_1v_2}\biggl(m^2_{12} + \text{sin}(\alpha)\text{cos}(\alpha)m^2_h - \text{sin}(\alpha)\text{cos}(\alpha)m^2_H    \biggr) - \lambda_4 - \lambda_5, \\
 \lambda_4 &= \frac{m^2_{12}}{v_1v_2} - 2\frac{m^2_C}{v^2_{sm}} + \frac{m^2_A}{v^2_{sm}},   \\
 \lambda_5 &= \frac{m^2_{12}}{v_1v_2} - \frac{m^2_A}{v^2_{sm}}, \\
 m^2_{11} &= m^2_{12}\text{tan}(\beta) - \frac{\lambda_1}{2}v^2_1 - \frac{\lambda_3 + \lambda_4 + \lambda_5}{2}v^2_2, \\
 m^2_{22} &=  m^2_{12}\text{tan}(\beta) - \frac{\lambda_2}{2}v^2_1 - \frac{\lambda_3 + \lambda_4 + \lambda_5}{2}v^2_1.
\end{align}
The term $m^2_{12}$ softly breaks the $Z_2$ symmetry leading the formed domain walls to be unstable and therefore annihilate some time after their formation. In this work we set $m^2_{12} = 0$ and leave the effects of a small non-vanishing value for future studies.

%% file: DWin2HDM.tex
\section{Domain walls in the 2HDM}
\begin{figure}[h]
     \centering
     \begin{subfigure}[b]{0.49\textwidth}
         \centering
         \includegraphics[width=\textwidth]{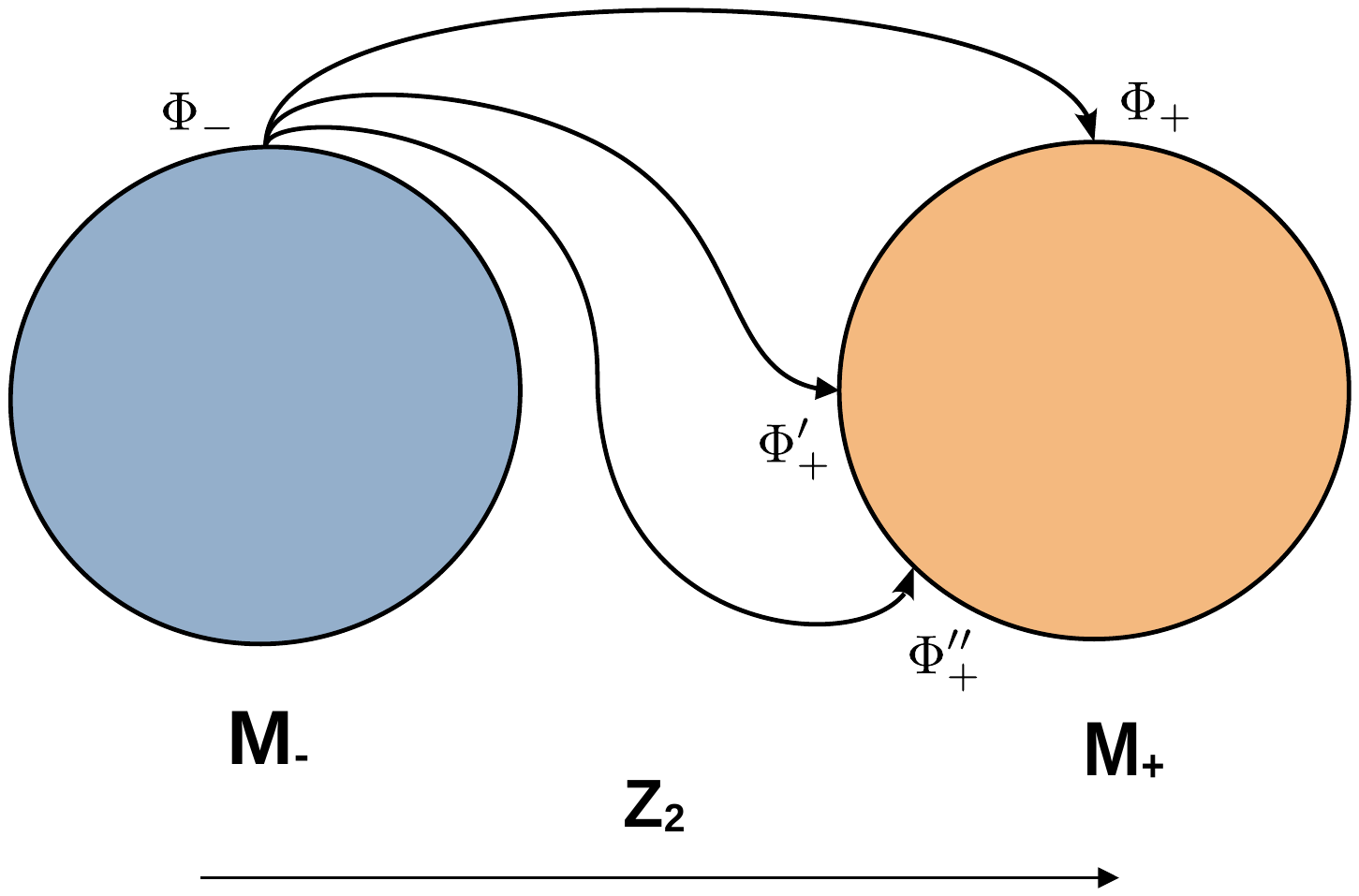}
         \subcaption{Vacuum manifold M of the model.}\label{subfig:vacmanifold}
     \end{subfigure}
     \hfill
     \begin{subfigure}[b]{0.49\textwidth}
         \centering
         \includegraphics[width=\textwidth]{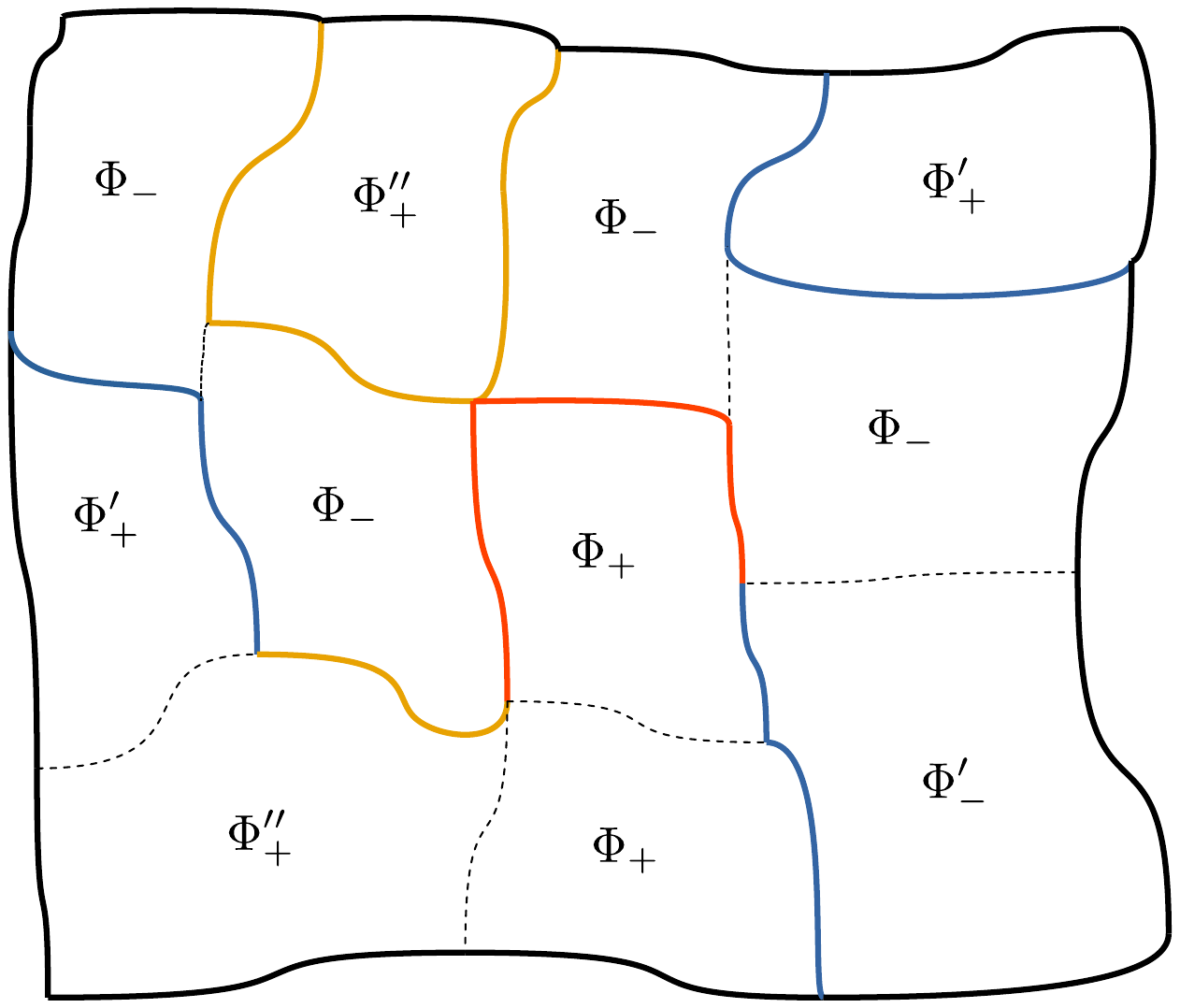}
         \subcaption{A patch of the universe after EWSB.}\label{subfig:earlyuniversevac}
     \end{subfigure}
\caption{(a) Vacuum manifold M of the model. In this case M consists of two disconnected sectors $M_-$ and $M_+$ related by the $Z_2$ symmetry and all the vacua in both sectors are degenerate. The elements of each sector are related by $\text{SU(2)}_L \times \text{U(1)}_Y$ transformations. $\Phi'_+$ and $\Phi''_+$ are related to $\Phi_+$ by different gauge transformations of $\text{SU(2)}_L \times \text{U(1)}_Y$.\\
(b) After EWSB, causally disconnected regions of the universe can end up in different vacua of the vacuum manifold. No topologically protected domain walls form between regions with vacua on the same spheres of M. Regions that end up with vacua in separate sectors of M can have different classes of domain walls depending on the Goldstone and Hypercharge angles they acquire.}
\label{fig:vacuummanifold}
\end{figure}

After electroweak symmetry breaking, the Higgs doublets acquire a neutral vev (\ref{eq:neutralvac}).
This vacuum breaks $\text{SU(2)}_L \times \text{U(1)}_Y \times Z_2$ into $\text{U(1)}_{em}$. In this case the vacuum manifold of the theory is homeomorphic to the coset space:
\begin{equation}
    M = (\text{SU(2)}_L \times \text{U(1)}_Y \times Z_2)/\text{ U(1)}_{em} \simeq Z_2 \times S^3,
\end{equation}
which is topologically equivalent to two disconnected 3-spheres \cite{Battye:2011jj} as depicted in Figure \ref{subfig:vacmanifold}. The vacuum manifold has then two disconnected sectors related by a $Z_2$ symmetry. These sectors are non-trivial (in contrast to the case when only $Z_2$ is broken) and consist of vacua which are related by $\text{SU(2)}_L \times \text{U(1)}_Y$ transformations. This leads to the formation of different classes of domain walls due to the multiple choices that can be taken for the electroweak matrix $U$ inside the two regions (see Figure \ref{subfig:vacmanifold}), in contrast to standard $Z_2$ domain wall solutions where the choice of the vacua inside the two domains is fixed to be vacua that are only related by a discrete symmetry \cite{Steer:2006ik}.
Figure \ref{subfig:earlyuniversevac} shows different possibilities for the boundary conditions after EWSB. In order to get a kink solution to the scalar field configuration, the boundary conditions at $\pm \infty$ need to lie on disconnected sectors of the vacuum manifold. 
Starting at $x \rightarrow -\infty$ with a vacuum $\Phi_-$ on the vacuum manifold sector $M_-$ corresponding to one 3-sphere:
\begin{equation}
    \Phi_- = \biggl\{ \dfrac{1}{\sqrt{2}}
    \begin{pmatrix}
          0 \\      v^*_1
     \end{pmatrix}  , \dfrac{1}{\sqrt{2}}
      \begin{pmatrix}
     0 \\
     -v^*_2
      \end{pmatrix}  \biggr\},
\end{equation}
we end up at $x \rightarrow +\infty$ with a vacuum $\Phi_+$ on the vacuum manifold $M_+$. Fixing our choice for $\Phi_-$, we have multiple choices for the vacuum $\Phi_+$:
\begin{equation}
    \Phi_+ = \biggl\{ \dfrac{1}{\sqrt{2}}
    U\begin{pmatrix}
          0 \\      v^*_1
     \end{pmatrix}  , \dfrac{1}{\sqrt{2}}
      U\begin{pmatrix}
     0 \\
     v^*_2
      \end{pmatrix}  \biggr\},
\end{equation}
where $U$ is an element of the broken electroweak symmetry group $\text{SU(2)}_L \times \text{U(1)}_Y$ and $\sqrt{(v^*_1)^2 + (v^*_2)^2} = 246 \text{ GeV}$. 
In order to compute the kink solution of the field configuration interpolating between those two vacua, we need to minimize the energy of such a field configuration.
In the 2HDM, the energy functional is given by:
\begin{equation}
    \mathcal{E}(x) = \dfrac{d\Phi^{\dag}_1}{dx} \dfrac{d\Phi_1}{dx} + \dfrac{d\Phi^{\dag}_2}{dx} \dfrac{d\Phi_2}{dx} + V_{\text{2HDM}}(\Phi_1, \Phi_2)
\label{eq:EF}    
\end{equation}
The first two terms describe the kinetic energy of the vacuum configuration, while the third is the potential of the scalar sector. There is an interplay between the kinetic energy that arises due to a changing profile of the field configuration as a function of $x$ and the potential energy of this field configuration.\\
Using the non-linear representation for the Higgs doublets (defined in (\ref{eq:nonlinrep1}), (\ref{eq:nonlinrep2}) and (\ref{eq:EWmatrix}))
in the expression of the energy functional (\ref{eq:EF}) we end up with:
\begin{align}
  \notag  \mathcal{E}(x) &= \dfrac{d\tilde{\Phi}^{\dag}_1}{dx} \dfrac{d\tilde{\Phi}_1}{dx} + \dfrac{d\tilde{\Phi}^{\dag}_2}{dx} \dfrac{d\tilde{\Phi}_2}{dx} + \biggl(\dfrac{d\tilde{\Phi}^{\dag}_{1,2}}{dx} U^{\dag}(x)\dfrac{dU}{dx}\tilde{\Phi}_{1,2}(x) + \text{h.c}\biggl) + \tilde{\Phi}^{\dag}_{1,2}(x)\dfrac{dU^{\dag}}{dx}\dfrac{dU}{dx}\tilde{\Phi}_{1,2}(x) \\
    & + V_{2HDM}(\Phi_1, \Phi_2),
\label{eq:energydensity}    
\end{align}
where, in terms of the vacuum manifold parameters:
\begin{align}
&\notag V_{2HDM}(v_1, v_2, v_+,\xi) = \dfrac{m^2_{11}}{2}v^2_1(x) + \dfrac{m^2_{22}}{2}\bigl(v^2_2(x) + v^2_+(x)\bigr)  + \dfrac{\lambda_1}{8}v^4_1(x) 
 + \dfrac{\lambda_2}{8}\bigl(v^2_2(x) + v^2_+(x)\bigr)^2 \\ & + \dfrac{\lambda_3 + \lambda_4 }{4}v^2_1(x)v^2_2(x)  +\dfrac{\lambda_3}{4}v^2_1(x)v^2_+(x)  + \dfrac{\lambda_5}{4}v^2_1(x)v^2_2(x)\cos(2\xi(x)).
\end{align}
Writing down $U(x)$ in terms of the Pauli matrices, we get:
\begin{equation}
U(x) = e^{i\theta(x)}\biggl[\cos(G(x)) \text{I}_2 + i\dfrac{g_1(x)}{2} \dfrac{\sin(G(x))}{G(x)}\sigma_1 + i\dfrac{g_2(x)}{2} \dfrac{\sin(G(x))}{G(x)}\sigma_2 + i\dfrac{g_3(x)}{2} \dfrac{\sin(G(x))}{G(x)}\sigma_3\biggr],
\end{equation}
where:
\begin{align}
G(x) &= \sqrt{\bigl(\dfrac{g_1(x)}{2}\bigr)^2 + \bigl(\dfrac{g_1(x)}{2}\bigr)^2 + \bigl(\dfrac{g_1(x)}{2}\bigr)^2}, &  g_i(x) = \tilde{g}_i(x)/v_{sm}.
\end{align}
In \cite{Law:2021ing}, the choice of the matrix $U(x)$ was simplified to the case where only a single Goldstone mode or the hypercharge angle $\theta$ was allowed to be non-zero and have asymmetric boundary conditions at $\pm \infty$. We will expand the results to the general case, where all modes in $U(x)$ can change. This will lead to more effects inside the domain walls compared to \cite{Law:2021ing}. \\
Since the calculation of all terms in \eqref{eq:energydensity} is straightforward but lengthy, we give the final expression of the energy functional in terms of the vacuum parametrization:
\begin{align}
\notag  \mathcal{E}(x) &= \dfrac{1}{2}\bigl(\dfrac{dv_1}{dx}\bigr)^2 + \dfrac{1}{2}\bigl(\dfrac{dv_2}{dx}\bigr)^2 + \dfrac{1}{2}\bigl(\dfrac{dv_+}{dx}\bigr)^2 + \dfrac{1}{2}v^2_2(x)\bigl(\dfrac{d\xi}{dx}\bigl)^2  +
  \dfrac{1}{2}v^2_1(x)\biggl[  \bigl(\dfrac{d\theta}{dx}\bigr)^2 + I_0(x) \\ \notag & +  2\dfrac{d\theta}{dx}I_3(x) \biggr]  + \dfrac{1}{2}v^2_2(x)\biggl[  \bigl(\dfrac{d\theta}{dx}\bigr)^2 + I_0(x) + 2\bigl(\dfrac{d\theta}{dx} + \dfrac{d\xi}{dx} \bigr)I_3(x) + 2\dfrac{d\theta}{dx}\dfrac{d\xi}{dx} \biggr] \\ \notag & + \dfrac{1}{2}v^2_+(x)\biggl[  \bigl(\dfrac{d\theta}{dx}\bigr)^2   + I_0(x) - 2\dfrac{d\theta}{dx}I_3(x) \biggr] + v_2(x) \biggl[   \text{ sin}(\xi)\dfrac{dv_+}{dx}I_1(x) - \text{ cos}(\xi)\dfrac{dv_+}{dx}I_2(x) \biggr] \\ \notag & + v_+(x) \biggl[   -\text{ sin}(\xi)\dfrac{dv_2}{dx}I_1(x) + \text{ cos}(\xi)\dfrac{dv_2}{dx}I_2(x) \biggr]
   + \dfrac{v_+(x)v_2(x)}{2}\biggl[  4\text{ sin}(\xi)I_2(x)\dfrac{d\theta}{dx} \\  & -  4\text{ cos}(\xi)I_1(x)\dfrac{d\theta}{dx} -2\text{ cos}(\xi)\dfrac{d\xi}{dx}I_1(x) -2\text{ sin}(\xi)\dfrac{d\xi}{dx}I_2(x) \biggr]  +  V_{2HDM}, 
   \label{eq:energyfunctionalgen}
\end{align}
where,
\begin{align}
   \notag I_0(x) &= G'^2(x)\text{ cos}^2(G(x)) + \biggl[ \bigl(g_1'^2(x) + g_2'^2(x) + g_3'^2(x) \bigr)G^2(x) - 2\bigl(g_1g_1' + g_2g_2' + g_3g_3'\bigr)^2 \\ &   + \bigl(G^2(x) + G^4(x)\bigr)G'^2(x) \biggr]\dfrac{\text{ sin}^2(G(x))}{G^4(x)}, \\ 
    I_1(x) &= B'(x)C(x) - B(x)C'(x) - A'(x)\text{ cos}(G(x)) - A(x)G'(x)\text{ sin}(G(x)),\\
    I_2(x) &= C'(x)A(x) -A'(x)C(x) - B'(x)\text{ cos}(G(x)) - B(x)G'(x)\text{ sin}(G(x)), \\
    I_3(x) &= A'(x)B(x) - A(x)B'(x) - C'(x)\text{ cos}(G(x)) - C(x)G'(x)\text{ sin}(G(x)), \\
    A(x) &= g_1(x) \dfrac{\text{ sin}(G(x))}{G(x)}, \\
    B(x) &= g_2(x) \dfrac{\text{ sin}(G(x))}{G(x)}, \\
    C(x) &= g_3(x) \dfrac{\text{ sin}(G(x))}{G(x)}, 
\end{align}
and the prime symbol denotes derivatives with respect to $x$.
In order to get non-trivial vacuum configurations corresponding to different vacua at $\pm \infty$,  we need to minimize the space integral of this energy functional with respect to small deviations in the fields:
\begin{equation}
    \delta E = \delta \biggl( \int dx \ \ \mathcal{E}(x) \biggr) = 0.
\label{eq:minimizationcond}    
\end{equation}
For static solutions, this leads to a system of differential equations analogous to the equations of motion:
\begin{equation}
     \dfrac{d}{dx}\biggl(\dfrac{d\mathcal{E}}{d(d\phi_i)}\biggr) - \dfrac{d\mathcal{E}}{d\phi_i} = 0,
\label{eq:eomi}     
\end{equation}
where $\phi_i$ denotes the 8 fields in the two doublets. \\
Since the energy functional and the equations of motion in the general case are very complicated, it is simpler to explain the behavior of the different fields inside the domain wall by choosing some special cases which make the energy functional and equations of motion considerably simpler. In the following sections we will consider a few special cases for the choice of $U(x)$ at the boundaries. These simplified cases capture the interesting effects that influence the fields inside the domain wall. We will first start by considering the case of standard domain walls where the matrix $U(x) = \text{I}_2$, constant everywhere in space, this means that we only consider vacua related by the $Z_2$ symmetry. The cases where only one mode of $U(x)$ is different in the two domains while taking all the other modes to be 0 was already considered in \cite{Law:2021ing}. Here we expand those results by first considering the case where more than one mode of $U(x)$ changes across the domains and later the general case where the hypercharge and the Goldstone modes are chosen arbitrarily. We also discuss the behavior of the fields that interact with the kink solution for $v_2$ inside the wall. \\
To solve the 8 differential equations describing the vacuum configuration as a function of $x$, we use the same numerical algorithm used in \cite{Battye:2020sxy,Law:2021ing,Battye:2011jj}, namely the gradient flow method. The gradient flow method introduces a fictitious time parameter to the vacuum profiles $\phi_i(x,t)$. We then modify the minimization condition (\ref{eq:minimizationcond}) to:
\begin{equation}
   \frac{\partial \phi_i }{\partial t} =  -\frac{\delta E}{\delta\phi_i} 
\end{equation}
Using this method, we find the solution of the system of differential equations by iteratively minimizing the energy functional at each time for the given boundary conditions until the vacuum configuration $\phi_i(x,t)$ leads to a minimum in the energy. As we approach such a minimal energy configuration, the derivative of $\phi_i(x,t)$ with respect to the fictitious time approaches zero and therefore we recover the static equations of motion in (\ref{eq:minimizationcond}). The solution is then declared as found if, after several iterations, the vacuum configuration stays the same. We also adopt the same rescaling of the dimensionful parameters that was used in \cite{Battye:2020sxy,Law:2021ing,Battye:2011jj}:
\begin{align}
 \hat{m}_i &= \frac{m_i}{m_h}, & \hat{v}_i = \frac{v_i}{v_{sm}}, && \hat{x} = x\cdot m_h, && \hat{\mathcal{E}} = \frac{\mathcal{E}}{m^2_hv^2_{sm}}, 
 \label{eq:rescaling}
\end{align}
where $m_h = 125 \text{GeV}$ denotes the mass of the SM Higgs particle. 
Such a rescaling is useful to get dimensionless space variable $\hat{x}$ and numerical values of the order 1. 
\subsection{Standard Domain Wall Solution}\label{standardDWsolution}
\begin{figure}[h]
\centering
\includegraphics[width=0.65\textwidth]{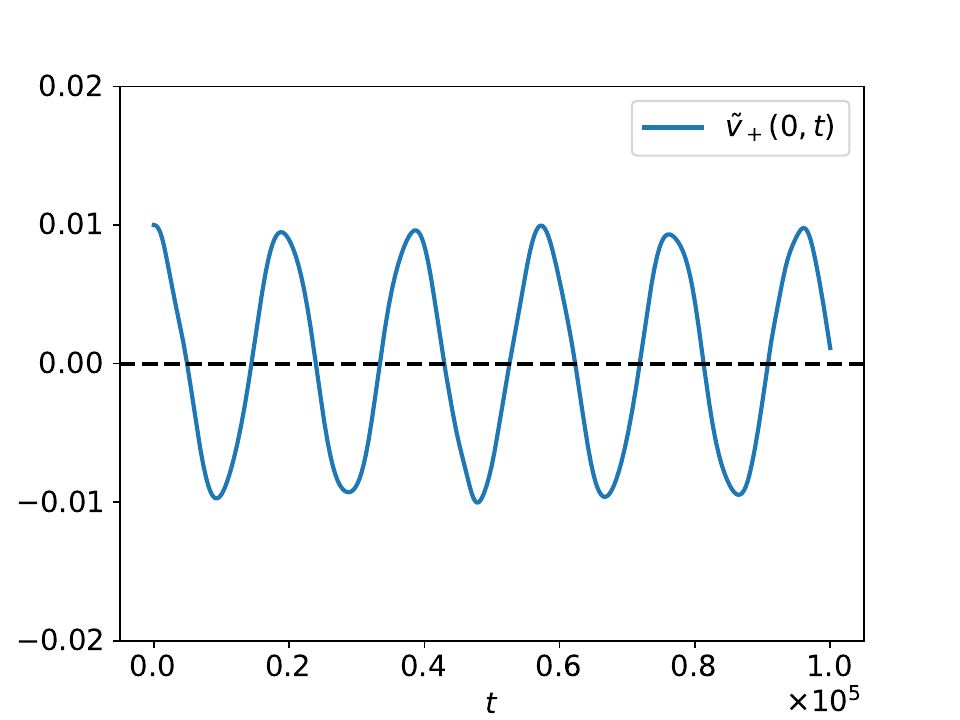}
\caption{Stability of the solution $v_+(x) = 0$ under a small fluctuation $\tilde{v}_+(x,t)$,  oscillating around $v_+ = 0$ with a real frequency $w$.}
\label{fig:stabilityvplus0}
\end{figure}

We start with the standard kink solution, where the two domains have the same and constant angles $\theta$ and $g_i$ while the vacuum expectation value of $v_2$ changes sign. The kink solution interpolates between the following two vacua:
\begin{align}
  \Phi_- = \biggl\{ \dfrac{1}{\sqrt{2}}
    \begin{pmatrix}
          0 \\      v^*_1
     \end{pmatrix}  , \dfrac{1}{\sqrt{2}}
      \begin{pmatrix}
     0 \\
     -v^*_2
      \end{pmatrix}  \biggr\},
  &&  \Phi_+ = \biggl\{ \dfrac{1}{\sqrt{2}}
    \begin{pmatrix}
          0 \\      v^*_1
     \end{pmatrix}  , \dfrac{1}{\sqrt{2}}
      \begin{pmatrix}
     0 \\
     v^*_2
      \end{pmatrix}  \biggr\}.
\end{align}
We use (\ref{eq:energyfunctionalgen}) and set the derivatives for $\theta$ and $g_i$ to zero. We obtain the following energy functional that has to be minimized: 
\begin{align}
  \notag  \mathcal{E}(x) &= \frac{1}{2} (\frac{dv_1}{dx})^2 +  \frac{1}{2} (\frac{dv_2}{dx})^2 + \frac{1}{2} (\frac{dv_+}{dx})^2 + \frac{1}{2}v^2_2(x)(\frac{d\xi}{dx})^2 + \frac{1}{2}m^2_{11}v^2_1(x) + \frac{1}{2}m^2_{22}(v^2_2(x) + v^2_+(x))  \\ & + \frac{1}{8}\lambda_1v^4_1(x) + \frac{1}{8}\lambda_2(v^2_2(x) + v^2_+(x))^2 + \frac{1}{4}\lambda_3 v^2_1(x) v^2_+(x) \notag  + \frac{1}{4}\biggl[\lambda_3 + \lambda_4  + \lambda_5 \text{cos}(2\xi(x))\biggr] \\ & \times v^2_1(x) v^2_2(x), 
\label{eq:ensta}  
\end{align}
leading to the following equations of motion for the field profiles: 
\begin{align}
 \notag   \dfrac{d}{dx}(\dfrac{d\mathcal{E}}{d(dv_1)}) - \dfrac{d\mathcal{E}}{dv_1} &= \frac{d^2v_1}{dx^2} - \frac{1}{2}\lambda_1v^3_1(x) + \lambda_3 v^2_+(x) - v_1(x)m^2_{11} \\ & - \frac{1}{2}v_1(x)\biggl(\lambda_3 + \lambda_4 + \lambda_5 \text{cos}(\xi(x))\biggr)v^2_2(x)  = 0, \\
    \notag   \dfrac{d}{dx}(\dfrac{d\mathcal{E}}{d(dv_2)}) - \dfrac{d\mathcal{E}}{dv_2} &= \frac{d^2v_2}{dx^2} - \frac{1}{2}\lambda_2v^3_2(x) - \frac{1}{2}v_2(x)\biggl[2(m^2_{22} + (\frac{d\xi}{dx})^2) + \lambda_2 v^2_+(x) \\ &  + \biggl(\lambda_3 + \lambda_4 + \lambda_5 \text{cos}(\xi(x))\biggr)v^2_1(x) \biggr] = 0,  \\
   \dfrac{d}{dx}(\dfrac{d\mathcal{E}}{d(dv_+)}) - \dfrac{d\mathcal{E}}{dv_+} &= -\frac{1}{2}\biggl[2m^2_{22} + \lambda_3 v^2_1(x) + \lambda_2 v^2_2(x)\biggr]v_+(x) - 
 \frac{1}{2} \lambda_2 v^3_+ + \frac{d^2v_+}{dx^2} = 0 \\
  \notag   \dfrac{d}{dx}(\dfrac{d\mathcal{E}}{d(d\xi)}) - \dfrac{d\mathcal{E}}{d\xi} &= v_2(x)\biggl[ \lambda_5 \text{cos}(\xi) \text{sin}(\xi) v^2_1(x) v_2(x) + 2\frac{dv_2}{dx} \frac{d\xi}{dx}  + 
   v_2(x) \frac{d^2\xi}{dx^2}\biggr] =0.
\end{align}
As $\mathcal{E}(x)$ is independent of $\theta$ and $g_i$, they therefore remain constant. 
\begin{figure}[h]
     \centering
     \begin{subfigure}[b]{0.49\textwidth}
         \centering
         \includegraphics[width=\textwidth]{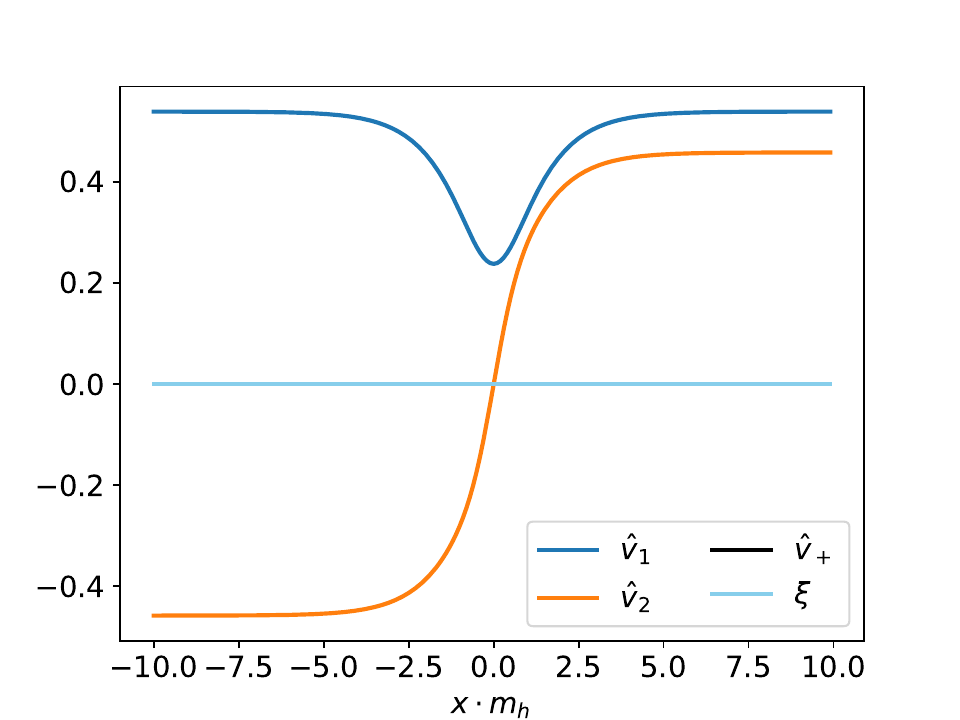}
         \subcaption{Domain wall solution for PP I.}\label{subfig:dwsolutionI}
     \end{subfigure}
     \hfill
     \begin{subfigure}[b]{0.49\textwidth}
         \centering
         \includegraphics[width=\textwidth]{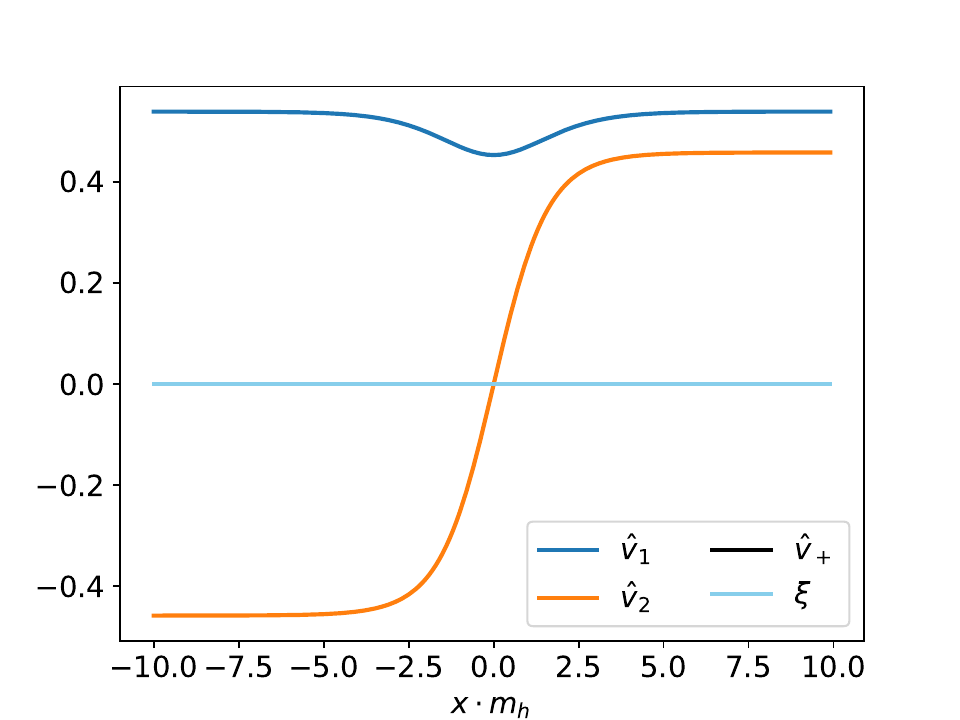}
         \subcaption{Solutions for 
 PP II.}\label{subfig:dwsolutionII}
     \end{subfigure}
\caption{Standard domain wall solution for different parameter points I (a) and II (b). We use the rescaled dimensionless vacuum parameters $\hat{v}_i = v_i/v_{sm} $ (cf. \ref{eq:rescaling}).}
\label{fig:stan}
\end{figure}
The system of differential equations is solved in an interval $-10 < \hat{x} < 10 $. In Figure \ref{fig:stan}  we plot the profile of the fields for parameter point (PP) I:
\begin{align}
m_H &= 800 \text{ GeV}, && m_A = 500 \text{ GeV}, && m_C = 400 \text{ GeV}, 
&& \text{tan}(\beta) = 0.85,
\label{eq:PP1}
\end{align}
and parameter point II:
\begin{align}
m_H &= 200 \text{ GeV}, && m_A = 200 \text{ GeV}, && m_C = 200 \text{ GeV}, && \text{tan}(\beta) = 0.85.
\label{eq:PP2}
\end{align}
This choice of parameter points is done for pedagogical reasons to show different properties in the kink solutions. A detailed phenomenological study involving constraints from experiments in colliders and cosmological observations is subject of future studies. The results show that the vacuum $v_2(x)$ interpolates from a negative value in the region on the left to a positive value in the region on the right while crossing the value 0. This is the usual behavior of a kink solution. Note that also the value of the vacuum expectation value of $v_1(x)$ is also affected as both $v_1$ and $v_2$ are coupled via the differential equations.\\
To understand this change in $v_1(x)$ inside the domain wall of $v_2$ (see Figures \ref{subfig:dwsolutionI} and \ref{subfig:dwsolutionII}), we derive the standard energy functional $\mathcal{E}_{standard}(x)$ in \ref{eq:ensta} taking the derivatives of all fields other than $v_1$ and $v_2$ to be 0 as they all vanish at all $x$:
\begin{align}
    \mathcal{E}_{standard}(x) &= \dfrac{1}{2}(\dfrac{dv_1}{dx})^2 + \dfrac{1}{2}(\dfrac{dv_2}{dx})^2 + \dfrac{m^2_{11}}{2}v^2_1(x) + \dfrac{m^2_{22}}{2}v^2_2(x)  + \dfrac{\lambda_1}{8}v^4_1(x) \\ \notag & + \dfrac{\lambda_2}{8}v^4_2(x) + \dfrac{\lambda_3 + \lambda_4 }{4}v^2_1(x)v^2_2(x)  + \dfrac{\lambda_5}{4}v^2_1(x)v^2_2(x) .
\end{align}
The effective mass term for $v_1$ is then given by:
\begin{equation}
    M_1(x) = \dfrac{m^2_{11}}{2} + \dfrac{(\lambda_3 + \lambda_4 + \lambda_5 )}{4}v^2_2(x).
\label{eq:effectivemass1}    
\end{equation}
\begin{figure}[h]
     \begin{subfigure}[b]{0.49\textwidth}
         \centering
     \includegraphics[width=\textwidth]{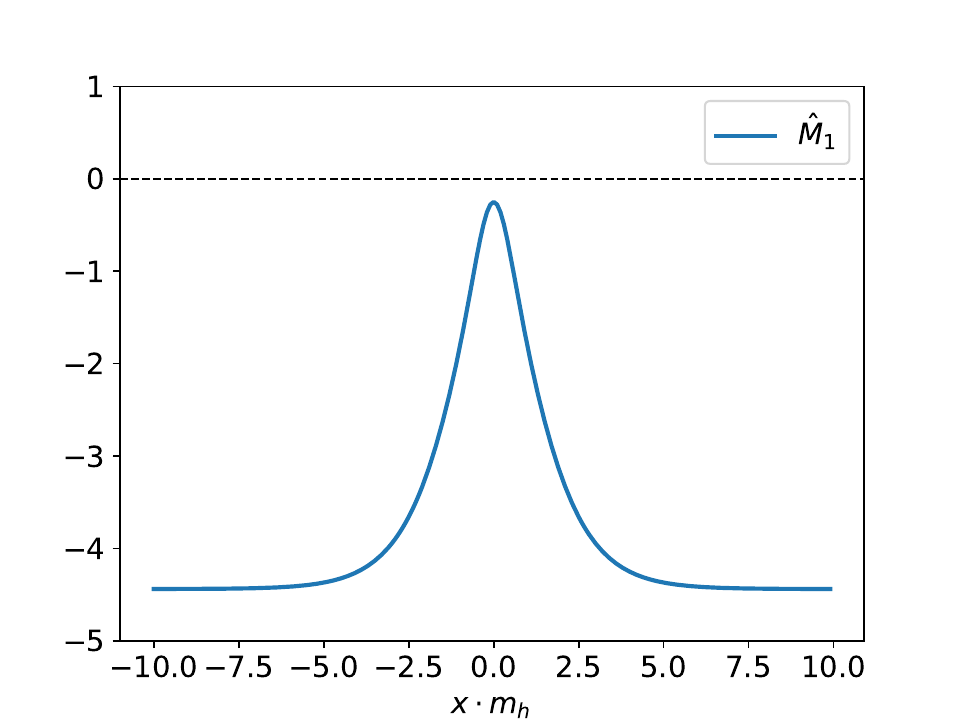}
         \subcaption{$\hat{M}_1$ for PP I.}\label{subfig:m1standard}
     \end{subfigure}
     \hfill
     \begin{subfigure}[b]{0.49\textwidth}
         \centering
         \includegraphics[width=\textwidth]{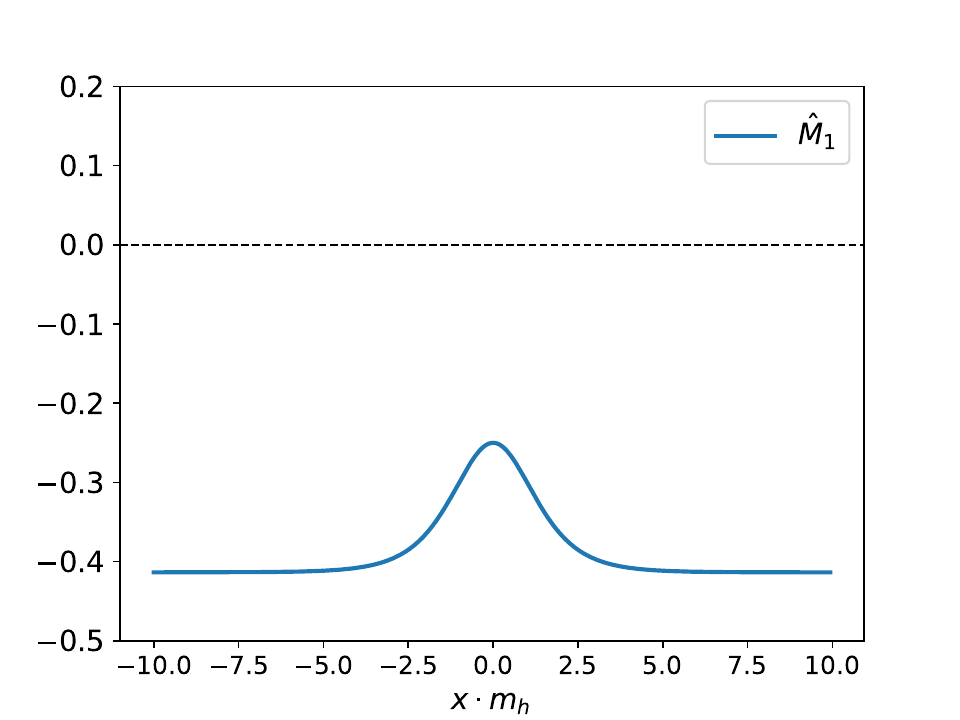}
         \subcaption{ $\hat{M}_1$ for PP II.}\label{subfig:m1standardother}
     \end{subfigure}
     \begin{subfigure}[b]{0.49\textwidth}
         \centering
         \includegraphics[width=\textwidth]{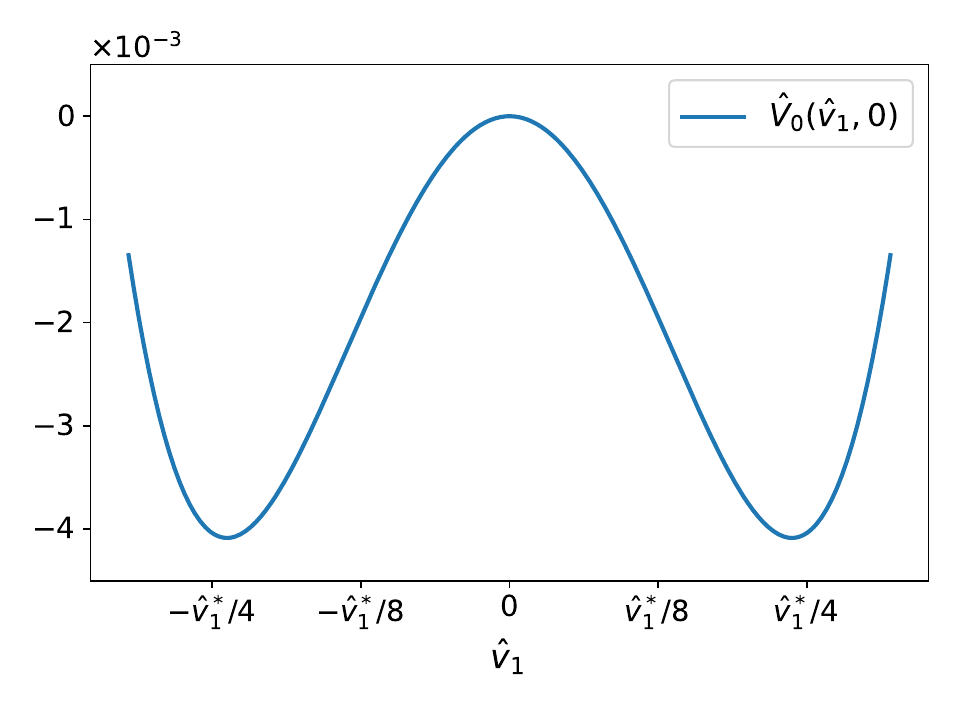}
         \subcaption{Potential inside the DW for PP I.}\label{subfig:potv1I}
     \end{subfigure}
     \hfill
     \begin{subfigure}[b]{0.49\textwidth}
         \centering
         \includegraphics[width=\textwidth]{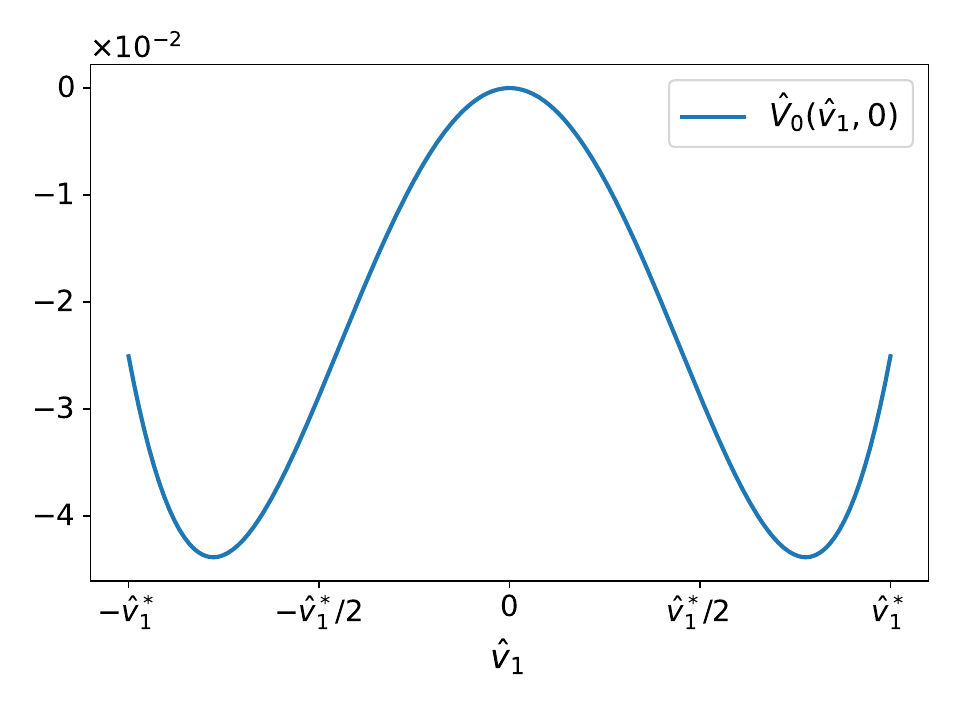}
         \subcaption{ Potential inside the DW for PP II.}\label{subfig:potv1II}
     \end{subfigure}     
\caption{(a) and (b): Rescaled effective mass term $\hat{M}_1$ of $v_1$ as a function of $x$. Notice that the effective mass becomes less negative inside the domain wall. This explains why the vacuum $v_1$ inside the wall gets smaller.(c) and (d): Rescaled potential $\hat{V}_0(\hat{v}_1,\hat{v}_+) = \hat{V}_{2HDM}(x=0)$ inside the domain wall $\bigl(v_2(0) = 0\bigr)$.}
\label{fig:M1}
\end{figure}
Inside the core of the wall ($x = 0$), $v_2 = 0$ and $M_1(0) = m^2_{11}/2$. When plotting the effective mass $M_1$ as function of $x$, we can see that the value becomes less negative inside the domain wall for both parameter points (see Figures \ref{subfig:m1standard} and \ref{subfig:m1standardother}). Consequently the scalar potential $V_{2HDM}(x=0) =V_0(v_1,v_+)$ inside the domain wall will have its minima for $v_1$ at a smaller value and the potential barrier between the minima will be lower. Note that for other parameter points, $(\lambda_3 + \lambda_4 + \lambda_5)$ can be positive and therefore the effective mass $M_1(0)$ inside the domain wall will be more negative than at the asymptotic values. This then corresponds to a bigger value for the minimum $v_1(0)$ of $V_0(v_1)$ and therefore $v_1$ inside the domain wall would be bigger than its asymptotic values at $\pm \infty$. 
\begin{figure}[h]
     \centering
     \begin{subfigure}[b]{0.49\textwidth}
         \centering
         \includegraphics[width=\textwidth]{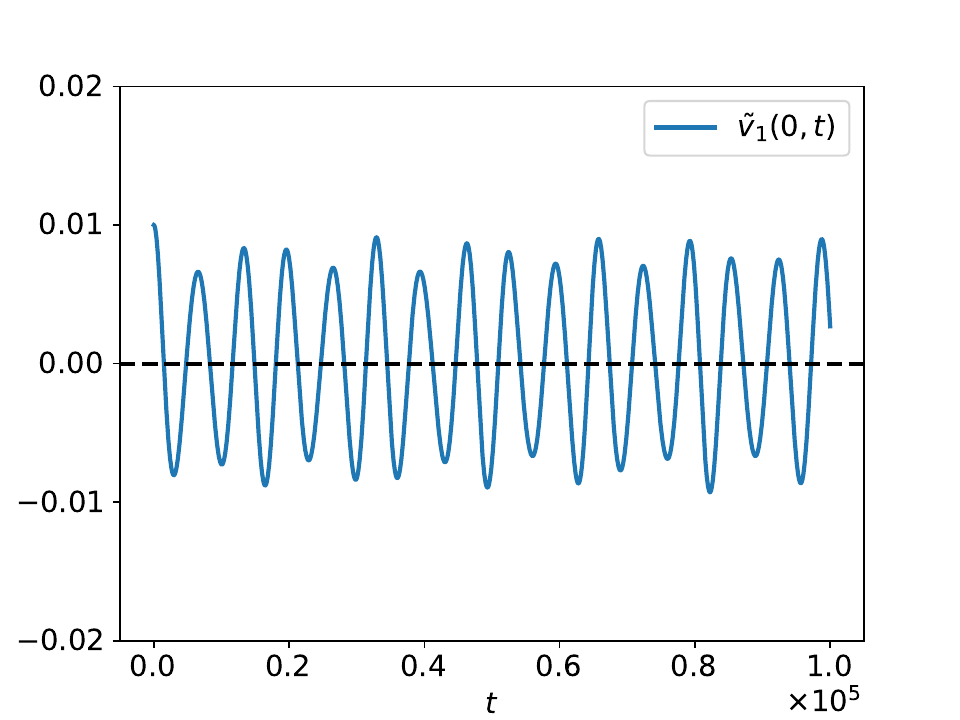}
         \subcaption{}\label{subfig:v1stable}
     \end{subfigure}
     \hfill
     \begin{subfigure}[b]{0.49\textwidth}
         \centering
         \includegraphics[width=\textwidth]{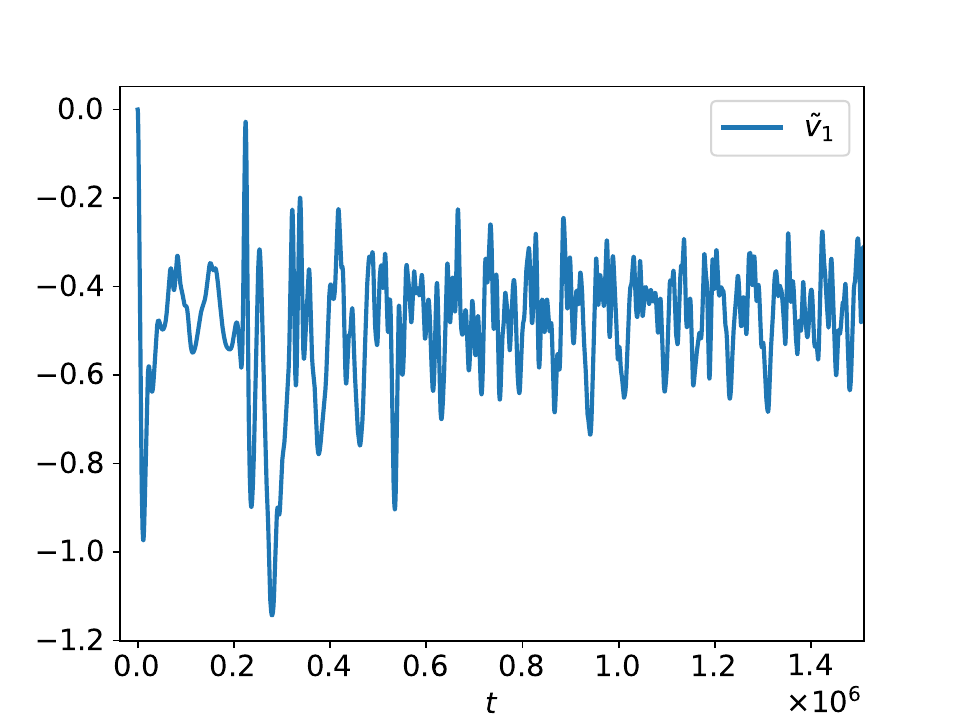}
         \subcaption{}\label{subfig:unstable}
     \end{subfigure}
\caption{(a) The solution found in Figure \ref{subfig:dwsolutionI} is stable as the fluctuation $\tilde{v}_1$ oscillates around 0. (b) The fluctuation $\tilde{v}_1(x,t)$ grows with time for the case when $v_1(x) = v_1$ indicating the instability of that solution. After some time the fluctuation oscillates around the lowest minimum corresponding to the obtained results for PP I.  }
\label{fig:stabilityv1}
\end{figure}
However, a changing profile for $v_1(x)$ inside the wall will always lead to a positive contribution to the kinetic part of $\mathcal{E}(x)$. Therefore, one needs to make sure that this solution for $v_1(x)$ is stable. This is done by considering small fluctuations $\tilde{v}_1(x,t)$ around the background kink solution $v_1(x)$. The equation of motion describing such fluctuations is given by:
\begin{equation}
    \partial^2_t\tilde{v}_1(x,t) - \partial^2_x\tilde{v}_1(x,t) + \frac{dV_{2HDM}}{d\tilde{v}_1} = 0.
\end{equation}
Taking small fluctuations up to first order, this reduces to:
\begin{align}
     - \partial^2_x \tilde{v}_1(x) + 2M_1(x)\tilde{v}_1(x) = w^2\tilde{v}_1(x),
\label{eq:stabilityv1}     
\end{align} 
for a fluctuation of the form $\tilde{v}_1(x,t) = e^{iwt}\tilde{v}_1(x)$.
If a solution with $w^2<0$ exists, then the fluctuation $v_1(x,t) \propto e^{\tilde{w}t}$ ,where $w = i\tilde{w}$, grows with time leading to the instability of the solution for $v_1(x)$. For $w^2>0$ the fluctuation keeps oscillating around the found solution for $v_1(x)$. The behavior of $\tilde{v}_1(x,t)$ for PP I is shown in Figure \ref{subfig:v1stable}, showing an oscillating fluctuation around 0, which means that the solution is stable.
Note that a vacuum configuration, where $v_1(x) = v_1$ is constant, will be unstable and the field $v_1(x)$ evolves to the lowest energy solution that we obtain from the numerical calculations (see Figures \ref{subfig:unstable} and \ref{subfig:dwsolutionI}).\\
\begin{figure}[h]
     \centering
     \begin{subfigure}[b]{0.49\textwidth}
         \centering
         \includegraphics[width=\textwidth]{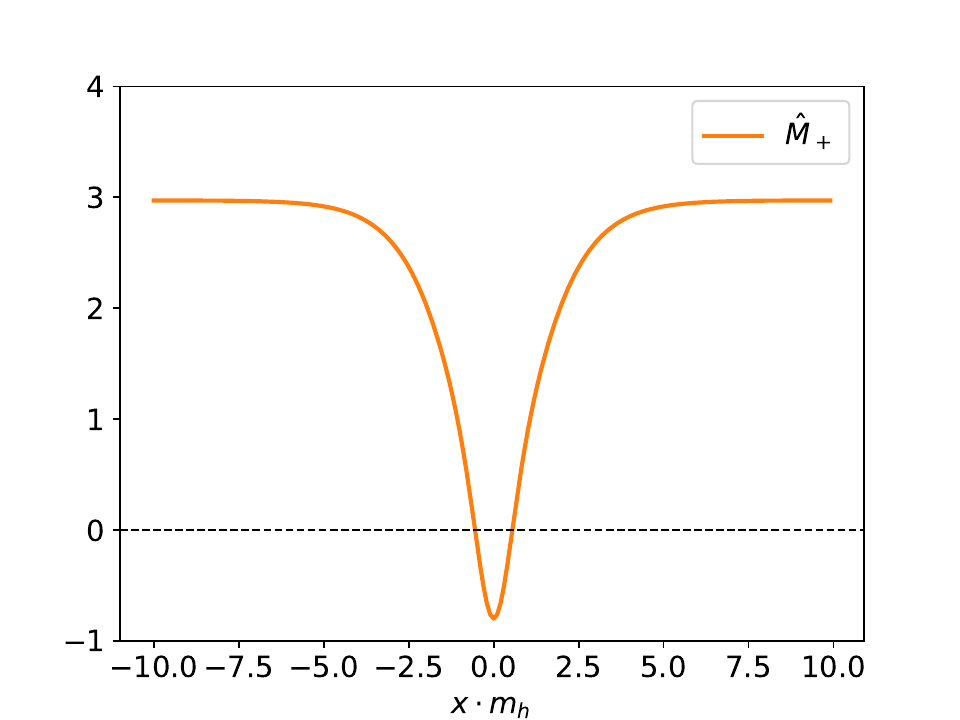}
         \subcaption{Negative $\hat{M}_+(0)$ (PP I).}\label{subfig:M+neg}
     \end{subfigure}
     \hfill
     \begin{subfigure}[b]{0.49\textwidth}
         \centering
         \includegraphics[width=\textwidth]{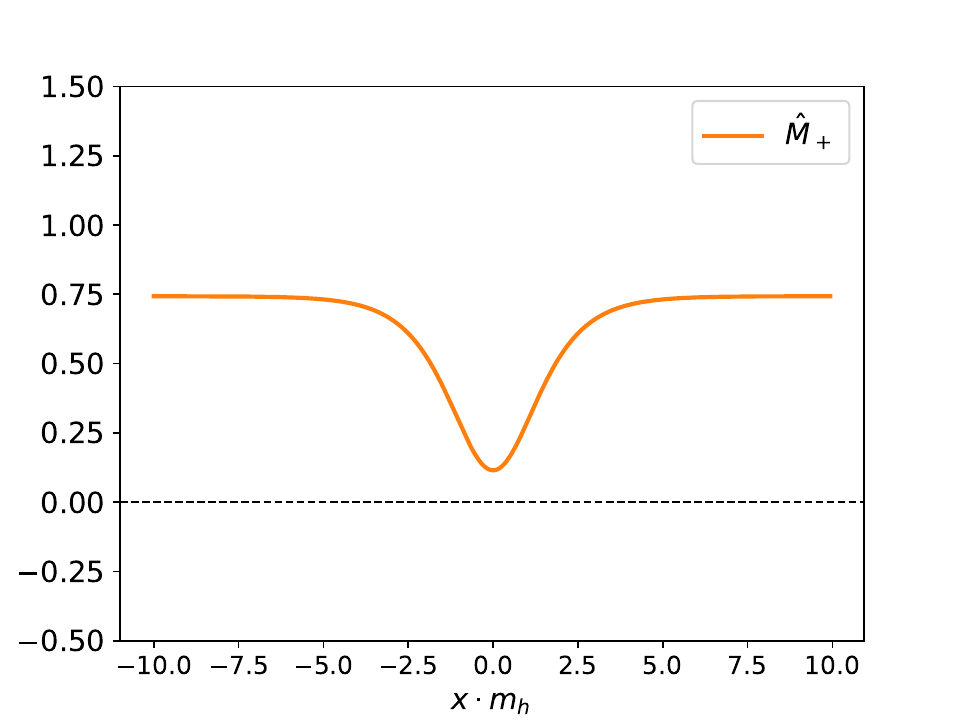}
        \subcaption{Positive $\hat{M}_+(0)$ (PP II).} \label{subfig:m+pos}
     \end{subfigure}
     \begin{subfigure}[b]{0.49\textwidth}
         \centering
         \includegraphics[width=\textwidth]{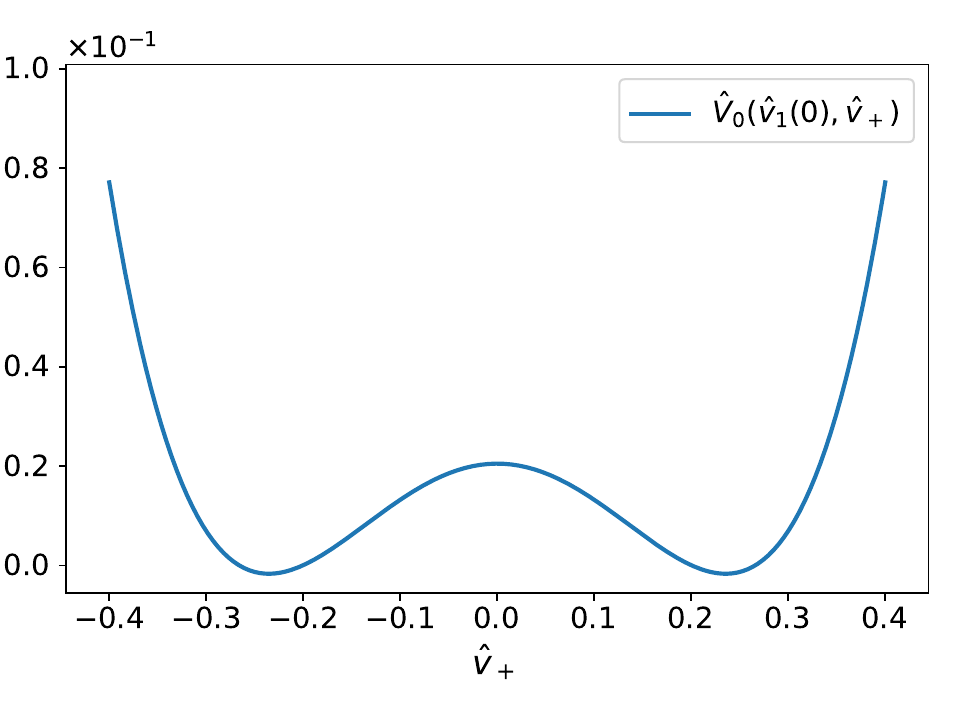}
         \subcaption{Potential for $v_+$ inside the DW (PP I).}\label{subfig:potv+I}
     \end{subfigure}
     \hfill
     \begin{subfigure}[b]{0.49\textwidth}
         \centering
         \includegraphics[width=\textwidth]{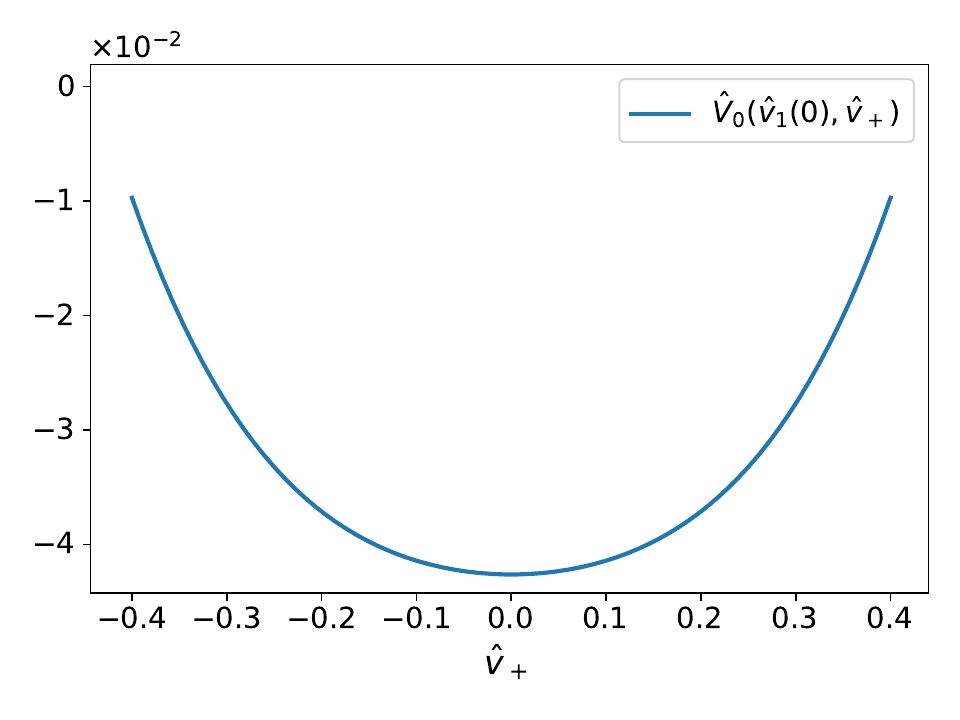}
         \subcaption{ Potential for $v_+$ inside the DW (PP II).}\label{subfig:potvplusII}
     \end{subfigure}      
\caption{Effective mass term of $v_+$ as a function of $x$ and the corresponding potential for different PP: (a) $\hat{M}_+$ for PP I, notice that the effective mass can become negative inside the wall, leading to the possibility of forming a condensate of $v_+$ localized on the wall; (b) shows $\hat{M}_+$ for PP II. Here the effective mass is positive everywhere and no charge violating vacua are expected inside the wall; (c) shows the potential inside the DW as a function of $v_+$ for PP I. In this case, the minimum is non-zero; (d) the same potential for PP II. In this case the global minimum of $v_+$ is zero.}
\label{fig:vplusbehaviour}
\end{figure}\\
For the fields $v_+(x)$ and $\xi(x)$, we observe that they stay equal to zero everywhere. A non-zero phase $\xi$ provides a positive contribution to the energy functional (\ref{eq:ensta}) leading to a higher energy solution. In other words, $\xi(x) = 0$ presents the lowest energy solution. Concerning $v_+(x)$, the situation is more complicated: in order to study its behavior inside the wall, we consider the terms in $\mathcal{E}$ that depend on $v_+(x)$, (cf. \ref{eq:ensta}):
\begin{align}
    \mathcal{E}_+(v_+) =  \frac{1}{2} (\frac{dv_+}{dx})^2 +  \frac{1}{2}m^2_{22}v^2_+(x) + \frac{1}{8}\lambda_2v^4_+(x) + \frac{1}{4}\lambda_2v^2_+(x)v^2_2(x) + \frac{1}{4}\lambda_3 v^2_1(x) v^2_+(x). 
\end{align}
The effective mass term for $v_+$ in the background of the DW is given by:
\begin{equation}
    M_+(x) = \frac{1}{2}m^2_{22} + \frac{1}{4}(\lambda_2 v^2_2(x) + \lambda_3 v^2_1(x)).
\label{eq:effectivemass+}    
\end{equation}
Outside the domain wall, the effective mass is obviously positive and the potential minimum for $v_+$ is 0. Inside the wall, one gets $$ M_+(0) = \frac{1}{2}m^2_{22} + \frac{1}{4}\lambda_3v^2_1(0),  $$ which, depending on the parameter point can also be negative (see Figures \ref{subfig:M+neg} and \ref{subfig:m+pos}). 
The field $v_+$ can develop a condensate inside the wall, i.e. when $M_+(0) < 0$. In that case the potential $V_{0}(v_1(0), v_+)$ will have two non zero minima for $v_+$ (see Figure \ref{subfig:potv+I}). If such parameter points exist, it is energetically more favorable for $v_+$ to get a condensate inside the wall. However, the kinetic energy part is minimized for $v_+(x) = 0$. Therefore, one should make sure that the kinetic contribution due to the spatial derivative of $v_+$ is not too large so that the solution for the non-zero condensate is stable inside the wall. In order to investigate the stability of the $v_+(x) = 0$ solution inside the wall, we consider the linearized time-dependent equation of motion for a small fluctuation $\tilde{v}_+(x,t)$ around $v_+(x)$ in the background of the domain wall:
\begin{equation}
    \partial^2_t\tilde{v}_+(x,t) - \partial^2_x\tilde{v}_+(x,t) + \frac{dV_{2HDM}}{d\tilde{v}_+} = 0.
\end{equation}\\
For a small fluctuation of the form \begin{equation}  
\tilde{v}_+(x,t)=e^{iwt}\tilde{v}_+(x),
\end{equation}
the evolution of the fluctuation follows the differential equation:
\begin{align}
     - \partial^2_x \tilde{v}_+(x) + 2M_+(x)\tilde{v}_+(x) = w^2\tilde{v}_+.
\label{eq:stabilityv+}     
\end{align} 
The small fluctuation around $v_+ =0$ is unstable if $w^2<0$, making $v_+(x,t) \propto e^{\tilde{w}t}$, where $w = i\tilde{w}$, growing with time and leading to the instability of the solution $v_+(x) = 0$. In case $w^2>0$, the fluctuation $\tilde{v}_+(x,t)$ oscillates around the stable solution. 
Figure \ref{subfig:M+neg} shows one parameter point where $M_+(0)$ is negative inside the wall. However, the lowest energy solution has $v_+(x) = 0$, meaning that the kinetic energy contribution from a $v_+(x)$ condensate inside the wall leads to a higher total energy than the solution with a vanishing $v_+(x)$ overall. We also verify numerically (see Figure \ref{fig:stabilityvplus0}) that the frequency $w$ of a fluctuation $\tilde{v}_+(x,t)$ to the $v_+(x) = 0$ solution is real as the fluctuation oscillates around $v_+(x) = 0$.
\subsection{Variation of a single angle across the wall}
We now  consider the effects of the Goldstone and hypercharge modes on the domain wall solution. We start by following the same approach in \cite{Law:2021ing} and simplify $\text{U}(x)$ (\ref{eq:EWmatrix}) by allowing the variation across the wall of either the hypercharge angle $\theta(x)$ or a single Goldstone mode $g_i(x)$ at a time. In this case, the vacua at $\pm \infty$ will be rotated relative to each other, by either a $\text{U}_Y(1)$ transformation or a transformation related to one Goldstone mode of $SU(2)_L$. For each case, we will discuss the solution of the equations of motion using either Dirichlet or von Neumann boundary conditions, where the former keeps the vacua at the boundaries fixed while the latter allows the dynamical variation of the vacua at the boundaries but keeps the spatial derivative of the vacua to be zero on the boundaries. For the following numerical solutions, we work in the alignment limit and fix the parameter point: 
\begin{equation}
m_H = 800 \text{ GeV, } m_A = 500 \text{ GeV, } m_C = 400 \text{ GeV, } \text{tan}(\beta) = 0.85 \text{ and } \alpha-\beta = 0.
\label{eq:parameterpoint}
\end{equation}
\subsubsection{Variation of hypercharge $\theta$}\label{theta}
\begin{figure}[H]
\centering
     \begin{subfigure}[b]{0.49\textwidth}
         \centering
         \includegraphics[width=\textwidth]{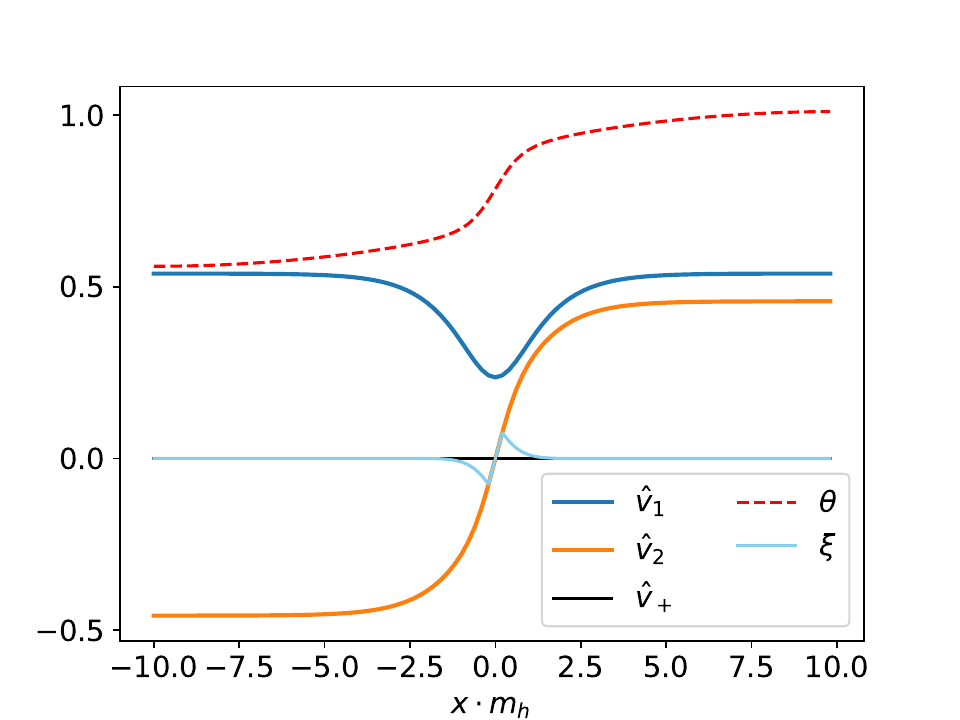}
         \subcaption{}\label{subfig:tneumaninter}
     \end{subfigure}
     \hfill
     \begin{subfigure}[b]{0.49\textwidth}
         \centering
         \includegraphics[width=\textwidth]{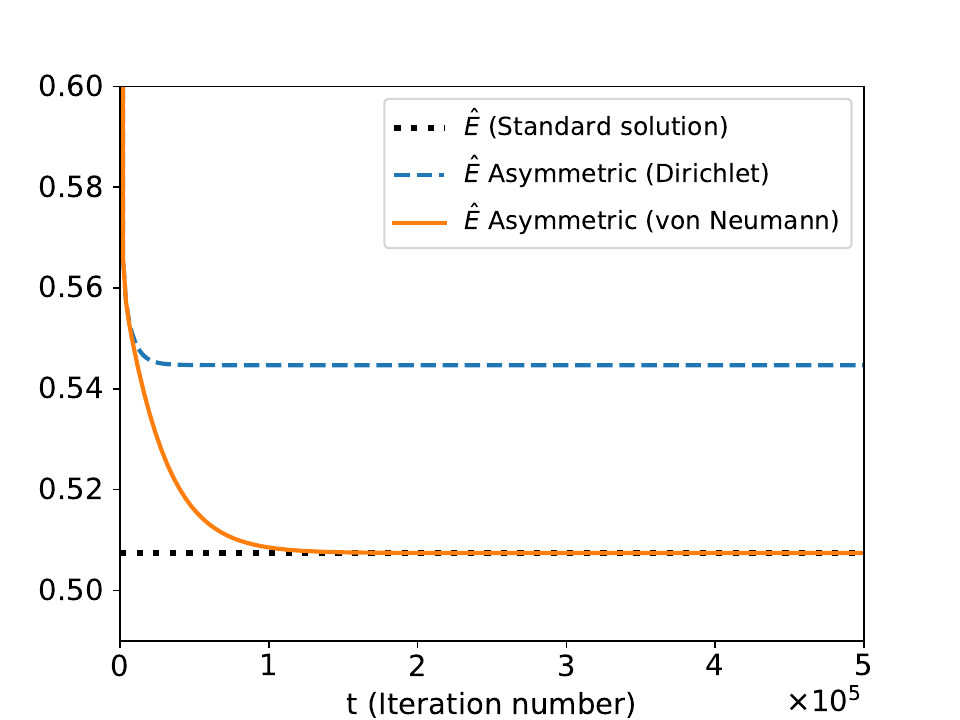}
         \subcaption{}\label{subfig:energyt}
     \end{subfigure}
\caption{(a) Evolution of the hypercharge angle $\theta$ at an intermediate stage of the calculations. Note that the CP-violating phase $\xi$ gets smaller as the difference in hypercharge in both domains gets smaller. (b) Evolution of the rescaled energy $\hat{E}$ (cf. \ref{eq:rescaling}) of the vacuum configuration using Dirichlet and von Neumann boundary conditions.}
\label{fig:interenergies}
\end{figure}

We first discuss the variation of the hypercharge angle $\theta(x)$ across the wall. In this case, the matrix $\text{U}(x)$ (\ref{eq:EWmatrix}) is given by:
$$ \text{U}(x) = e^{i\theta(x)}I_2, $$
which leads to the energy functional $\mathcal{E}(x)$:
\begin{align}
    \notag & \mathcal{E}(x) = \dfrac{1}{2}\biggl(\dfrac{dv_1}{dx}\biggr)^2 + \dfrac{1}{2}\biggl(\dfrac{dv_2}{dx}\biggr)^2 + \dfrac{1}{2}\biggl(\dfrac{dv_+}{dx}\biggr)^2 + \dfrac{1}{2}v^2_2(x)\biggl(\dfrac{d\xi}{dx}\biggr)^2  +
  \dfrac{1}{2}v^2_1(x)\biggl(\dfrac{d\theta}{dx}\biggr)^2  \\ & + \dfrac{1}{2}v^2_2(x)\biggl[  \biggl(\dfrac{d\theta}{dx}\biggr)^2 + 2\dfrac{d\theta}{dx}\dfrac{d\xi}{dx} \biggr]  + \dfrac{1}{2}v^2_+(x)  \biggl(\dfrac{d\theta}{dx}\biggr)^2  +  V_{2HDM}. 
\label{eq:energytheta}  
\end{align}
\begin{figure}[H]
     \centering
     \begin{subfigure}[b]{0.49\textwidth}
         \centering
         \includegraphics[width=\textwidth]{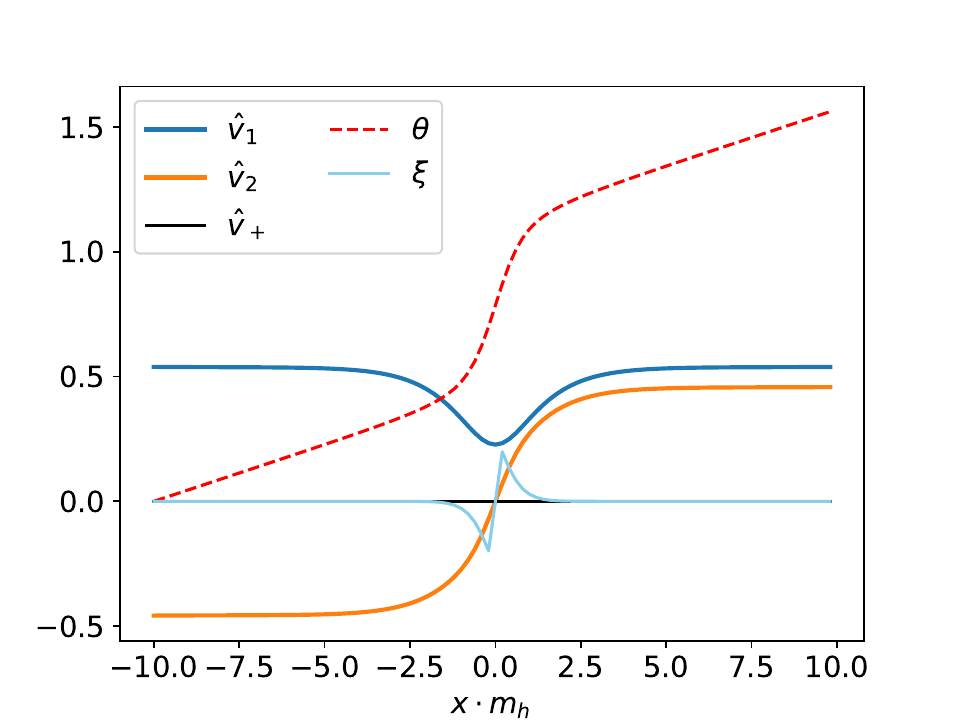}
         \subcaption{Dirichlet boundary conditions.}\label{subfig:tdirichlet}
     \end{subfigure}
     \hfill
     \begin{subfigure}[b]{0.49\textwidth}
         \centering
         \includegraphics[width=\textwidth]{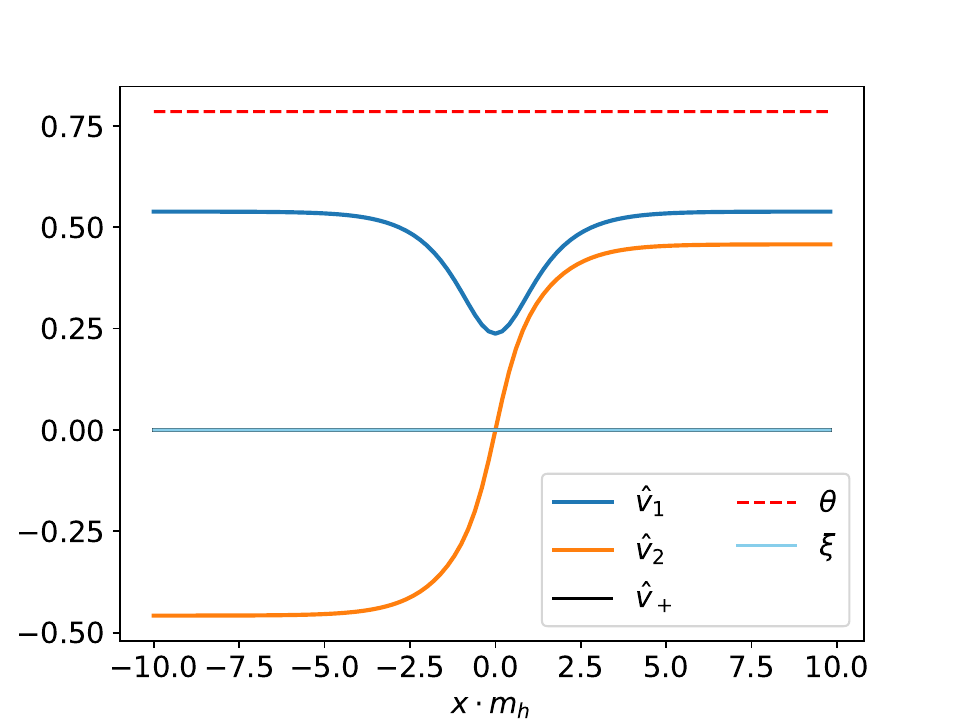}
         \subcaption{von Neumann boundary conditions.}\label{subfig:tneumann}
     \end{subfigure}
\caption{Numerical solutions of the DW equations of motion for vacua on the boundaries having different hypercharge angle $\theta$. (a) Using Dirichlet boundary conditions. (b) Using von Neumann boundary conditions. $\hat{v}_i$ are the rescaled vacuum parameters, cf. (\ref{eq:rescaling}). We observe a non zero phase $\xi(x)$ inside the wall when using Dirichlet boundary condition. This means that the vacua inside the wall are CP-violating (see (\ref{eq:CPviolatingvacuum})). }
\label{fig:resultt}
\end{figure}
\noindent
One can see immediately from (\ref{eq:energytheta}), that a change in the hypercharge across the wall cannot lead to a charge breaking solution $v_+(0) \neq 0$ inside the wall. The term $\frac{1}{2}v^2_+(x)  (d\theta/dx)^2$ in (\ref{eq:energytheta}) always leads to a positive contribution to the energy and therefore it only minimizes the energy of the vacuum configuration when $v_+(0) = 0$.
Using the equation of motion for the hypercharge $\theta(x)$, one can derive a relation between the change in the hypercharge $\theta(x)$ and the derivative of the CP-violating phase $\xi(x)$ \cite{Law:2021ing}:
\begin{equation}
    \frac{d\theta}{dx} = \frac{-v^2_2}{v^2_1 + v^2_2 + v^2_+} \frac{d\xi}{dx}.
\label{eq:relationthetaxi}    
\end{equation}
Such an equation is only valid for finite energy solutions where the spatial derivatives of the vacua at the boundaries vanish.
From (\ref{eq:relationthetaxi}) one would expect that, as the hypercharge angle starts to change from the value it has in one domain to the value in the  other domain, the value of the phase $\xi(x)$ will become non zero. Figure \ref{subfig:tdirichlet} shows the numerical solution of the equations of motion using Dirichlet boundary conditions  with $\theta(-\infty) = 0$ and $\theta(+\infty) =\pi/2$. The initial guess for the profile of $\theta(x)$ was taken to be a hyperbolic tangent function interpolating between 0 and $\pi/2$. The numerical results point to a non-vanishing phase $\xi(x)$ localized inside the wall, where the hypercharge changes significantly. However, this vacuum configuration has a higher energy than the vacuum configuration of the standard domain walls with $\xi(x) = 0$ (see Figure \ref{subfig:energyt}). Therefore, for these parameter points, such domain walls are unstable and should decay into the standard domain wall configuration. One can also notice that the derivative in $\theta(x)$ outside the wall are non zero due to the choice of Dirichlet boundary conditions. This shows that minimizing the energy of the wall tends to change the hypercharge angle at the boundaries. For this reason, it is more advantageous to use von Neumann boundary conditions when solving the differential equations, so that the hypercharge angle in each domain can change dynamically to minimize the energy of the wall. This can be seen in Figure \ref{subfig:tneumann}, where the hypercharge in both regions evolve to be equal to each other. This will then lead to $d\theta/dx = 0$ and therefore to a vanishing $\xi(x)$ for all $x$.
\noindent
In Figure \ref{subfig:tneumaninter} we plot the profile of the solution at an intermediate time step using von Neumann boundary conditions. This shows how the hypercharge angle $\theta(x)$ at the boundaries change dynamically to minimize the energy and we see that the value of the CP-violating phase $\xi(x)$ inside the wall gets smaller. Even though the CP-violating domain wall solution is unstable, it is expected that after EWSB, the hypercharge angles on causally disconnected regions of the universe can be different. Therefore the early stages of the formation of the domain wall network would exhibit such CP-violating vacua inside the wall until the profile of the hypercharge angle $\theta(x)$ relaxes to a solution where it is constant for all $x$. We will also see later that in the realistic scenario, where we consider $\text{U}(x)$ to be a general $\text{SU}(2) \times \text{U}_Y(1)$ matrix, the stable domain wall solution will exhibit a non zero (albeit small) CP-violating vacua because the hypercharge angle will be different on both domains. 
\subsubsection{Variation of $g_1$}
\begin{figure}[H]
\centering
\includegraphics[width=0.65\textwidth]{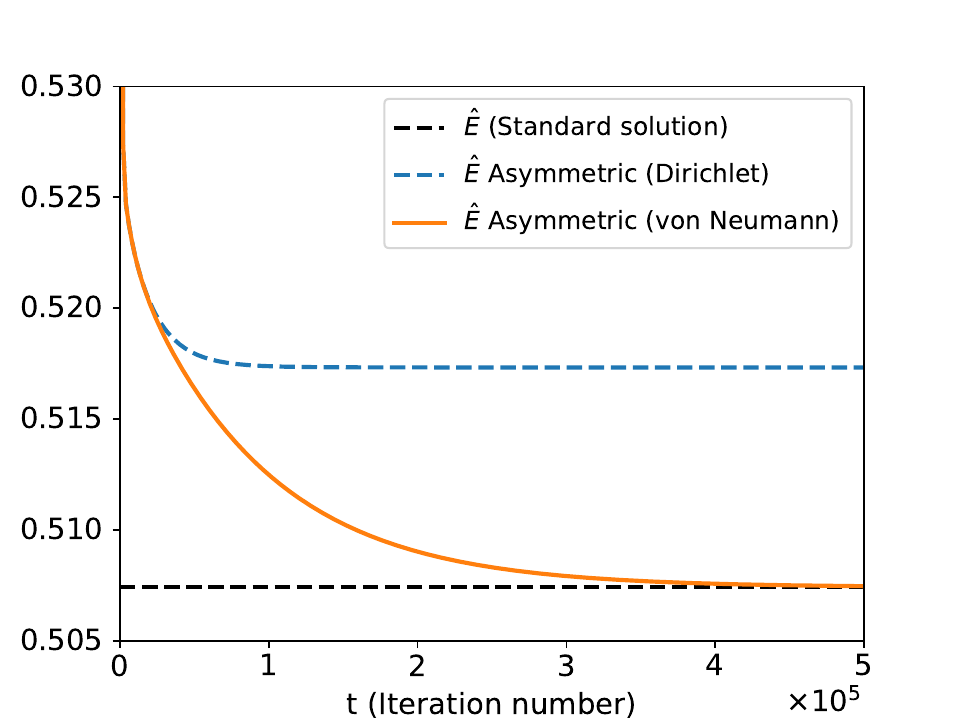}
\caption{Evolution of the energy of the vacuum configuration with iteration time for Dirichlet and von Neumann boundary conditions. For $g_1$ having asymmetric boundaries, the energy is higher than the energy of the standard vacuum configuration.}
\label{fig:g1energy}
\end{figure}

We now discuss the case when only the Goldstone mode $g_1$ is non-zero and different on both domains. The matrix $U(x)$ (\ref{eq:EWmatrix}) in such a case is given by:
\begin{equation}
    U(x) = \cos\bigl(g_1(x)/2\bigr)I_2 + i\sin\bigl(g_1(x)/2\bigr)\sigma_1 = \begin{pmatrix}
\cos\bigl(g_1(x)/2\bigr) & i\sin\bigl(g_1(x)/2\bigr) \\
i\sin\bigl(g_1(x)/2\bigr) & \cos\bigl(g_1(x)/2\bigr) 
\end{pmatrix}.
\end{equation}
The corresponding energy functional $\mathcal{E}$ (\ref{eq:energyfunctionalgen}) is simplified to:
\begin{align}
     \notag \mathcal{E}(x) &= \dfrac{1}{2}\biggl(\dfrac{dv_1}{dx}\biggr)^2 + \dfrac{1}{2}\biggl(\dfrac{dv_2}{dx}\biggr)^2 + \dfrac{1}{2}\biggl(\dfrac{dv_+}{dx}\biggr)^2 + \dfrac{1}{2}v^2_2(x)\biggl(\dfrac{d\xi}{dx}\biggr)^2  + \frac{1}{8}v^2_+\biggl(\frac{dg_1}{dx}\biggr)^2  \\ & -\frac{1}{2}v_2\sin(\xi)\frac{dv_+}{dx}\frac{dg_1}{dx}  +\frac{1}{2}v_+\frac{dg_1}{dx}\biggl(\sin(\xi)\frac{dv_2}{dx} + v_2\cos(\xi) \frac{d\xi}{dx}   \biggr)  + V_{2HDM}.
\end{align}
In \cite{Law:2021ing}, by using the equation of motion for $g_1(x)$, an expression relating how the change in $g_1$ will affect $v_+$ and $\xi$ was derived, namely:
\begin{equation}
       \frac{dg_1}{dx} = \frac{2}{v^2_{1} + v^2_{2} + v^2_{+}}\biggl( v_{2}\sin(\xi)\frac{dv_{+}}{dx} - v_{+}\sin(\xi)\frac{dv_{2}}{dx} - v_{2}v_+\cos(\xi)\frac{d\xi}{dx}  \biggr).
\label{eq:condg1}       
\end{equation}
This would then imply that a change in $g_1$ across the wall will lead to a non-zero $v_+(x)$ and $\xi(x)$ inside the wall.
\begin{figure}[H]
     \centering
     \begin{subfigure}[b]{0.49\textwidth}
         \centering
         \includegraphics[width=\textwidth]{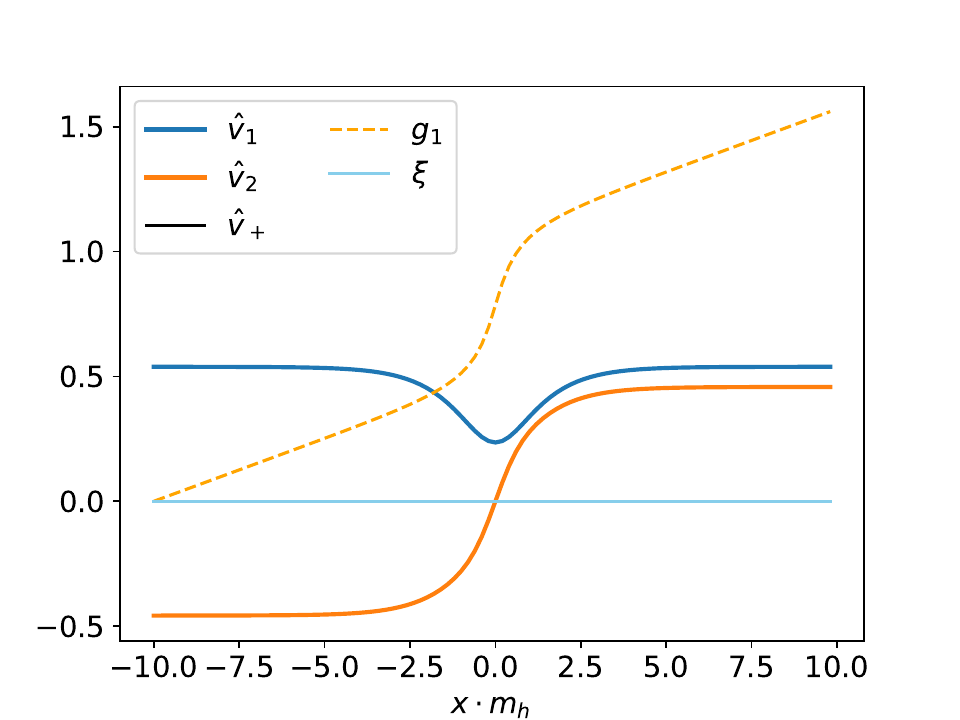}
         \subcaption{Dirichlet boundary conditions.}\label{subfig:g1dirichlet}
     \end{subfigure}
     \hfill
     \begin{subfigure}[b]{0.49\textwidth}
         \centering
         \includegraphics[width=\textwidth]{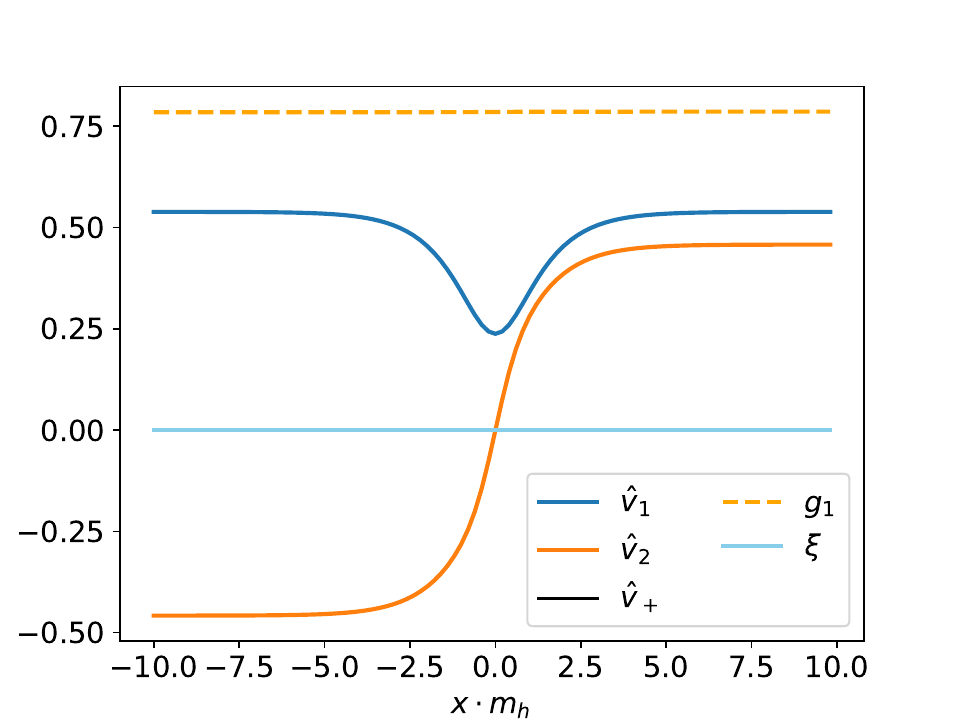}
         \subcaption{von Neumann boundary conditions.}\label{subfig:g1neumann}
     \end{subfigure}
\caption{Numerical solutions of the DW equations of motion for vacua having a different Goldstone mode $g_1$ using: (a) Dirichlet boundary conditions and (b) von Neumann boundary conditions. The solutions do not exhibit charge or CP-violating vacua inside the wall. Only the solutions using von Neumann boundary conditions satisfies the relation (\ref{eq:condg1}). }
\label{fig:resultg1}
\end{figure}
\noindent
The numerical solution of the equations of motion using Dirichlet boundary conditions, however, points to a vanishing $v_+(x)$ and $\xi(x)$ (see Figure \ref{subfig:g1dirichlet}), such a behavior was also found in \cite{Law:2021ing}. This can be attributed to the fact that using Dirichlet boundary conditions for this system of differential equations is not the correct approach. The equation (\ref{eq:condg1}) is valid for a static field configuration that satisfies the equations of motion, has a vanishing derivative for the Goldstone mode at the boundaries and minimizes the energy functional $\int dx\, \mathcal{E}(x)$. However it is clear that the condition of vanishing derivative at the boundaries is not satisfied for our solution using Dirichlet boundary conditions, even though the solution is static ($dg_1/dt = 0$).\\
In Figure \ref{fig:g1energy}, we compare the energies of domain wall solutions using different boundary conditions. We observe that the energy of the solution using the Dirichlet boundary conditions is only a local minimum and has a higher energy than the standard domain wall solution. When using von Neumann boundary conditions (see Figure \ref{subfig:g1neumann}), the Goldstone modes $g_1$ in both domains change dynamically to become the same value, which eventually leads to $dg_1/dx = 0$. Such a field configuration is the correct solution: it explains the vanishing values of $v_+(x)$ and $\xi(x)$ inside the wall, satisfies the relation (\ref{eq:condg1}) and minimizes the energy.
\subsubsection{Variation of $g_2$}
We now discuss the case when only $g_2$ changes across the wall. The matrix $U(x)$ (\ref{eq:EWmatrix}) is given by:
\begin{equation}
    U(x) = \cos\bigl(g_2(x)/2\bigr)I_2 + i\sin\bigl(g_2(x)/2\bigr)\sigma_2 = \begin{pmatrix}
\cos\bigl(g_2(x)/2\bigr) & \sin\bigl(g_2(x)/2\bigr) \\
-\sin\bigl(g_2(x)/2\bigr) & 
 \cos\bigl(g_2(x)/2\bigr) 
\end{pmatrix}.
\end{equation}
The energy functional $\mathcal{E}$ (\ref{eq:energyfunctionalgen}) simplifies to:
\begin{align}
   \notag  \mathcal{E}(x) &= \dfrac{1}{2}\biggl(\dfrac{dv_1}{dx}\biggr)^2 + \dfrac{1}{2}\biggl(\dfrac{dv_2}{dx}\biggr)^2 + \dfrac{1}{2}\biggl(\dfrac{dv_+}{dx}\biggr)^2 + \dfrac{1}{2}v^2_2(x)\biggl(\dfrac{d\xi}{dx}\biggr)^2 + \frac{1}{8}(v^2_+ + v^2_1 + v^2_2)\biggl(\frac{dg_2}{dx}\biggr)^2  \\  & + \frac{1}{2}\frac{dg_2}{dx}\biggl(-v_+\text{cos}(\xi)\frac{dv_2}{dx} + v_2\text{cos}(\xi)\frac{dv_+}{dx} + v_2v_+\text{sin}(\xi)\frac{d\xi}{dx} \biggr) + V_{2HDM}.  
\label{eq:potentialg2}   
\end{align}
One major difference in comparison with the analysis of the standard domain wall solution in section \ref{standardDWsolution} is that $\dd g_2/\dd x \neq 0$ induces a linear term for $v_+(x)$ in the potential (\ref{eq:potentialg2}):
$$-\frac{1}{2}\frac{dg_2}{dx}v_+(x)\text{cos}(\xi)\frac{dv_2}{dx}, $$
which can give a negative contribution to the energy of the vacuum configuration depending on the sign of $\dd g_2/\dd x$ and $\dd v_2/\dd x$. 
Using the equation of motion for $g_2(x)$, one derives an expression relating the change in $g_2(x)$ to the derivative of $v_+$ inside the wall \cite{Law:2021ing}:
\begin{equation}
    \frac{dg_2}{dx} = \frac{-2v^2_2\cos^2(\xi)}{v^2_1 + v^2_2 + v^2_+}\frac{d}{dx}\biggl( \frac{v_+}{v_2\cos(\xi)} \biggr).
\label{eq:relationg2}    
\end{equation}
This implies that a variation in the Goldstone mode $g_2(x)$ will lead to a non-vanishing $v_+(x)$.
In this case, even if $\xi(x)=0$, it is possible to get a negative contribution to the energy of the wall by having a non vanishing $v_+(0)$, in contrast to the previous case where the dependence was on $\text{sin}(\xi)$. A non-zero derivative for $g_2(x)$ leads to the creation of a stable $v_+(x)$ condensate inside the wall, if the energy $E$ of such a solution is smaller than the energy of the standard domain wall solution.
\begin{figure}[H]
     \centering
     \begin{subfigure}[b]{0.49\textwidth}
         \centering
         \includegraphics[width=\textwidth]{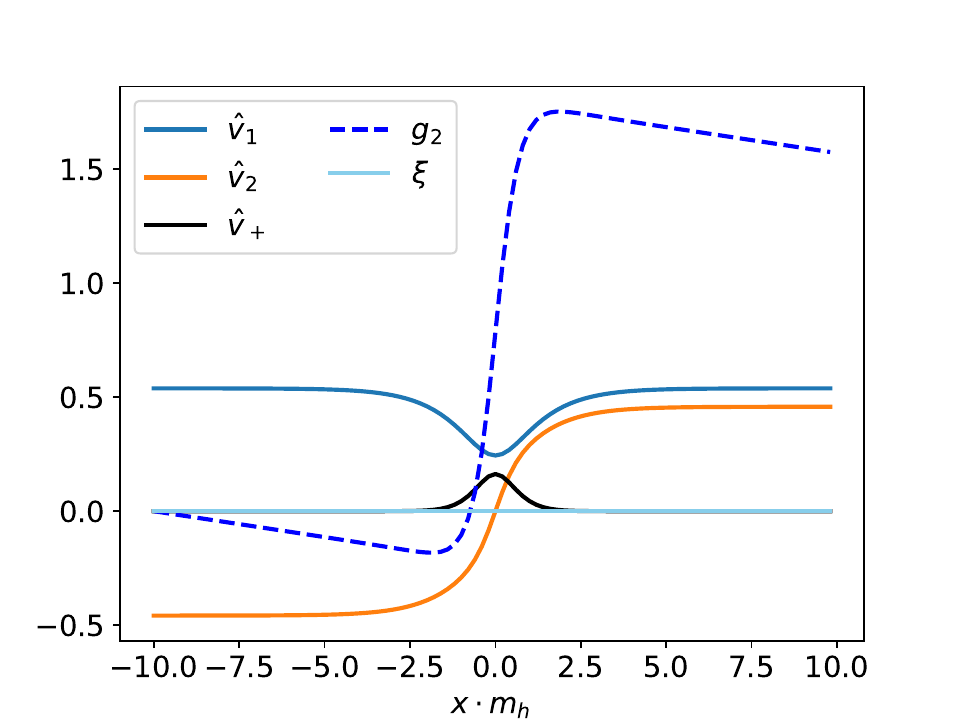}
         \subcaption{Dirichlet boundary conditions.}\label{subfig:g2dirichlet}
     \end{subfigure}
     \hfill
     \begin{subfigure}[b]{0.49\textwidth}
         \centering
         \includegraphics[width=\textwidth]{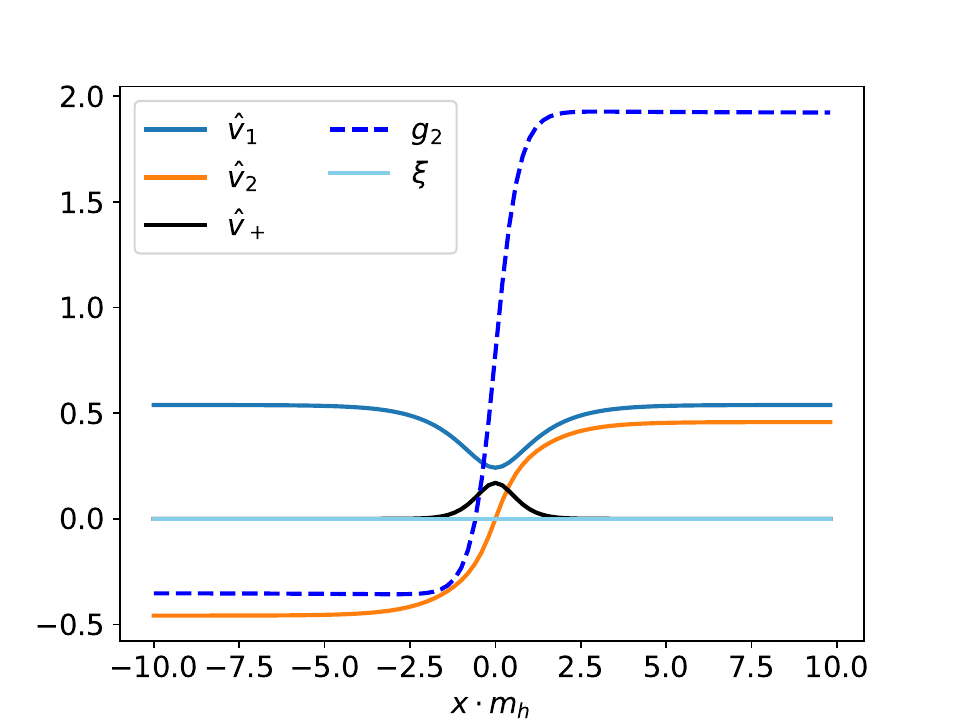}
         \subcaption{von Neumann boundary conditions.}\label{subfig:g2neumann}
     \end{subfigure}
\caption{Numerical solutions of the DW equations of motion in the case of variation of $g_2$. (a) Using Dirichlet boundary conditions and (b) using von Neumann boundary conditions.}
\label{fig:resultg2}
\end{figure}
\noindent
\begin{figure}[h]
     \centering
     \begin{subfigure}[b]{0.49\textwidth}
         \centering
         \includegraphics[width=\textwidth]{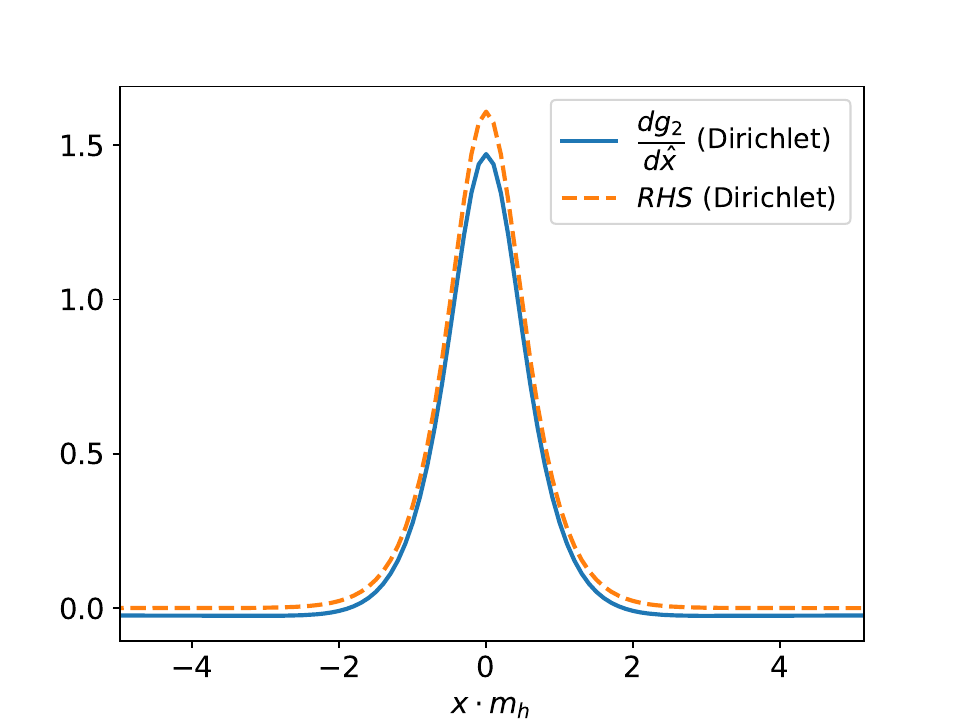}
         \subcaption{Condition \ref{eq:relationg2} for Dirichlet boundaries.}\label{subfig:relationg2dirichlet}
     \end{subfigure}
     \hfill
     \begin{subfigure}[b]{0.49\textwidth}
         \centering
         \includegraphics[width=\textwidth]{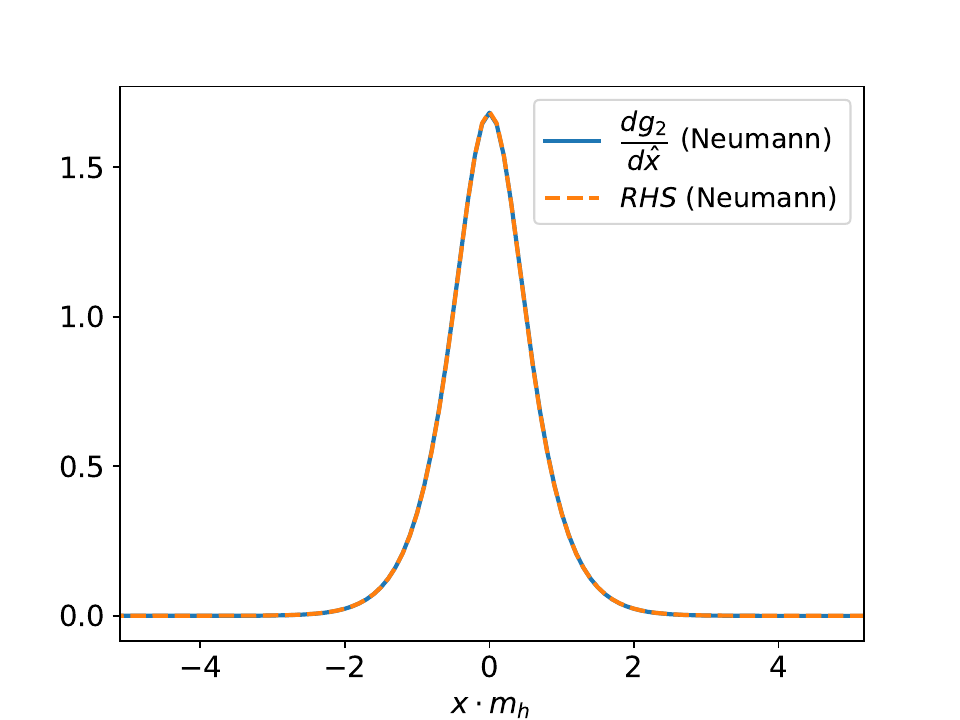}
         \subcaption{Condition \ref{eq:relationg2} for von Neumann boundaries.}\label{subfig:relationg2neumann}
     \end{subfigure}
\caption{Verification of the relation (\ref{eq:relationg2}) for Dirichlet and von Neumann boundary conditions. "RHS" denotes the right hand side of the equation (\ref{eq:relationg2}), namely the quantity $(-2v^2_2\cos^2(\xi))/(v^2_1 + v^2_2 + v^2_+)\frac{d}{dx}\bigl( \frac{v_+}{v_2\cos(\xi)} \bigr)$. Using Dirichlet boundary conditions, the relation is not fulfilled because the used boundaries at $\pm \infty$ give an unstable solution. Using von Neumann boundary condition, we get a perfect agreement.}
\label{fig:relationg2}
\end{figure}
\noindent
Figure \ref{fig:resultg2} shows the numerical solutions to the equations of motion using Dirichlet (Fig.\ref{subfig:g2dirichlet}) and von Neumann boundary conditions (Fig.\ref{subfig:g2neumann}). Like in the previous cases, we take the boundaries of the Goldstone mode to be: $g_2(-\infty) = 0$ and $g_2(+\infty) = \pi/2$. For the initial guess, we use a hyperbolic tangent interpolating between both values. We observe a non-vanishing value for $v_+(x)$ inside the wall. The $U(1)_{em}$ is therefore broken inside the wall leading to exotic phenomena such as charge breaking processes and the photon getting a mass \cite{Viatic:2020yme,Battye:2021dyq}. When using Dirichlet boundary conditions we observe again that the spatial derivative of $g_2(x)$ outside the domain wall is not vanishing. We can see in Figure \ref{subfig:relationg2dirichlet} that the relation (\ref{eq:relationg2}) is not exactly fulfilled using these boundary conditions, reflecting the instability of the solution with Dirichlet boundary conditions. Using von Neumann boundary conditions, we can verify that the non-vanishing $v_+(x)$ inside the wall is stable and that the change in $g_2(x)$ between the two regions gets enhanced, leading to a slightly higher value for $v_+(0)$ inside the wall compared with the solution using Dirichlet boundary conditions. We also verified that the relation (\ref{eq:relationg2}) is satisfied for this choice of boundary condition (see Figure \ref{subfig:relationg2neumann}).
\begin{figure}[h]
     \centering
     \begin{subfigure}[b]{0.49\textwidth}
         \centering
         \includegraphics[width=\textwidth]{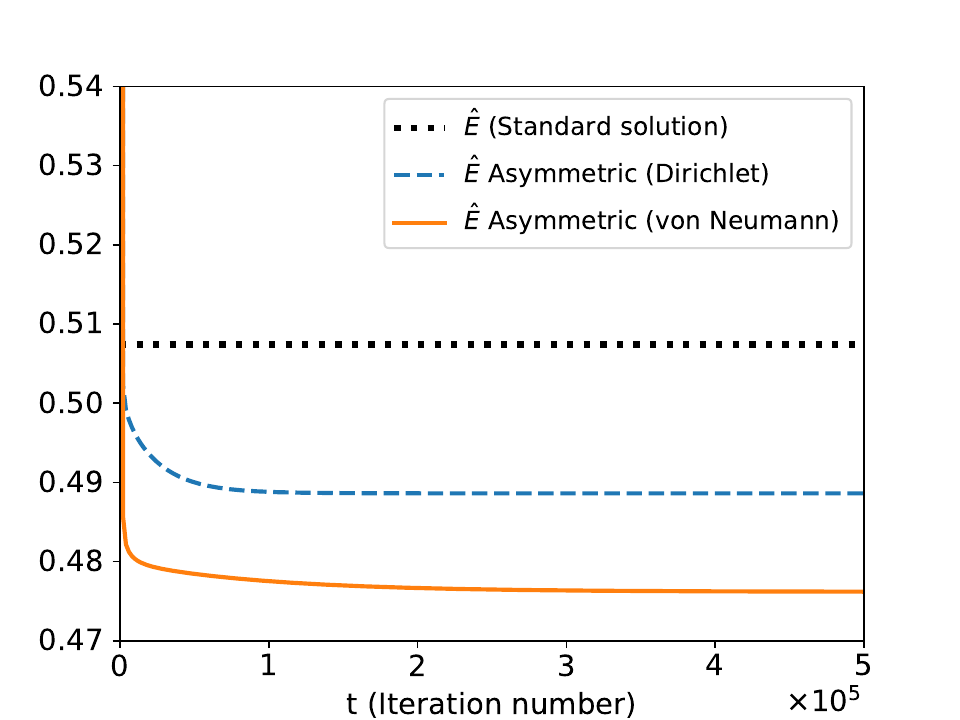}
         \subcaption{Stable charge violating kink (PP I).}\label{subfig:g2ener}
     \end{subfigure}
     \hfill
     \begin{subfigure}[b]{0.49\textwidth}
         \centering
         \includegraphics[width=\textwidth]{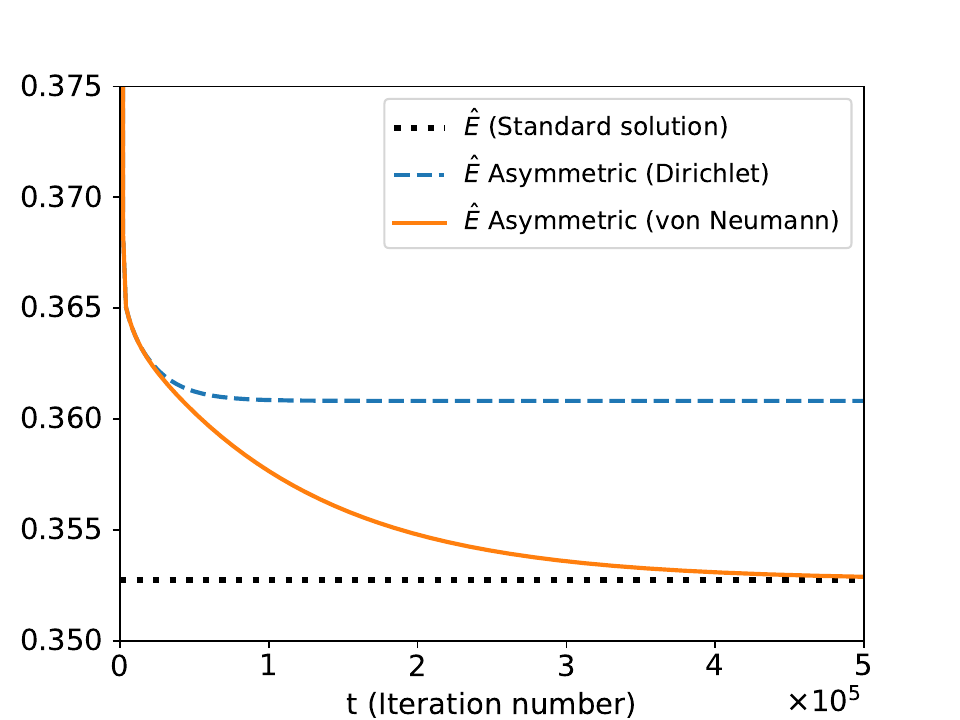}
         \subcaption{Unstable charge violating kink (PP II).}\label{subfig:g2enerunst}
     \end{subfigure}
\caption{Rescaled energies of kink solutions with standard and asymmetric boundary conditions. (a) The charged condensate $v_+(x)$ is stable because the energy of the charge violating kink solution is lower than the energy of the kink with vanishing $v_+(x)$. (b) The charge violating kink solution is unstable because the energy is higher than the solution with $v_+(x) = 0$. In this case the kink solution will decay into the standard solution. Such a solution was found for the parameter point II (cf. \ref{eq:PP2}). }
\label{fig:resultenergyg2}
\end{figure}
\noindent
Even though the choice of von Neumann boundary condition is the correct choice to get stable solutions for our system of differential equations, we can still use Dirichlet boundary condition in order to get an approximate solution for a fixed choice of $g_2$ between the two regions, as this type of boundary condition fixes those values. In the case of von Neumann boundary conditions, if we choose different initial values for $g_2$ on the boundaries, we always end up with a solution for $g_2(x)$ where the difference $\Delta(g_2)_{stable} =g_2(+\infty) - g_2(-\infty)$ is fixed. This means that when starting with random $\Delta(g_2)_{initial}$  between $[0,2\pi]$, $\Delta(g_2)$ always relaxes to a fixed value $\Delta(g_2)_{stable}$ that only depends on the mass parameters and $\text{tan}(\beta)$.
For the used parameter point (\ref{eq:parameterpoint}), the $v_+(x)$ condensate is stable and the vacuum configuration has a lower energy than the standard domain wall solution (see Figure \ref{subfig:g2ener}).
There are also other parameter points where the contributions from the derivative $dg_2/dx$ and a non-vanishing condensate $v_+(x)$ leads to a higher energy than the standard domain wall solution with $v_+(x) = 0$. In such a case, the Goldstone mode $g_2$ dynamically changes its values until it becomes equal in both domains leading to the charge breaking domain wall to decay into the standard domain wall solution as the latter has a lower energy (see Figure \ref{subfig:g2enerunst}). 
\begin{figure}[H]
\centering
\includegraphics[width=0.65\textwidth]{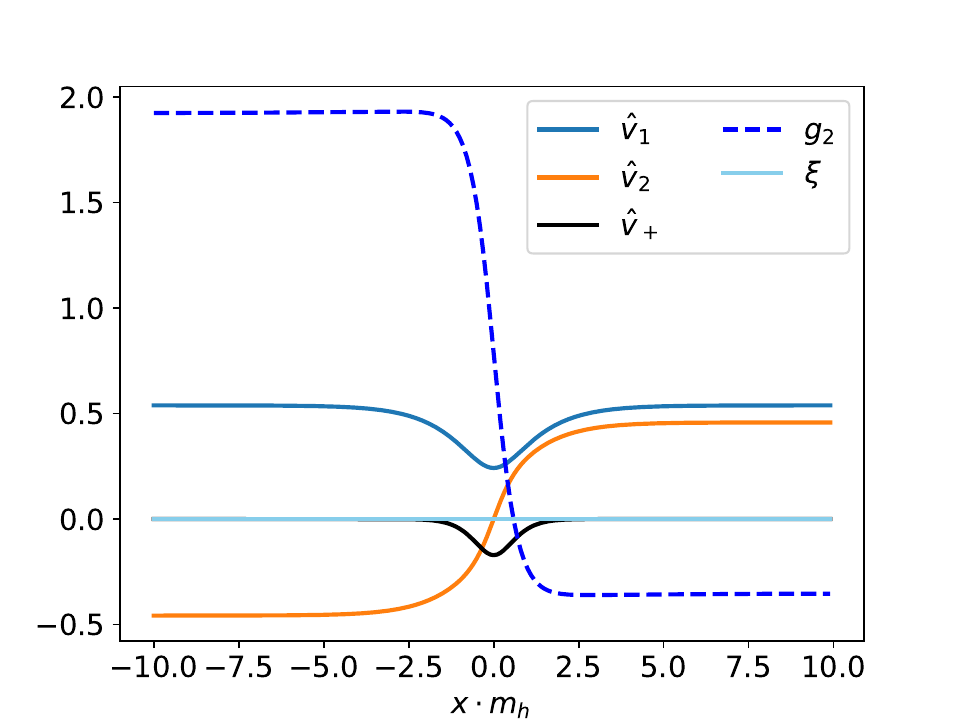}
\caption{Numerical solution of the DW equations of motion in the case of variation of $g_2$. In contrast to the previous case, the derivative $dg_2/dx$ is negative, leading to a negative condensate $v_+(x)$ inside the wall. }
\label{fig:g2flipped}
\end{figure}
\noindent
For the case when $g_2(x)$ decreases when going from the region $v_2<0$ to the region $v_2>0$, for example, when taking $g_2(-\infty) = \pi/2$ and $g_2(+\infty) = 0$  (see Figure \ref{fig:g2flipped}), we obtain a negative value for the condensate $v_+(x)$. 
\subsubsection{Variation of $g_3$}\label{g3}
We now discuss the case when we allow $g_3$ to have different values on the boundaries. The matrix $U(x)$ (\ref{eq:EWmatrix}) is given by:
\begin{equation}
U(x) = \cos\bigl(g_3(x)/2\bigr)I_2 + i\sin\bigl(g_3(x)/2\bigr)\sigma_3 = \begin{pmatrix}
e^{(ig_3(x)/2)} & 0 \\
0 & 
e^{(-ig_3(x)/2)}
\end{pmatrix}.
\end{equation}
The energy functional $\mathcal{E}$ (\ref{eq:energyfunctionalgen}) simplifies to:
\begin{align}
   \notag \mathcal{E} &= \dfrac{1}{2}\biggl(\dfrac{dv_1}{dx}\biggr)^2 + \dfrac{1}{2}\biggl(\dfrac{dv_2}{dx}\biggr)^2 + \dfrac{1}{2}\biggl(\dfrac{dv_+}{dx}\biggr)^2 + \dfrac{1}{2}v^2_2(x)\biggl(\dfrac{d\xi}{dx}\biggr)^2 + \frac{1}{8}(v^2_+ + v^2_1 + v^2_2)\biggl(\frac{dg_3}{dx}\biggr)^2 \\  & -\frac{1}{2}v^2_2\frac{d\xi}{dx}\frac{dg_3}{dx} + V_{2HDM}.
\end{align}
The change in $g_3(x)$ will lead to a change in the phase $\xi(x)$ as was derived in \cite{Law:2021ing} using the equation of motion for $g_3(x)$:
\begin{equation}
    \frac{dg_3}{dx} = \frac{2v^2_2}{v^2_1 + v^2_2 + v^2_+}\frac{d\xi}{dx}.
\end{equation}
\begin{figure}[h]
     \centering
     \begin{subfigure}[b]{0.49\textwidth}
         \centering
         \includegraphics[width=\textwidth]{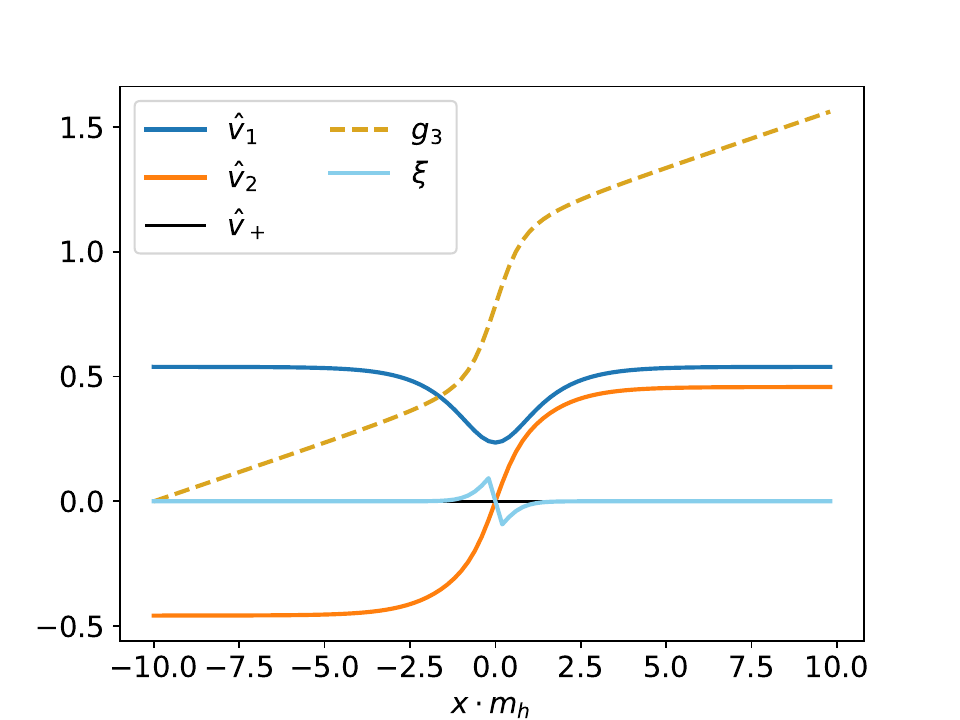}
         \subcaption{Dirichlet boundary conditions.}\label{subfig:g3dirichlet}
     \end{subfigure}
     \hfill
     \begin{subfigure}[b]{0.49\textwidth}
         \centering
         \includegraphics[width=\textwidth]{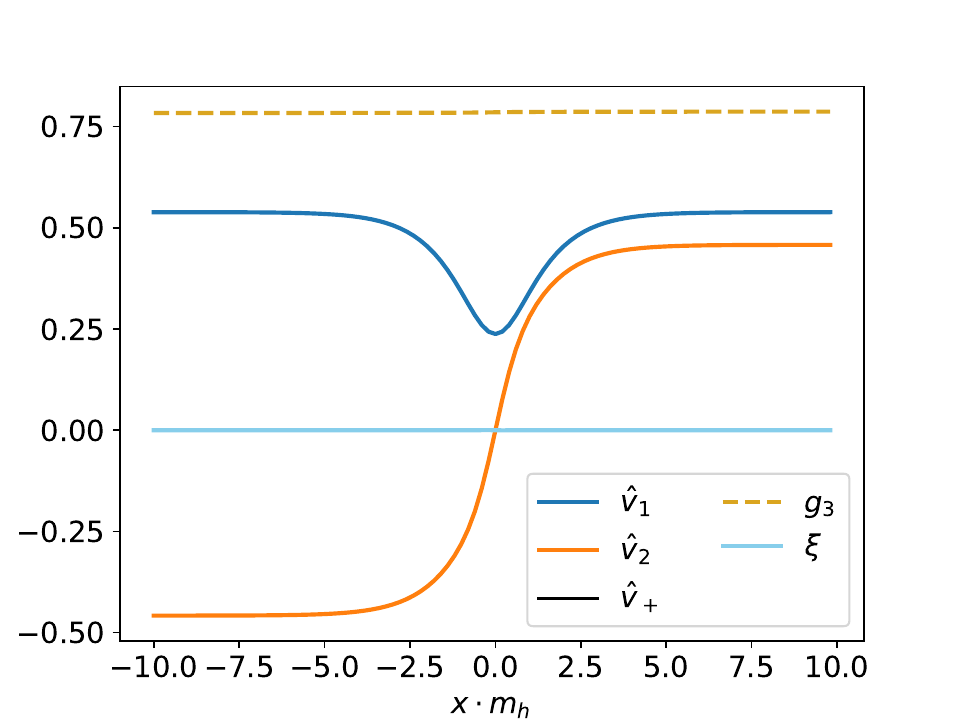}
         \subcaption{von Neumann boundary conditions.}\label{subfig:g3neumann}
     \end{subfigure}
\caption{Solutions of the equations of motion for the case when $g_3$ is different on both domains. (a) Using Dirichlet boundary conditions. (b) Using von Neumann boundary conditions. }
\label{fig:resultg3}
\end{figure}
\begin{figure}[h]
\centering
\includegraphics[width=0.60\textwidth]{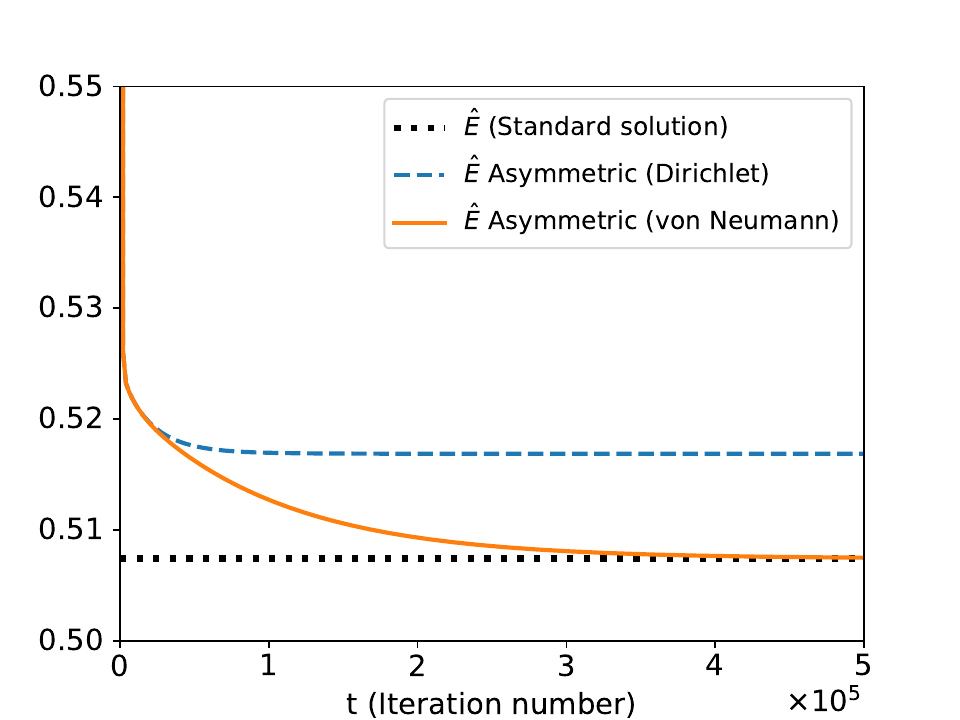}
\caption{Evolution of the rescaled energy of the vacuum configuration using different boundary conditions. Notice that the energy using Dirichlet boundary conditions is higher than the energy using von Neumann boundary conditions. The class of domain walls that are CP-violating due to a variation in $g_3$ decays after some time to the class of standard domain walls.}
\label{fig:energyg3}
\end{figure}
Figure \ref{subfig:g3dirichlet} shows the solution for asymmetric boundary conditions in $g_3$ using Dirichlet boundary conditions. We observe that the phase $\xi(x)$ is non zero inside the wall, which means that the vacuum inside the domain wall is CP-violating. Using Dirichlet boundary conditions, we see again that the derivative of the Goldstone mode $g_3(x)$ is non zero at the boundaries: such a CP-breaking solution has a higher energy than the standard domain wall solution (see Figure \ref{fig:energyg3}) and is therefore unstable for this parameter point. The gradient flow method gives us a solution where the Goldstone field tries to change its value at the boundaries, reflecting the instability of the solution. However, if we use von Neumann boundary conditions, the Goldstone mode $g_3$ dynamically changes its value at the boundaries in such a way that, after some time, both regions end up having the same $g_3$ (see Figure \ref{subfig:g3neumann}). The CP-breaking solution at the wall will then decay and we end up with a standard domain wall. Nevertheless, the Dirichlet boundary condition provides a good approximation for determining the amount of CP-violation inside the wall just after the formation of the defect.
\subsection{Variation of the hypercharge angle $\theta$ and the Goldstone modes}
We now turn to the case where we allow the hypercharge angle $\theta$ of $U(1)_Y$ to vary on both domains alongside the Goldstone modes of $SU(2)_L$. In this case, we expect the formation of domain walls which are both CP and charge violating at the same time. We first discuss the case when the hypercharge angle $\theta$ and one single Goldstone mode $g_i$ are different on both domains. We then discuss the case of pure $SU(2)_L$ and finish with considering the general case when $U(x)$ is a general electroweak symmetry matrix of $SU(2)_L \times U(1)_Y$. We provide the numerical solutions for the equations of motion using both Dirichlet and von Neumann boundary conditions.
\subsubsection{Variation of $\theta$ and $g_1$}
\begin{figure}[H]
     \centering
     \begin{subfigure}[b]{0.49\textwidth}
         \centering
    \includegraphics[width=\textwidth]{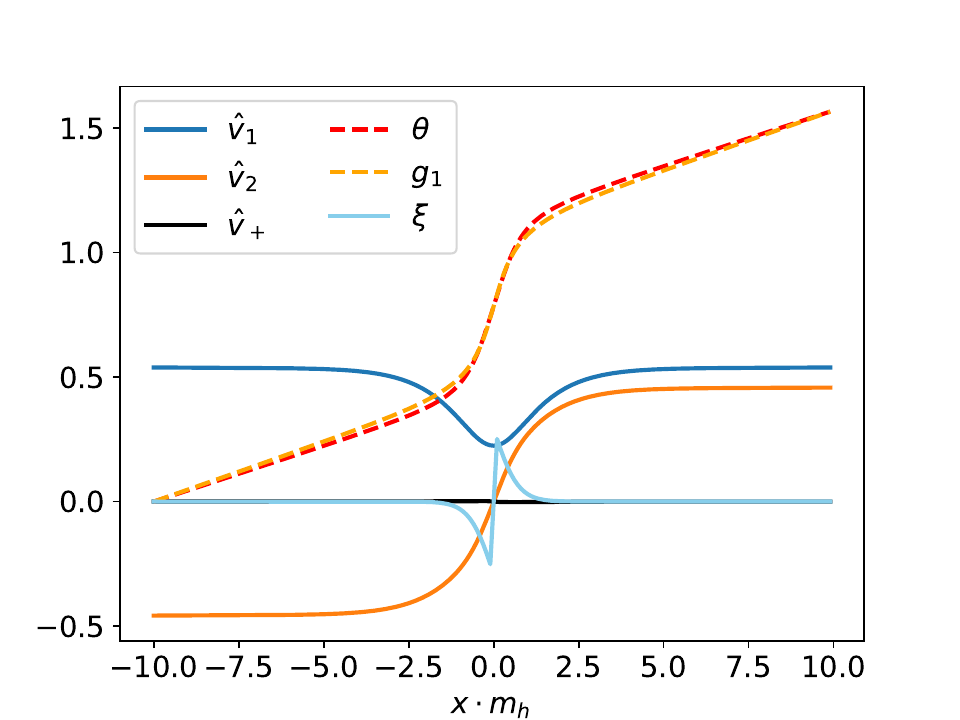}
         \subcaption{Dirichlet boundary conditions.}\label{subfig:tg1dirichlet}
     \end{subfigure}
      \begin{subfigure}[b]{0.49\textwidth}
         \centering     \includegraphics[width=\textwidth]{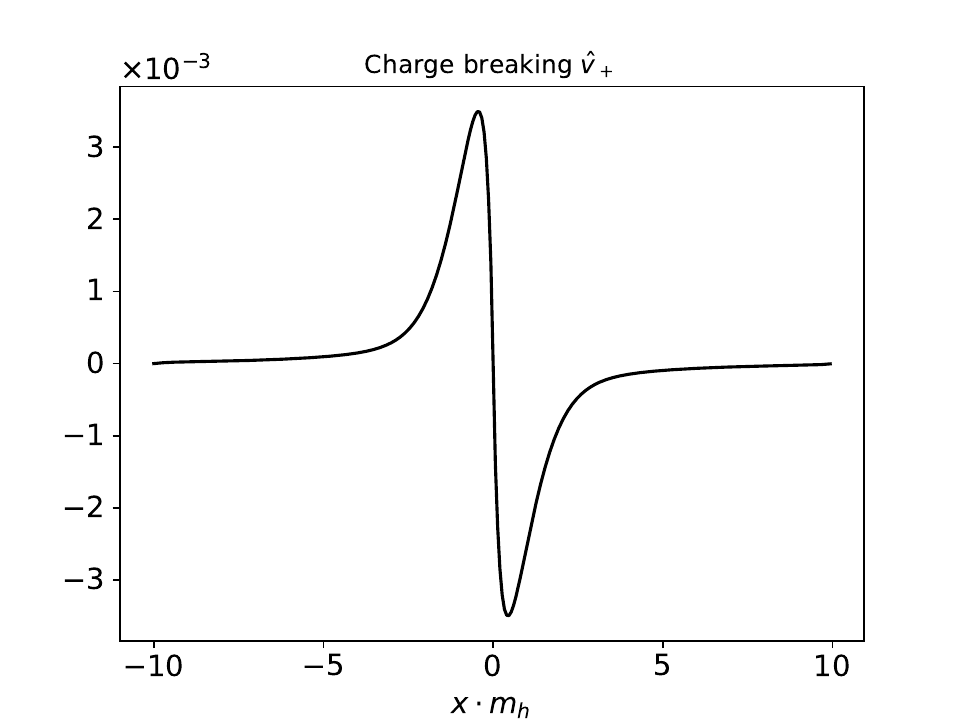}
         \subcaption{$v_+$ using Dirichlet boundary conditions.}
         \label{subfig:chargetg1}
     \end{subfigure}
\caption{(a) Solutions of the equations of motion when varying both the hypercharge angle $\theta$ and Goldstone mode $g_1$ using Dirichlet boundary conditions. (b) Zoom on the solution for $v_+(x)$.}
\label{fig:resulttg1}
\end{figure}

\begin{figure}
     \centering
    \includegraphics[width=0.6\textwidth]{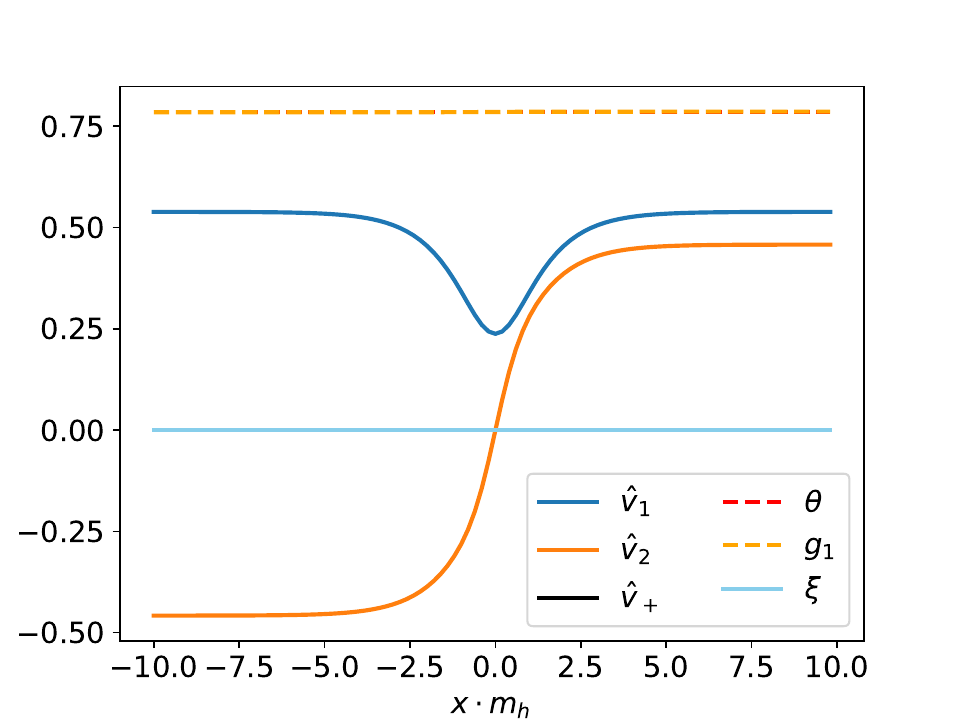}
         \caption{Solutions for the equations of motion using von Neumann Boundary conditions. Notice that the hypercharge angle and the Goldstone mode $g_1$ are the same on both domains.}\label{fig:tg1neumann}
\end{figure}

We start by considering the case when the vacua of the two domains have different hypercharge angle $\theta$ and Goldstone mode $g_1$. The matrix $U(x)$ (\ref{eq:EWmatrix}) simplifies to:
\begin{equation}
    U(x) = e^{i\theta(x)}\begin{pmatrix}
\cos\bigl(g_1(x)/2\bigr) & i\sin\bigl(g_1(x)/2\bigr) \\
i\sin\bigl(g_1(x)/2\bigr) & \cos\bigl(g_1(x)/2\bigr) 
\end{pmatrix}.
\end{equation}
Such a case is relevant to see whether a solution $\xi(x) \neq 0 $ inside the wall will lead to a non-vanishing $v_+(x)$ condensate on the wall. The energy functional (\ref{eq:energyfunctionalgen}) is given by: 
\begin{align}
   \notag \mathcal{E}(x) &= \frac{1}{2}\biggl(\frac{dv_1}{dx}\biggr)^2 + \frac{1}{2}\biggl(\frac{dv_2}{dx}\biggr)^2 + \frac{1}{2}\biggl(\frac{dv_+}{dx}\biggr)^2 + \frac{1}{2}v^2_2(x)\biggl(\frac{d\xi}{dx}\biggr)^2  \\ \notag & + \frac{1}{2}\biggl(\frac{d\theta}{dx}\biggr)^2\biggl[v^2_1(x) + v^2_2(x) + v^2_+(x)\biggr] + \frac{1}{8}\biggl(\frac{dg_1}{dx}\biggr)^2\biggl[v^2_1(x) + v^2_2(x) + v^2_+(x)\biggr]   \\ \notag & + \frac{1}{2}\frac{dg_1}{dx}\biggl[ v_+(x)\sin(\xi)\frac{dv_2}{dx} -  v_2(x)\sin(\xi)\frac{dv_+}{dx}
 + 2v_+v_2\cos(\xi)\frac{d\theta}{dx} +  v_+v_2\cos(\xi)\frac{d\xi}{dx}  \biggr] \\ & + V_{2HDM}.
\end{align}
Figure \ref{subfig:tg1dirichlet} shows the numerical results for the vacuum configuration using Dirichlet boundary conditions. We take the values of the hypercharge angle $\theta$ and Goldstone mode $g_1$ to be: $\theta(-\infty) = 0$, $g_1(-\infty) = 0$ for the vacuum $\Phi_-$ on the left and $\theta(+\infty) = \pi/2$, $g_1(+\infty) = \pi/2$ for the vacuum $\Phi_+$ on the right. In this case we find a very small non-vanishing condensate $v_+(x)$ inside the wall. We also observe that the CP-violating phase $\xi(x)$ is slightly enhanced inside the wall compared to the case where only the hypercharge varies across the wall (cf. Figure \ref{subfig:tdirichlet}). However, this CP and charge breaking solution is unstable as it has a higher energy than the standard domain wall solution.
\noindent
When using von Neumann boundary conditions (see Figure \ref{fig:tg1neumann}), both the hypercharge angle $\theta$ and the Goldstone mode $g_1$ evolve to become equal on both domains and the solution does not exhibit CP or charge violation. 
     
\subsubsection{Variation of $\theta$ and $g_2$}
\begin{figure}[H]
     \centering
     \begin{subfigure}[b]{0.49\textwidth}
         \centering
         \includegraphics[width=\textwidth]{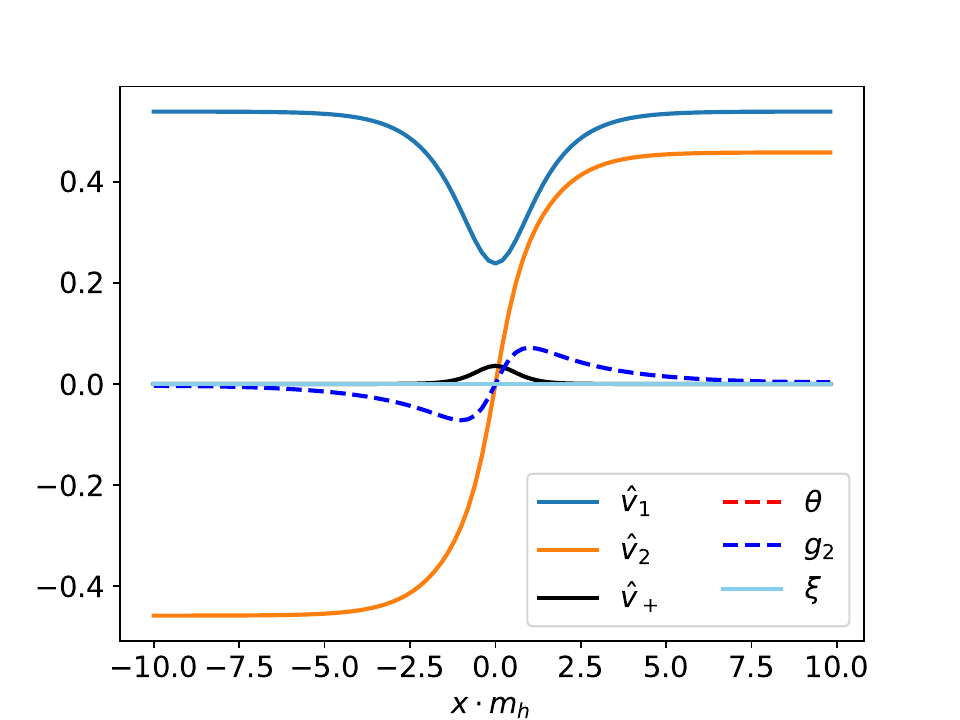}
         \subcaption{} \label{subfig:standardg2decaystart}
     \end{subfigure}
     \begin{subfigure}[b]{0.49\textwidth}
         \centering
         \includegraphics[width=\textwidth]{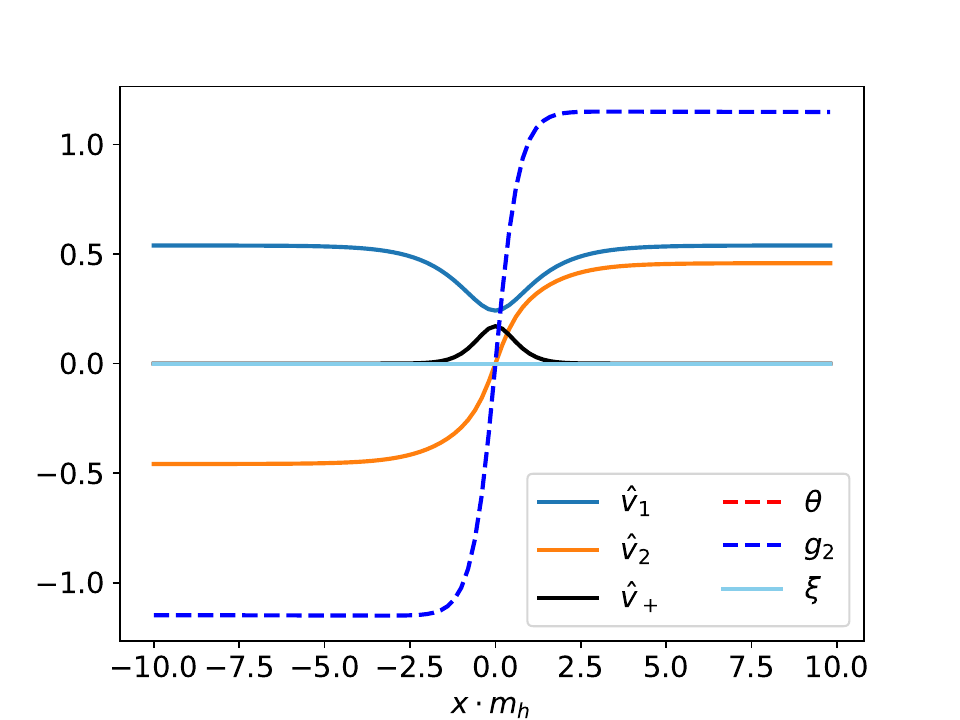}
         \subcaption{}\label{subfig:standardg2decayend}
     \end{subfigure}
\caption{Evolution of the $v_+(x)$ condensate inside the wall. (a) Starting from a standard domain wall solution. (b) After some time the standard domain wall solution decays into a stable charge-breaking solution without CP-violation.}
\label{fig:decaystandardg2t}
\end{figure}

We now vary both $\theta$ and $g_2$. The matrix $U(x)$ (\ref{eq:EWmatrix}) simplifies to:
\begin{equation}
    U(x)  = e^{i\theta(x)} \begin{pmatrix}
\cos\bigl(g_2(x)/2\bigr) & \sin\bigl(g_2(x)/2\bigr) \\
-\sin\bigl(g_2(x)/2\bigr) & 
 \cos\bigl(g_2(x)/2\bigr) 
\end{pmatrix}.
\end{equation}
In this case one would expect that we get both effects of charge and CP-violation inside the wall. The energy functional (\ref{eq:energyfunctionalgen}) is:
\begin{align}
   \notag \mathcal{E} &= \frac{1}{2}\biggl(\frac{dv_1}{dx}\biggr)^2 + \frac{1}{2}\biggl(\frac{dv_2}{dx}\biggr)^2 + \frac{1}{2}\biggl(\frac{dv_+}{dx}\biggr)^2 + \frac{1}{2}v^2_2(x)\biggl(\frac{d\xi}{dx}\biggr)^2 + \frac{1}{2}\biggl(\frac{d\theta}{dx}\biggr)^2\biggl[v^2_1(x) + v^2_2(x) + v^2_+(x)\biggr] \\ \notag & + \frac{1}{8}\biggl(\frac{dg_2}{dx}\biggr)^2\biggl[v^2_1(x) + v^2_2(x) + v^2_+(x)\biggr] + \frac{1}{2}\frac{dg_2}{dx}\biggl[ -v_+(x)\cos(\xi)\frac{dv_2}{dx} +  v_2(x)\cos(\xi)\frac{dv_+}{dx} \\ & 
 - 2v_+v_2\sin(\xi)\frac{d\theta}{dx} +  v_+v_2\sin(\xi)\frac{d\xi}{dx}  \biggr] + V_{2HDM}.
\label{eq:energyfunctionalthetag2} 
\end{align}
\begin{figure}[h]
     \centering
     \begin{subfigure}[b]{0.49\textwidth}
         \centering
         \includegraphics[width=\textwidth]{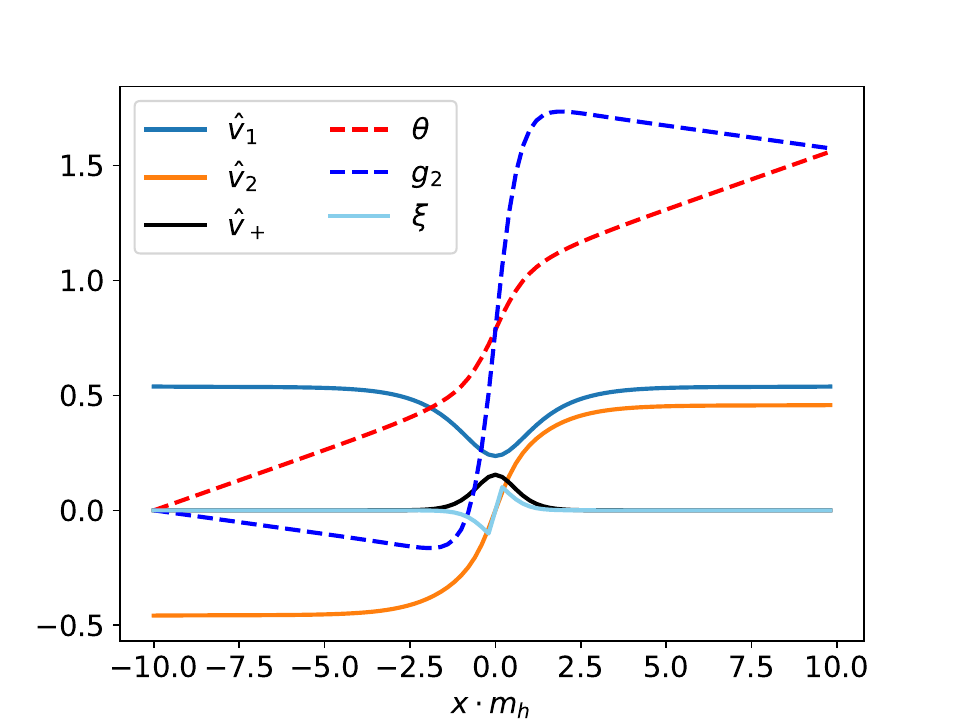}
         \subcaption{Dirichlet boundary conditions.}
         \label{subfig:tg2dirichlet}
     \end{subfigure}
     \hfill
     \begin{subfigure}[b]{0.49\textwidth}
         \centering
         \includegraphics[width=\textwidth]{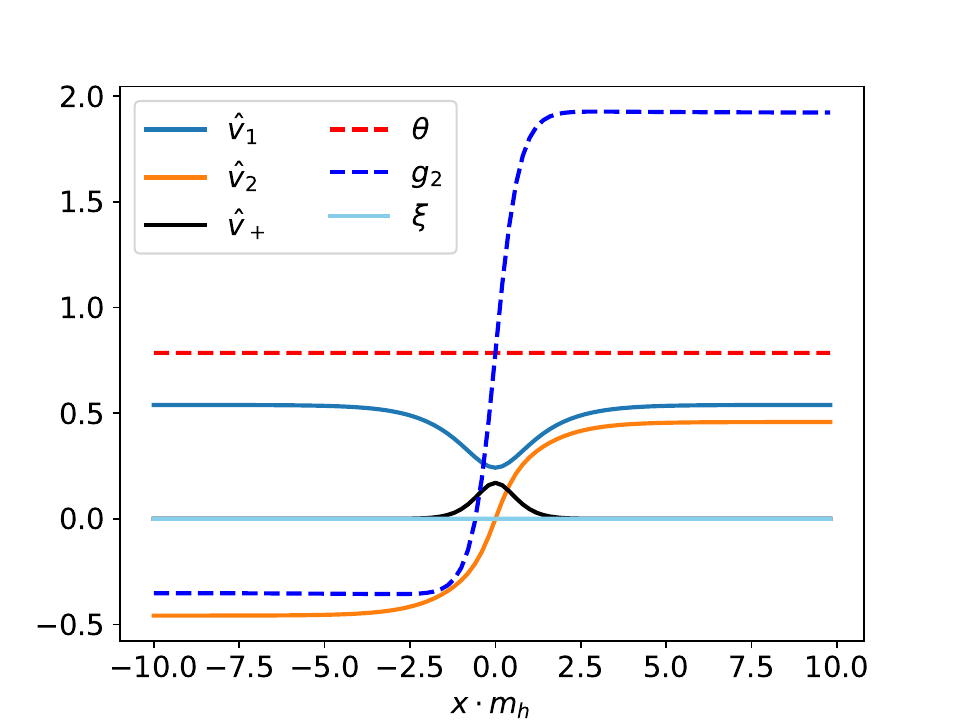}
         \subcaption{von Neumann boundary conditions.}
         \label{subfig:tg2neumann}
     \end{subfigure}
     \begin{subfigure}[b]{0.49\textwidth}
         \centering
         \includegraphics[width=\textwidth]{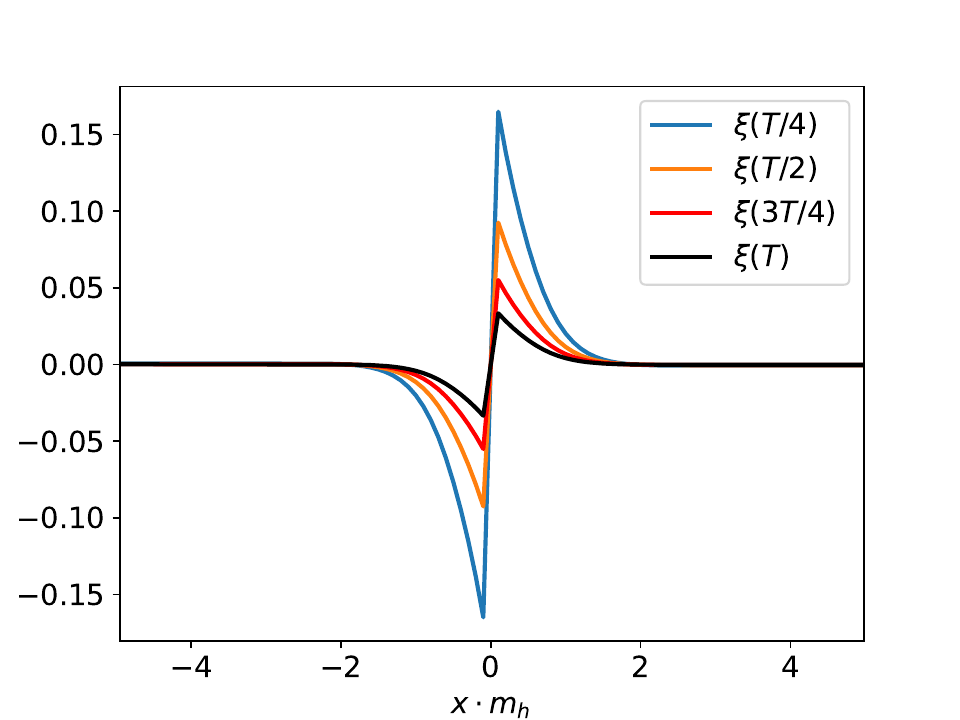}
         \subcaption{Evolution of $\xi(x)$.}
         \label{subfig:evolutionxidiff}
     \end{subfigure}
     \hfill
     \begin{subfigure}[b]{0.49\textwidth}
         \centering
         \includegraphics[width=\textwidth]{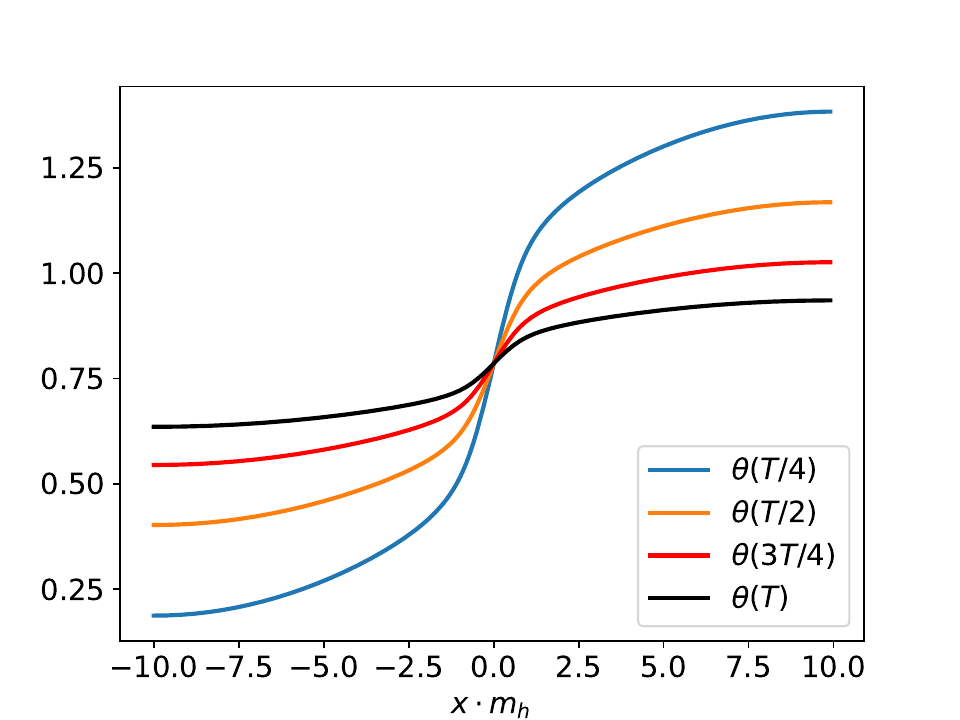}
         \subcaption{Evolution of $\theta(x)$.}
         \label{subfig:evolutionthetadiff}
     \end{subfigure}
\caption{Solutions of the equations of motion in the case of different values of the hypercharge angle $\theta$ and Goldstone mode $g_2$ in the two regions. (a) Using Dirichlet boundary conditions. (b) Using von Neumann boundary conditions. Note that when using the von Neumann boundary condition, the CP-violating effect vanishes after some time as the hypercharge angle $\theta(x)$ becomes constant for all $x$, while the condensate $v_+(x) \neq 0$ stays stable. (c) Evolution of $\xi(x)$ at different iteration times with $T = 3\times 10^5$, notice that the CP-violating phase $\xi(x)$ decreases with time. (d) Evolution of $\theta(x)$ at different iteration times with $T = 3\times 10^5$, notice that $\theta(x)$ tend to dynamically change to become the same value on both domains. }
\label{fig:resulttg2}
\end{figure}
\noindent
Figure \ref{fig:resulttg2} shows the solution for both Dirichlet (in Figure \ref{subfig:tg2dirichlet}) and von Neumann (Figure \ref{subfig:tg2neumann}) boundary conditions. The initial boundary conditions for $\theta$ and $g_2$ ($\theta(-\infty) = g_2(-\infty) = 0$, $\theta(+\infty) = g_2(+\infty) = \pi/2$) lead to a charge and CP-breaking vacuum inside the wall when using Dirichlet boundary conditions. However, such a solution is energetically unstable since a non-zero $\xi(x)$ gives a positive contribution to the energy of the domain wall. We also observe a non-zero derivative for $\theta(x)$ and $g_2(x)$ at the boundaries. This behavior reflects the instability of this solution, since the values of $\theta$ and $g_2$ on both domains try to change in order to further minimize the energy of the vacuum configuration. Using von Neumann boundary conditions (\ref{subfig:tg2neumann}), where the boundaries can change dynamically to minimize the energy of the vacuum configuration, the derivatives of $\theta(x)$ and $g_2(x)$ at the boundaries vanish. One notices that for the solution which minimizes the energy the most, the charge breaking vacuum inside the wall gets enhanced while the CP-breaking phase $\xi(x)$ inside the wall will start decreasing and eventually vanishes once the values for $\theta$ on both domains become equal to each other (as is shown in Figures \ref{subfig:evolutionthetadiff} and \ref{subfig:evolutionxidiff}). This vacuum configuration is stable, since its dimensionless energy $\hat{E} = 0.476$ is lower than the standard domain wall's energy $\hat{E}_{standard} = 0.507$. This means that the standard domain wall solution will decay into the charge-breaking domain wall solution as can be seen in Figure \ref{fig:decaystandardg2t}. The Goldstone mode $g_2$ changes dynamically from a vacuum configuration where it is equal on both domains to a lower energy configuration with different values for $g_2$ on both domains. The hypercharge angle $\theta$, however, stays zero on both domains and a CP-violating phase $\xi(x)$ does not develop inside the wall, as such a solution would otherwise give a positive contribution to the energy of the defect.
\begin{figure}[H]
\centering
   \begin{subfigure}[b]{0.49\textwidth}
         \centering
         \includegraphics[width=\textwidth]{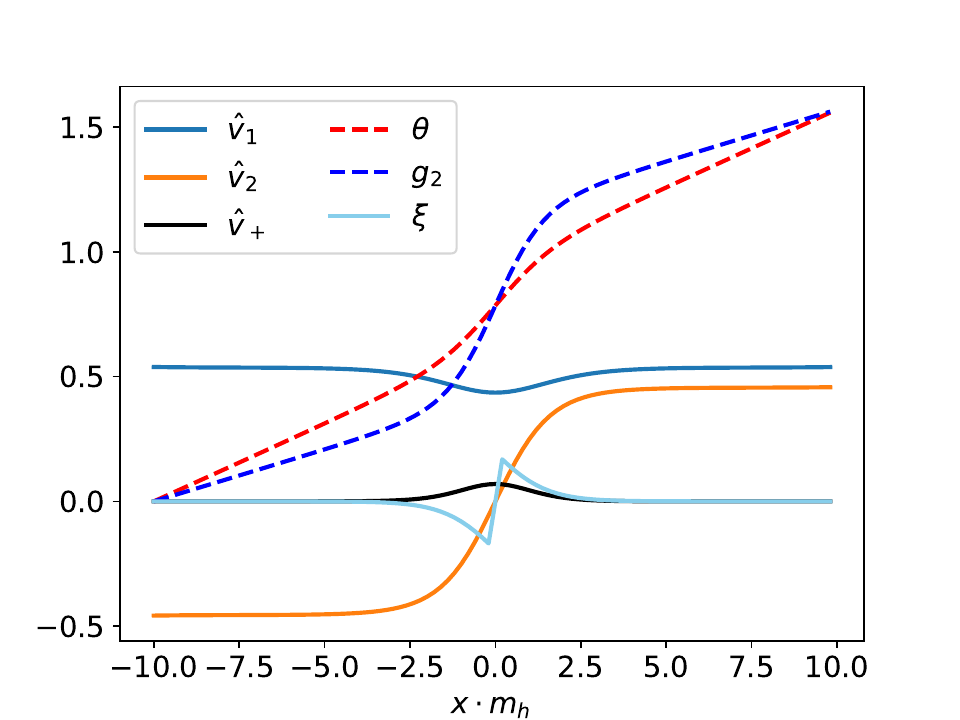}
          \subcaption{Dirichlet boundary conditions.} \label{subfig:Dirichlet boundary conditions g2theta}
     \end{subfigure}
     \begin{subfigure}[b]{0.49\textwidth}
         \centering
         \includegraphics[width=\textwidth]{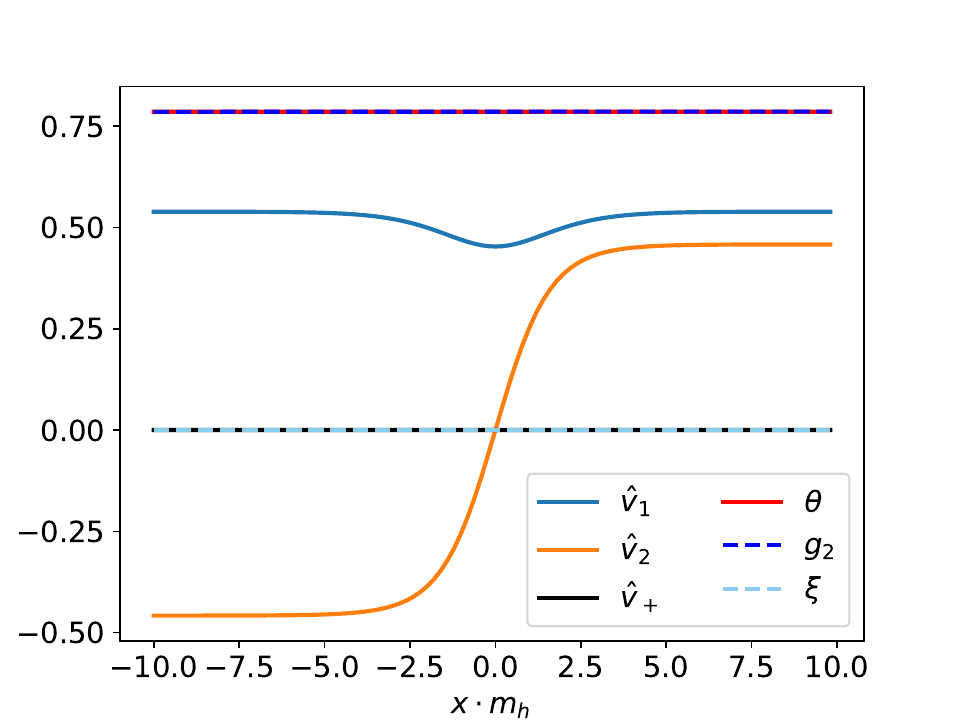}
          \subcaption{von Neumann boundary conditions.} \label{subfig:Neumann boundary conditions g2theta}
     \end{subfigure}
\caption{(a) Solutions of the equations of motion for PP II (\ref{eq:PP2}) using  Dirichlet boundary conditions, this field configuration is unstable. (b) Solutions using von Neumann boundary conditions. The CP and charge-breaking vacua inside the wall vanish and we end with a standard domain wall solution.}
\label{fig:PP2g2theta}
\end{figure}
\noindent
For other parameter points, such as PP II (\ref{eq:PP2}), we observe the opposite behavior: the energy of the vacuum configuration for the CP and charge-breaking wall is higher than the energy of the standard wall. Such a scenario is shown in Figure \ref{fig:PP2g2theta} using Dirichlet (Figure \ref{subfig:Dirichlet boundary conditions g2theta}) and von Neumann (Figure \ref{subfig:Neumann boundary conditions g2theta}) boundary conditions.  The Dirichlet solution exhibits a CP and charge-breaking vacuum which is energetically unstable. Using von Neumann boundary conditions, the values of $v_+(0)$ and $\xi(x)$ decrease until they vanish (both regions end up with the same value for $g_2$ and $\theta$) and we recover the standard domain wall solution. 
\subsubsection{Variation of $\theta$ and $g_3$}
We now discuss the variation of $\theta$ alongside $g_3$. In this case one expects only a CP-violating solution and no charge violation inside the wall. The matrix $U(x)$ (\ref{eq:EWmatrix}) simplifies to:
\begin{equation}
    U(x)  = e^{i\theta(x)} \begin{pmatrix}
e^{ig_3(x)/2} & 0 \\
0 & e^{-ig_3(x)/2}
\end{pmatrix}.
\end{equation}
The energy functional (\ref{eq:energyfunctionalgen}) is given by:
\begin{align}
     \notag \mathcal{E} &= \frac{1}{2}\biggl(\frac{dv_1}{dx}\biggr)^2 + \frac{1}{2}\biggl(\frac{dv_2}{dx}\biggr)^2 + \frac{1}{2}\biggl(\frac{dv_+}{dx}\biggr)^2 + \frac{1}{2}v^2_2(x)\biggl(\frac{d\xi}{dx}\biggr)^2 + \frac{1}{2}\biggl[v^2_1(x) + v^2_2(x) + v^2_+(x)\biggr]\biggl(\frac{d\theta}{dx}\biggr)^2 \\ \notag & + \frac{1}{8}\biggl(\frac{dg_3}{dx}\biggr)^2\biggl[v^2_1(x) + v^2_2(x) + v^2_+(x)\biggr] - \frac{1}{2}\frac{dg_3}{dx}\biggl[ \frac{d\theta}{dx}\biggl(v^2_1(x) + v^2_2(x) - v^2_+(x)\biggr) + v^2_2(x)\frac{d\xi}{dx} \biggr] \\  & + v^2_2(x)\frac{d\theta}{dx}\frac{d\xi}{dx} + V_{2HDM}.
\end{align}
Using the equations of motion for $\theta(x)$ and $g_3(x)$, we can derive an equation that describes how the change in $\theta(x)$ and $g_3(x)$ causes a change in $\xi(x)$:
\begin{equation}
    \frac{d\theta}{dx} - \frac{1}{2}\frac{dg_3}{dx} = - \frac{v^2_2}{v^2_1 + v^2_2}\frac{d\xi}{dx}.
\label{eq:thetag3}    
\end{equation}
From such a relation, one expects the possibility of having an interference in the contributions of $\theta$ and $g_3$.
\begin{figure}[h]
     \centering
     \begin{subfigure}[b]{0.49\textwidth}
         \centering
         \includegraphics[width=\textwidth]{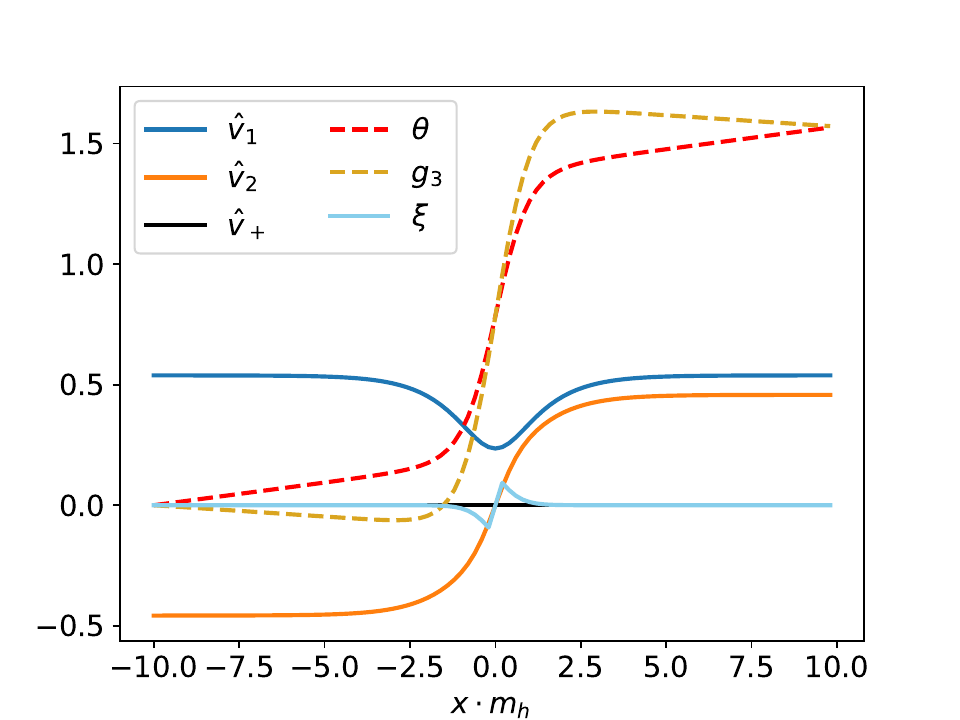}
         \subcaption{Dirichlet boundary conditions.}
         \label{subfig:thetag3dirichlet}
     \end{subfigure}
     \hfill
     \begin{subfigure}[b]{0.49\textwidth}
         \centering
         \includegraphics[width=\textwidth]{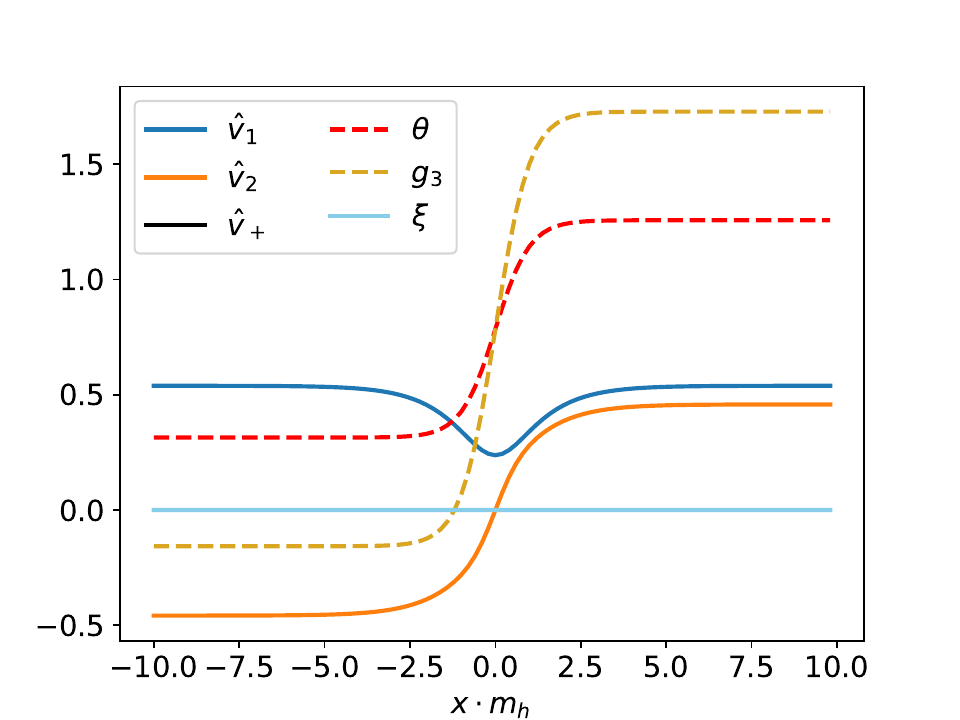}
         \subcaption{von Neumann boundary conditions.}
         \label{subfig:thetag3neumann}
     \end{subfigure}
\caption{Solutions of the equations of motion in the case when $\theta$ and $g_3$ are different on both domains. (a) Using Dirichlet boundary conditions. (b) Using von Neumann boundary conditions. The CP-violating phase $\xi(x)$ is zero even though the Goldstone mode $g_3$ and hypercharge angle $\theta$ are different on both domains. This behavior is explained by the relation (\ref{eq:thetag3}).}
\label{fig:resulttg3}
\end{figure}
\noindent
Figure \ref{fig:resulttg3} shows the numerical solution to the equations of motion. The initial condition were taken to be $\theta(-\infty) = g_3(-\infty) = 0$, $\theta(+\infty) = g_3(+\infty) = \pi/2$.
In this case, we again see that the Dirichlet boundary condition leads to a localized CP-violating phase $\xi(x)$ inside the wall and that the derivatives of the Goldstone modes do not vanish at the boundaries. This vacuum field configuration is energetically unstable and decays to the standard vacuum configuration with $\xi(x) = 0$ for all $x$. In contrast to the results from the variation of $\theta$ and $g_3$ individually, the minimum vacuum configuration gets two different values for $\theta$ and $g_3$ in the two domains (see Figure \ref{subfig:thetag3neumann}).
\begin{figure}[H]
\centering
\includegraphics[width=0.60\textwidth]{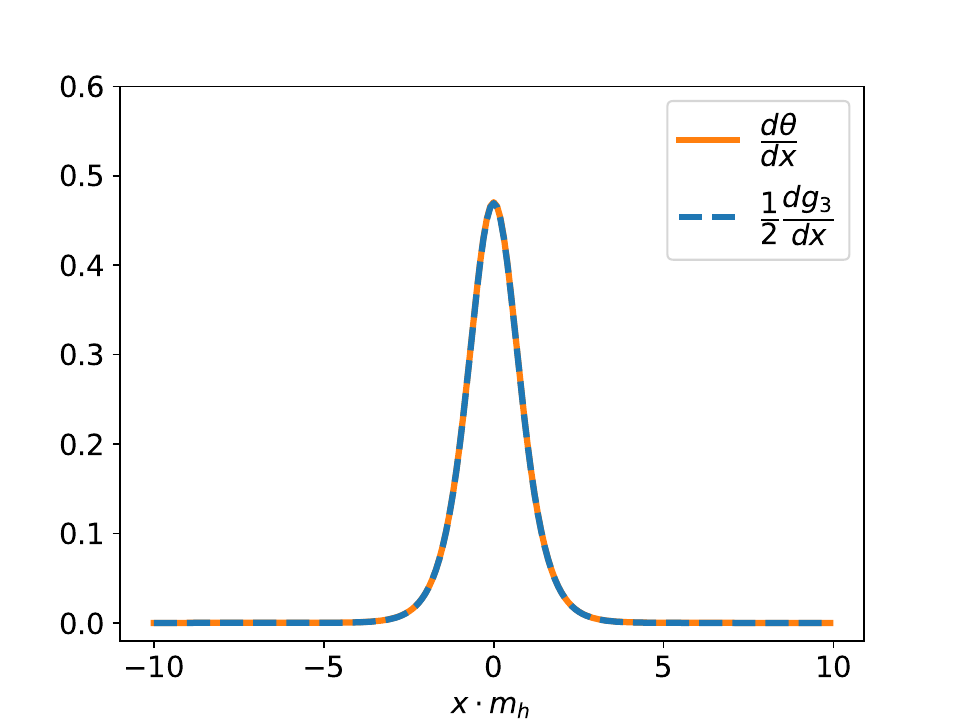}
\caption{Values of the derivative of $\theta(x)$ and $g_3(x)$ using von Neumann boundary condition. The numerical solution satisfies (\ref{eq:thetag3}) leading to the vanishing of the phase $\xi(x)$ inside the wall even though the Goldstone modes are different on both sides of the wall.}
\label{fig:derivativeg3theta}
\end{figure}
\noindent
Using (\ref{eq:thetag3}), one sees that if the derivative of $\theta(x)$ is equal to half the derivative of $g_3(x)$, one obtains $d\xi/dx = 0$. This clarifies why the values of $\xi(x)$ are vanishingly small despite the different Goldstone modes on both domains (see Figure \ref{subfig:thetag3neumann}). This condition can be verified in Figure \ref{fig:derivativeg3theta}, where we see that both expressions agree numerically.
\subsubsection{Variation of $SU(2)_L$ modes $g_1$, $g_2$ and $g_3$}\label{su2}
\begin{figure}[H]
         \centering
         \includegraphics[width=0.6\textwidth]{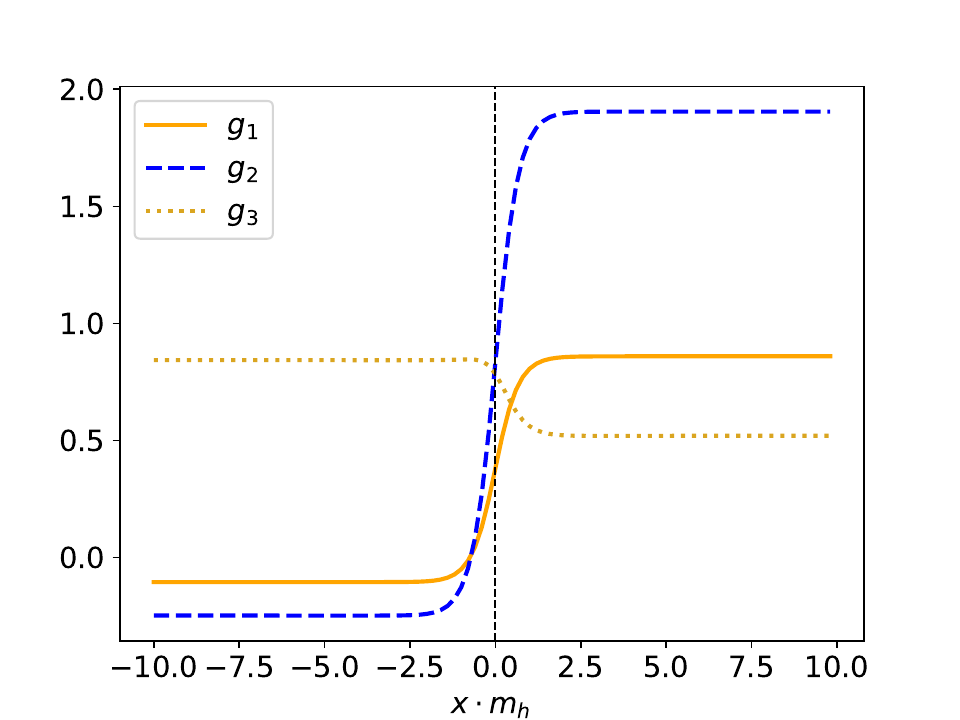}
         \caption{Goldstone modes $g_1(x)$, $g_2(x)$ and $g_3(x)$. Notice that the profiles of the Goldstone modes are asymmetric inside the wall. This suggests an interference between the different Goldstone modes.}\label{fig:goldstonesu2}
\end{figure}

In this special case the vacua in both regions are related by an $\text{SU(2)}_L$ gauge transformation and we ignore the effects coming from the change in the hypercharge angle $\theta$. The matrix $U(x)$ (\ref{eq:EWmatrix}) simplifies to:
\begin{equation}
 U(x) = \begin{pmatrix}
\cos\bigl(G(x)\bigr) + ig_3(x)\sin\bigl(G(x)\bigr)/G(x) & \bigl(g_2(x) +ig_1(x)\bigr)\sin\bigl(G(x)\bigr)/G(x) \\
-\bigl(g_2(x) -ig_1(x)\bigr)\sin\bigl(G(x)\bigr)/G(x) & \cos\bigl(G(x)\bigr) - ig_3(x)\sin\bigl(G(x)\bigr)/G(x)
\end{pmatrix}.
\end{equation}
The energy functional (\ref{eq:energyfunctionalgen}) is  given by:
\begin{align}
\notag  \mathcal{E}(x) &= \dfrac{1}{2}\biggl(\dfrac{dv_1}{dx}\biggr)^2 + \dfrac{1}{2}\biggl(\dfrac{dv_2}{dx}\biggr)^2 + \dfrac{1}{2}\biggl(\dfrac{dv_+}{dx}\biggr)^2 + \dfrac{1}{2}v^2_2(x)\biggl(\dfrac{d\xi}{dx}\biggl)^2  +
  \dfrac{1}{2}v^2_1(x) I_0(x) + \dfrac{1}{2}v^2_+(x)I_0(x)  \\ \notag & + \dfrac{1}{2}v^2_2(x)\biggl[I_0(x)  + 2\dfrac{d\xi}{dx}I_3(x) \biggr]  + v_2(x) \biggl[ sin(\xi)\dfrac{dv_+}{dx}I_1(x) - cos(\xi)\dfrac{dv_+}{dx}I_2(x) \biggr]  \\ \notag  &  + v_+(x) \biggl[ -\sin(\xi)\dfrac{dv_2}{dx}I_1(x) + \cos(\xi)\dfrac{dv_2}{dx}I_2(x) \biggr] - v_+(x)v_2(x)\biggl[ \cos(\xi)\dfrac{d\xi}{dx}I_1(x) \\  &  + \sin(\xi)\dfrac{d\xi}{dx}I_2(x) \biggr] +  V_{2HDM}.
\end{align} 
\begin{figure}[H]
     \centering
     \begin{subfigure}[b]{0.49\textwidth}
         \centering
         \includegraphics[width=\textwidth]{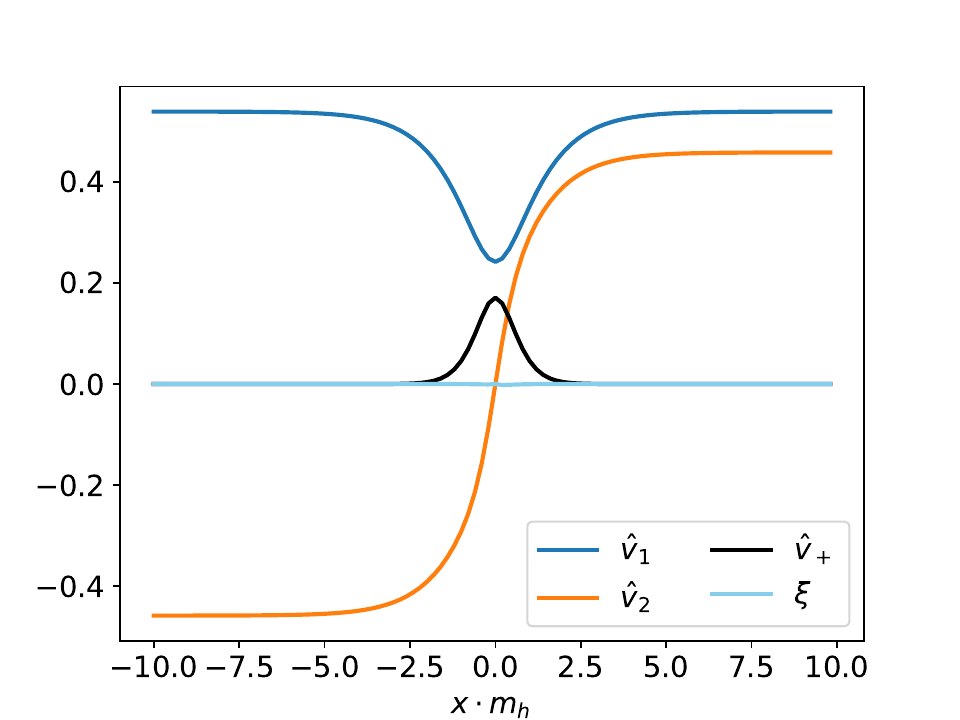}
         \subcaption{Physical vacuum parameters.}\label{subfig:su2higgs}
     \end{subfigure}
     \hfill
    \begin{subfigure}[b]
     {0.49\textwidth}
         \centering
         \includegraphics[width=\textwidth]{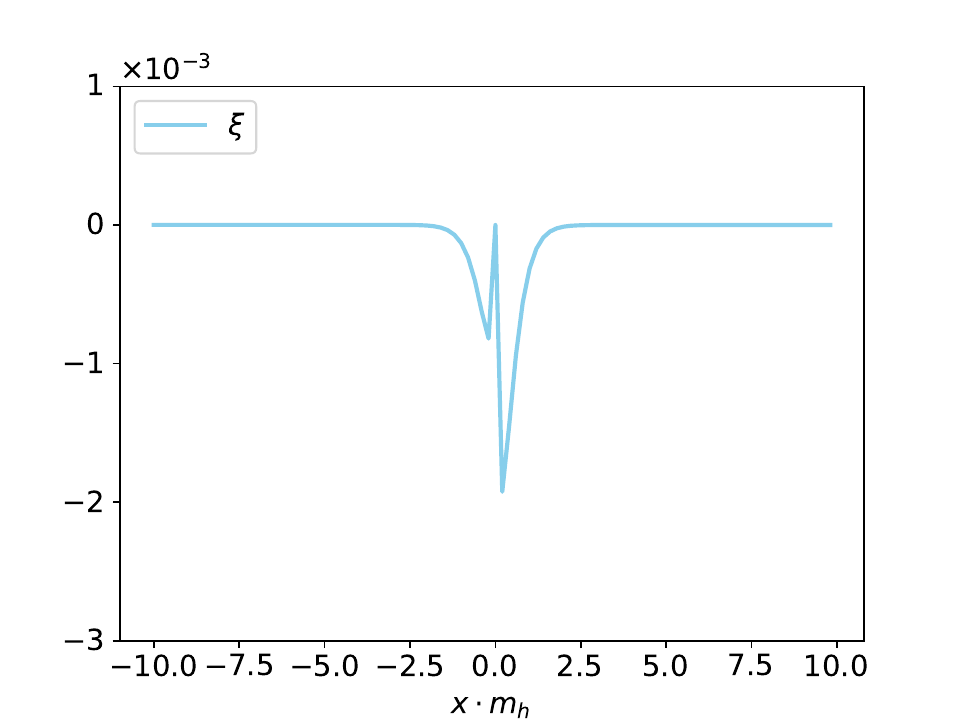}
         \subcaption{CP-Violating phase $\xi(x)$.}\label{subfig:xisu2}
     \end{subfigure}
\caption{Vacuum field configuration for domains with different $\text{SU}(2)_L$ modes $g_1$, $g_2$ and $g_3$ using von Neumann boundary conditions. (a) The solution for the vacuum parameters $v_i(x)$ and $\xi(x)$, we find that the DW is charge breaking ($v_+(0) \neq 0$). (b) Zoom on the CP-violating phase $\xi(x)$, we see that for this particular case $\xi(x)$ is asymmetric.}
\label{fig:su2}
\end{figure} 
\noindent
The numerical results using von Neumann boundary conditions are shown in Figure \ref{fig:su2}. The initial profile of the Goldstone modes were chosen to be $g_{i}(-\infty) = 0$ and $g_i(+\infty) = \pi/2$, (where $i$ denotes {1,2,3}) with a tangent hyperbolic function interpolating both boundaries. The solution has a stable charge-violating vacuum $v_+(x)$ inside the domain wall, as well as a small and stable CP-violating phase $\xi(x)$. In contrast to the previous results, $\xi(x)$ in this particular case is asymmetric. This behavior could be attributed to the fact that the profiles of the Goldstone modes $g_i(x)$ are not symmetrical (see Figure \ref{fig:goldstonesu2}). There is also an interference between the Goldstone modes as some have negative derivatives while the others have a positive derivative inside the wall. Note that, in this case, the Goldstone modes keep having a different value on both domains.

\subsection{General Domain Wall Solution}\label{general}
\begin{figure}[H]
     \centering
     \begin{subfigure}[b]{0.49\textwidth}
         \centering
         \includegraphics[width=\textwidth]{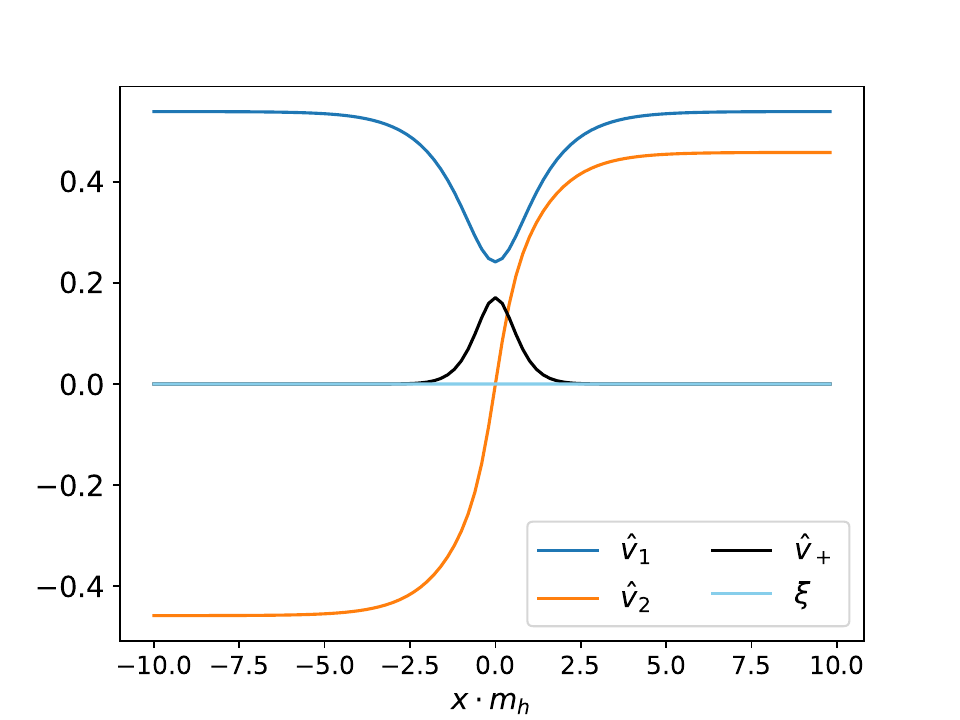}
         \subcaption{Higgs vacuum parameters.}\label{subfig:vacpar}
     \end{subfigure}
     \begin{subfigure}[b]{0.49\textwidth}
         \centering
         \includegraphics[width=\textwidth]{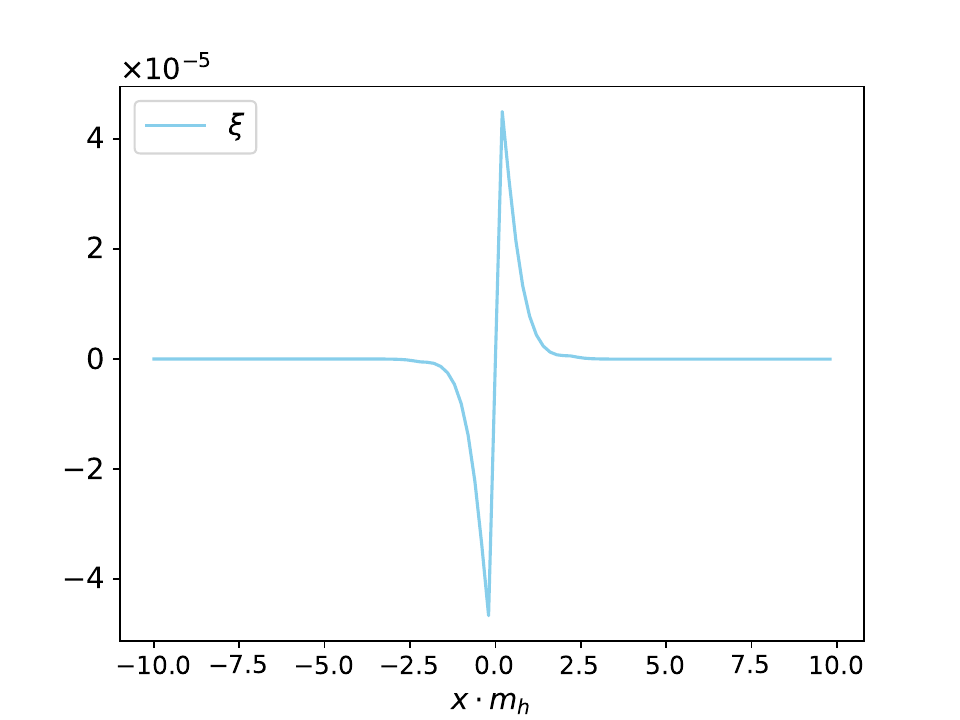}
         \subcaption{CP-violating phase $\xi$.}\label{subfig:xigeneral}
     \end{subfigure} 
\caption{Domain wall solution for the vacuum parameters in the case when the hypercharge angle $\theta$ and the Goldstone modes $g_i$ can change on both regions. The solution, using von Neumann boundary conditions, exhibits a stable charge breaking condensate $v_+$ and also a small but stable CP-violating phase $\xi(x)$.  }
\label{fig:vacuumparametersgeneral}
\end{figure}

\begin{figure}[H]
     \centering     
     \begin{subfigure}[b]{0.49\textwidth}
         \centering
         \includegraphics[width=\textwidth]{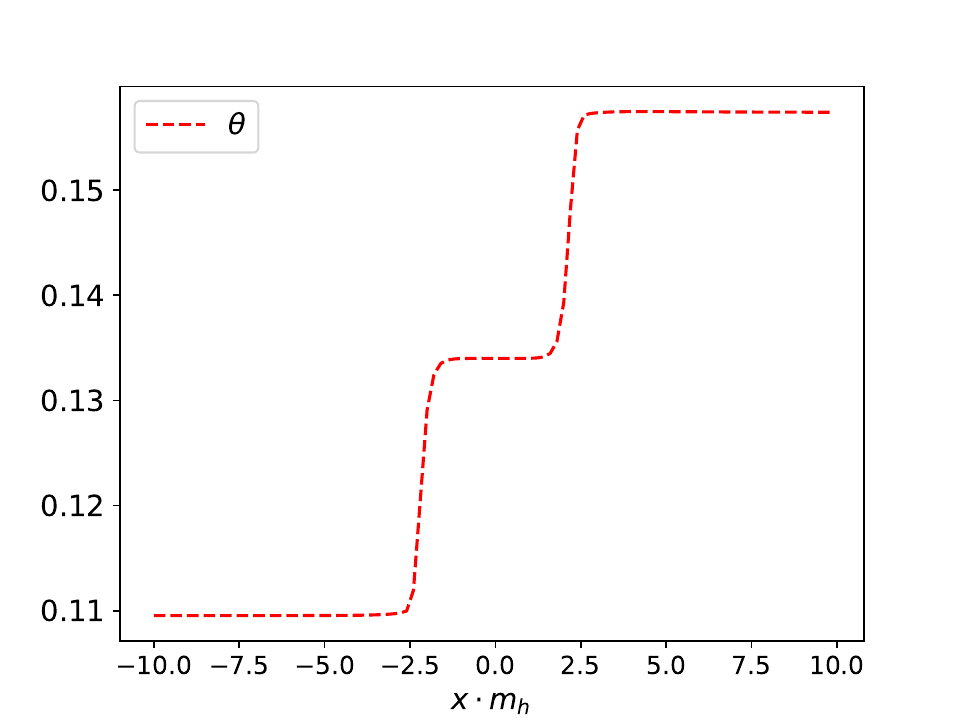}
         \subcaption{Hypercharge angle $\theta$.}\label{subfig:thetagen}
     \end{subfigure}
     \begin{subfigure}[b]{0.49\textwidth}
         \centering
         \includegraphics[width=\textwidth]{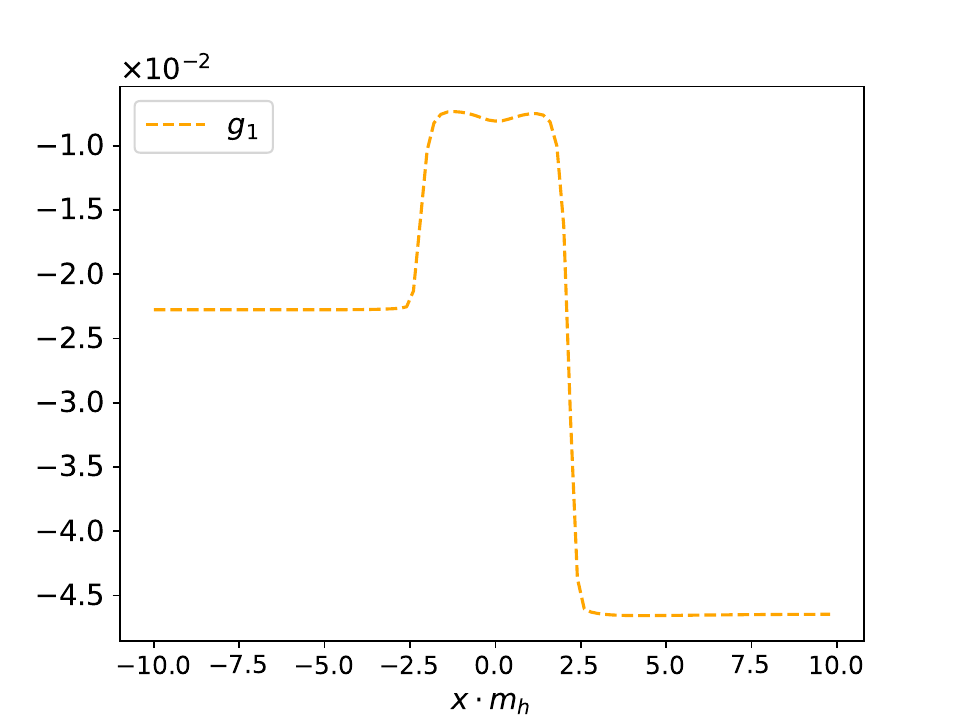}
         \subcaption{Goldstone mode $g_1$.}\label{subfig:g1gen}
     \end{subfigure}    
     \begin{subfigure}[b]{0.49\textwidth}
         \centering
         \includegraphics[width=\textwidth]{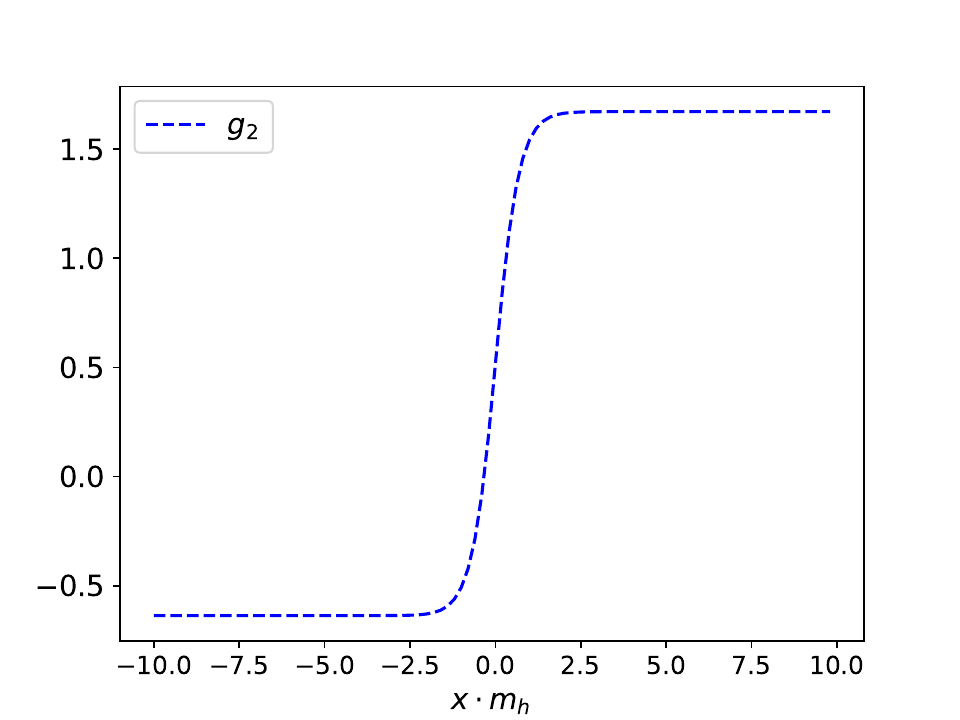}
         \subcaption{Goldstone mode $g_2$.}\label{subfig:g2gen}
     \end{subfigure}
     \begin{subfigure}[b]{0.49\textwidth}
         \centering
         \includegraphics[width=\textwidth]{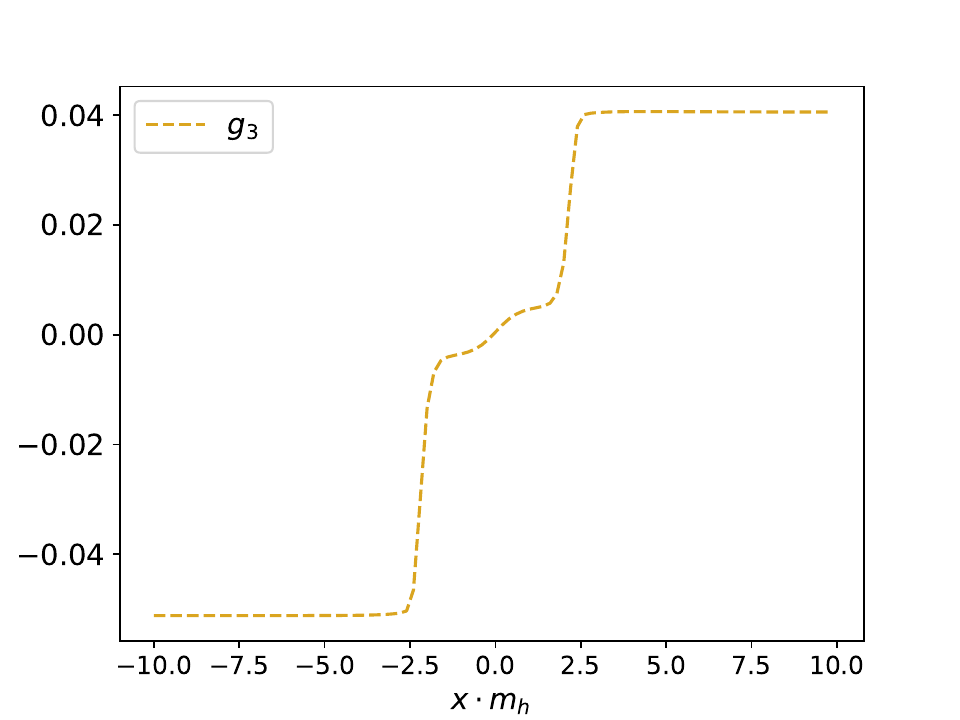}
         \subcaption{Goldstone mode $g_3$.}\label{subfig:g3gen}
     \end{subfigure}
\caption{Domain wall solution for the hypercharge angle $\theta$ and Goldstone modes $g_i$. One notices the non-trivial behavior of the hypercharge angle and Goldstone modes $g_1(x)$ and $g_3(x)$.}
\label{fig:electroweakmodes}
\end{figure}

We finish the discussion of domain wall solutions in the 2HDM by considering the case with a general matrix $U(x)$ (\ref{eq:EWmatrix}):
\begin{equation}
      \text{U}(x) = e^{i\theta(x)} \text{exp}\biggl(i\dfrac{g_i(x)}{2}\sigma_i\biggr),
\end{equation}
We recall the general formula for the energy functional \ref{eq:energyfunctionalgen}:
\begin{align}
\notag  \mathcal{E}(x) &= \dfrac{1}{2}\bigl(\dfrac{dv_1}{dx}\bigr)^2 + \dfrac{1}{2}\bigl(\dfrac{dv_2}{dx}\bigr)^2 + \dfrac{1}{2}\bigl(\dfrac{dv_+}{dx}\bigr)^2 + \dfrac{1}{2}v^2_2(x)\bigl(\dfrac{d\xi}{dx}\bigl)^2  +
  \dfrac{1}{2}v^2_1(x)\biggl[  \bigl(\dfrac{d\theta}{dx}\bigr)^2 + I_0(x) \\ \notag & +  2\dfrac{d\theta}{dx}I_3(x) \biggr]  + \dfrac{1}{2}v^2_2(x)\biggl[  \bigl(\dfrac{d\theta}{dx}\bigr)^2 + I_0(x) + 2\bigl(\dfrac{d\theta}{dx} + \dfrac{d\xi}{dx} \bigr)I_3(x) + 2\dfrac{d\theta}{dx}\dfrac{d\xi}{dx} \biggr] \\ \notag & + \dfrac{1}{2}v^2_+(x)\biggl[  \bigl(\dfrac{d\theta}{dx}\bigr)^2   + I_0(x) - 2\dfrac{d\theta}{dx}I_3(x) \biggr] + v_2(x) \biggl[   \text{ sin}(\xi)\dfrac{dv_+}{dx}I_1(x) - \text{ cos}(\xi)\dfrac{dv_+}{dx}I_2(x) \biggr] \\ \notag & + v_+(x) \biggl[   -\text{ sin}(\xi)\dfrac{dv_2}{dx}I_1(x) + \text{ cos}(\xi)\dfrac{dv_2}{dx}I_2(x) \biggr]
   + \dfrac{v_+(x)v_2(x)}{2}\biggl[  4\text{ sin}(\xi)I_2(x)\dfrac{d\theta}{dx} \\  & -  4\text{ cos}(\xi)I_1(x)\dfrac{d\theta}{dx} -2\text{ cos}(\xi)\dfrac{d\xi}{dx}I_1(x) -2\text{ sin}(\xi)\dfrac{d\xi}{dx}I_2(x) \biggr]  +  V_{2HDM}, 
\end{align}
\\\\
Here, the Goldstone modes $g_i$ and hypercharge angle $\theta$ are chosen randomly and can be different on both domains. 
\begin{table}[H]
\centering
\begin{tabular}{ |c||c|c|c|c|c|c|c|c|  }
% \multicolumn{4}{|c|}{Country List} \\
 \hline
 Field & $v_1$ & $v_2$ & $v_+$ & $\xi$ &  $\theta$ & $g_1$ & $g_2$ & $g_3$\\
 \hline \hline
Boundary at $+\infty$  & positive  & positive & 0 & 0  & $\pi/2$ & 0 & $\pi/6$ & 0 \\

Boundary at  $-\infty$ & positive & negative  & 0 & 0 & 0 & 0 & $\pi/6$ & 0 \\
 \hline
\end{tabular}
\caption{Asymptotic values of the fields at the boundaries. The initial profile for $\theta(x)$ is taken to be a tangent hyperbolic function interpolating between $0$ at $-\infty$ and $\pi/2$ at $+\infty$. We use von Neumann boundary condition to get the lowest energy solution. }
\label{Tab:boundariescond}
\end{table}
\noindent
Table \ref{Tab:boundariescond} shows an example of the initial asymptotic values for the fields at $\pm \infty$ that we use for the numerical calculations with $\theta$ having a tangent hyperbolic profile interpolating the values on the two boundaries. We use von Neumann boundary conditions to get the numerical solution of the 8 equations of motions describing the profiles of the fields.\\\\
Figures \ref{fig:vacuumparametersgeneral} and \ref{fig:electroweakmodes} show the numerical solution of the vacuum configuration for this choice of hypercharge angle $\theta$ and Goldstone modes. The solution features a stable charge-violation as well as a small but stable CP-violating phase $\xi(x)$ inside the wall. In contrast to all previous cases, the behavior of the hypercharge angle $\theta(x)$ and the Goldstone modes $g_1(x)$ and $g_3(x)$ is non-trivial inside the wall. We also note that, even though we started with the Goldstone modes $g_i$ being the same on both domains, the lowest energy solution has different values for $g_i(\pm \infty)$.

\subsection{Dependence of the kink solution on the parameter points of the 2HDM}
\begin{figure}[H]
     \centering
     \begin{subfigure}[b]{0.49\textwidth}
         \centering
         \includegraphics[width=\textwidth]{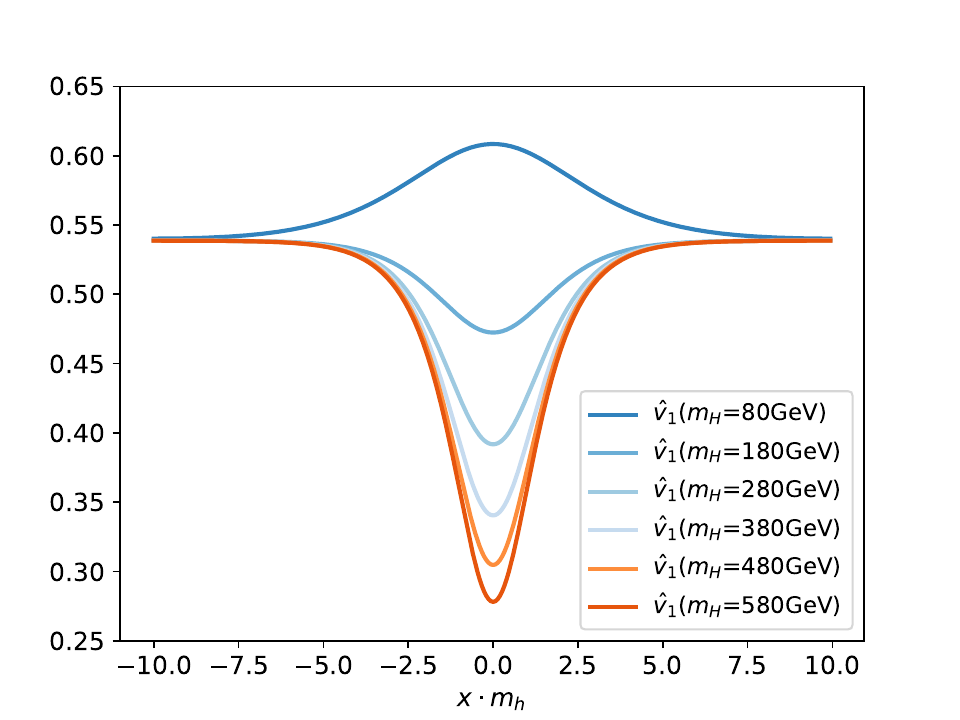}
         \subcaption{Dependence of $v_1$ on $m_H$.}\label{subfig:v1mH}
     \end{subfigure}
     \begin{subfigure}[b]{0.49\textwidth}
         \centering
         \includegraphics[width=\textwidth]{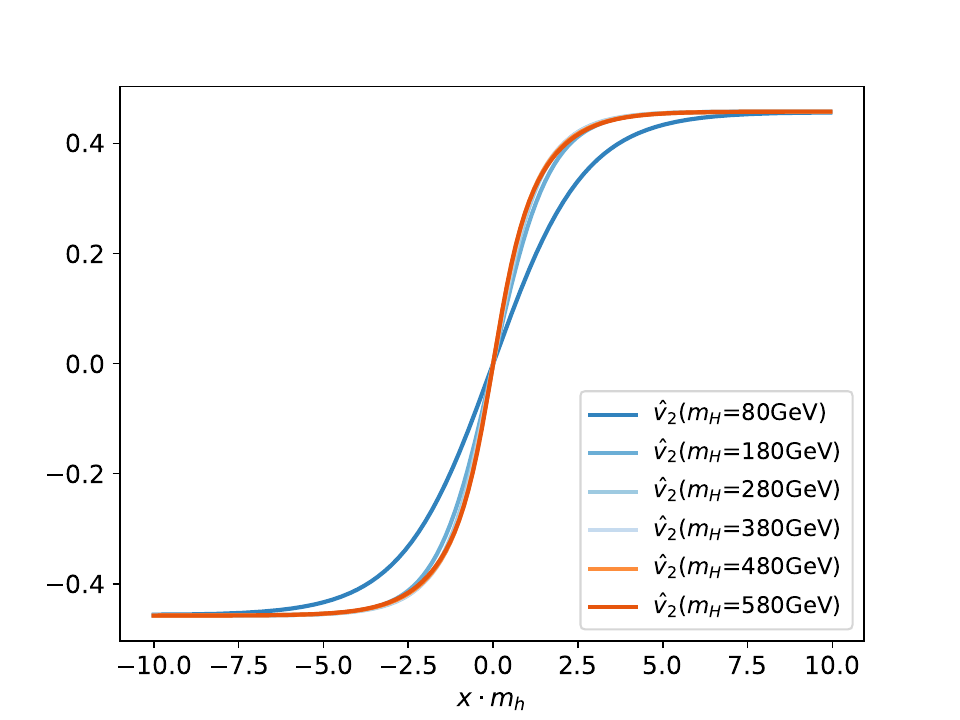}
        \subcaption{Dependence of $v_2$ on $m_H$.} \label{subfig:v2mH}
     \end{subfigure}
      \begin{subfigure}[b]{0.49\textwidth}
         \centering
         \includegraphics[width=\textwidth]{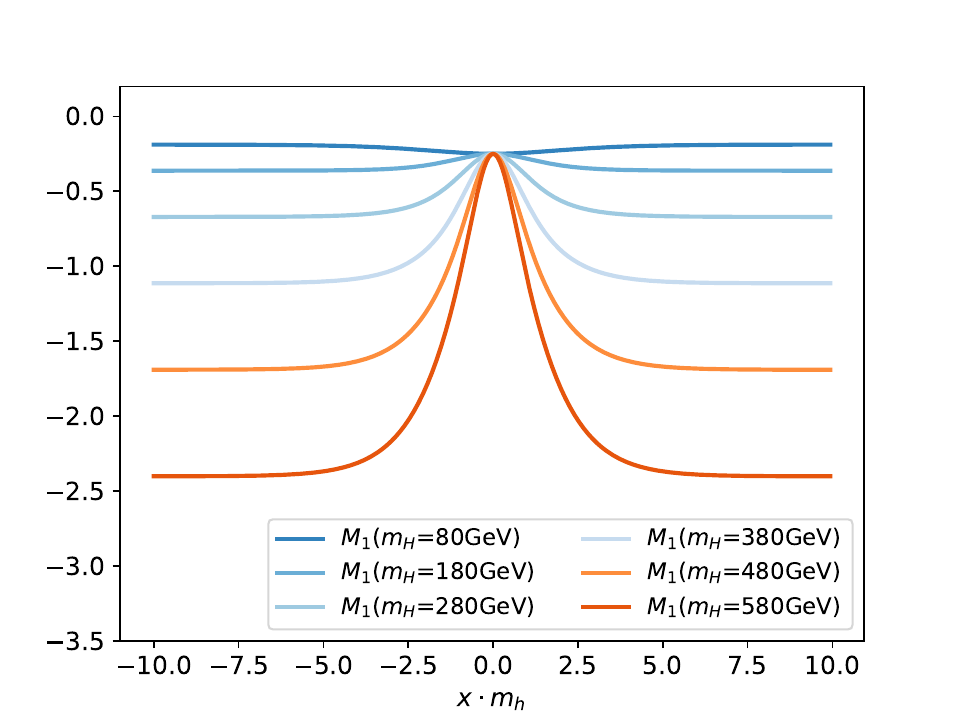}
         \subcaption{Dependence of $M_1$ on $m_H$.}\label{subfig:m1mH}
     \end{subfigure}
     \begin{subfigure}[b]{0.49\textwidth}
         \centering
         \includegraphics[width=\textwidth]{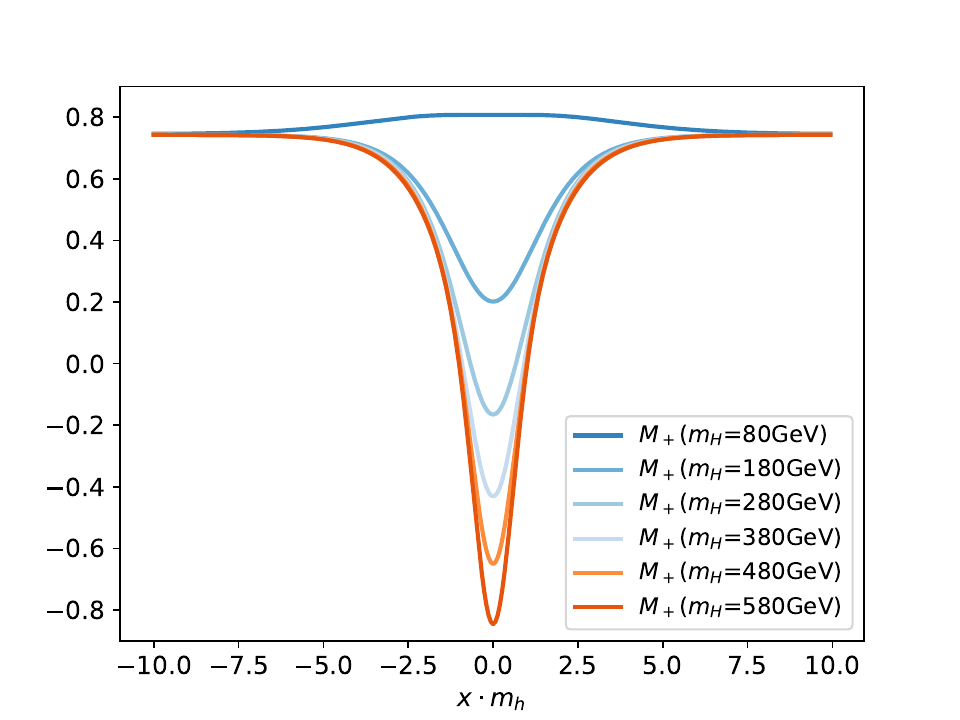}
        \subcaption{Dependence of $M_+$ on $m_H$.} \label{subfig:m+mH}
     \end{subfigure}
 \caption{Dependence of the vacuum expectation values $v_1(x)$ and $v_2(x)$ and effective masses $M_1(x)$ and $M_+(x)$ on the variation of the CP-even Higgs mass $m_H$. The masses of the other Higgs scalars are fixed to be $200 \text{ GeV}$ and $\tan(\beta) = 0.85$. (a) Profile of $v_1(x)$ inside the wall, notice that for small $m_H$, $v_1(x)$ becomes bigger inside the wall. (b) Profile of $v_2(x)$, we observe that a higher $m_H$ leads to a bigger width for the wall. (c) Effective mass $M_1(x)$ for $v_1(x)$, notice that for higher masses, this value becomes less negative inside the wall leading to a smaller $v_1(0)$. (d) Effective mass $M_+(x)$ for $v_+(x)$. We observe that outside the wall, this quantity does not depend on $m_H$. However higher $m_H$ lead to a smaller and even negative values $M_+(0)$ inside the DW.}
\label{fig:variationmH}
\end{figure}

\begin{figure}[H]
     \centering
     \begin{subfigure}[b]{0.49\textwidth}
         \centering
         \includegraphics[width=\textwidth]{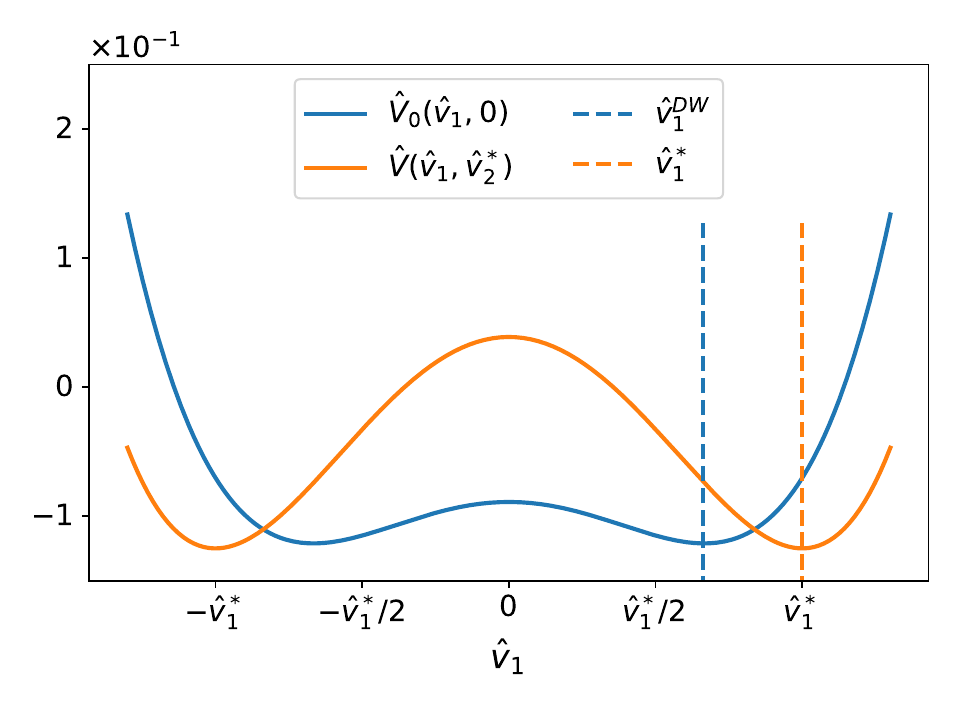}
\subcaption{}\label{subfig:v1behavioursmaller}
     \end{subfigure}
     \hfill
     \begin{subfigure}[b]{0.49\textwidth}
         \centering
    \includegraphics[width=\textwidth]{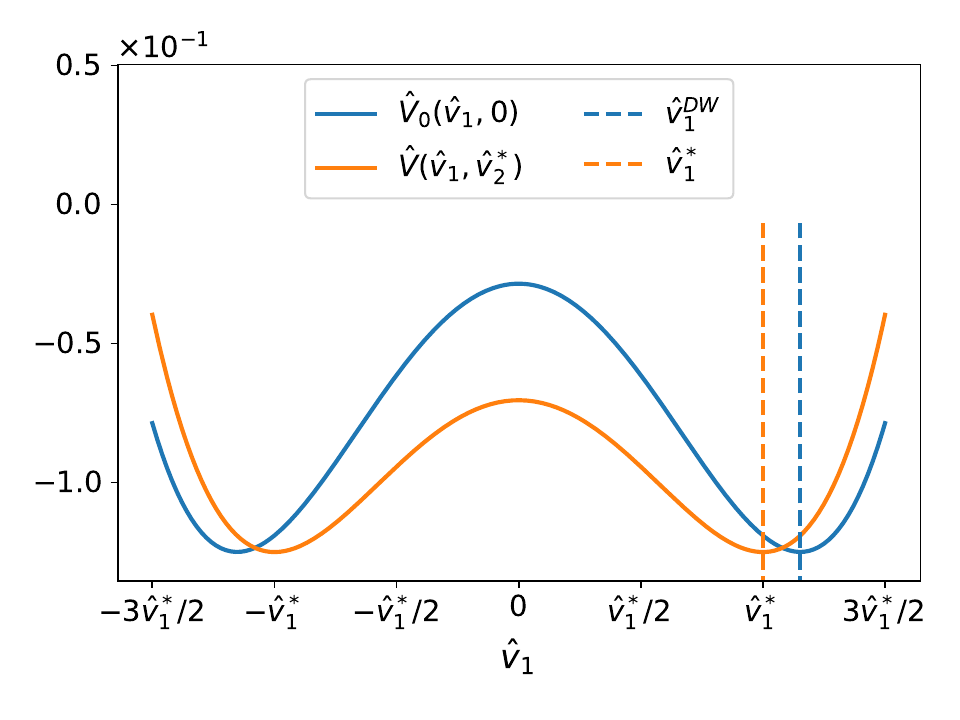}
         \subcaption{}\label{subfig:v1behaviourbigger}
     \end{subfigure}
\caption{Comparison of the potential for $v_1$ inside (blue) and outside (orange) the domain wall. (a) For $m_H = 250 \text{ GeV}$ and (b) For $m_H = 80 \text{ GeV}$. The other mass parameters are fixed to $m_A = m_C = 200 \text{ GeV}$. The potential $\hat{V}_0(\hat{v}_1,0)$ is shifted to have the same height as $\hat{V}(\hat{v}_1, \hat{v}_2)$.  }
\label{fig:v1behaviour}
\end{figure}

\begin{figure}[H]
     \centering
     \begin{subfigure}[b]{0.49\textwidth}
         \centering
         \includegraphics[width=\textwidth]{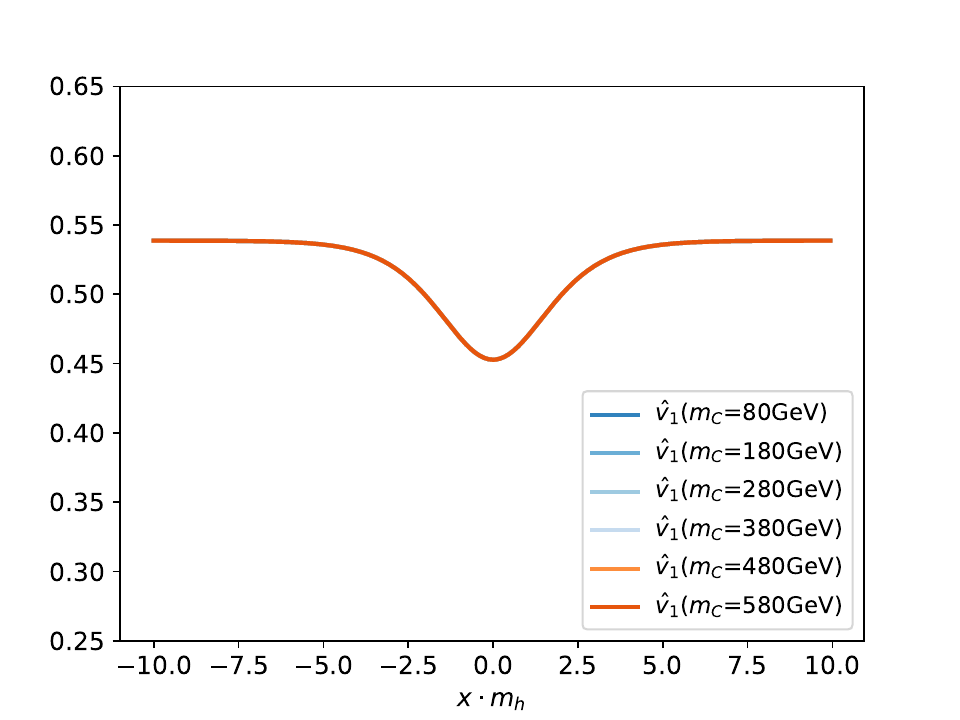}
         \subcaption{Dependence of $v_1$ on $m_C$.}\label{subfig:v1mC}
     \end{subfigure}
     \begin{subfigure}[b]{0.49\textwidth}
         \centering
         \includegraphics[width=\textwidth]{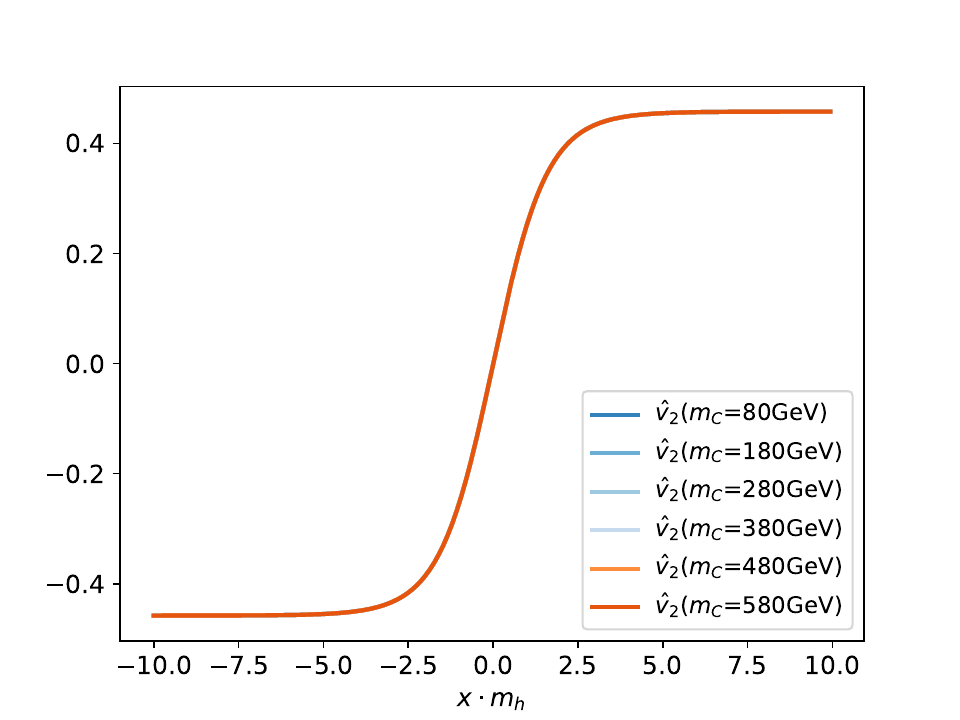}
        \subcaption{Dependence of $v_2$ on $m_C$.} \label{subfig:v2mC}
     \end{subfigure}
     \begin{subfigure}[b]{0.49\textwidth}
         \centering
         \includegraphics[width=\textwidth]{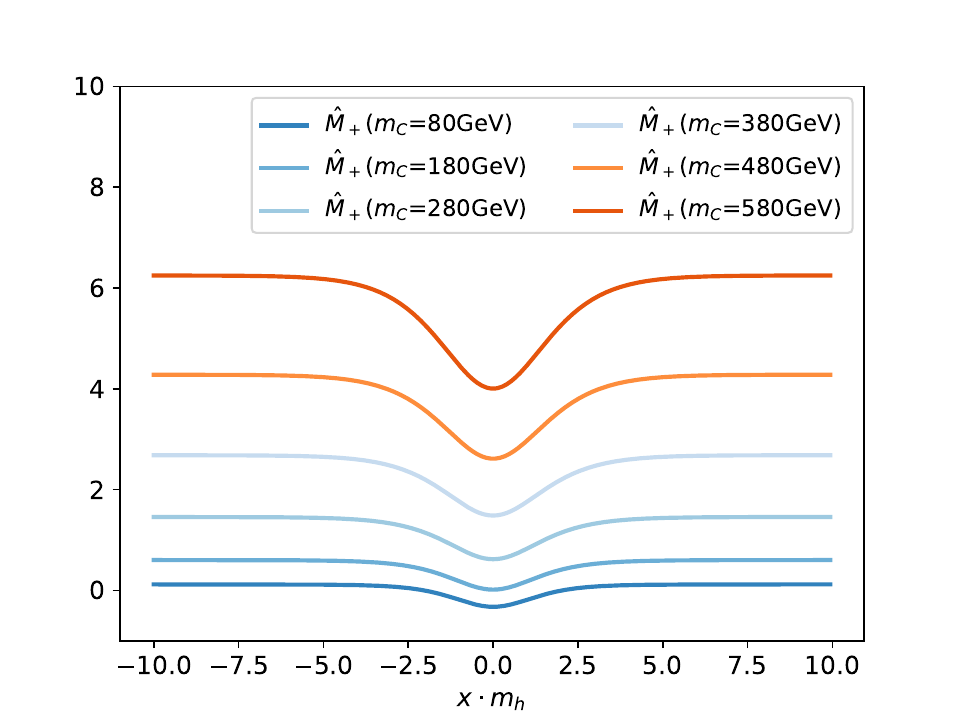}
        \subcaption{Dependence of $M_+$ on $m_C$.} \label{subfig:m+mC}
     \end{subfigure}
     \begin{subfigure}[b]{0.49\textwidth}
         \centering
         \includegraphics[width=\textwidth]{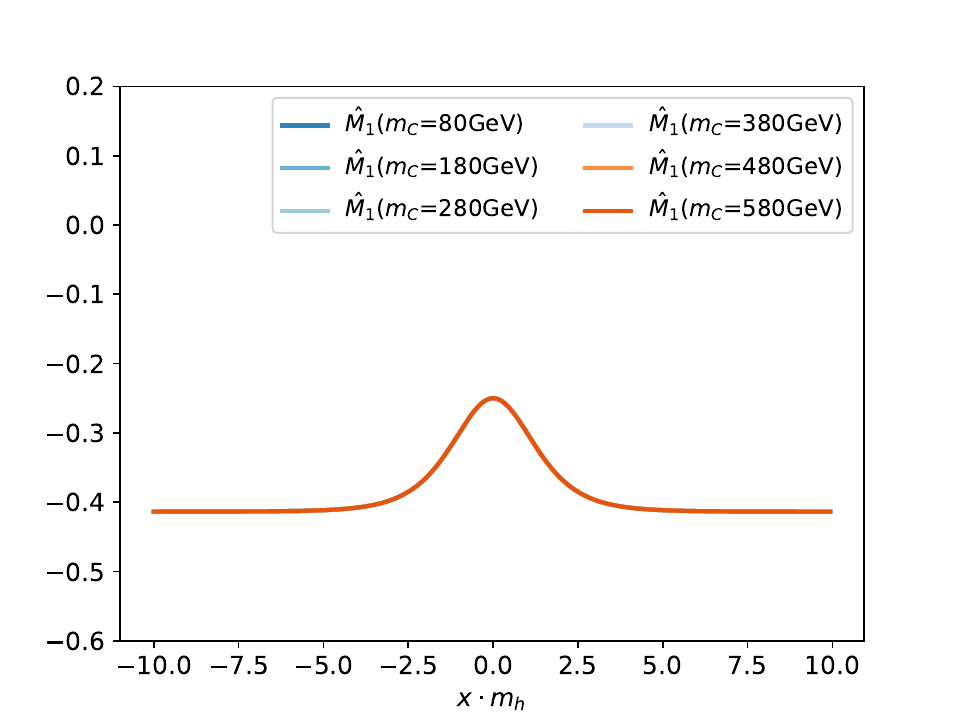}
        \subcaption{Dependence of $\hat{M}_1$ on $m_C$.} \label{subfig:m1mC}
     \end{subfigure}
\caption{Dependence of the vacuum expectation values $v_1(x)$ (a) and $v_2(x)$ (b) and effective masses $M_+(x)$ (c) and $M_1(x)$ (d) on the change of the charged Higgs mass. These results are obtained by fixing $m_{H,A} = 200 \text{ GeV}$ and $\tan(\beta) = 0.85$. The profiles of the vacua $v_1(x)$ and $v_2(x)$ are independent on $m_C$ (see (a) and (b)). The variation of $m_C$ has an impact on the effective mass for $v_+(x)$ and lower values of $m_C$ can lead to negative $M_+(0)$ inside the wall.}
\label{fig:variationmC}
\end{figure}

\begin{figure}[H]
  \centering
  \begin{subfigure}[b]
   {0.49\textwidth}
         \centering
    \includegraphics[width=\textwidth]{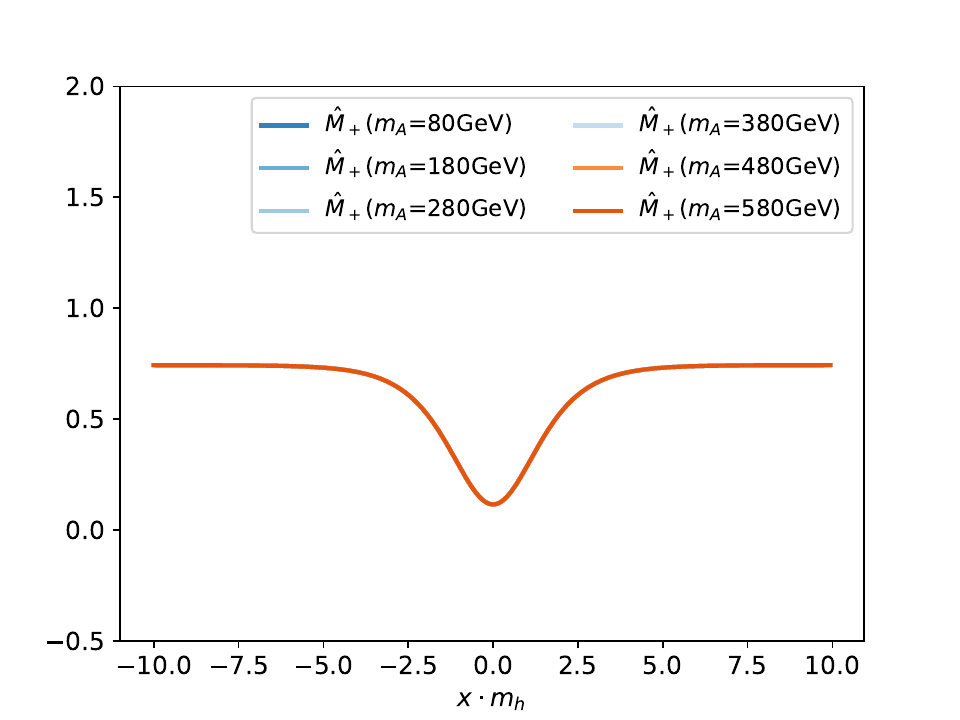}
        \subcaption{Dependence of $\hat{M}_+$ on $m_A$.} \label{subfig:m+mA}
     \end{subfigure}
     \begin{subfigure}[b]{0.49\textwidth}
         \centering
\includegraphics[width=\textwidth]{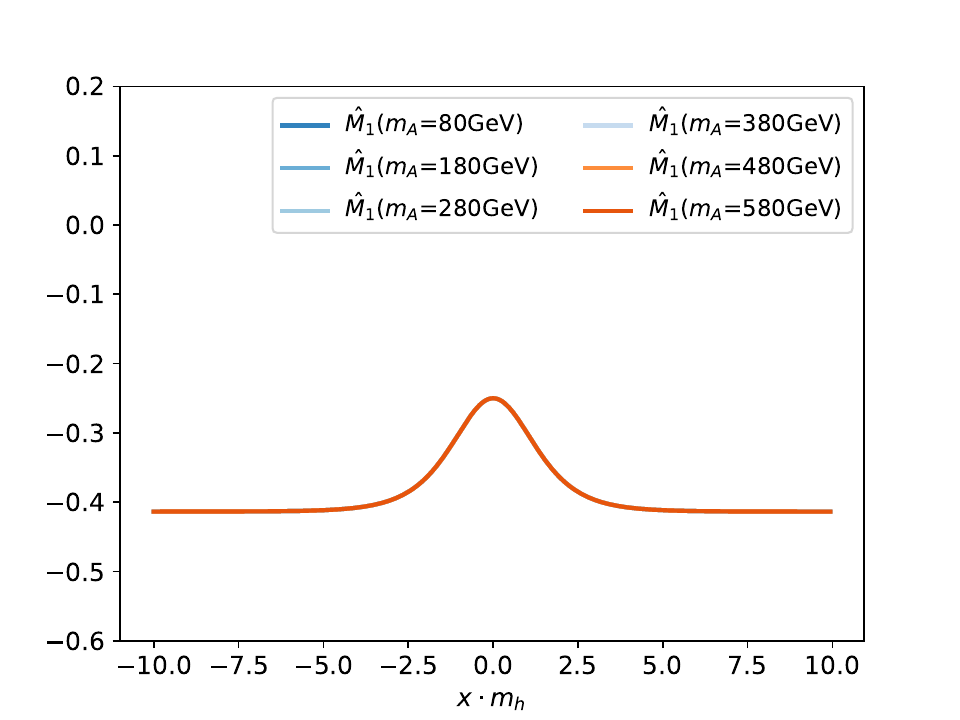}
        \subcaption{Dependence of $\hat{M}_1$ on $m_A$.} \label{subfig:m1mA}
     \end{subfigure}
 \caption{Dependence of the effective masses $M_1(x)$ and $M_+(x)$ on the variation of the CP-odd Higgs mass. We find that the properties of the domain wall solution are independent of $m_A$.}
\label{fig:variationmA}
\end{figure}

\begin{figure}[h]
\centering
 \begin{subfigure}[b]{0.49\textwidth}
         \centering
 \includegraphics[width=\textwidth]{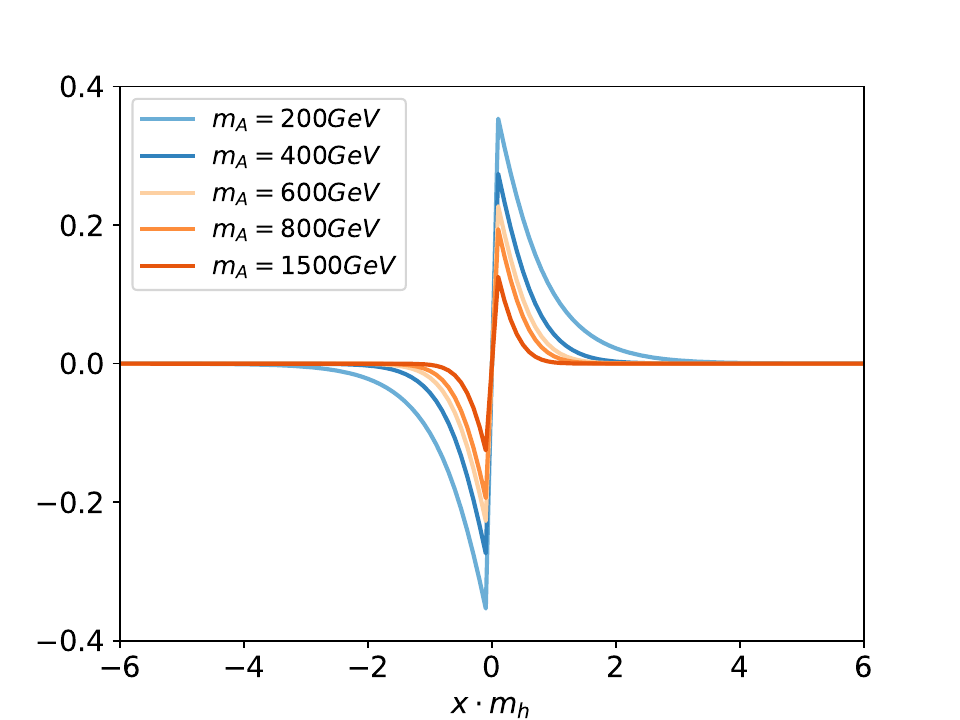}
 \subcaption{Dependence of $\xi(x)$ on $m_A$.} \label{subfig:ximadependence}
     \end{subfigure} 
     \begin{subfigure}[b]{0.49\textwidth}
         \centering
 \includegraphics[width=\textwidth]{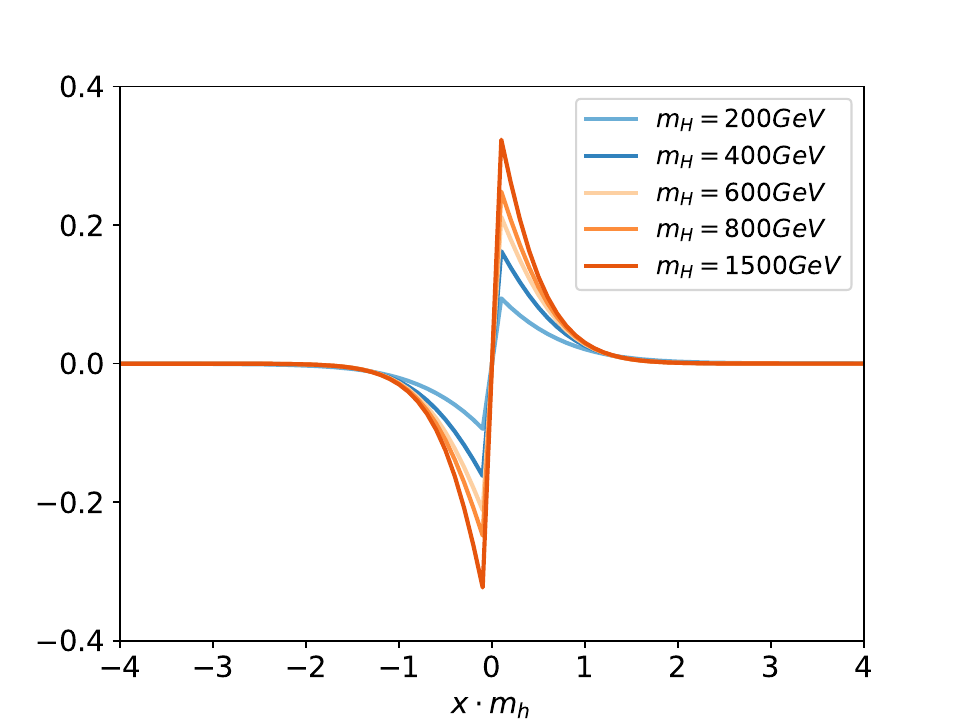}
 \subcaption{Dependence of $\xi(x)$ on $m_H$.} \label{subfig:ximhdependence}
     \end{subfigure}
\caption{Dependence of $\xi(x)$ on $m_A$ and $m_H$. We use the Dirichlet boundary condition in order to get an estimate for the CP-violating phase $\xi(x)$ at the formation stage of the DW. (a) We fix the masses $m_H = 800 \text{ GeV}$, $m_C = 400 \text{ GeV}$ and $\tan(\beta) = 0.85$. We observe that $\xi(x)$ gets smaller with higher $m_A$. (b) We fix the masses $m_A = 500 \text{ GeV}$, $m_C = 400 \text{ GeV}$ and $\tan(\beta) = 0.85$. We observe that $\xi(x)$ gets bigger with higher $m_H$.}
\label{fig:ximadep}
\end{figure}

\begin{figure}[H]
     \begin{subfigure}[b]{0.49\textwidth}
         \centering
         \includegraphics[width=\textwidth]{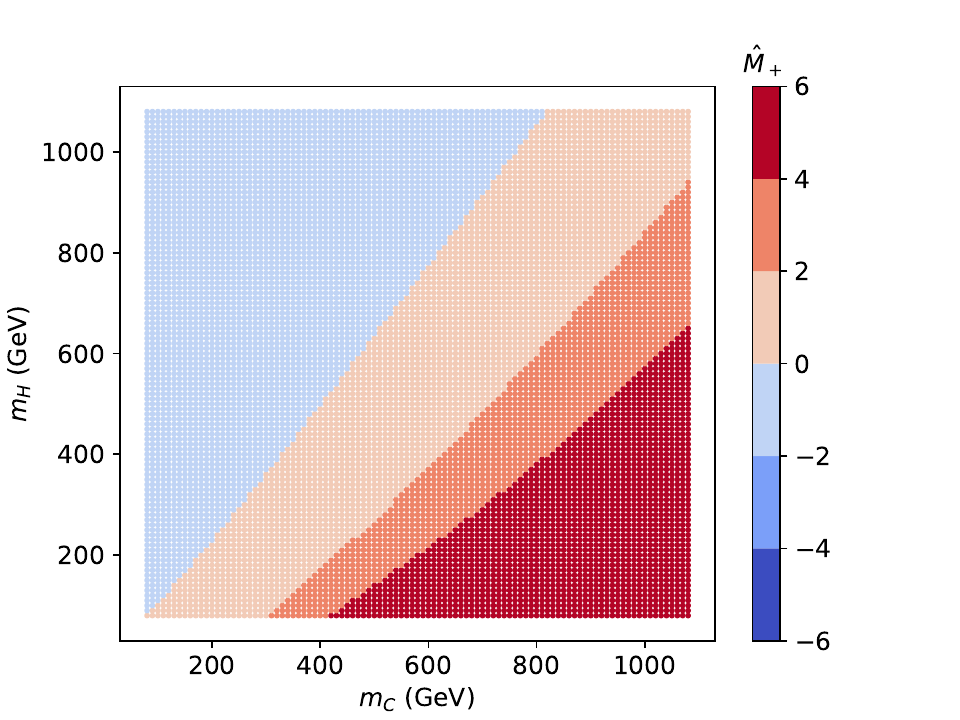}
         \subcaption{tan($\beta$) = 0.85}\label{subfig:tan085}
     \end{subfigure}
     \begin{subfigure}[b]{0.49\textwidth}
         \centering
         \includegraphics[width=\textwidth]{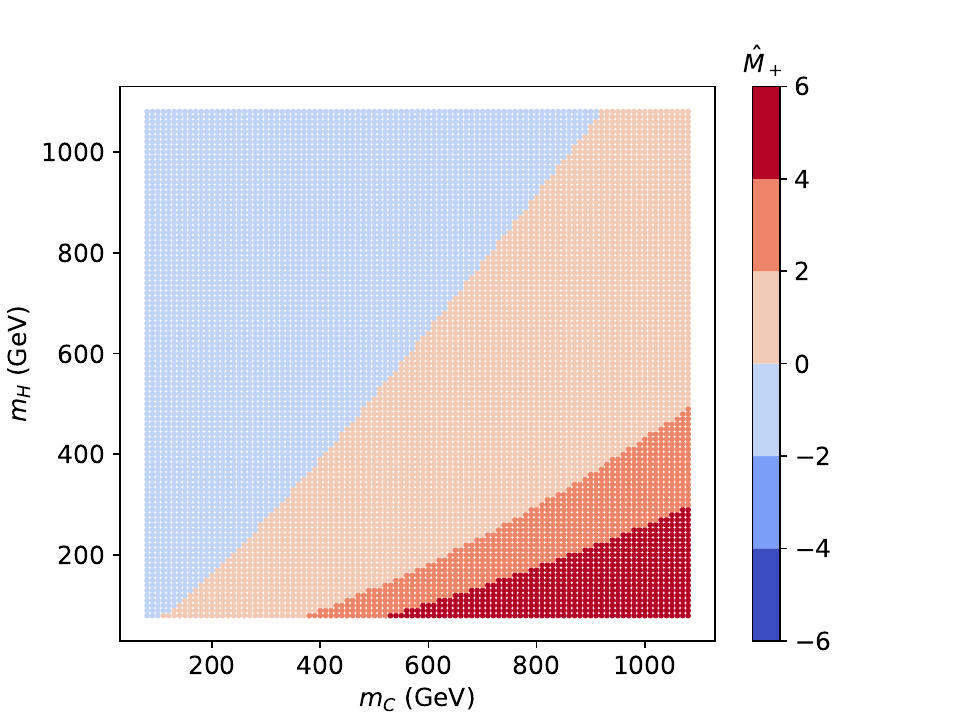}
         \subcaption{tan($\beta$) = 1.5}\label{subfig:tan15}
     \end{subfigure}
     \begin{subfigure}[b]{0.49\textwidth}
         \centering
         \includegraphics[width=\textwidth]{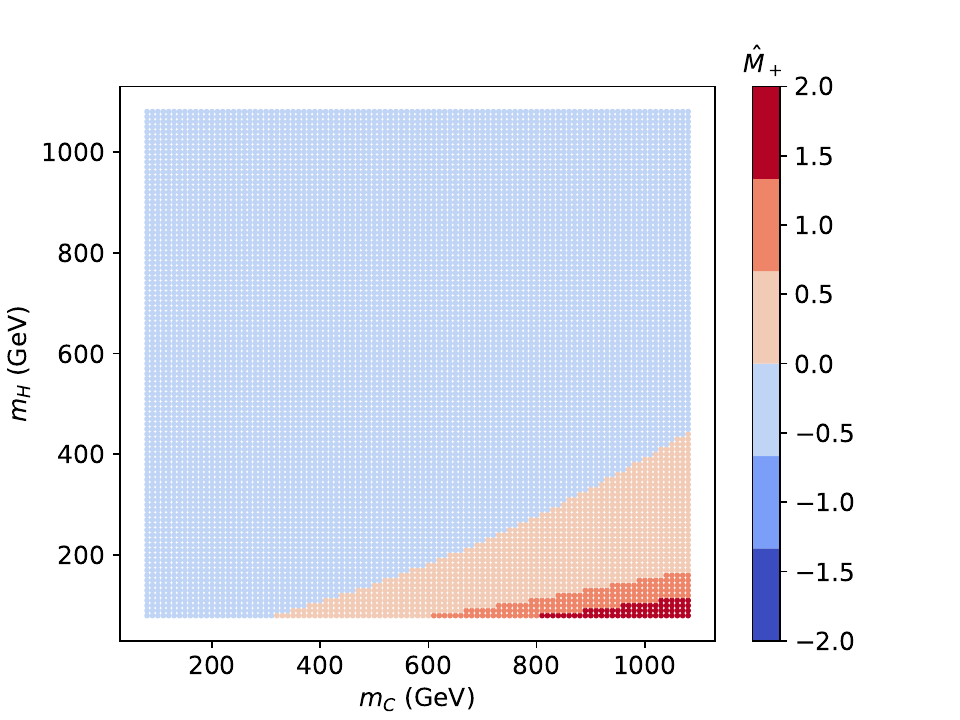}
         \subcaption{tan($\beta$) = 5}\label{subfig:tan050}
     \end{subfigure}
     \begin{subfigure}[b]{0.49\textwidth}
         \centering
         \includegraphics[width=\textwidth]{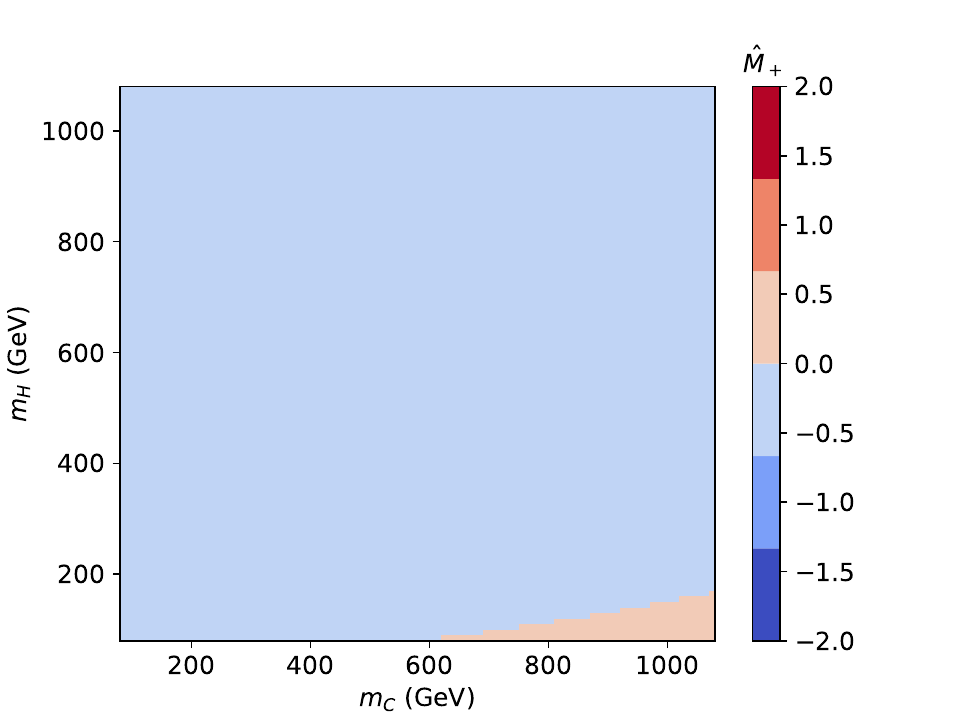}
         \subcaption{tan($\beta$) = 10}\label{subfig:tan100}
     \end{subfigure}
\caption{Dependence of $\hat{M}_+(0)$ inside the domain wall on the masses $m_C$ and $m_H$ for different values of $\tan(\beta)$. Negative values lead to the possibility of generating a charged vacuum inside the wall.}
\label{fig:mplusmassdep}
\end{figure}

\begin{figure}[H]
     \centering
     \begin{subfigure}[b]{0.49\textwidth}
         \centering
    \includegraphics[width=\textwidth]{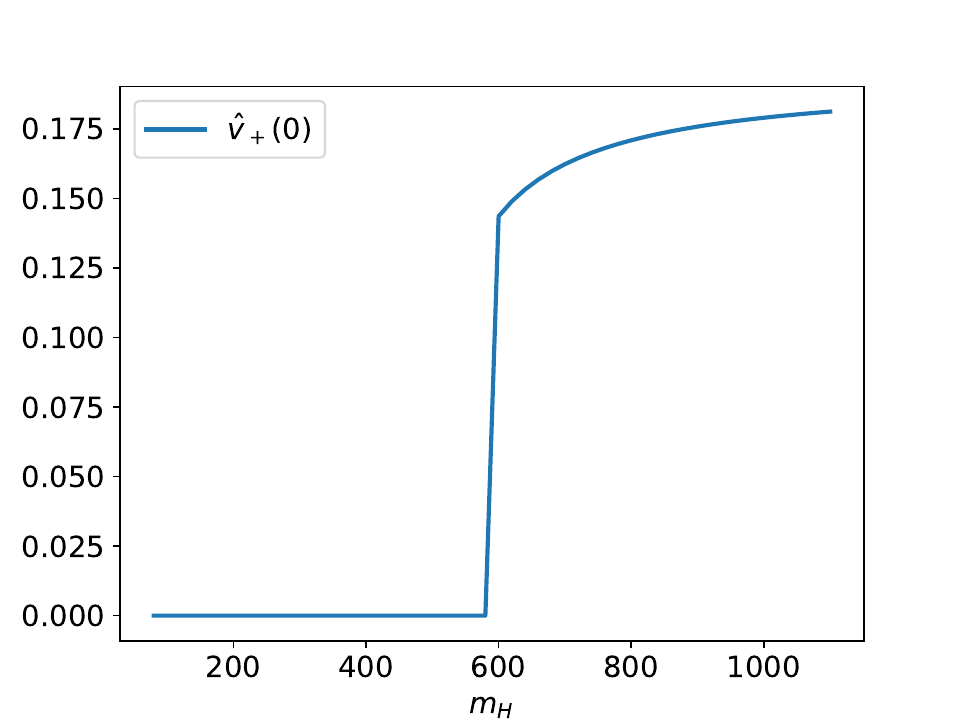}
         \subcaption{Dependence on $m_H$.}\label{subfig:chargedmH}
     \end{subfigure}
     \hfill
     \begin{subfigure}[b]{0.49\textwidth}
         \centering
    \includegraphics[width=\textwidth]{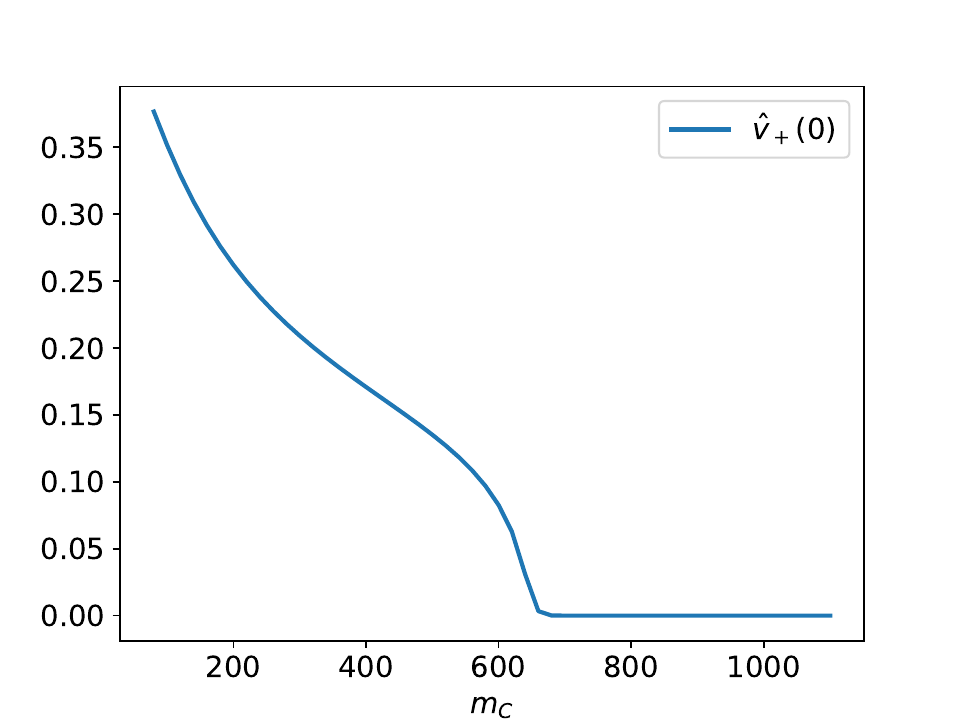}
         \subcaption{Dependence on $m_C$.}\label{subfig:chargedmC}
     \end{subfigure}
\caption{Dependence of the charged condensate $\hat{v}_+(0)$ on the mass parameters of the model. (a) We vary $m_H$ and fix $m_C = 400 \text{ GeV}$. (b) We vary $m_C$ and fix $m_H = 800 \text{ GeV}$.}
\label{fig:vplusdepmhmc}
\end{figure}

In this subsection, we briefly describe the dependence of the kink solution on the masses of the Higgs bosons of the 2HDM as well as $\tan(\beta)$. For all the discussed parameters, we use the alignment limit: $\alpha = \beta$. In this work we do not take into account experimental constraints on the 2HDM. Our choice of the parameter points discussed is done in order to give an overview of the different properties that arise for domain walls in the 2HDM. \\\\ 
We start by analyzing the effects of the parameter points on the standard domain wall solution. We therefore take the matrix $U(x)$ to be the identity.
First, we consider varying the mass of the CP-even Higgs $m_H$ with values between 80 GeV and 580 GeV, while the other parameters are fixed to:
\begin{align}
    m_A &= 200 \text{ GeV}, && m_C = 200 \text{ GeV}, && \tan(\beta) = 0.85. 
\end{align}
The profiles of the vacuum expectation values $v_1(x)$ and $v_2(x)$ for different $m_H$ are shown in Figures \ref{subfig:v1mH} and \ref{subfig:v2mH}.
\noindent
We observe that for higher $m_H$, the value of $v_1$ inside the domain wall ($v_1(x=0)$) becomes smaller. This is explained by the behavior of the effective mass $M_1(x)$ inside the domain wall (see \ref{eq:effectivemass1}) as shown in Figure \ref{subfig:m1mH}. Consider a potential of the form:
\begin{align}
    V(v_1(x), v_2(x)) = \dfrac{m_{22}}{2}v^2_2(x) + \dfrac{\lambda_2}{8}v^4_2(x) +M_1(x)v^2_1(x) + \dfrac{\lambda_1}{8}v^4_1(x).
\end{align}
Outside the wall and for parameter points with masses $m_H > 80\text{ GeV}$, the effective mass term $M_1$ is smaller (more negative) than inside the domain wall. Therefore, the minimum $v^{DW}_1 = v_1(x=0)$ of the potential $V_0(v_1, v_2 = 0)$ inside the wall (depicted in blue) is smaller than $v^*_1$, the minimum of the potential $V(v_1, v_2 = v^*_2)$ outside the wall (depicted in orange) as is shown in Figure \ref{subfig:v1behavioursmaller}.
\noindent
Note that for $m_H = 80\text{ GeV}$, the opposite effect occurs and $v_1(x=0)$ inside the wall is bigger than outside of it (see Figure \ref{subfig:v1behaviourbigger}). This is due to the fact that for $m_H = 80 \text{ GeV}$, $M_1(x)$ becomes more negative inside the wall than outside of it (see Figure \ref{subfig:m1mH}). This leads the minimum $v^{DW}_1 = v_1(x=0)$ of the potential $V_0(v_1, v_2 = 0)$ inside the wall to be bigger than $v^*_1$, the minimum of the potential $V(v_1, v_2 = v^*_2)$ outside the wall as is shown in Figure \ref{subfig:v1behaviourbigger}.
Notice also that the effective mass term $M_1(x=0)$ is the same for all mass parameters. This is because the parameter $m_{11}$ does not depend on $m_H$. As for the dependence of the kink solution for $v_2(x)$ on $m_H$, we observe that increasing the mass leads to a thinner profile for the kink as can be seen in Figure $\ref{subfig:v2mH}$. Figure \ref{subfig:m+mH} shows $M_+(x)$ for different $m_H$. A negative effective mass $M_+(0)$ inside the wall and therefore the possibility of having a charged condensate $v_+(x)$ localized on the wall is possible for higher masses of $H$. \\\\
We now vary the mass of the charged Higgs $m_C$ while keeping all other masses fixed to $200 \text{ GeV}$ and $\tan(\beta) = 0.85$. We do not observe any change in the profiles of the vacua $v_1(x)$ and $v_2(x)$ (see Figure \ref{subfig:v1mC} and \ref{subfig:v2mC}). However, we observe a change in the effective mass $M_+(x)$ for different masses $m_C$ and it becomes smaller inside the wall as shown in Figure \ref{subfig:m+mC}. However, the value $\hat{M}_+(x=0)$ is only negative for small masses of the charged Higgs. Therefore, if the mass $m_H$ of the CP-even Higgs is small, it is possible to have a stable charged condensate in the wall only if the charged Higgs masses $m_C$ is very low.
\noindent
To investigate the dependence of the domain wall properties on the mass $m_A$ of the CP-odd Higgs we vary $m_A$ from $80 \text{ GeV}$ to $580 \text{ GeV}$. We keep the masses of the other scalars to be $200 \text{ GeV}$ and $\tan(\beta) = 0.85$. The results for $M_1(x)$ and $M_+(x)$ are shown in Figure \ref{fig:variationmA} and we do not observe a variation in the properties of the domain walls. \\\\
We also study the effect of $m_A$ and $m_H$ on the CP-violating phase $\xi(x)$. In this case we study the scenario when the two domains have different hypercharge angle $\theta$ and use the Dirichlet boundary condition with $\theta(-\infty) = 0$ and $\theta(+\infty) = \pi/2$. We first fix the masses $m_H = 800 \text{ GeV}$, $m_C = 400 \text{ GeV}$ and $\tan(\beta) = 0.85$ and vary $m_A$. The results are shown in Figure \ref{subfig:ximadependence} and we observe that increasing $m_A$ leads to a smaller phase $\xi(x)$ inside the DW. However, when fixing $m_A = 500 \text{ GeV}$ and varying $m_H$, we observe the opposite behavior: increasing $m_H$ leads to a higher CP-violating phase $\xi(x)$ (see Figure \ref{subfig:ximhdependence}).
\noindent
Finally, we plot in Figure \ref{fig:mplusmassdep} the dependence of $\hat{M}_+(0)$ on both $m_H$ and $m_C$ for several values of tan($\beta$). $\hat{M}_+(0)$ does not depend on the mass of the CP-odd Higgs (as seen in Figure \ref{subfig:m+mA}), therefore we fixed $m_A = 300 \text{ GeV}$. For low $\tan(\beta) = 0.85$, a large number of parameter points have positive $\hat{M}_+(0)$ (see Figure \ref{subfig:tan085}). This leads the scalar potential inside the wall $V_0(v_1,v_+)$ to have its minimum at $v_+(0)=0$ (see Figure \ref{subfig:potvplusII}). This means that any charge violating solution for those domain walls is unstable. In order to get a stable $v_+(x)$ condensate inside the wall for this low value of $\tan(\beta)$, we need to choose high values for $m_H$. As we increase tan($\beta$), the fraction of parameter points with a negative $\hat{M}_+(0)$ increases (as is shown in Figure \ref{subfig:tan15}) and a charged condensate inside the domain wall can be stable (in case of different $g_2$ Goldstone mode on both domains). For high values of tan($\beta$), most parameter points have a negative $\hat{M}_+(0)$ as can be seen in Figures \ref{subfig:tan100} and \ref{subfig:tan050}.  
\noindent
In the case that $v_+$ develops a stable condensate inside the wall, the masses $m_H$ and $m_C$ can have a sizable effect on the maximum value $v_+(x=0)$ of such a condensate. In order to study this, we solved the equations of motion for the case when the vacua have different values of $g_2$ using von Neumann boundary conditions. This was done for two scenarios: 
\begin{itemize}
    \item To study the effect of $m_H$, we fix $m_C = 400 \text{ GeV}$, $\tan(\beta) = 0.85$ and vary $m_H$ between 80 GeV and 1100 GeV. The results are shown in Figure \ref{subfig:chargedmH}. We observe that the $v_+$ condensate is unstable for $m_H < 580 \text{ GeV}$. The value of $v_+(0)$ inside the wall increases with the mass $m_H$.
    \item To study the effect of of $m_C$, we fix $m_H = 800 \text{ GeV}$, $\tan(\beta) = 0.85$ and vary $m_C$ between 80 GeV and 1100 GeV. The results are shown in Figure \ref{subfig:chargedmC}. In this case, $v_+(0)$ gets smaller with heavier $m_C$. For $m_C>680 GeV$ the condensate becomes unstable. 
\end{itemize}
\subsection{Non-Topological kink solutions in the 2HDM}
\begin{figure}[h]
\centering
\includegraphics[width=0.65\textwidth]{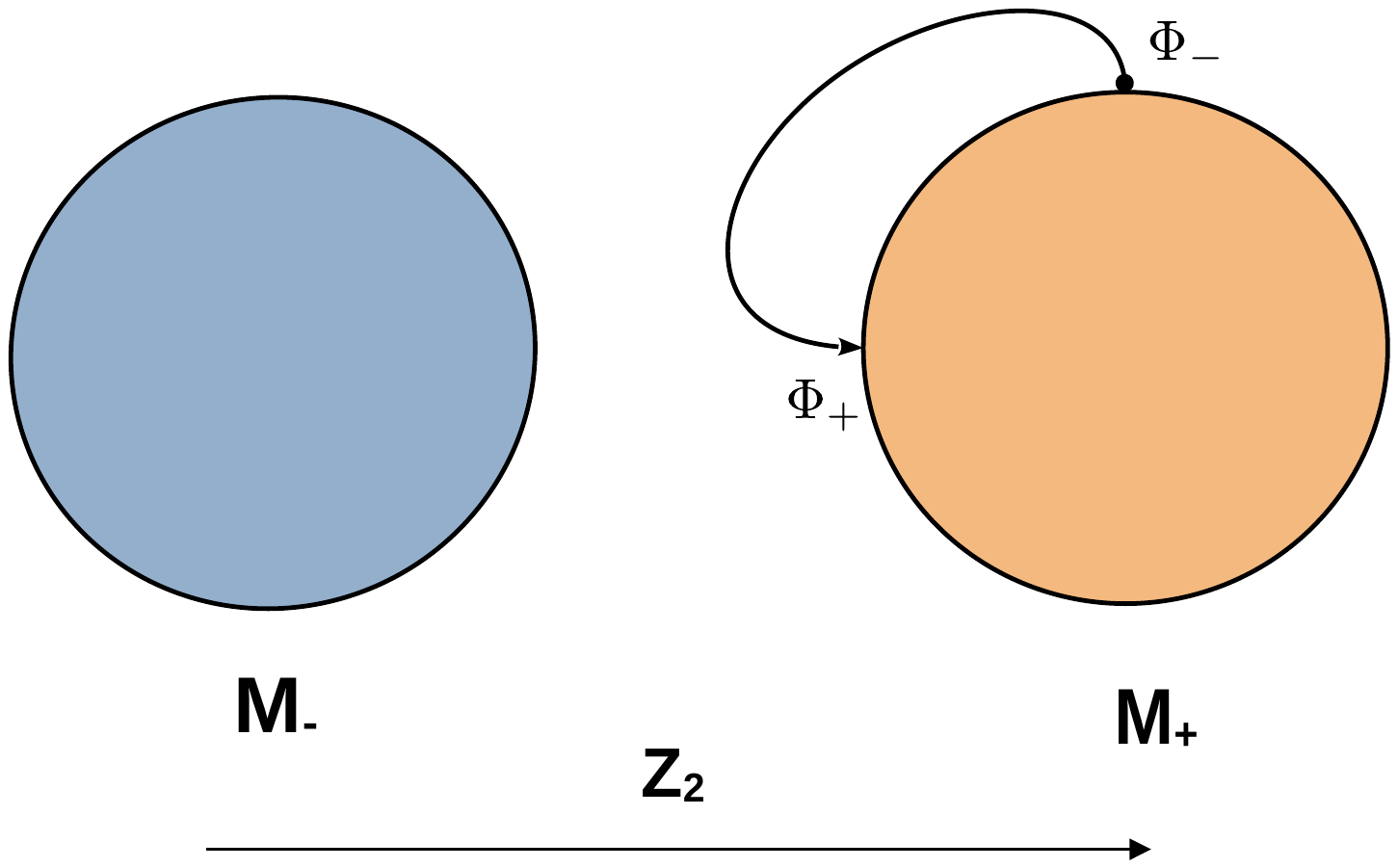}
\caption{Non-topological domain wall vacuum configuration. The vacua on both regions in space lie on the same sector of the vacuum manifold.}
\label{fig:nontopmanifold}
\end{figure}

\begin{figure}[h]
     \centering
     \begin{subfigure}[b]{0.49\textwidth}
         \centering
         \includegraphics[width=\textwidth]{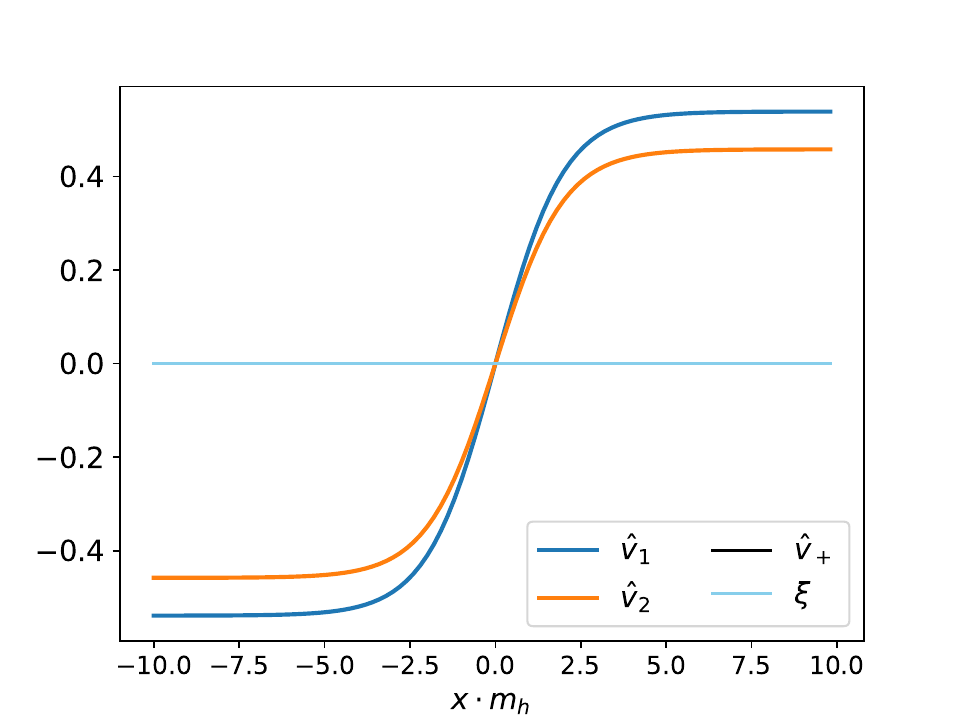}
          \subcaption{}\label{subfig:nontopkink}
     \end{subfigure}
     \hfill
     \begin{subfigure}[b]{0.49\textwidth}
         \centering
    \includegraphics[width=\textwidth]{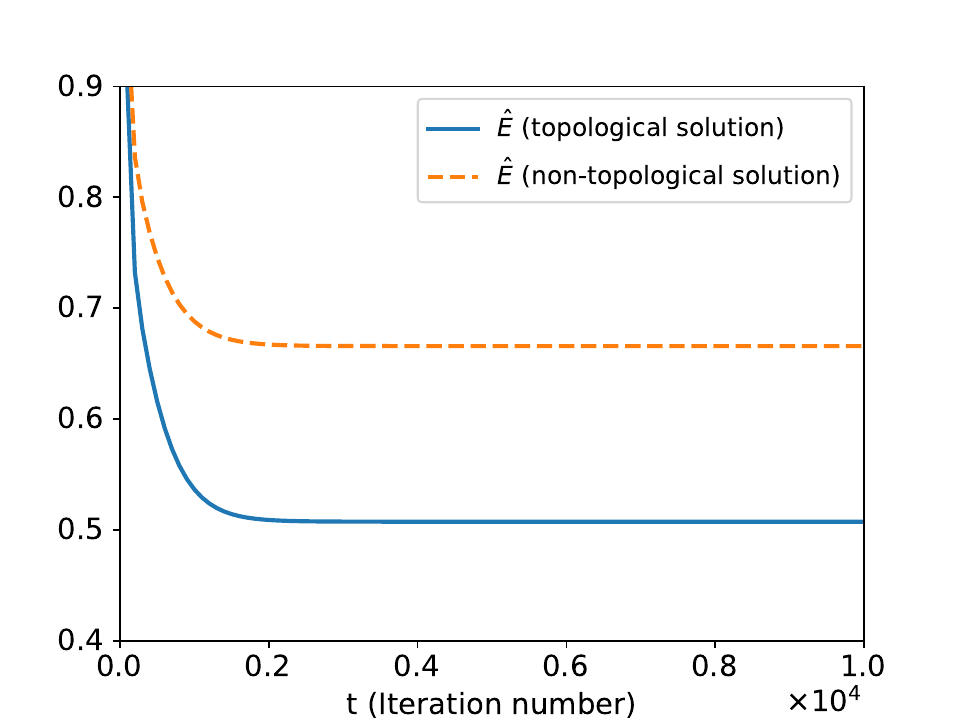}
         \subcaption{}\label{subfig:nontopenergy}
     \end{subfigure}
\caption{(a) Non-topological kink solution for the vacuum configuration. (b) Energies of the topological solution (interpolating vacua related by a $Z_2$ symmetry) in blue and non-topological kink solution (\ref{eq:nontopconfig}) in dashed orange. }
\label{fig:nontopological}
\end{figure}

We discussed in the previous sections the possibility of having topological kink solutions that interpolate between vacua belonging to disconnected sectors of the vacuum manifold. We now briefly discuss the scenario when both vacua at  $ x \to \pm \infty$ belong to the same sector (in the case of the 2HDM, the same 3-sphere $M_+$ or $M_-$) as shown in Figure \ref{fig:nontopmanifold}. 
\noindent
We consider the following two degenerate vacua:
\begin{align}
    \Phi_- = \biggl\{ \dfrac{1}{\sqrt{2}}
    \begin{pmatrix}
          0 \\      -v^*_1
     \end{pmatrix}  , \dfrac{1}{\sqrt{2}}
      \begin{pmatrix}
     0 \\
     -v^*_2
      \end{pmatrix}  \biggr\}, && 
    \Phi_+ = \biggl\{ \dfrac{1}{\sqrt{2}}
    \begin{pmatrix}
          0 \\      v^*_1
     \end{pmatrix}  , \dfrac{1}{\sqrt{2}}
      \begin{pmatrix}
     0 \\
     v^*_2
      \end{pmatrix}  \biggr\}.
\label{eq:nontopconfig}      
\end{align}
One can obtain $\Phi_-$ by performing a $U(1)_Y$ transformation on $\Phi_+$. Therefore, both these vacua belong to the same sector $M_+$. A kink solution for such a vacuum configuration is not topologically protected against a variation in the fields and therefore any such solution should, in principle, be unstable \cite{Pogosian:2001fm}. Indeed we obtain a kink solution (see Figure \ref{subfig:nontopkink}) for this configuration using von Neumann boundary conditions. The solution has a higher energy than the topological kink solution, where the vacua are related by a $Z_2$ transformation (see Figure $\ref{subfig:nontopenergy}$).
However the solution seems to be stable, even when using von Neumann boundary conditions. This might be explained by the fact that a potential energy barrier exists between the vacua $\Phi_+$ and $\Phi_-$ (see Figure \ref{subfig:potential}). \\
If we transform the vacuum $\Phi_+$ with an $\text{SU}(2) \times \text{U}(1)_Y$ gauge transformation U (\ref{eq:EWmatrix}), the kink solution can develop charge and/or CP-violating vacua localized inside the wall but such a vacuum configuration is no longer stable. In the 1D simulation, one vacuum grows in the region of the other vacuum. The charge and/or CP-violating vacua also move alongside the wall. Therefore, after some time, the kink solution dynamically decays into the trivial  vacuum configuration, where one vacuum occupies the whole space.

%% file: FermionScattering.tex
\section{SM fermions scattering off the domain walls}
The existence of different types of domain walls in the 2HDM can have profound implication on the physics of the early universe. For instance, it was shown in \cite{Battye:2021dyq} that photons with small frequencies such as in the CMB will scatter off superconducting domain walls with a large reflection coefficient. In \cite{Steer:2006ik}, it was shown that domain walls in Grand Unified Theories arising after the spontaneous breaking of $\text{SU}(5)\times Z_2$, interact with the Higgs scalar field. This interaction induces exotic scattering phenomena of fermions off the domain wall via the Yukawa sector, such as neutrinos being reflected as d-quarks, with the electric and color charges being absorbed by gauge fields living on the wall.    
In our work, we study the interactions of SM fermions with 2HDM domain walls via the Yukawa sector by considering the scattering solution of the Dirac equation of SM fermions within a domain wall background. We leave the case of fermion zero modes and bound states solutions \cite{vachaspati_2023} on the walls, which might be relevant for baryogenesis scenarios, for future work.  \\
The solution for the scattering of fermions off standard domain walls generated by the spontaneous breaking of a discrete symmetry can be found in \cite{Steer:2006ik,Campanelli:2004si}. One finds that for thin walls, the rate of reflection and transmission of fermions off the walls is:
\begin{align}
    R(p) &= \frac{m^2}{m^2 + p^2}, & T(p) = \frac{p^2}{m^2 + p^2} 
\label{eq:standardrates}    
\end{align}
In this work, we want to get analytical solutions for the Dirac equations within the background of  different types of domain walls arising in the 2HDM. As the functions describing the spatial dependence of these vacuum configurations are non-trivial, it is appropriate to use a thin-wall approximation to simplify the form of the different vacuum configurations inside the wall. The thin-wall approximation is valid for the scattering of fermions which have wavelengths larger than the width of the wall. As another simplification, we do not consider the back-reaction effects of fermions on the vacuum configurations in our study, which could change the spatial kink profile of the vacuum configuration \cite{Klimashonok:2019iya}.
\subsection{Thin-Wall approximation}
In this section, we briefly discuss the validity of this approximation for domain walls in the 2HDM as well as describe the vacuum configuration in this approximation. \\
From results of the last chapter we can infer a typical width of the domain walls to be approximately $L_w \approx 4/m_h$. The typical wavelength $\lambda_f$ of a particle in the thermal plasma in the early universe is proportional to $2\pi/T$, where T is the temperature of the thermal plasma. The thin wall approximation is valid when $L_w < \lambda_f$ corresponding to particle momenta smaller than $  200 \text{ GeV}$, which is sufficient to describe the momenta of most particles existing in the thermal plasma after EWSB.\\ 
Taking a thin-wall profile for the domain walls, we can approximate the kink solution of $v_2(x)$ to be a step-function:
\begin{align}
    v_2(x) = -\tilde{v}_2\Theta(x) + \tilde{v}_2\Theta(-x).
\end{align}
As for the vacua $v_1(x)$, $v_+(x)$ and $\text{Im}(v_2(x)e^{i\xi(x)})$, it is possible to approximate them with a delta distribution:
\begin{align}
    v_1(x) &= v_1 + \tilde{v}_1 \delta(x), \\
    v_+(x) &= \tilde{v}_+ \delta(x), \\
    \tilde{v}_2(x) &= \text{Im}(v_2(x)e^{i\xi(x)}) =  \tilde{v}_2 \delta(x-a) +  \tilde{v}_2 \delta(x+a), \label{eq:v2tilde}   
\end{align}
where $\tilde{v}_{1,2,+}$ are dimensionless parameters defined as $\tilde{v}_{1,2,+} = v_{1,2,+}/ \text{GeV}$.
\begin{figure}[h]
\centering
\includegraphics[width=0.65\textwidth]{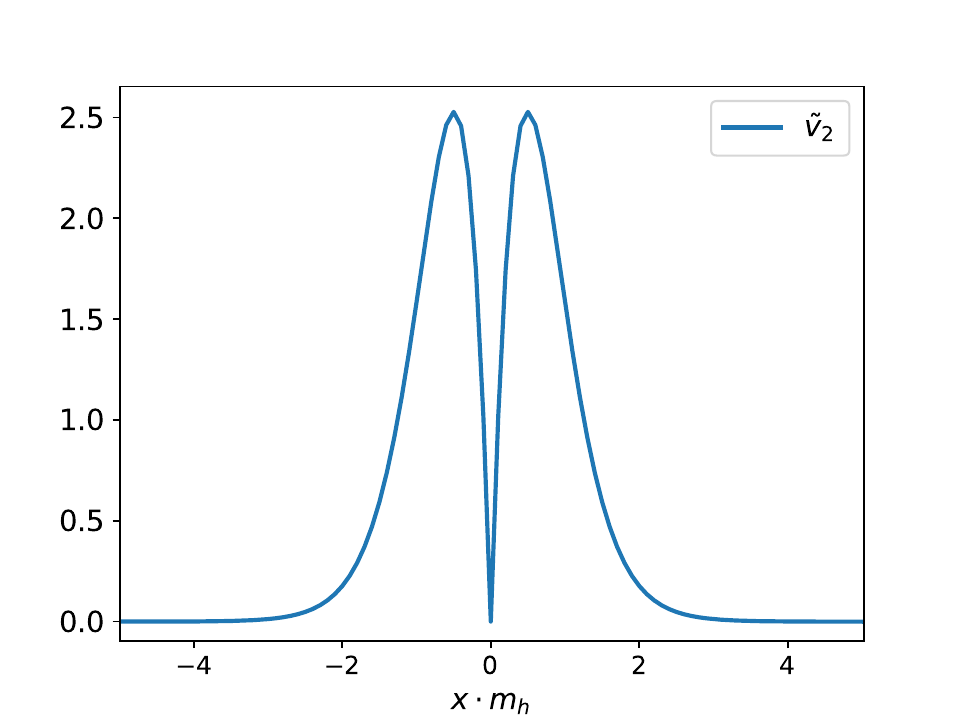}
\caption{Profile of $\tilde{v}_2(x)$ using Dirichlet boundary conditions. When using von Neumann boundary conditions, $\tilde{v}_2$ gets smaller with time as the CP-violating phase $\xi(x)$ inside the wall vanishes. }
\label{fig:v2tilde}
\end{figure}
\noindent
The parameter $\tilde{v}_2(x)$ describes the profile of the imaginary part of the CP-violating mass term that appears in the Dirac equations (see Figure \ref{fig:v2tilde}). We saw in the last chapter that the phase $\xi(x)$ vanishes (in the simplified cases) when using von Neumann boundary conditions (see e.g sections [\ref{theta}] and [\ref{g3}]). However, the vanishing of $\xi(x)$ inside the wall only occurs after the dynamical evolution of the Goldstone modes or hypercharge angle $\theta$ inside the different domains, making them equal to each other. Therefore, the CP-violating phase should still be substantial at the moment of the formation of the domain wall network and the study of CP-violating scattering of fermions off the wall is relevant for that period, when considering the simplified cases. However, as shown in sections [\ref{su2}] and [\ref{general}], for domains rotated by a relative $SU(2)_L$ or an electroweak symmetry $SU(2)_L \times U_Y(1)$ the CP-violating phase $\xi(x)$ of the kink solution is small but stable. For simplicity, we consider the simplified cases in this work and leave the scattering of fermions off general domain walls in the 2HDM for future investigations.
\subsection{CP-Violating interactions of fermions with the domain walls}\label{CPviolatingscattering}
\begin{figure}[H]
\centering
\begin{subfigure}[b]{0.49\textwidth}
\centering
\includegraphics[width=\textwidth]{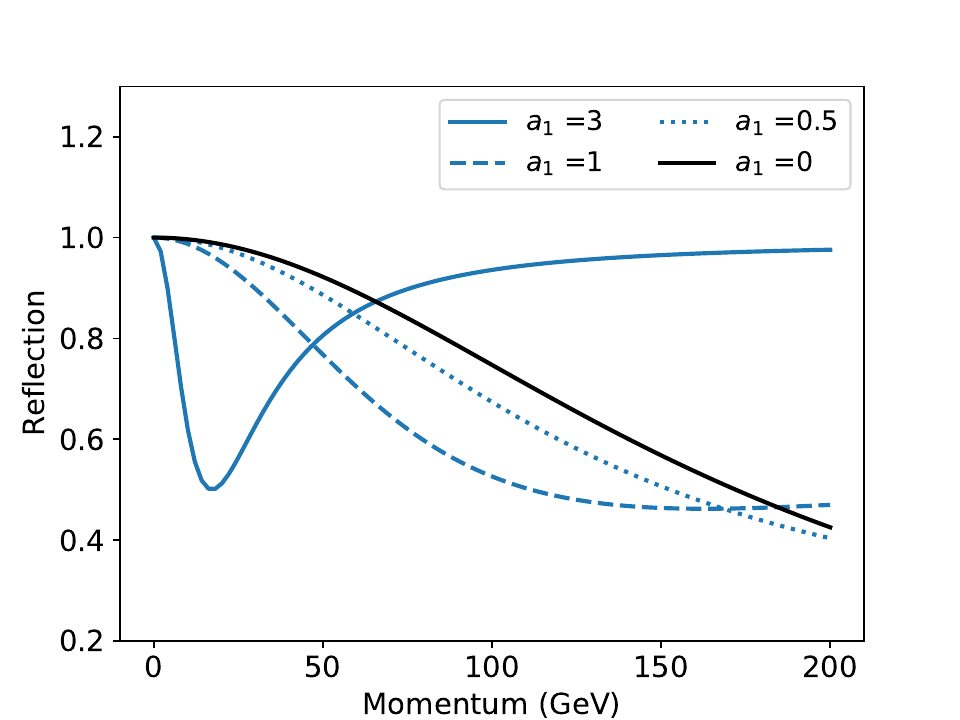}
\subcaption{$\hat{R}$ as a function of $a_1$ and momenta}  \label{subfig:variationa1ref}
\end{subfigure}
\hfill
\begin{subfigure}[b]{0.49\textwidth}
\centering
\includegraphics[width=\textwidth]{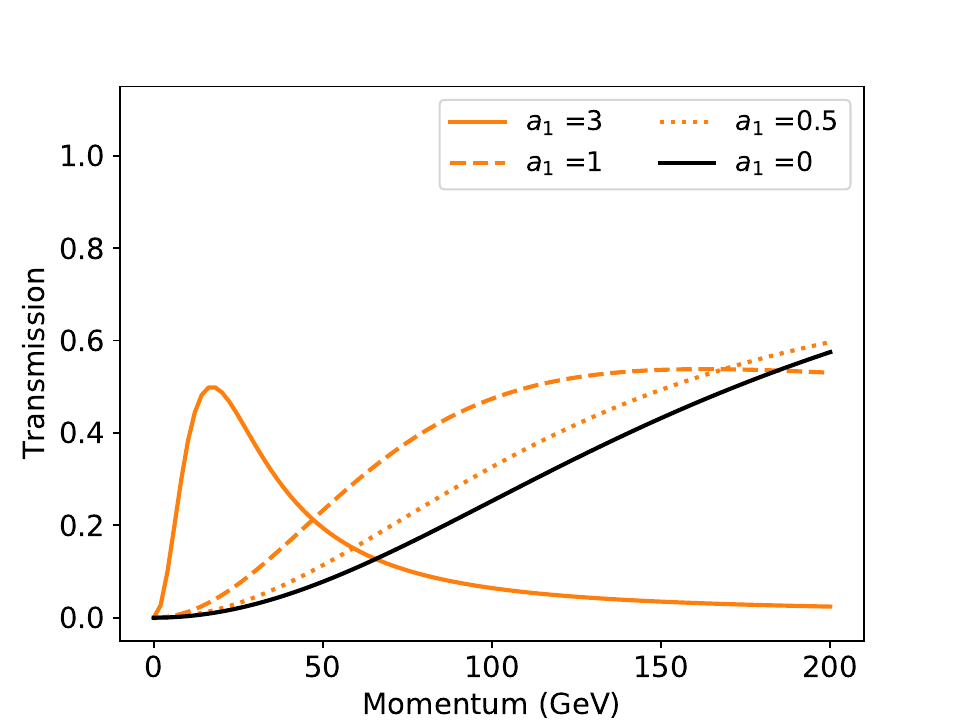}
\subcaption{$\hat{T}$ as a function of $a_1$ and momenta} \label{subfig:variationa1tra}
\end{subfigure}
\caption{Results for the reflection (a) and transmission (b) coefficients for the case when the numerical value of the Yukawa contribution $a_1$ is much higher than the numerical value from the gauge field contribution $b_2$. Notice that for higher values of CP-violation $a_1$, the reflection coefficient increases with momentum while $\hat{T}$ decreases. }
\label{fig:yukawadominant}
\end{figure}

We start with the case of fermions scattering off CP-violating domain walls induced by a difference in the hypercharge angle $\theta$. 
Recall that in this case, the Higgs doublets are given by: 
\begin{align}
\Phi_1 (x) &= \frac{U(x)}{\sqrt{2}} \begin{pmatrix} 0  \\  v_1(x) \end{pmatrix},  & \Phi_2 (x) &= \frac{U(x)}{\sqrt{2}} \begin{pmatrix} 0  \\  v_2(x)e^{i\xi(x)} \end{pmatrix},
\end{align}
where the matrix $U(x)$ is given by:
\begin{equation}
U(x) = e^{i\theta(x)} \text{I}_2.
\end{equation}
We can remove this matrix $U(x)$ from the Yukawa sector by performing a gauge transformation, leading to a pure gauge term for the hypercharge gauge field $B^{\mu}$: 
\begin{align}
    & \Phi_1(x) \xrightarrow{U_Y(1)} U^{-1}(x) \Phi_1(x) = \tilde{\Phi}_1(x), \\
    & \Phi_2(x) \xrightarrow{U_Y(1)} U^{-1}(x) \Phi_2(x) = \tilde{\Phi}_2(x), \\
    & B_{\mu} \xrightarrow{U_Y(1)} \frac{i}{g}U(x)\partial_{\mu}(U^{-1}(x)) = \frac{1}{g}\partial_{\mu}\theta(x).
\end{align}
The Yukawa Lagrangian for the up-type quarks in the type-2 2HDM is then given by:
\begin{align}
  \notag  \mathcal{L}_{Fermion} &= i\bar{u}_L\bigl(\slashed{\partial} + iY_{u,L}\slashed{\partial}\theta(x)\bigr)u_L + i\bar{u}_R\bigl(\slashed{\partial} + iY_{u,R}\slashed{\partial}\theta(x)\bigr)u_R - y_uv_2(x)e^{i\xi(x)}\bar{u}_Lu_R \\ & - y_uv_2(x)e^{-i\xi(x)}\bar{u}_Ru_L,
\end{align}
where $Y_{u,L}$ and $Y_{u,R}$ are the hypercharges of the left and right-handed up-type quarks respectively, $y_u$ is the Yukawa coupling of the Higgs doublet to the up-type quark and $u_L$, $u_R$ are the left-handed and right-handed components of the up-type quark respectively.\\
One can then derive the Dirac equation for up-type quarks:
\begin{align}
    \biggl( i\slashed{\partial} - \slashed{\partial}\theta(x)(Y_{u,L}P_L + Y_{u,R}P_R) - m_{R}(x) -im_I(x)\gamma_5   \biggr)u(x) = 0,
\label{eq:diraccp}    
\end{align}
where $P_L$ and $P_R$ are the left and right- handed projector operators and $m_R(x)$, $m_I(x)$ are the real and imaginary parts of the fermion mass, respectively:
\begin{align}
    m_u(x) &= y_uv_2(x)e^{i\xi(x)},  \\
    m_R(x) &= y_uv_2(x)\cos(\xi(x)) \approx y_uv_2(x),  \\
    m_I(x) &= y_uv_2(x)\sin(\xi(x)) \approx y_uv_2(x)\xi(x) = y_u\tilde{v}_2(x),
\end{align}
where we apply a small angle approximation for the CP-violating angle $\xi(x)$, which is motivated by the simulations of the domain walls in the previous chapter. After multiplying the left hand side of (\ref{eq:diraccp}) with $i\gamma_1$, we can rearrange this equation and get:
\begin{align}
  \notag  \partial_x u(x,t) & \coloneqq \hat{G}(x,t) u(x,t) \\
    &=\biggl[ -i\gamma_1\gamma_0\partial_t -i\partial_x\theta(x)(Y_LP_L + Y_RP_R) + i\gamma_1m_R(x) - \gamma_1\gamma_5m_I(x) \biggr]u(x,t).
\label{eq:matrixnotationdiraccp}    
\end{align}
The solution to this equation can be calculated in an analogous way to the case of fermion scattering off a CP-violating bubble wall \cite{Joyce:1994zn}:
\begin{align}
  u(x,t) = \hat{P}\text{ exp}\biggl( \int^{x}_{x_0} dx' \hat{G}(x')  \biggr) u(x_0,t),
  \label{eq:matchingcp}
\end{align}
where $\hat{P}$ is an ordering operator. We consider the case of a plane wave solution of a quark scattering off the domain wall from the left ($x<0$), that can either be reflected or transmitted to the other region ($x>0$):
\begin{align}
    u(t,x) &= e^{-iEt+ip_ux}u_{inc} + e^{-iEt-ip_ux}u_{ref} \text{    for $x<0$,} \\ 
    u(t,x) &= e^{-iEt+ip_ux}u_{tra} \text{    for $x>0$,} 
\end{align}
where $E$ denotes the energy of the incoming quark and $u_{inc}, u_{ref} \text{ and } u_{tra}$ are 4-component spinors describing respectively the incident, reflected and transmitted fermion.
\begin{align}
  u_{inc} = \begin{pmatrix} u_{1i} \\ u_{2i} \\ u_{3i} \\ u_{4i}\end{pmatrix} , &&  u_{ref} = \begin{pmatrix} u_{1m} \\ u_{2m} \\ u_{3m} \\ u_{4m}\end{pmatrix} , && u_{tra} = \begin{pmatrix} u_{1p} \\ u_{2p} \\ u_{3p} \\ u_{4p}\end{pmatrix}.
\end{align}
By plugging this ansatz into the Dirac equation (\ref{eq:diraccp}) for the regions far away from the wall (where the vacua $v_1(x)$ and $v_2(x)$ take on their asymptotic values and $\xi(x)=\theta(x)=0$), we can derive relations between the different components of the spinors:
\begin{align}
    u_{4m} &= \frac{-p_u}{E+m_u}u_{1m}, && u_{3m} = \frac{-p_u}{E+m_u}u_{2m}, && u_{4p} = \frac{p_u}{E-m_u}u_{1p}, && u_{3p} = \frac{p_u}{E-m_u}u_{2p}.
\label{eq:fardw}    
\end{align}
The matching condition of the solution $u(x)$ at $x=0$ is calculated using (\ref{eq:matchingcp}):
\begin{align}
  u(+\epsilon,t) = \hat{P}\text{ exp}\biggl( \int^{+\epsilon}_{-\epsilon} dx' \hat{G}(x')  \biggr) u(-\epsilon,t),
  \label{eq:matchingcp2}
\end{align}
where $\epsilon$ is a small number taken to $\epsilon \rightarrow 0$. The final result is given by:
\begin{align}
    u(+\epsilon) = \text{ exp}\biggl(-i\frac{\Delta\theta}{2}(Y_L+ Y_R)\biggr)\biggl[ \text{cosh}(a)\text{ I}_4 - \frac{\text{sinh}(a)}{a}\hat{A} \biggr] u(-\epsilon),
\label{eq:connectingmatrixcp}    
\end{align}
where:
\begin{align}
    \hat{A} &= \begin{pmatrix} 0 & 2\tilde{v}_2 & i\frac{\Delta\theta(Y_R-Y_L)}{2} & 0  \\ 2\tilde{v}_2  & 0 & 0 & i\frac{\Delta\theta(Y_R-Y_L)}{2}, \\
    i\frac{\Delta\theta(Y_R-Y_L)}{2} & 0 & 0 & -2\tilde{v}_2 \\
    0 & i\frac{\Delta\theta(Y_R-Y_L)}{2} & -2\tilde{v}_2 & 0
    \end{pmatrix}, &&
    a &= \sqrt{4\tilde{v}^2_2 - \frac{\Delta\theta^2(Y_R-Y_L)^2}{4}} .
\end{align}
Note that the spinor $u(x)$ is not continuous at $x=0$. This is due to the delta-functions in $\tilde{v}_2(x)$ (see (\ref{eq:v2tilde}) and Figure \ref{fig:v2tilde}). It is a known issue that the presence of delta distributions in the Dirac equation leads to a discontinuity of the spinor's wave function \cite{10.1119/1.19283,1987AmJPh..55..737C,PhysRevA.24.1194,PhysRevA.47.3417} (in contrast to the discontinuity in the derivative of the wave function when dealing with delta-distribution potentials in the Schrödinger equation). In the limit of a standard domain wall, $a \rightarrow 0 $ and $\hat{A}$ is a zero matrix, therefore we recover the continuity condition of the quark's spinor at $x=0$ given by $u(-\epsilon) = u(+\epsilon)$. \\
The Dirac spinor of an incident particle moving in the positive $x$-direction is given by:
\begin{align}
    u_{inc}(x,t) =  e^{(-iEt + ipx)}\begin{pmatrix} \sqrt{E+m_u} \\ 0 \\ 0 \\ \dfrac{p}{\sqrt{E+m_u}}\end{pmatrix}.
\label{eq:incident}    
\end{align}
Using (\ref{eq:incident}) alongside the equations (\ref{eq:fardw}) and (\ref{eq:connectingmatrixcp}), we can find the solution for the spinor components:
\begin{align}
    u_{1p} &= e^{-ib_1} \biggl(\frac{1}{\sqrt{E+m_u}}\biggr) \frac{4a^2p^2E\text{ cosh}(a)}{4a^2E^2 \text{ cosh}^2(a) - (2a_1m_u + ib_2p)^2\text{ sinh}^2(a)}, \label{eq:spinor1p} \\
    u_{2p} &= e^{-ib_1} \biggl(\frac{1}{4a\sqrt{E+m_u}}\biggr) \frac{-8a^2p^2(2a_1m_u+ib_2p)\text{ sinh}(a)}{4a^2E^2 \text{ cosh}^2(a) - (2a_1m_u + ib_2p)^2\text{ sinh}^2(a)}, \label{eq:spinor2p} \\
    u_{1m} &= \biggl(2E\sqrt{E+m_u}\biggr)\frac{-2a^2m_u\text{ cosh}^2(a) + a_1(2a_1m_u + ib_2p)\text{ sinh}^2(a)}{4a^2E^2 \text{ cosh}^2(a) - (2a_1m_u + ib_2p)^2\text{ sinh}^2(a)},\label{eq:spinor1m} \\
    u_{2m} &= \frac{-ap\sqrt{E+m_u}(2a_1p - ib_2m_u)\text{ sinh}(2a)}{4a^2E^2 \text{ cosh}^2(a) - (2a_1m_u + ib_2p)^2\text{ sinh}^2(a)}, 
\label{eq:spinor2m}    
\end{align}
with:
\begin{align}
    a_1 &= 2\tilde{v}_2, && b_1 = (Y_L + Y_R)\dfrac{\Delta\theta}{2}, && b_2 =  (Y_R - Y_L)\Delta\theta.
\end{align}
In order to get the transmission and reflection coefficients of particles scattering off the wall, we need to calculate the fermion currents on both regions:
\begin{align}
    \mathcal{J}_{inc} &= u^{\dag}_{inc}\gamma_0\gamma_1u_{inc}, \\
    \mathcal{J}_{tra} &= u^{\dag}_{tra}\gamma_0\gamma_1u_{tra}, \\
    \mathcal{J}_{ref} &= u^{\dag}_{ref}\gamma_0\gamma_1u_{ref}.
\end{align}
Using the expression (\ref{eq:incident}) for the incident spinor, we can derive the transmission and reflection coefficients for the up-type quark scattering off the wall:
\begin{align}
    \hat{\text{R}}\text{(p)} &= -\frac{\mathcal{J}_{ref}}{\mathcal{J}_{inc}} = \frac{1}{E+m_u}\biggl( \abs{u_{1m}}^2 + \abs{u_{2m}}^2  \biggr),\\
    \hat{\text{T}}\text{(p)} &= \frac{\mathcal{J}_{tra}}{\mathcal{J}_{inc}} = \frac{1}{E-m_u}\biggl( \abs{u_{1p}}^2 + \abs{u_{2p}}^2  \biggr).
\end{align}
\begin{figure}[H]
     \centering
     \begin{subfigure}[b]{0.49\textwidth}
         \centering
         \includegraphics[width=\textwidth]{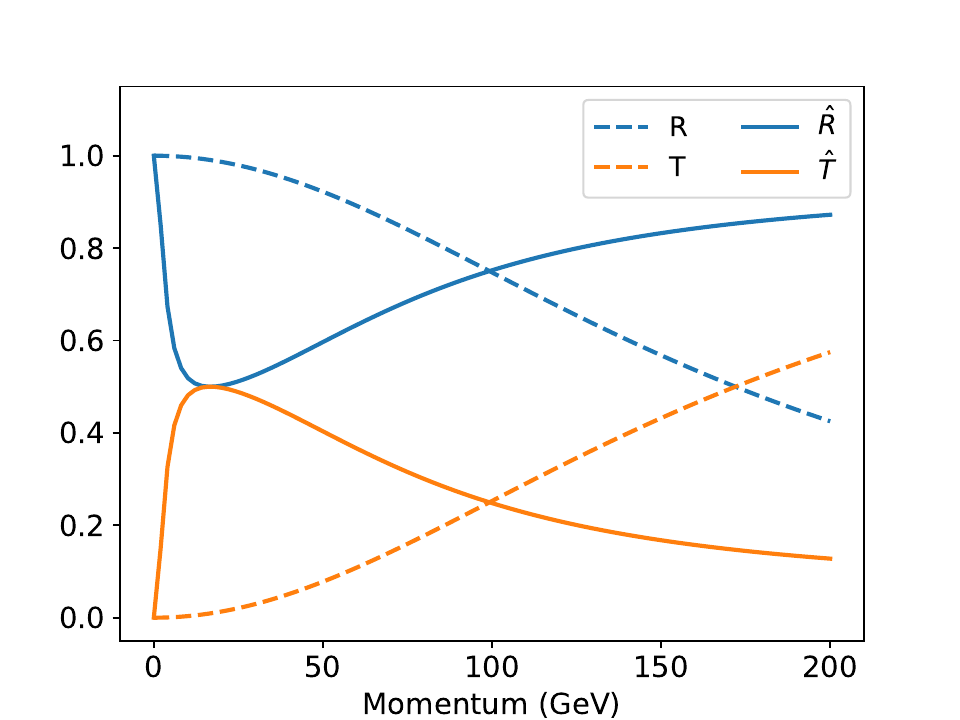}
       \subcaption{}  \label{subfig:cpgeneral1}
     \end{subfigure}
     \hfill
     \begin{subfigure}[b]{0.49\textwidth}
         \centering
         \includegraphics[width=\textwidth]{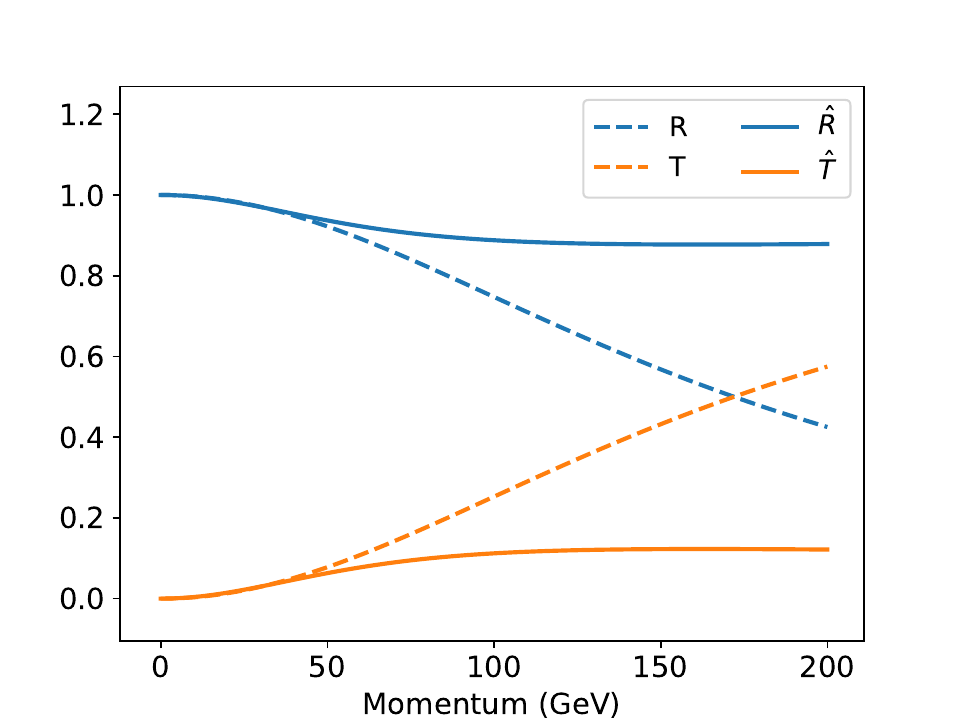}
        \subcaption{} \label{subfig:cpgeneral2}
     \end{subfigure}
\caption{General results for the reflection $\hat{R}$ and transmission $\hat{T}$ coefficients of a top quark scattering off the CP-violating wall (solid lines). We also plot the reflection R and transmission coefficient T for the standard case (no CP-violation, dashed lines). (a) Reflection and transmission with $a_1 = 2.1$ and $b_2$ = 1. (b) Reflection and transmission with $a_1 = 2.1$ and $b_2$ = 3.} 
\label{fig:generalcpresults}
\end{figure}
\noindent
In Figure \ref{fig:generalcpresults}, we show a comparison between the reflection $\hat{R}$ and transmission $\hat{T}$ coefficients of quarks scattering off the CP-violating DW and the coefficients $R$ and $T$ for quarks scattering off standard CP-conserving DW for two different cases. The first case depicted in Figure $\ref{subfig:cpgeneral1}$ shows the results for $a_1 = 2.1$ and $b_2 = 1$, while for the second case (shown in Figure \ref{subfig:cpgeneral2}) the parameters are: $a_1 = 2.1$ and $b_2 = 3$. We see that in both cases, the reflection and transmission coefficients differ a lot from the standard reflection and transmission coefficients. In particular, we see that the reflection coefficient for incident particles with higher momenta grows and stays the dominant process.  
Since the analytical results for the general case are quite complicated, we consider some special cases in order to understand the physical interpretation of the solution. We first consider the case where $b_2$ is small compared to $a_1$. 
In such a scenario, the reflection and transmission coefficients simplify to:
\begin{align}
    \hat{\text{R}}\text{(p)} &= \frac{4m^2_uE^2 + p^4(\text{ cosh}^2(2a_1) -1)}{[(E^2 + m^2_u) + p^2\text{ cosh}(2a_1)]^2} \label{eq:reflectionni} \\
    \hat{\text{T}}\text{(p)} &= \frac{2p^2(p^2 + (E^2 + m^2)\text{ cosh}(2a_1))}{[(E^2 + m^2_u) + p^2\text{ cosh}(2a_1)]^2} 
\end{align}
As shown in Figure \ref{subfig:variationa1ref}, in this case and for high values of $a_1$, the reflection coefficient increases with momentum. This seems counter-intuitive as particles reacting with a potential barrier should have a higher probability of crossing when they have higher energies. One can also deduce from (\ref{eq:reflectionni}) that for big values of CP-violation ($a_1$ is big), the reflection coefficient approaches 1 for all momenta, as the terms proportional to $\cosh^2(2a_1)$ will be dominant.
\noindent
For the opposite case $b_2 >> 2\tilde{v}_2$, the transmission and reflection coefficients are:
\begin{align}
    \hat{\text{R}}\text{(p)} &= \frac{2m_u^2\text{cos}^2(\frac{b_2}{2})}{m_u^2 + 2p^2 + m_u^2\text{cos}(b_2)}, \\
    \hat{\text{T}}\text{(p)} &= \frac{2p^2}{m_u^2 + 2p^2 + m_u^2\text{cos}(b_2)}.
\end{align}
\begin{figure}[H]
\centering
\begin{subfigure}[b]{0.49\textwidth}
\centering
\includegraphics[width=\textwidth]{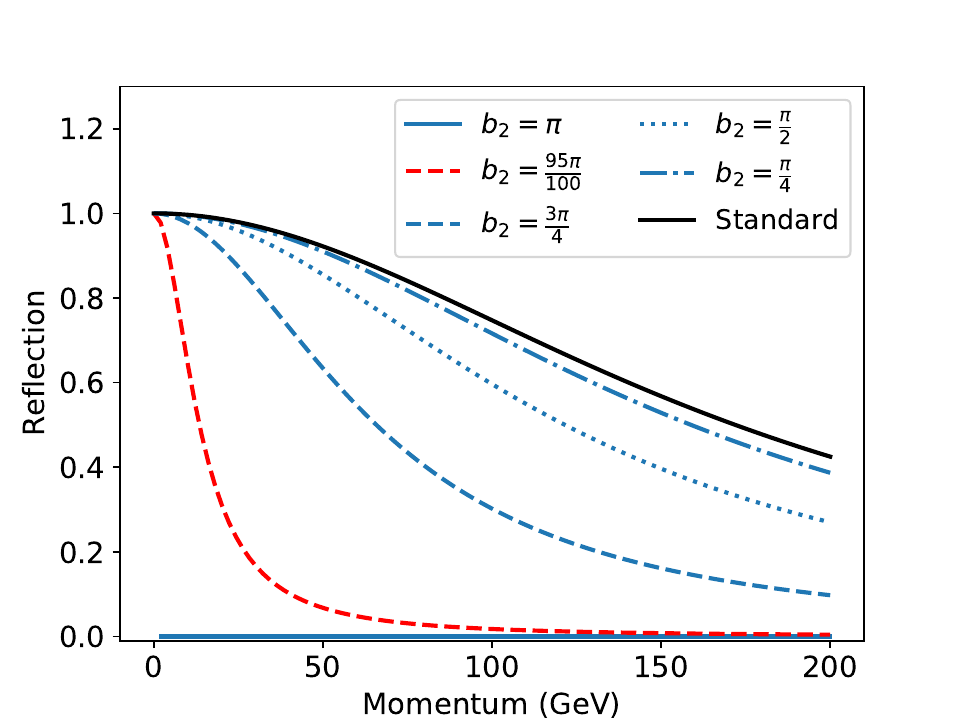}
\subcaption{$\hat{R}$ as a function of $b_2$ and momenta}  \label{subfig:variationb2ref}
\end{subfigure}
\hfill
\begin{subfigure}[b]{0.49\textwidth}
\centering
\includegraphics[width=\textwidth]{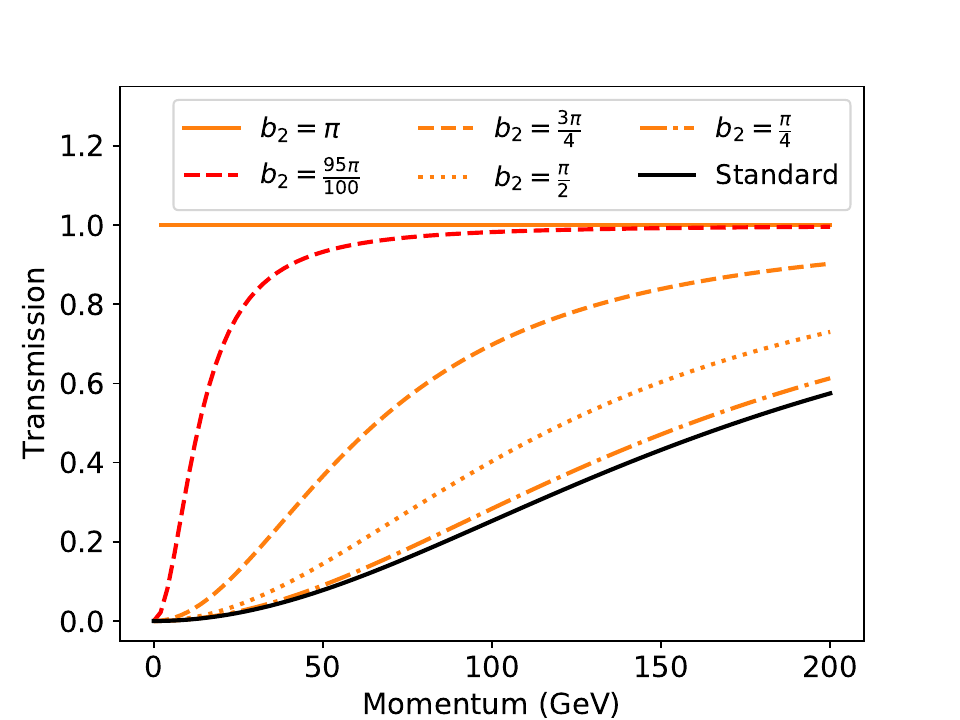}
\subcaption{$\hat{T}$ as a function of $b_2$ and momenta} \label{subfig:variationb2tra}
\end{subfigure}
\caption{Results for the reflection (a) and transmission (b) coefficients for the case when the numerical value of the gauge contribution $b_2$ is much higher than the numerical value of $a_1$. These rates will oscillate depending on $b_2$. The standard reflection and transmission rates (black lines) refer to the scattering off standard domain walls (\ref{eq:standardrates}).}
\label{fig:b2dominant}
\end{figure}
\noindent
For this case, the reflection and transmission coefficients will oscillate with a $b_2$ dependence (see Figure \ref{fig:b2dominant}). Notice that for $b_2 = \pi$ all particles will be transmitted, irrespective of their incoming momentum.
\begin{figure}[H]
     \centering
     \begin{subfigure}[b]{0.49\textwidth}
         \centering
         \includegraphics[width=\textwidth]{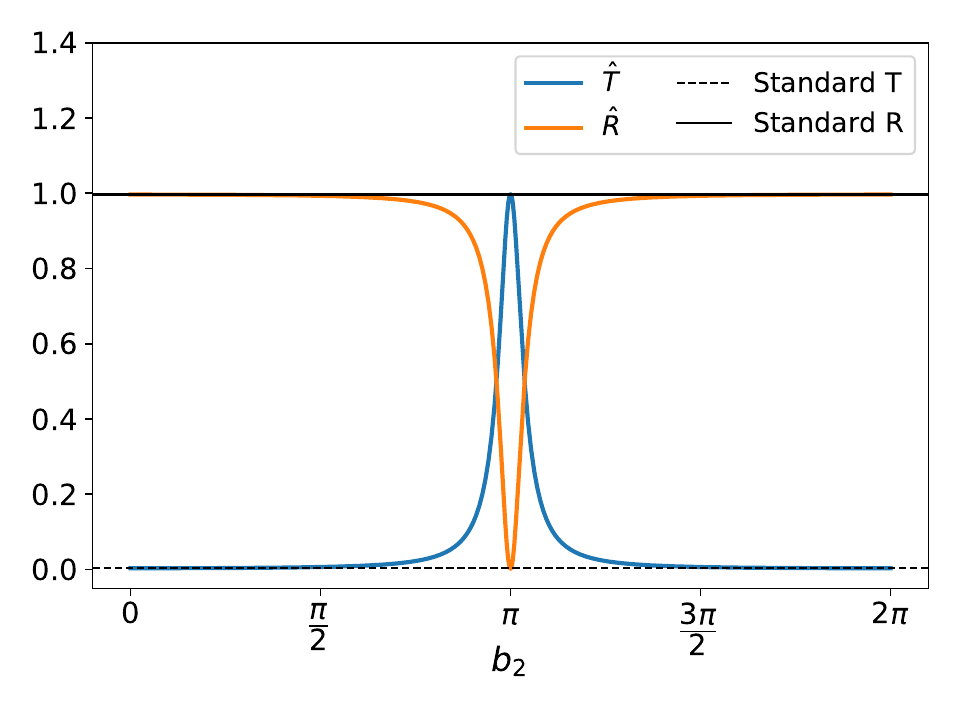}
       \subcaption{p = 10 GeV}  \label{subfig:b2dep10}
     \end{subfigure}
     \hfill
     \begin{subfigure}[b]{0.49\textwidth}
         \centering
         \includegraphics[width=\textwidth]{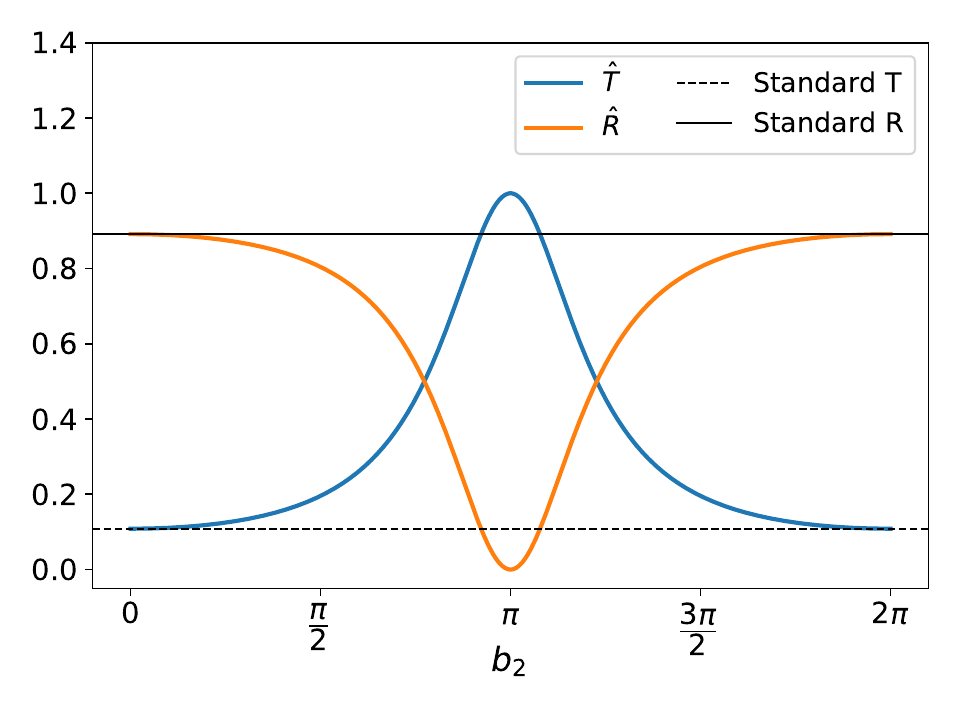}
        \subcaption{p = 60 GeV} \label{subfig:b2dep60}
     \end{subfigure}
\caption{Variation of the reflection and transmission coefficient for a top quark as a function of $b_2$ for different fixed momenta: (a) $p=10 \text{ GeV}$ and (b) $p=60 \text{ GeV}$ . The standard reflection and transmission rates refer to the scattering off standard domain walls (\ref{eq:standardrates}).}
\label{fig:b2dep}
\end{figure}
\noindent
In Figure \ref{fig:b2dep}, we show the oscillatory behavior of the reflection and transmission coefficients for this special case as a function of $b_2$ for different momenta of the incident particle. We see that for small momenta (see Figure \ref{subfig:b2dep10}), the deviations of the scattering rates from those of the standard DW are only relevant for $b_2$ in the vicinity of $b_2 = \pi$. In contrast with particles with higher momenta (see Figure \ref{subfig:b2dep60}).\\\\
We now calculate the rate of reflection and transmission of the left handed (LH) and right handed (RH) components, derived from the currents of left and right-handed particles: 
\begin{align}
    \mathcal{J}^L_{tra} &= \bar{u}_{tra} \gamma_1 P_Lu_{tra}, \\
    \mathcal{J}^R_{tra} &= \bar{u}_{tra} \gamma_1 P_R u_{tra}, \\
    \mathcal{J}^L_{ref} &= \bar{u}_{ref} \gamma_1 P_L u_{ref}, \\
    \mathcal{J}^R_{ref} &= \bar{u}_{ref} \gamma_1 P_R u_{ref}.
\end{align}
leading to the transmission and reflection coefficients:
\begin{align}
\text{T}_L &= \frac{\mathcal{J}^L_{tra}}{\mathcal{J}_{inc}} = \frac{1 }{4p}\biggl[ 2d\biggl( \abs{u_{1p}}^2 + \abs{u_{2p}}^2 \biggr) - (1+d^2)\biggl(u^*_{1p}u_{2p} + u_{1p}u^*_{2p} \biggr) \biggr],  \\
\text{T}_R &= \frac{\mathcal{J}^R_{tra}}{\mathcal{J}_{inc}} = \frac{1}{4p}\biggl[ 2d\biggl( \abs{u_{1p}}^2 + \abs{u_{2p}}^2 \biggr) + (1+d^2)\biggl(u^*_{1p}u_{2p} + u_{1p}u^*_{2p} \biggr) \biggr],  \\
\text{R}_L &= -\frac{\mathcal{J}^L_{ref}}{\mathcal{J}_{inc}} = \frac{1}{4p}\biggl[ 2g\biggl( \abs{u_{1m}}^2 + \abs{u_{2m}}^2 \biggr) - (1+g^2)\biggl(u^*_{1m}u_{2m} + u_{1m}u^*_{2m} \biggr) \biggr], \\
\text{R}_R &= -\frac{\mathcal{J}^R_{ref}}{\mathcal{J}_{inc}} = \frac{1 }{4p}\biggl[ 2g\biggl( \abs{u_{1m}}^2 + \abs{u_{2m}}^2 \biggr) + (1+g^2)\biggl(u^*_{1m}u_{2m} + u_{1m}u^*_{2m} \biggr) \biggr],
\end{align}
where:
\begin{align}
    d &= \dfrac{p}{E-m_u}, && g = \dfrac{p}{E+m_u}. 
\end{align}
From equations (\ref{eq:spinor1p})-(\ref{eq:spinor2m}) it is clear that for $a_1 = 0$ (as in the case of the standard domain wall), $(u^*_{1p}u_{2p})$ and $(u^*_{1m}u_{2m})$ are purely imaginary expressions leading to the vanishing of the second terms in the equations for $T_{L,R}$ and $R_{L,R}$. In such a case the reflection and transmission coefficient do not depend on the chirality of the particle.\\
Because analytical expressions for the general case are lengthy and complicated,  we give for simplicity the analytical expressions in the limit when $b_2 \rightarrow 0$ :
\begin{align}
\text{T}_L &= \frac{p(p^3 + p(2m^2 + p^2)\cosh(2a) + 
       2m(m^2 + p^2)\sinh(2a))}{(2m^2 + p^2 + p^2\cosh(2a))^2}, \label{eq:TL} \\
\text{T}_R &= \frac{p(p^3 + p(2m^2 + p^2)\cosh(2a) - 
       2m(m^2 + p^2)\sinh(2a))}{(2m^2 + p^2 + p^2\cosh(2a))^2}, \label{eq:TR} \\
\text{R}_R &= \frac{4m^2E^2 - 8mpE^2\cosh(a)\sinh(a) + p^4\sinh^2(2a)}{2(2m^2 + p^2 + p^2\cosh(2a))^2)} \label{eq:RR}, \\
\text{R}_L &=\frac{4m^2E^2 + 8mpE^2\cosh(a)\sinh(a) + p^4\sinh^2(2a)}{2(2m^2 + p^2 + p^2\cosh(2a))^2}. \label{eq:RL}
\end{align}
\begin{figure}[H]
     \centering
     \begin{subfigure}[b]{0.49\textwidth}
         \centering
         \includegraphics[width=\textwidth]{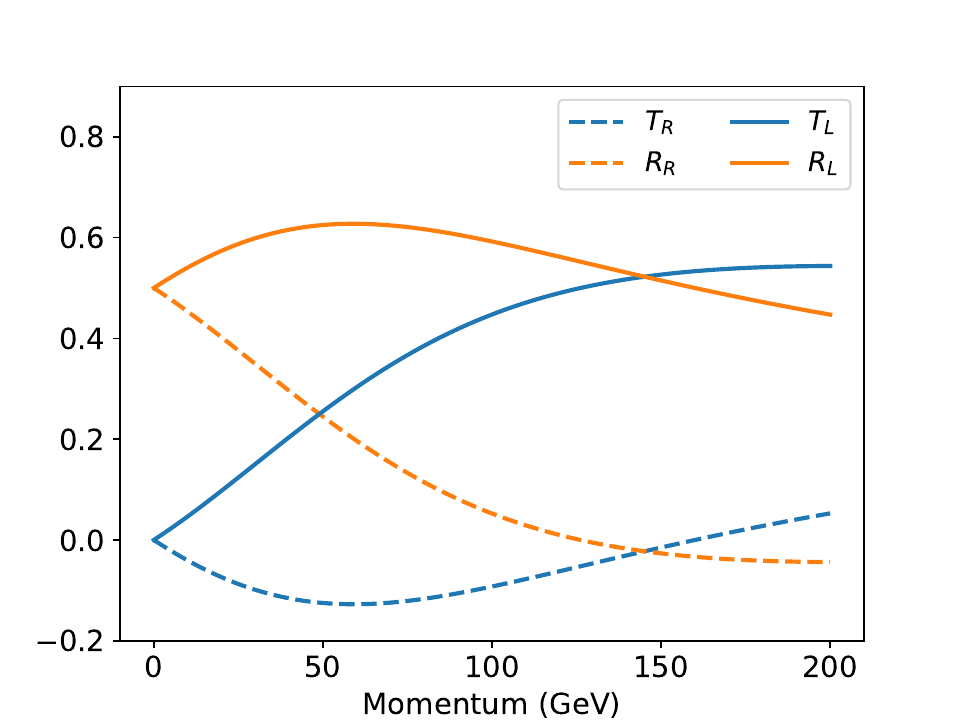}
       \subcaption{}  \label{subfig:tltrtop}
     \end{subfigure}
     \hfill
     \begin{subfigure}[b]{0.49\textwidth}
         \centering
         \includegraphics[width=\textwidth]{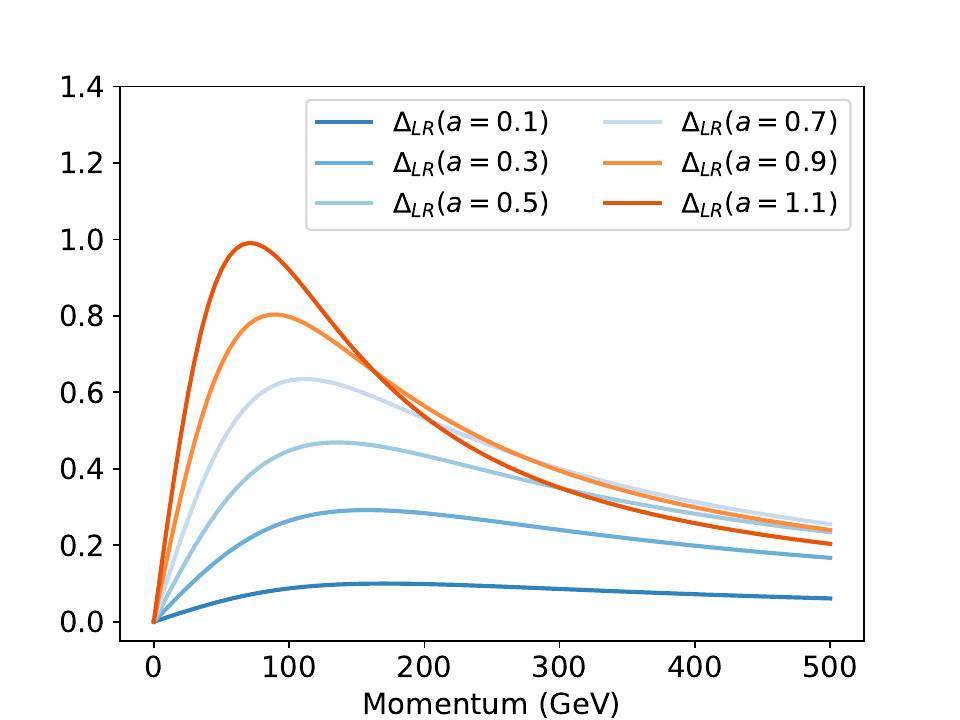}
        \subcaption{} \label{subfig:deltarltop}
     \end{subfigure}
\caption{(a) Left and right-handed reflection and transmission currents normalized to the incident current. We take $a_1 = 0.5$. (b) The difference $\Delta_{LR}$ between the reflection of left and right-handed particles as a function of the CP-violating rate $a_1$.}
\label{fig:leftrighttop}
\end{figure}
\noindent
Note that for small momenta, these rates can become negative (see Figure \ref{subfig:tltrtop} for $a_1 = 0.5$). This is not an issue as these rates are not describing probability amplitudes for the particles and only $\hat{R}$ and $\hat{T}$ should be, in principle, positive numbers between 0 and 1. In case $a_1 \rightarrow 0$, we do not observe CP-violation in the scattering rates of the particle. In order to get a rate for the CP-violation, we calculate the difference $\Delta_{LR} = \text{R}_L - \text{R}_R$ . The results are shown in Figure \ref{subfig:deltarltop}. We find from (\ref{eq:TL})-(\ref{eq:RL}), that when the momentum of the incoming particles gets larger, the left and right-handed rates should converge to the same value. This explains the behavior of $\Delta_{LR}$ becoming smaller at higher momenta. Since we get different rates for $\text{R}_L$ and $\text{R}_R$, the motion of the wall will generate an axial asymmetry in front of it. 
\subsection{Charge violating interactions of fermions with the domain walls}
\begin{figure}[H]
     \centering
     \begin{subfigure}[b]{0.49\textwidth}
         \centering
         \captionsetup{justification=centering}
         \includegraphics[width=\textwidth]{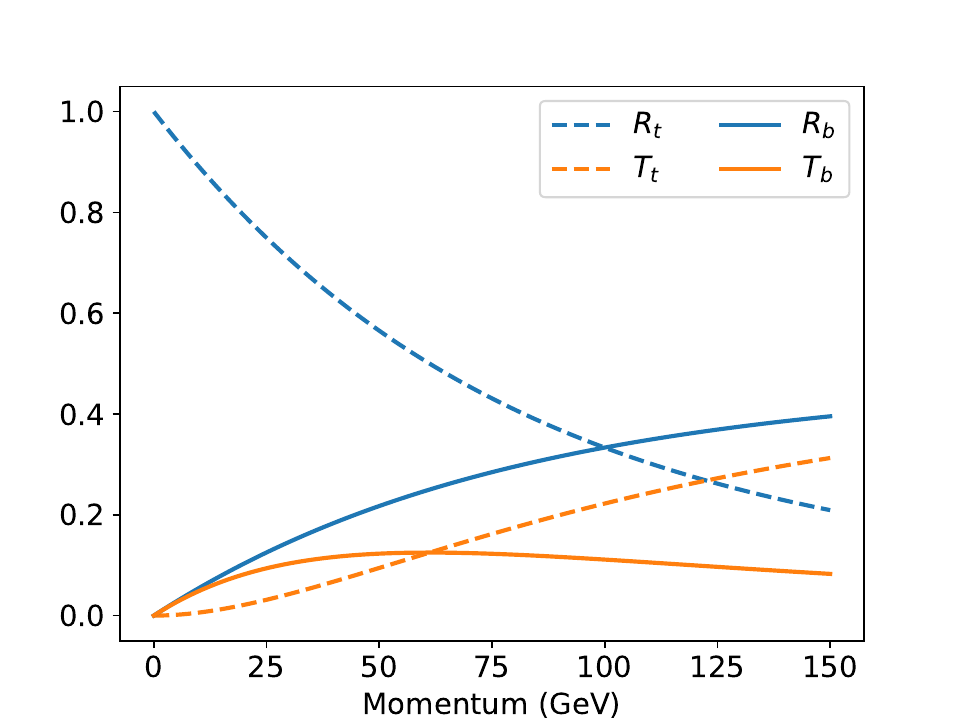}
        \subcaption{Reflection/Transmission coefficients for \\ top and bottom quarks.} \label{subfig:generalcasecharged}
     \end{subfigure}
     \hfill
     \begin{subfigure}[b]{0.49\textwidth}
         \centering
           \captionsetup{justification=centering}
         \includegraphics[width=\textwidth]{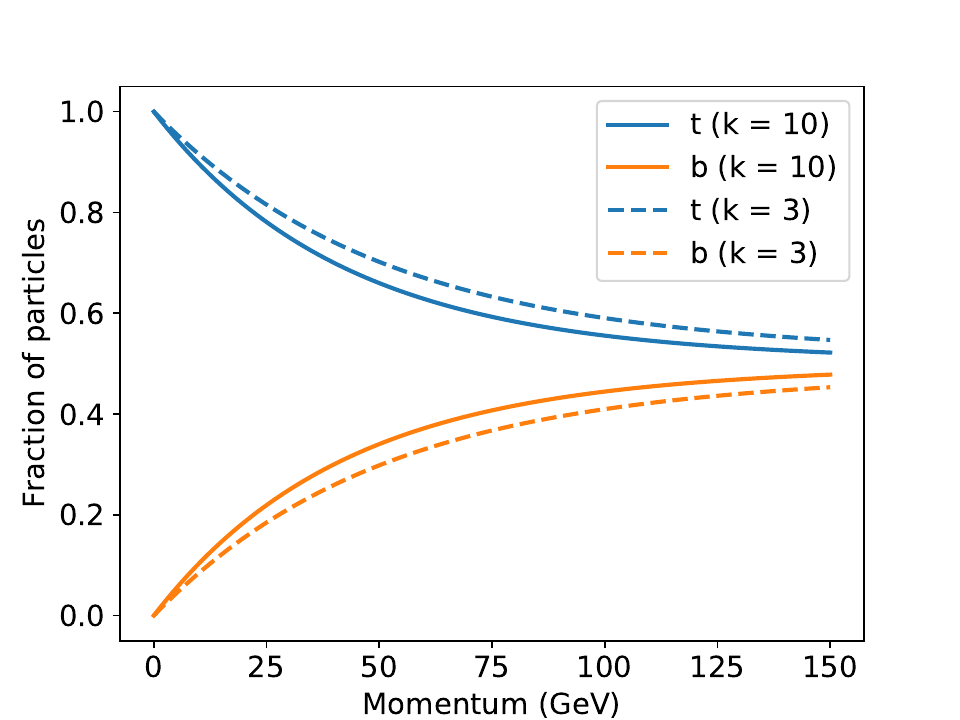}
         \subcaption{Fraction of top (t) and bottom (b) quarks after scattering.}\label{subfig:chargedamount}
     \end{subfigure}
\caption{(a) Reflection and transmission coefficients for top and bottom quarks after the interaction of an initial top quark with the charge breaking domain wall. Note that the sum of reflection and transmission coefficients for both particles adds up to 1. The parameters used are $k=10$ and $m=0.5$ (b) Rate of top quarks being transformed into bottom quarks or kept as top quarks. Notice that, for high momenta, the process of reflection a bottom quark has the highest probability. The amount of produced bottom quarks gets higher with increasing $k$.}     
\end{figure}

\begin{figure}[H]
     \centering
     \begin{subfigure}[b]{0.49\textwidth}
         \centering
         \includegraphics[width=\textwidth]{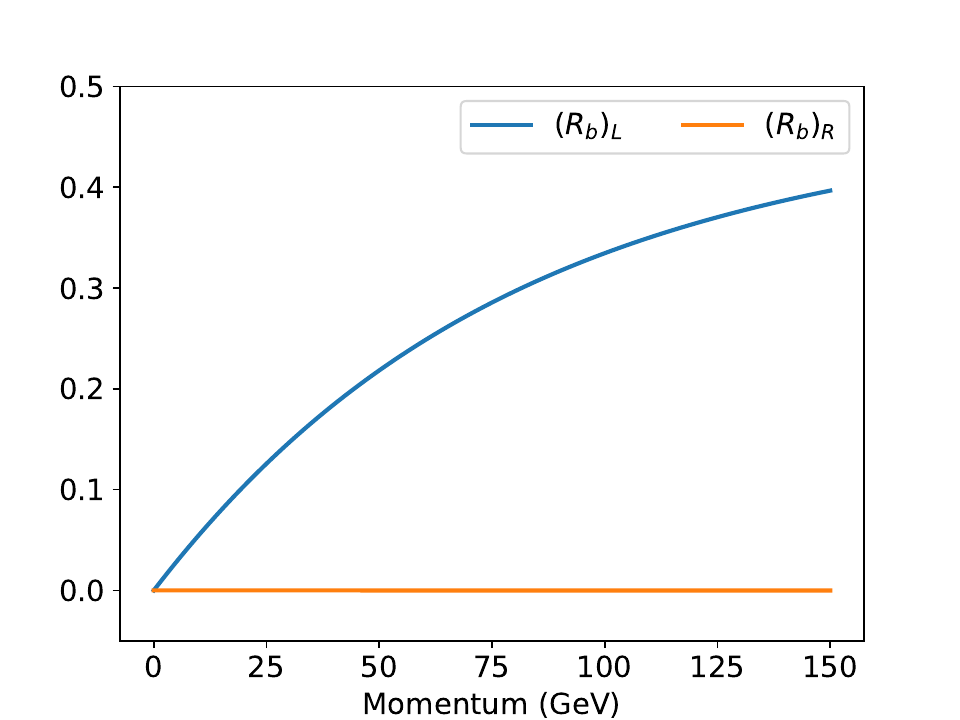}
         \subcaption{}\label{subfig:RDLRDR}
     \end{subfigure}
     \hfill
     \begin{subfigure}[b]{0.49\textwidth}
         \centering
         \includegraphics[width=\textwidth]{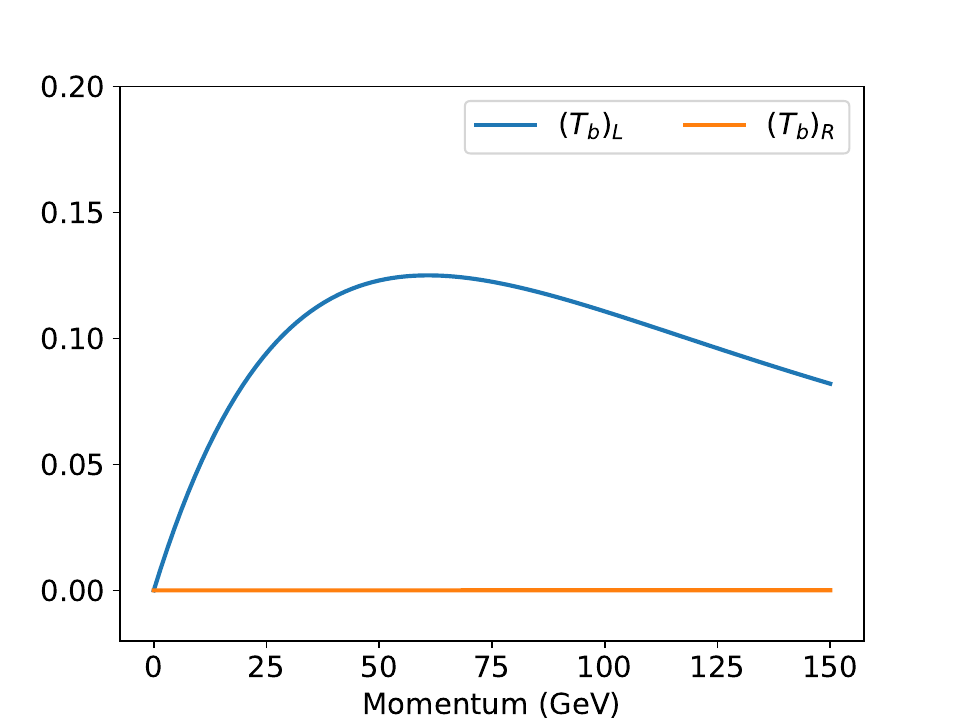}
          \subcaption{} \label{subfig:TDLTDR}
   \end{subfigure}
        \centering
     \begin{subfigure}[b]{0.49\textwidth}
         \centering
         \includegraphics[width=\textwidth]{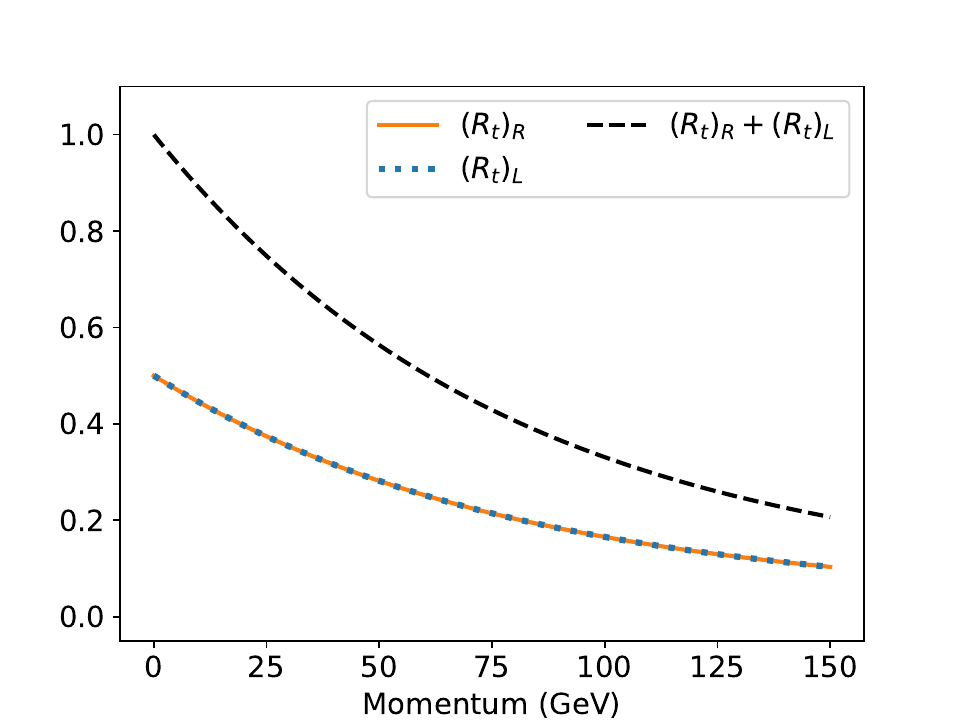}
         \subcaption{}\label{subfig:RULRUR}
     \end{subfigure}
     \hfill
     \begin{subfigure}[b]{0.49\textwidth}
         \centering
         \includegraphics[width=\textwidth]{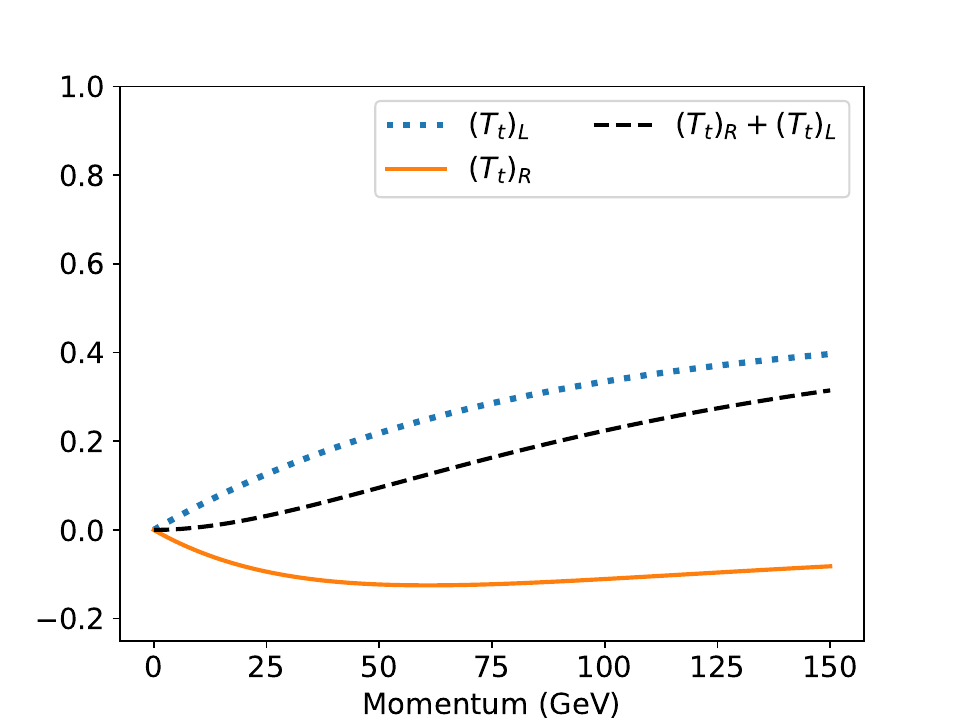}
          \subcaption{} \label{subfig:TULTUR}
   \end{subfigure}
\caption{Chiralities of the reflected and transmitted particles: (a) Reflection coefficients for $\mathbf{LH}$ and $\mathbf{RH}$ bottom quarks. (b) Transmission coefficients for $\mathbf{LH}$ and $\mathbf{RH}$ bottom quarks. Notice that all the produced bottom quarks are left handed. (c) Reflection coefficients for $\mathbf{LH}$ and $\mathbf{RH}$ top quarks. (d) Transmission coefficients for $\mathbf{LH}$ and $\mathbf{RH}$ top quarks. We observe that the reflection rate as a top quark does not depend on the chirality of the particle, while the transmission for the top quark is chirality-dependent. }
\label{fig:lefthanded}
\end{figure}

\begin{figure}[H]
     \centering
     \begin{subfigure}[b]{0.49\textwidth}
         \centering
         \includegraphics[width=\textwidth]{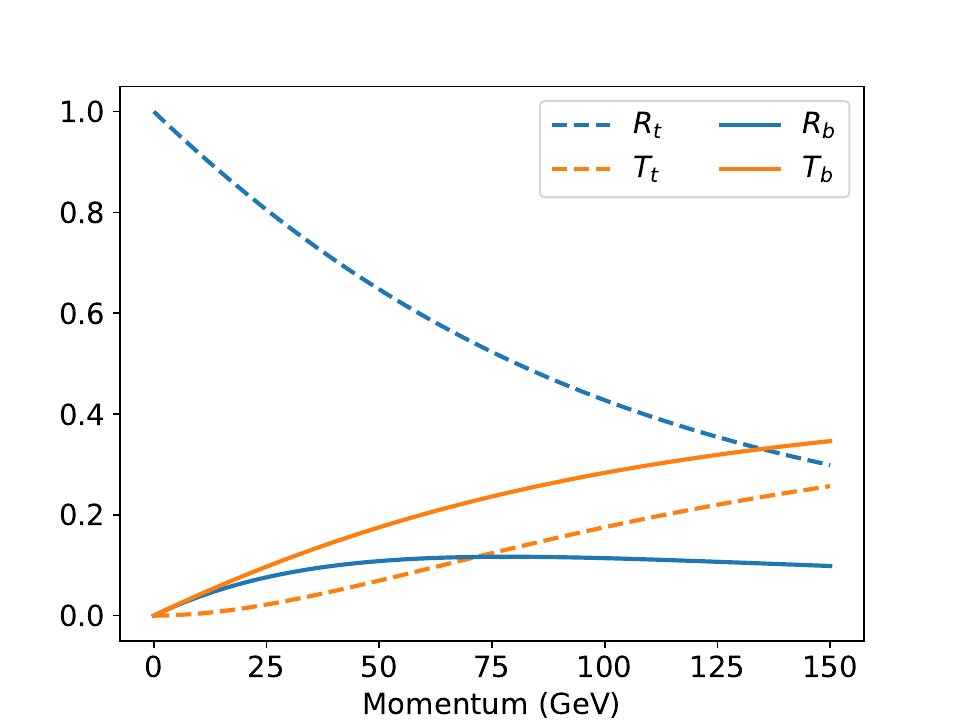}
        \subcaption{} \label{subfig:generalcasechargedmbigger}
     \end{subfigure}
     \hfill
     \begin{subfigure}[b]{0.49\textwidth}
         \centering
         \includegraphics[width=\textwidth]{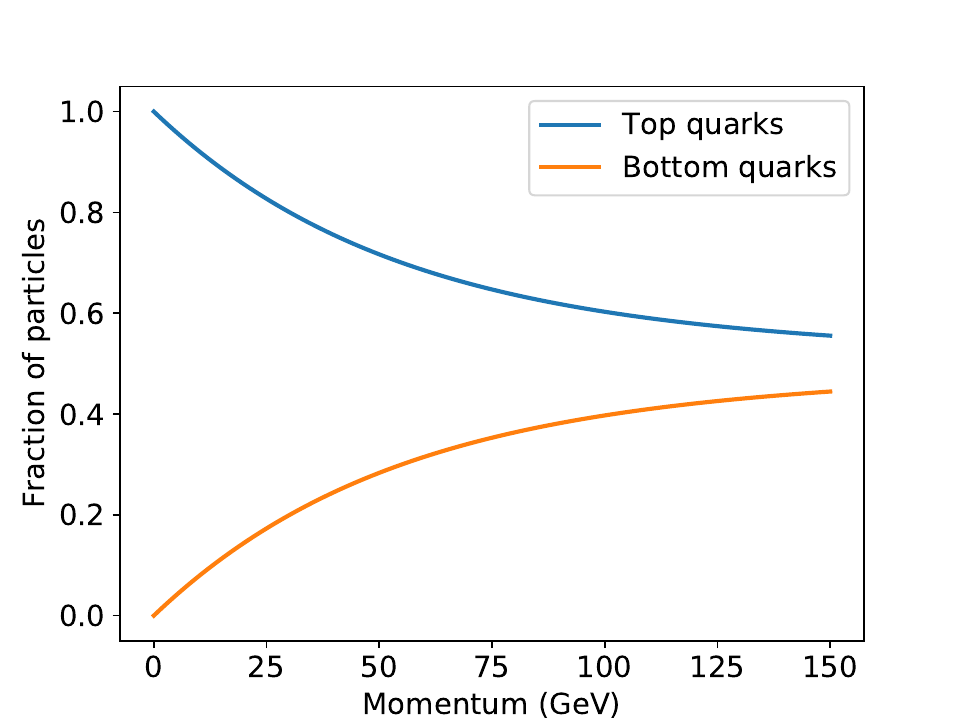}
        \subcaption{} \label{subfig:chargedamountmbigger}
     \end{subfigure}
          \centering
     \begin{subfigure}[b]{0.49\textwidth}
         \centering
         \includegraphics[width=\textwidth]{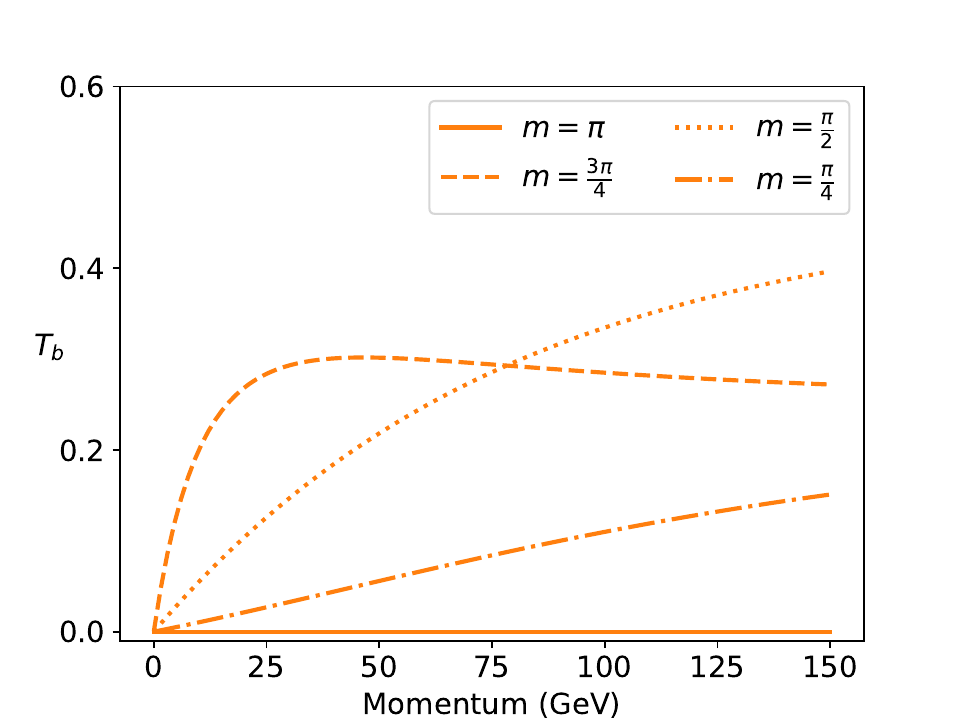}
        \subcaption{} \label{subfig:variationtbm}
     \end{subfigure}
     \hfill
     \begin{subfigure}[b]{0.49\textwidth}
         \centering
         \includegraphics[width=\textwidth]{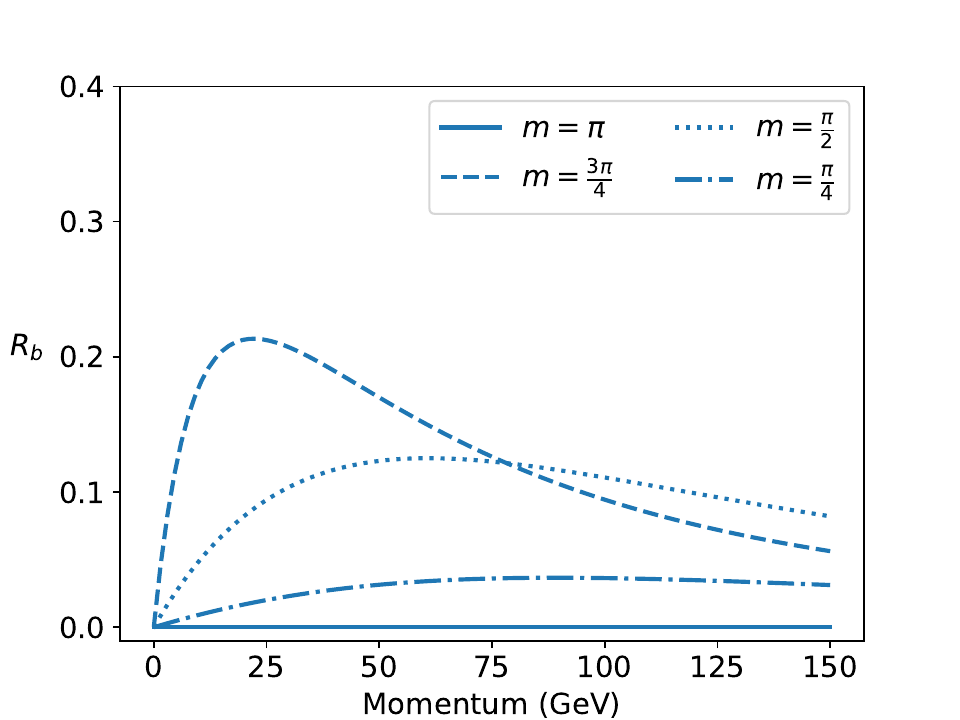}
        \subcaption{} \label{subfig:variationrbm}
     \end{subfigure}
\caption{(a) Reflection and transmission coefficients for top and bottom quarks in the case of $m>k$. (b) Rate of top quarks being transformed into bottom quarks or kept as top quarks. (c) and (d) Transmission and reflection coefficients for different values of $m$.} 
\label{fig:mbigger}
\end{figure}

\begin{figure}[H]
\centering
\includegraphics[width=0.65\textwidth]{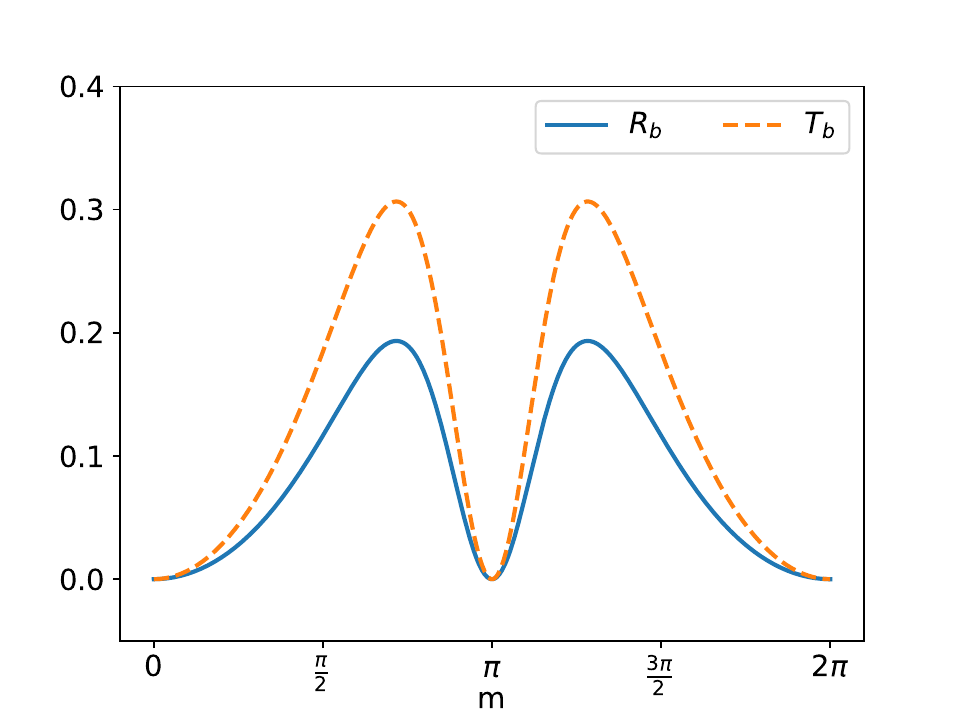}
\caption{Reflection and transmission coefficients as a bottom quark for an incident top quark with momentum $p=40 \text{ GeV}$ as a function of $m$, cf.(\ref{eq:parameterscharge}).}
\label{fig:variationm}
\end{figure}

\begin{figure}[h]
     \centering
     \begin{subfigure}[b]{0.49\textwidth}
         \centering
         \includegraphics[width=\textwidth]{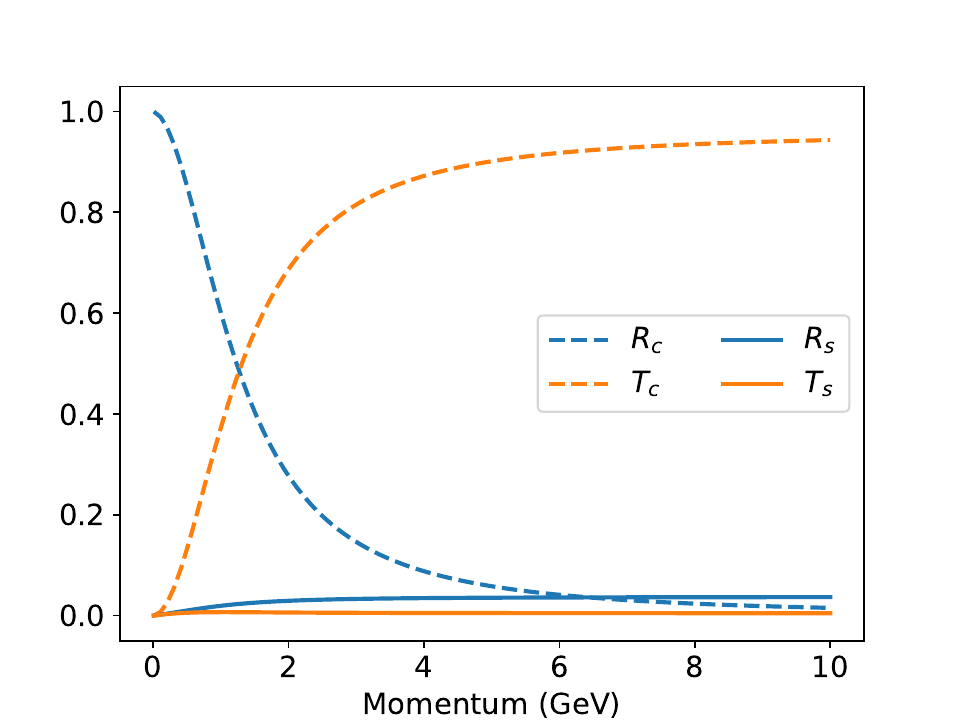}
        \subcaption{} \label{fig:generalcharm}
     \end{subfigure}
     \hfill
     \begin{subfigure}[b]{0.49\textwidth}
         \centering
         \includegraphics[width=\textwidth]{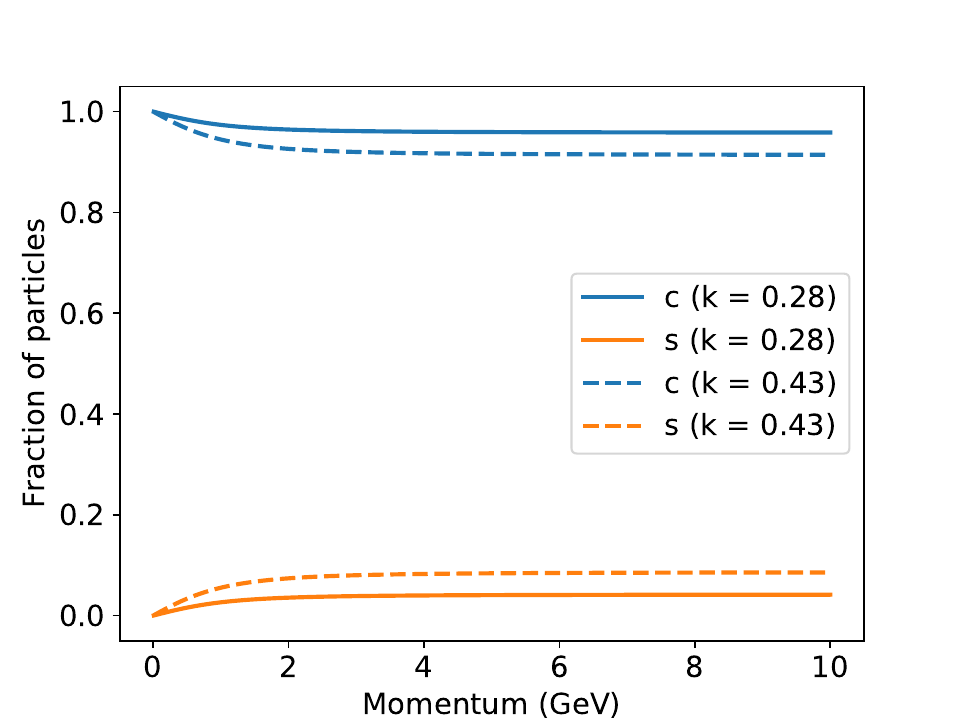}
        \subcaption{} \label{fig:fractioncharm}
     \end{subfigure}
\caption{(a) Reflection and transmission coefficients for second generation quarks. We take $k = 0.28$, $m=0.1$, $m_c = 1.27 \text{ GeV}$ and $m_s = 0.095 \text{ GeV}$ corresponding to charm (denoted c) and strange quarks (denoted s). Note that the conversion rate to strange quarks is very small even for $v_+ = 65 \text{ GeV}$ corresponding to $k = 0.43$. (b) Ratio of charm and strange quarks after the scattering for different values of $k$.}
\label{fig:lowkcharm}
\end{figure}

\begin{figure}[H]
     \centering
     \begin{subfigure}[b]{0.49\textwidth}
         \centering
         \includegraphics[width=\textwidth]{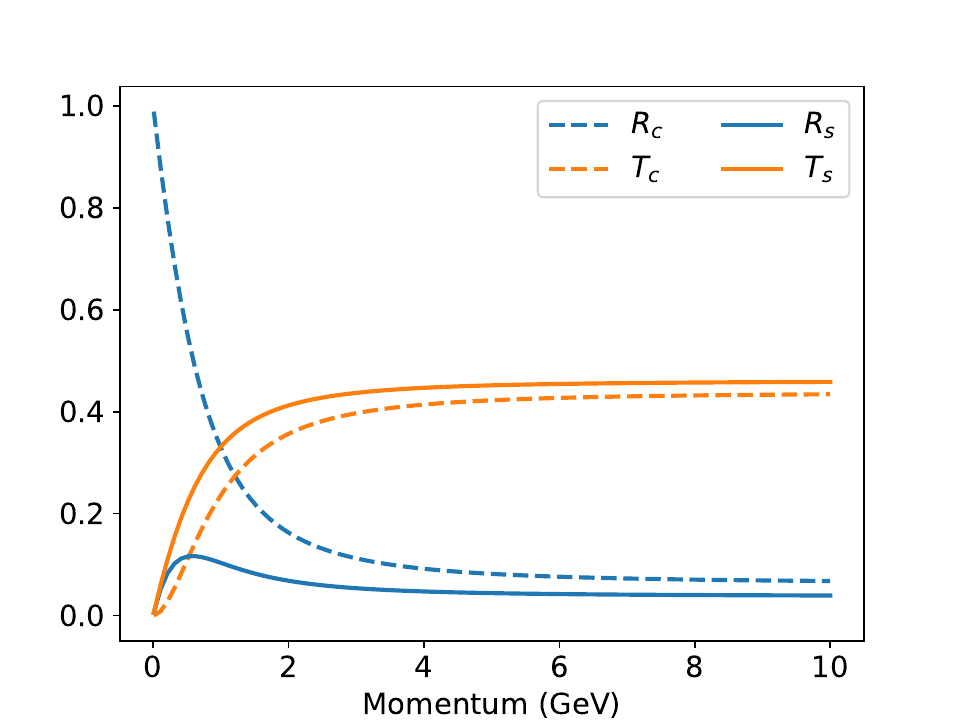}
        \subcaption{} \label{subfig:charmm}
     \end{subfigure}
     \hfill
     \begin{subfigure}[b]{0.49\textwidth}
         \centering
         \includegraphics[width=\textwidth]{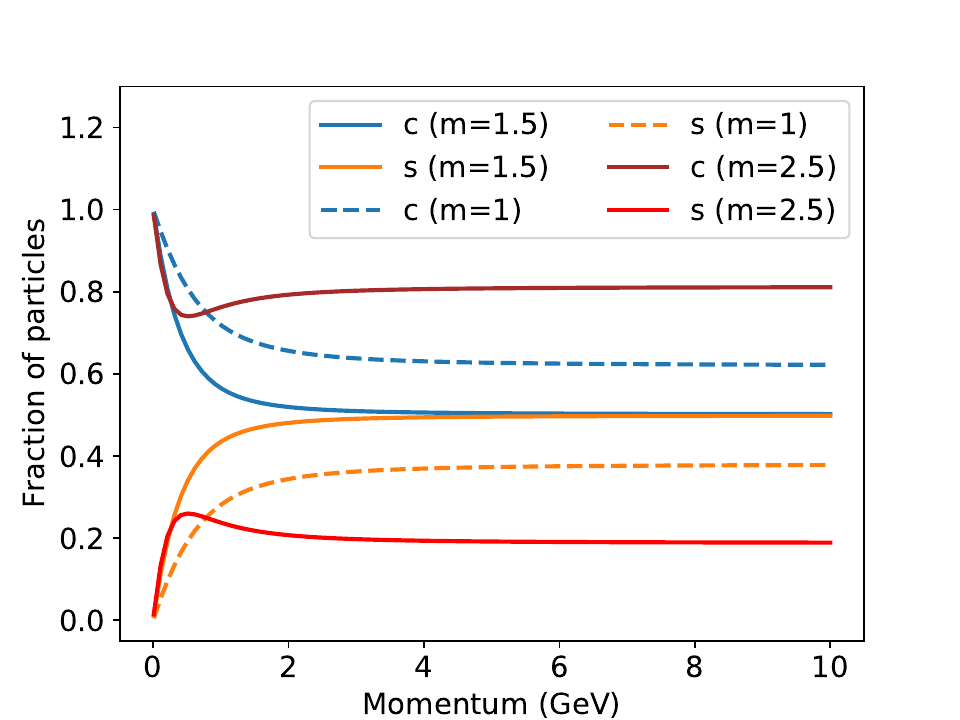}
        \subcaption{} \label{fig:fractioncharmm}
     \end{subfigure}
\caption{(a) Reflection and transmission coefficients for second generation quarks. We take $k = 0.43$, $m=1.5$, $m_c = 1.27 \text{ GeV}$ and $m_s = 0.095 \text{ GeV}$ corresponding to charm (denoted c) and strange quarks (denoted s). (b) Fraction of particles after scattering with the wall for different values of $m$.  }
\label{fig:highmcharm}
\end{figure}

We now discuss the case when the domain wall exhibits a non negligible value for $v_+$ inside its core. In this case the vacuum inside the domain wall is charge-breaking and the Yukawa sector inside the wall now includes couplings between fermions with different charges. \\
In the following, we consider the case of a Yukawa sector in the type-2 2HDM. The relevant terms are:
\begin{equation}
\mathcal{L}_{Yukawa} = -y_u \bar{Q}_L\Phi^{c}_2u_R - y_d \bar{Q}_L\Phi_1d_R - y_l\bar{L}_L\Phi_1e_R +h.c,
\end{equation}
where $\Phi^{c}_2 = i\sigma_2\Phi_2$. $y_u$, $y_d$ and $y_l$ denote the Yukawa couplings for the up-type quarks, down-type quarks and leptons respectively. $Q_L$ and $L_L$ denote the left-handed quark and lepton doublets charged under $SU_L(2)$ and finally $u_R$, $d_R$ and $e_R$ denote the right-handed up-type, down-type quark and lepton singlet under $SU_L(2)$.
We take the simplification that only one of the Goldstone modes changes between the two domains. In this case, it is the second generator of SU(2) which leads to charge violating effects inside the domain wall and the CP-violating phase $\xi(x)$ vanishes (see \ref{subfig:g2neumann}). The Higgs doublets are: 
\begin{align}
\Phi_1 (x) &= \frac{U(x)}{\sqrt{2}} \begin{pmatrix} 0  \\  v_1(x) \end{pmatrix},  & \Phi_2 (x) &= \frac{U(x)}{\sqrt{2}} \begin{pmatrix} v_+(x)  \\  v_2(x) \end{pmatrix}.
\end{align}
The matrix $U(x)$ in this case is given by:
\begin{equation}
U(x) = \begin{pmatrix} \cos(\frac{g_2(x)}{2}) & \sin(\frac{g_2(x)}{2})  \\  -\sin(\frac{g_2(x)}{2}) & \cos(\frac{g_2(x)}{2}) \end{pmatrix}.
\end{equation}
We can remove this matrix from the Yukawa sector by performing a gauge transformation, leading to a pure gauge term for the gauge field $W^{\mu}_2$ confined inside the wall: 
\begin{align}
    & \Phi_1(x) \xrightarrow{SU(2)} \text{U}^{-1}(x) \Phi_1(x) = \tilde{\Phi}_1(x), \\
    & \Phi_2(x) \xrightarrow{SU(2)} \text{U}^{-1}(x) \Phi_2(x) = \tilde{\Phi}_2(x), \\
    & W^{\mu}_2 \frac{\sigma_2}{2} \xrightarrow{SU(2)}  \frac{i}{g}\text{U}(x)\partial_{\mu}(\text{U}^{-1}(x)), 
\end{align}
Writing the Lagrangian in terms of the individual quark fields, we get:
\begin{align}
   \notag \mathcal{L}_{F} =& \ \ i\bar{u}\slashed{\partial}u +  i\bar{d}\slashed{\partial}d  + \frac{1}{2}\biggl[ -i\bar{u}_{L}\biggl(\slashed{\partial}_xg_2(x)\biggr)d_L + i\bar{d}_{L}\biggl(\slashed{\partial}_xg_2(x)\biggr)u_L \biggr] - y_uv_2(x)\bar{u}u - y_dv_1(x)\bar{d}d \\ &
    + y_uv_+(x)(\bar{d}_Lu_R + \bar{u}_R\text{d}_L ).
\label{eq:lagrangiancharged}    
\end{align}
We then derive the Dirac equation for the up and down-type quarks:
\begin{align}
   & i\slashed{\partial}d + \frac{i}{2}\biggl(\slashed{\partial}_xg_2(x)\biggr)P_Lu - y_d v_1(x)d + y_uv_+(x)P_Ru  = 0, 
\label{eq:diraccharged1}   
   \\ &
   i\slashed{\partial}u - \frac{i}{2}\biggl(\slashed{\partial}_xg_2(x)\biggr)P_Ld - y_u v_2(x)u + y_uv_+(x)P_Ld  = 0. 
\label{eq:diraccharged2}   
\end{align}
To solve this system of equations, we first re-write it in the matrix form introduced in the previous chapter, cf.(\ref{eq:matrixnotationdiraccp}):
\begin{equation}
    \begin{pmatrix}
        i\slashed{\partial} - y_uv_2(x) & -\hat{G}_2(x) \\
        -\hat{G}_1(x) & i\slashed{\partial} - y_dv_1(x) 
    \end{pmatrix} \begin{pmatrix}
        u \\ d
    \end{pmatrix} = 0,
\label{eq:matrixchargedgeneral}    
\end{equation}
where:
\begin{align}
    \hat{G}_1(x) &= -\frac{i}{2}\biggl(\slashed{\partial}_xg_2(x)\biggr)P_L - y_uv_+(x)P_R, \\ 
    \hat{G}_2(x) &= \frac{i}{2}\biggl(\slashed{\partial}_xg_2(x)\biggr)P_L - y_uv_+(x)P_L.
\end{align}
Notice that all terms that mix the up and down-type quark only include the left-handed component of the down-type quark. This will induce a chiral asymmetry as we will discuss later.
We consider the scattering of an incident up-type quark off the domain wall. 
In the standard case, when $v_+ = 0$, the particle can be either reflected or transmitted as an up-type quark. However, due to the mixing between the up and down-type quarks in the Dirac equations (\ref{eq:diraccharged1}) and (\ref{eq:diraccharged2}), there is now the possibility that the incoming up-type quark gets also reflected or transmitted as a down-type quark after its interaction with the domain wall, inducing a charge violating interaction and the difference in charge between the up and down-type quark will be absorbed by the gauge bosons living inside the wall \cite{vachaspati_2023, Steer:2006ik}.
Using a plane wave solution for the spinor fields,
\begin{align}
    u(t,x) &= e^{-iEt}u_{inc}(x) + e^{-iEt}u_{ref}(x) \text{    for $x<0$,} \\ 
    u(t,x) &= e^{-iEt}u_{tra}(x) \text{    for $x>0$,} \\
    d(t,x) &= e^{-iEt }d_{ref}(x) \text{    for $x<0$,} \\ 
    d(t,x) &= e^{-iEt}d_{tra}(x) \text{    for $x>0$},
\end{align}
and inserting these expressions in (\ref{eq:matrixchargedgeneral}), we get:
\begin{equation}
    \partial_x \begin{pmatrix} u(x) \\ d(x) \end{pmatrix} = \hat{G}(x) \begin{pmatrix}
        u(x) \\ d(x)
    \end{pmatrix} = \begin{pmatrix} -iE \gamma_1\gamma_0 + i\gamma_1m_u(x) & i\gamma_1\hat{G}_2(x) \\
   i\gamma_1\hat{G}_1(x) & iE \gamma_1\gamma_0 + i\gamma_1m_d(x) \end{pmatrix} \begin{pmatrix} u(x) \\ d(x) \end{pmatrix} ,
\end{equation}
which can be solved by taking:
\begin{equation}
    \begin{pmatrix}
        u(x) \\ d(x)
    \end{pmatrix} = \hat{P} \text{ exp}\biggl( \int^x_{x_0} dx' \ \ \hat{G}(x') \biggr) \begin{pmatrix}
        u(x_0) \\ d(x_0)
    \end{pmatrix},
    \label{eq:MatrixCharge}
\end{equation}
where $\hat{P}$ is, as in the previous chapter, taken to be an ordering operator.\\
We now derive the expressions for the spinors in the different regions $x<0$ and $x>0$. \\
Far from the wall, the Dirac equations are: 
\begin{align}
   & i\slashed{\partial}d - y_d v_1d = 0, \label{eq:diracoutsided} \\ &
   i\slashed{\partial}u - y_u v_2u  = 0. \label{eq:diracoutsideu} 
\end{align}
In the region $x<0$, we parameterize (\ref{eq:diracoutsided}) and (\ref{eq:diracoutsideu}) by:
\begin{align}
  & i\gamma_0(-iE)\begin{pmatrix} d_{1m} \\ d_{2m} \\ d_{3m} \\ d_{4m}\end{pmatrix} +i\gamma_1(-ip_d)\begin{pmatrix} d_{1m} \\ d_{2m} \\ d_{3m} \\ d_{4m}\end{pmatrix} -y_dv_1\begin{pmatrix} d_{1m} \\ d_{2m} \\ d_{3m} \\ d_{4m}\end{pmatrix} = 0,\\
   & i\gamma_0(-iE)\begin{pmatrix} u_{1m} \\ u_{2m} \\ u_{3m} \\ u_{4m}\end{pmatrix} +i\gamma_1(-ip_u)\begin{pmatrix} u_{1m} \\ u_{2m} \\ u_{3m} \\ u_{4m}\end{pmatrix} -y_uv_2\begin{pmatrix} u_{1m} \\ u_{2m} \\ u_{3m} \\ u_{4m}\end{pmatrix} = 0, 
\end{align}
leading to the following relations between the components of the spinors:
\begin{align}
    u_{4m} = \frac{-p_u}{E+m_u}u_{1m}, && u_{3m} = \frac{-p_u}{E+m_u}u_{2m}, \label{eq:m1}  \\
    d_{4m} = \frac{-p_d}{E+m_d}d_{1m}, && d_{3m} = \frac{-p_u}{E+m_u}u_{2m},     
\label{eq:m2}    
\end{align}
where $p_u$ denotes the momentum of the incident up-type quark and $p_d$ denotes the momentum of the produced down-type quark.\\
For the region $x>0$, we get:
\begin{align}
   & i\slashed{\partial}d - y_d v_1d = 0, \\ &
   i\slashed{\partial}u + y_u v_2u  = 0, 
\end{align}
leading to:
\begin{align}
    u_{4p} = \frac{p_u}{E-m_u}u_{1p}, && u_{3p} = \frac{p_u}{E-m_u}u_{2p}, \label{eq:p1}  \\
    d_{4p} = \frac{p_d}{E+m_d}d_{1p}, && d_{3p} = \frac{p_d}{E+m_d}d_{2p}.
\label{eq:p2}    
\end{align}
We obtain for the complete spinor:
\begin{align}
   & u(-\epsilon) = \begin{pmatrix} u_{1i} + u_{1m} \\ u_{2i} + u_{2m} \\ u_{3i} - \frac{p_u}{E+m_u}u_{2m} \\ u_{4i} - \frac{p_u}{E+m_u}u_{1m} \end{pmatrix}, &&
    d(-\epsilon) = \begin{pmatrix} d_{1m} \\ d_{2m} \\ -\frac{p_d}{E+m_d}d_{2m} \\ -\frac{p_d}{E+m_d}d_{1m} \end{pmatrix}, \\ \notag \\
   & u(+\epsilon) = \begin{pmatrix}  u_{1p} \\ u_{2p} \\ \frac{p_u}{E-m_u}u_{2p} \\ \frac{p_u}{E-m_u}u_{1p} \end{pmatrix}, &&
    d(+\epsilon) = \begin{pmatrix} d_{1p} \\ d_{2p} \\ \frac{p_d}{E+m_d}d_{2p} \\ \frac{p_d}{E+m_d}d_{1p} \end{pmatrix}.
\end{align}
The results have to be matched at the boundary $x=0$, using (\ref{eq:MatrixCharge}) where $x = +\epsilon$ and $x_0 = -\epsilon$:
\begin{equation}
    \begin{pmatrix}
        u(+\epsilon) \\ d(+\epsilon)
    \end{pmatrix} = \hat{P} \text{ exp}\biggl( \int^{+\epsilon}_{-\epsilon} dx \ \ \hat{G}(x) \biggr) \begin{pmatrix}
        u(-\epsilon) \\ d(-\epsilon)
    \end{pmatrix}.
    \label{eq:MatrixChargecomponents}
\end{equation}
In order to get the matching conditions at $x=0$ for the Dirac spinors, we need to calculate the exponential matrix of the integral of $\hat{G}(x)$. For simplicity, we took $v_1(x) = v_1$ as we want to isolate the effects of a non-zero $v_+$. We obtain the final result:
\begin{align}
     \begin{pmatrix}
        u(+\epsilon) \\ d(+\epsilon)
    \end{pmatrix} = \begin{pmatrix}
        M_1 && M_2 \\
        M_3 && M_4
    \end{pmatrix} \begin{pmatrix}
        u(-\epsilon) \\ d(-\epsilon)
    \end{pmatrix},
\label{eq:e}
\end{align}
where:
\begin{align}
    M_1 =& \text{ cosh}^2\biggl(\frac{1}{2}\sqrt{k^2-m^2}\biggr)\text{ I}_2 + \biggl(\frac{k^2 + m^2}{k^2-m^2}\biggr) \text{ sinh}^2\biggl(\frac{1}{2}\sqrt{k^2-m^2}\biggr) \\& \notag +i\frac{km}{k^2-m^2}\biggl[-1 + \text{ cosh}\biggl(\sqrt{k^2-m^2}\biggr)\biggr]  \gamma_5\gamma_1 ,\\
    M_2 =& \frac{m}{\sqrt{k^2-m^2}}\text{ sinh}\biggl(\sqrt{k^2-m^2}\biggr)P_L - \frac{ik}{\sqrt{k^2-m^2}}\text{ sinh}\biggl(\sqrt{k^2-m^2}\biggr)\gamma_1P_L, \\
    M_3 =& -\frac{m}{\sqrt{k^2-m^2}}\text{sinh}\biggl(\sqrt{k^2-m^2}\biggr)P_L - \frac{ik}{\sqrt{k^2-m^2}}\text{ sinh}\biggl(\sqrt{k^2-m^2}\biggr)\gamma_1P_R, \\
    M_4 = &  \text{ cosh}^2\biggl(\frac{1}{2}\sqrt{k^2-m^2}\biggr) \text{ I}_2  - \text{ sinh}^2\biggl(\frac{1}{2}\sqrt{k^2-m^2}\biggr) \gamma_5,
\end{align}
with:
\begin{align}
    m &= \frac{\Delta g_2}{2}, & k = y_u\tilde{v}_+.
\label{eq:parameterscharge}    
\end{align}
We therefore have 16 variables (4 for each spinor at $x>0$ and $x<0$) and 16 equations (8 from (\ref{eq:e}) and 8 from (\ref{eq:m1}), (\ref{eq:m2}), (\ref{eq:p1}) and (\ref{eq:p2})). 
The solution of such a system of equation gives very lengthy analytical results. After finding the solution for the spinor components, we calculate the transmission and reflection coefficients of the up and down-type quarks corresponding to the scattering of an initial up-type quark scattering off the domain wall:
\begin{align}
    \text{R}_u(p_u) &= -\frac{\mathcal{J}^{ref}_u}{\mathcal{J}_{inc}} = -\frac{u^{\dag}_{ref}\gamma_0\gamma_1 u_{ref}}{2p_u}, & \text{R}_d(p_u) &= -\frac{\mathcal{J}^{ref}_d}{\mathcal{J}_{inc}} = -\frac{d^{\dag}_{ref}\gamma_0\gamma_1 d_{ref}}{2p_u}, \\
    \text{T}_u(p_u) &= -\frac{\mathcal{J}^{tra}_u}{\mathcal{J}_{inc}} = \frac{u^{\dag}_{tra}\gamma_0\gamma_1 u_{tra}}{2p_u}, & \text{T}_d(p_u) &= \frac{\mathcal{J}^{tra}_d}{\mathcal{J}_{inc}} = \frac{d^{\dag}_{tra}\gamma_0\gamma_1 d_{tra}}{2p_u},
\end{align}
where we used a plane wave solution for the incident up-type quark moving in the positive $x$-direction:
\begin{align}
    u_{inc}(x,t) =  e^{(-iEt + ipx)}\begin{pmatrix} \sqrt{E+m_u} \\ 0 \\ 0 \\ \dfrac{p_u}{\sqrt{E+m_u}}\end{pmatrix},    
\end{align}
Since the analytical formulas in the general case are very complicated and lengthy, we present the results in Figure \ref{subfig:generalcasecharged} for the numerical values of $m=0.5$, $k = 10$ and a mass $m_u = 172.76 \text{ GeV}$ for the top quark and $m_d = 4.2 \text{ GeV}$ for the bottom quark. We observe that the reflection and transmission probabilities as a bottom quark are non-zero. In Figure \ref{subfig:chargedamount}, we can see that as we increase the momentum of the incoming particle, the rate of top quarks being transformed into bottom quarks after the interaction with the domain wall becomes higher, while the probability of the quark staying a top quark decreases.
\noindent
We also verify that all the reflected or transmitted bottom quarks are left-handed as it can be already deduced from equations (\ref{eq:diraccharged1}) and (\ref{eq:diraccharged2}), where the coupling to the bottom quark includes only the left handed projector on the spinor. As can be seen in Figure \ref{fig:lefthanded} the rate of reflection and transmission coefficient as right-handed bottom quarks is zero. Concerning the chirality of the top quark after the interaction with the wall, we observe a difference in the transmission rate between the left-handed and right-handed components of the top quark, while the reflected top quarks do not show a chiral asymmetry (see Figures \ref{subfig:RULRUR} and \ref{subfig:TULTUR}). Therefore the scattering of top quarks on the charge breaking wall will generate a chiral asymmetry in front as well as behind the wall.
\noindent
For the case $k>>m$ and $m \rightarrow 0$ corresponding to having the numerical value of $v_+(0)$ inside the wall much bigger than the change in $g_2$ between the two domains, the formulas become considerably simpler. Using (\ref{eq:incident}) for the spinor components of the incident top quark we get:
\begin{align}
    \text{R}_u(p_u) &= \frac{(d^2_u - 1)^2 g^2_d (u_{1i})^2}{p_ud_u\bigl[ g_d(1+d^2_u) + d_u(1+g^2_d)\tanh^2(\frac{k}{2}) \bigr]^2},  \\ \text{R}_d(p_u) &= \frac{g_d\bigl[1+4d_ug_d + g^2_d + d^2_u(1+g^2_d)\bigr]u^2_{1i}\sinh^2(k)}{p_u\bigl[(d_u -g_d)(-1 + d_ug_d) + (d_u + g_d)(1+d_ug_d)\cosh(k)\bigr]^2},  \\
    \text{T}_u(p_u) &= \frac{2d_u\bigl[-4(g^2_d -1)^2\text{cosh}(k) + (1+6g^2_d + g^2_d)(3+\text{cosh}(2k)) \bigr](u_{1i})^2}{p_u\bigl[4d_u(1+g^2_d)\sinh^2(\frac{k}{2}) + 4g_d(1+d^2_u)\cosh^2(\frac{k}{2}) \bigr]^2},  \\ \text{T}_d(p_u) &= \frac{4g_d\bigl[1-4d_ug_d +g^2_d + d^2_u(1+g^2_d) \bigr](u_{1i})^2\sinh^2(k)}{p_u\bigl[4d_u(1+g^2_d)\sinh^2(\frac{k}{2}) + 4g_d(1+d^2_u)\cosh^2(\frac{k}{2}) \bigr]^2},
\end{align}
where:
\begin{align*}
    g_u &= \frac{p_u}{E+m_u},  && g_d = \frac{p_d}{E+m_d},
   && d_u = \frac{p_u}{E-m_u},  && d_d = \frac{p_d}{E-m_d},
    && u^2_{1i} = E+m_u .
\end{align*}
\begin{figure}[h]
      \centering
     \begin{subfigure}[b]{0.49\textwidth}
         \centering
         \includegraphics[width=\textwidth]{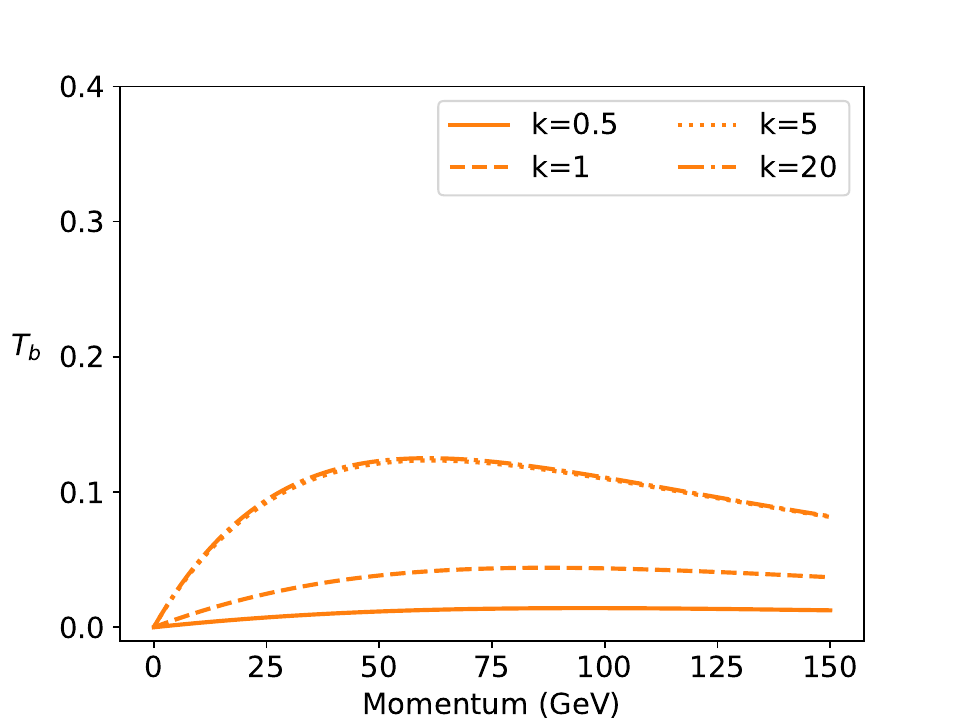}
        \subcaption{Variation of $T_b$ with $k$.} \label{subfig:variationtbk}
     \end{subfigure}
     \hfill
     \begin{subfigure}[b]{0.49\textwidth}
         \centering
         \includegraphics[width=\textwidth]{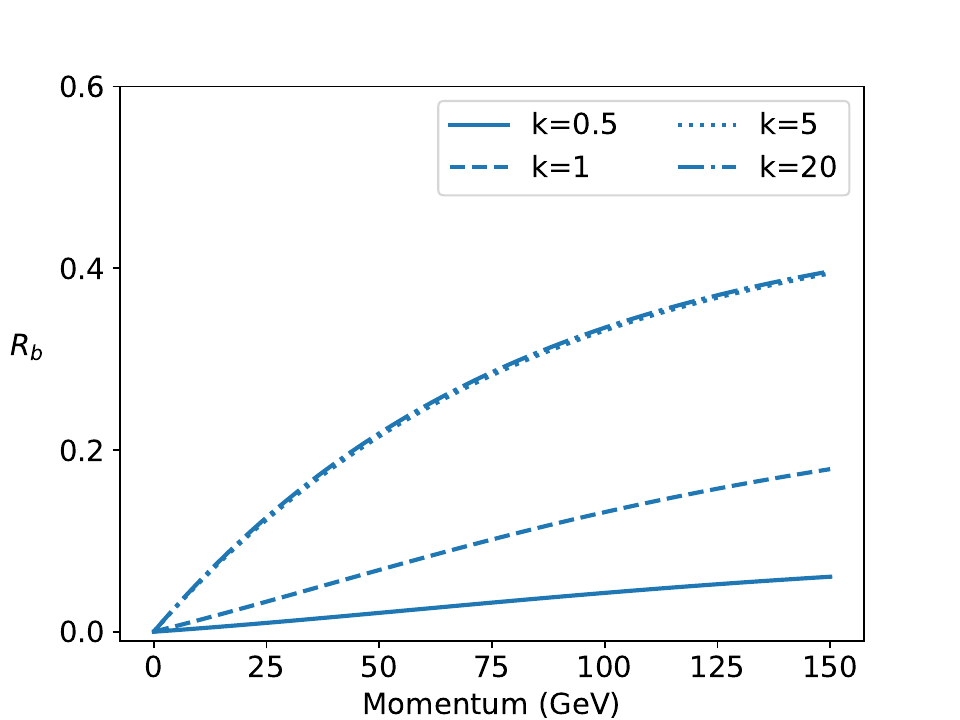}
         \subcaption{Variation of $R_b$ with $k$.}\label{subfig:variationrbk}
     \end{subfigure}
\caption{(a) Transmission rate as bottom quarks for different values of $k$ and $m=0$. (b) Reflection rate as a bottom quark for different values of $k$ and $m=0$. Notice that the reflection rate as a bottom quark is in this case higher than the transmission rate even for high momenta of the incoming incident top quark. We also observe that the rates are almost the same for $k>5$.}     
\end{figure}
\noindent
The results for the transmission $T_{t,b}$ and reflection $R_{t,b}$ coefficients as top or bottom quarks are shown in Figures $\ref{subfig:variationtbk}$ and $\ref{subfig:variationrbk}$. An interesting feature is that the rate of reflection as a bottom quark is higher than the rate of transmission as a bottom quark, even for higher momenta. Recall that we observed the same behavior in the previous section [\ref{CPviolatingscattering}] when the numerical value of the imaginary mass $\tilde{v}_2$ is dominant and the reflection rate for particles becomes higher than the transmission rate.\\\\
We now look at the case when the charge-breaking term $k = y_uv_+$ is small compared with the change in the Goldstone modes between the two domains ($m>k$), cf.(\ref{eq:parameterscharge}).
The boundary condition at $x=0$ is given by:
\begin{align}
   M_1 =& \cos^2\biggl(\frac{1}{2}\sqrt{m^2-k^2}\biggr)\text{ I}_2 + \biggl(\frac{k^2 + m^2}{m^2-k^2}\biggr) \sin^2\biggl(\frac{1}{2}\sqrt{m^2-k^2}\biggr) \\& \notag  +i\frac{2km}{m^2-k^2}\sinh^2\biggl(\sqrt{k^2-m^2}\biggr)  \gamma_5\gamma_1, \\
    M_2 =& \frac{m}{\sqrt{m^2-k^2}}\sinh\biggl(\sqrt{m^2-k^2}\biggr)P_L - \frac{ik}{\sqrt{m^2-k^2}}\sinh\biggl(\sqrt{m^2-k^2}\biggr)\gamma_1P_L, \\
    M_3 =& -\frac{m}{\sqrt{m^2-k^2}}\sin\biggl(\sqrt{m^2-k^2}\biggr)P_L - \frac{ik}{\sqrt{m^2-k^2}}\sin\biggl(\sqrt{m^2-k^2}\biggr)\gamma_1P_R, \\
    M_4 = &  \cos^2\biggl(\frac{1}{2}\sqrt{m^2-k^2}\biggr) \text{ I}_2  + \sin^2(\frac{1}{2}\sqrt{m^2-k^2}) \gamma_5.
\end{align}
Figure \ref{subfig:generalcasechargedmbigger} gives the results for numerical values $m = 1.5 $, $k= 0.1$, $m_u = 172.76 \text{ GeV}$ and $m_d = 4.2  \text{ GeV}$.
\noindent
One can see that, in this case, the transmission coefficient as a bottom quark gets higher as we increase the momentum of the incoming particle. The difference between this case and the one where the Yukawa term is dominant $(k>>m)$ is that there is no chirality flipping of the particle in the coupling between the gauge fields living inside the wall and the fermions (see (\ref{eq:lagrangiancharged})).\\
For $m >> k$ one can get rather simple analytical expressions:
\begin{align}
R_u(p_u) &= \dfrac{g^2_d(1-g^2_u)^2u^2_{1i}}{p_ug_u[g_d(d_u+g_u) + (1+g^2_d)\tan^2(\frac{m}{2})]^2}, \\
R_d(p_u) &= \dfrac{g_d[1-4g_dg_u + g^2_u + g^2_d(1+g^2_u)]u^2_{1i}\tan^2(\frac{m}{2}) }{p_u[g_d(d_u+g_u) + (1+g^2_d)\tan^2(\frac{m}{2})]^2}, \\
T_u(p_u) &= \dfrac{-4\cos(m)(-1+g^2_d)^2 + (1+6g^2_d+g^4_d)(3+\cos(2m))}{8\cos^4(\frac{m}{2})[g_d(d_u+g_u) + (1+g^2_d)\tan^2(\frac{m}{2})]^2}, \\
T_d(p_u) &= \dfrac{g_du^2_{1i}[1+4g_dg_u + g^2_u + g^2_d(1+g^2_u)]\tan^2(\frac{m}{2})}{p_u[g_d(d_u+g_u) + (1+g^2_d)\tan^2(\frac{m}{2})]^2}.
\end{align}
In this case the rate of reflection and transmission coefficients will oscillate with increasing $m$. Such a behavior is shown in Figures \ref{subfig:variationtbm} and \ref{subfig:variationrbm}. In order to study this oscillating behavior in more detail, we fix the momentum of the incoming top quark and vary $m$ between $[0,2\pi]$. The results are shown in Figure \ref{fig:variationm}. We observe that the rate of conversion of top quarks into bottom quarks also vanishes for $m=\pi$. 
\noindent
We now consider the scattering of the second generation quarks off the wall. We take, as an example, the scattering of charm quarks off the domain wall. In this case the charge breaking parameter $k$ gets smaller due to the small Yukawa coupling of the charm quark to the Higgs doublet.
Figure \ref{fig:lowkcharm} shows the results for $k>m$. In this case we observe that the charge breaking effect is very small and most charm quarks scatter off the wall as charm quarks, even for high values of $v_+ = 65\text{ GeV}$ corresponding to $k = 0.43$. However, when we consider the case $m>k$, the charge breaking effect can be quite high depending on the value of $m$ as is shown in Figure \ref{fig:highmcharm}.
\noindent
Finally, we also mention that anti up-type quarks scattering off the domain walls will lead to exactly the same rate of transformation into anti down-type as the rate of the up-type quarks transformed into down-type quarks. Therefore, the interaction of fermions with these types of domain walls does not lead to a net generation of charge in the early universe, a strongly constrained phenomenon \cite{Caprini:2003gz}. However, as we saw in the previous chapter, domain walls solution in 2HDM can exhibit simultaneously both a charge and a CP violation inside the wall. This will lead to the generation of a non-zero CP-violating phase $\xi$ inside the domain wall. This phase, along with a non-zero $v_+$ inside the wall, might lead to the generation of a net charge in the early universe, as particles and antiparticles will interact with the wall with different rates. In such a case, the domain wall network has to annihilate very quickly in order to avoid generating a charge asymmetry higher than the observed cosmological constraints. \\
Another possible problem with these types of domain walls is that they could efficiently deplete the number of up-type quarks into down-type quarks. For example, bottom and anti-bottom quarks could be generated from the interaction of top and anti-top quarks with the wall. Due to the large difference of the masses between the two flavors, the inverse reaction would be suppressed and we would end up with a surplus of bottom and anti-bottom quarks well before the usual annihilation time of top quarks in the thermal plasma. This phenomenon might then have consequences on Big Bang Nucleosynthesis (BBN).

%% file: Summary_and_conclusions.tex
\section{Summary and conclusions}
In this paper we investigated the different classes of domain walls arising in the 2HDM after EWSB. We extended the work done in \cite{Law:2021ing} and included the variation of all the angles of the $\text{SU}(2)_L\times \text{U}(1)_Y$ symmetry. In contrast to the standard domain wall solution, where only a discrete symmetry such as $Z_2$ gets spontaneously broken, we saw that the breaking of abelian and non-abelian symmetries alongside the discrete symmetry leads to the formation of kink solutions with non-trivial effects in the core of the defect, such as CP and charge-violating vacua. We have found that these different classes of kink solutions can be unstable and decay to the standard kink solution if their energy is higher than the energy of the standard kink solution. We demonstrated this behavior using von Neumann boundary conditions, where the vacua, the hypercharge angle $\theta$ and the Goldstone modes $g_i$ can change their boundary values dynamically in order to minimize the energy of the field configuration. When discussing the simplified cases (where only the hypercharge or/and a single Goldstone mode $g_i$ were allowed to change on both domains) we found that for the CP-breaking kink solution, the CP-violating phase $\xi(x)$ inside the defect decays after some time and we recover again the standard kink solutions. Nevertheless, such CP-violating effects can be quite sizable at the time of the domain walls formation as demonstrated when using Dirichlet boundary conditions instead of von Neumann conditions. In the case of charge breaking kink solutions, we showed that the stability of the solution depends on the sign of the effective mass $M_+$ of $v_+$ inside the wall. \\
When investigating the general case where all the equations of motions of the Goldstone modes as well as hypercharge angle are taken into account, we saw in particular, that these modes had a non-trivial profile inside the wall in contrast to a kink-like profile that was found in all other simplified cases. We also found that the CP-violating phase $\xi(x)$ in that case is stable even though it had a small value. \\
We also focused on the interaction of Standard Model fermions with different classes of kink solutions. In the case of CP-violating kink solutions, the fermions scatter in a CP-violating manner on the wall. In particular, we observed that the transmission and reflection coefficients can deviate substantially from the ones found for the case of standard kink solutions. This is in contrast with results found for CP-violating bubble walls \cite{Ayala:2017gqa}, where the transmission and reflection coefficients T and R do not depend on the CP-violating phase. However, in a realistic scenario, the phase $\xi(x)$ is dynamically vanishing and the scattering of fermions off these CP-violating simplified kink solutions is only relevant at the first stage of the evolution of the domain wall network after its formation. The CP-violating phase eventually decays with time as the Goldstone modes and hypercharge angle $\theta$ change dynamically, becoming equal in both domains and the scattering of fermions off the walls eventually occurs in a standard way.
Even though we showed that the CP-violating phase $\xi(x)$ inside the wall (for the simplified cases) is unstable, the initial value just after the formation of the domain wall network can be large and the CP-violating scattering of fermions could have an important impact for the initial phases of the domain wall's evolution. This effect generates a chiral asymmetry in front of the wall, which could serve to achieve electroweak baryogenesis via domain walls. In fact, due to $v_2(x=0) = 0$ and the possibility of $v_1$ becoming smaller inside the wall, the sphaleron rate inside the wall is less suppressed than outside of it and the scenario of the annihilation of the domain wall network would satisfy the last Sakharov condition, i.e. the departure from the thermal equilibrium \cite{Sakharov:1967dj}. Electroweak baryogenesis using topological defects such as cosmic strings was already investigated in \cite{Cline:1998rc, Espinosa:1999jm}, where it was shown that the volume suppression factor due to the 1D nature of cosmic strings is a huge constraint. As domain walls are 2D sheets in space, the volume suppression factor is, in principle, much less constraining and studying electroweak baryogenesis using 2HDM domain walls is an interesting scenario. A model-independent investigation of such a possibility was done in \cite{Brandenberger:1994mq, Trodden:1994ve} where the topological defects were formed before EWSB. Studying this formalism in the context of 2HDM, where the domain walls form just after EWSB is subject of future work.   \\
Furthermore, we considered the scattering of fermions off charge-violating kink solutions. We took, as an example, the scattering case of incident top quarks and showed that they can also be reflected or transmitted off the wall as bottom quarks. Such a scattering is charge-violating and the charge is absorbed by the gauge fields and bosons living on the wall. The rate of such charge-violation phenomenon is the same for the antiparticles. This means that we do not expect a global charge asymmetry from such phenomena. This might be different when considering the scattering off both CP and charge-breaking DW. As a charge asymmetry in the universe is heavily constrained \cite{Caprini:2003gz}, this might provide strong constraints on the 2HDM. Such a scattering would also, in principal, deplete the population of up-type quarks in the early universe, which might have some consequences on BBN. \\
The consequences of such phenomena in the context of early universe cosmology are subjects of future work.